\documentclass[twocolumn,onecolappendix]{aastex62}

\usepackage{listings}
\usepackage{color}

\definecolor{dkgreen}{rgb}{0,0.6,0}
\definecolor{gray}{rgb}{0.5,0.5,0.5}
\definecolor{mauve}{rgb}{0.58,0,0.82}

\newcommand{\vect}[1]{\boldsymbol{#1}}

\usepackage{savesym}
\savesymbol{tablenum}
\usepackage[group-separator={,}]{siunitx}
\restoresymbol{SIX}{tablenum}
\usepackage{xspace}
\usepackage{graphicx}
\usepackage{natbib}
\usepackage{amsmath}
\usepackage{nccmath}
\usepackage{booktabs}
\usepackage{multirow}
\usepackage[caption=false]{subfig}
\usepackage{upgreek}
\usepackage[figuresright]{rotating}
\graphicspath{{./figures/}}
\usepackage{array}
\newcolumntype{C}{>{$}c<{$}}
\usepackage{longtable}

\newcommand{\hide}[1]{}

\newcommand{\expo}[1]{\ensuremath{10^{#1}}\xspace}

\newcommand{\kms}{\ensuremath{\,{\rm km\,s^{-1}}}\xspace}

\newcommand{\microns}{\ensuremath{\,\mu{\rm m}}\xspace}

\newcommand{\cm}{\ensuremath{\,{\rm cm}}\xspace}
\newcommand{\pc}{\ensuremath{\,{\rm pc}}\xspace}
\newcommand{\kpc}{\ensuremath{\,{\rm kpc}}\xspace}
\newcommand{\K}{\ensuremath{\,{\rm K}}\xspace}

\newcommand{\degree}{\ensuremath{^\circ}\xspace}

\newcommand{\jy}{\ensuremath{\,{\rm Jy}}\xspace}

\newcommand{\ev}{\ensuremath{\,{\rm eV}}\xspace}
\newcommand{\rgal}{\ensuremath{\,R_G}\xspace}   

\newcommand{\hii}{{\rm H\,{\footnotesize II}}\xspace}

\def\fd{{}^{\circ}\mskip-9mu.\,}          

\shorttitle{The Galactic \hii Region Luminosity Function}
\shortauthors{Mascoop et al.}
\begin{document}

\title{The Galactic \hii Region Luminosity Function at Radio and Infrared Wavelengths}

\correspondingauthor{J.~L.~Mascoop}
\email{jlmascoop@mix.wvu.edu}

\author[0000-0002-3758-2492]{J.~L.~Mascoop}
\affiliation{Department of Physics and Astronomy, West Virginia University, Morgantown, WV 26506, USA}
\affiliation{Center for Gravitational Waves and Cosmology, West Virginia University, Chestnut Ridge Research Building, Morgantown, WV 26505, USA}

\author[0000-0001-8800-1793]{L.~D.~Anderson}
\affiliation{Department of Physics and Astronomy, West Virginia University, Morgantown, WV 26506, USA}
\affiliation{Center for Gravitational Waves and Cosmology, West Virginia University, Chestnut Ridge Research Building, Morgantown, WV 26505, USA}
\affiliation{Adjunct Astronomer at the Green Bank Observatory, P.O. Box 2, Green Bank, WV 24944, USA}

\author[0000-0003-0640-7787]{Trey.~V.~Wenger}
\affiliation{Dominion Radio Astrophysical Observatory, Herzberg Astronomy and Astrophysics Research Centre, National Research Council, P.O. Box 248, Penticton, BC V2A 6J9, Canada}

\author[0000-0002-1397-546X]{Z.~Makai}
\affiliation{Department of Physics and Astronomy, West Virginia University, Morgantown, WV 26506, USA}
\affiliation{Center for Gravitational Waves and Cosmology, West Virginia University, Chestnut Ridge Research Building, Morgantown, WV 26505, USA}

\author[0000-0002-7045-9277]{W.~P.~Armentrout}
\affiliation{Department of Physics and Astronomy, West Virginia University, Morgantown, WV 26506, USA}
\affiliation{Center for Gravitational Waves and Cosmology, West Virginia University, Chestnut Ridge Research Building, Morgantown, WV 26505, USA}
\affiliation{Green Bank Observatory, P.O. Box 2, Green Bank, WV 24944, USA}

\author[0000-0002-2465-7803]{Dana.~S.~Balser}
\affiliation{National Radio Astronomy Observatory, 520 Edgemont Road, Charlottesville, VA 22903, USA}

\author[0000-0003-4866-460X]{T.~M.~Bania}
\affiliation{Institute for Astrophysical Research, Astronomy Department, Boston University, 725 Commonwealth Ave., Boston, MA 02215, USA}

\begin{abstract}
The Galactic \hii region luminosity function (LF) is an important metric for understanding global star formation properties of the Milky Way, but only a few studies have been done and all use relatively small numbers of \hii regions. We use a sample of 797 first Galactic quadrant \hii regions compiled from the \textit{WISE} Catalog of Galactic \hii Regions to examine the form of the LF at multiple infrared and radio wavelengths. Our sample is statistically complete for all regions powered by single stars of type O9.5V and earlier. We fit the LF at each wavelength with single and double power laws. Averaging the results from all wavelengths, the mean of the best-fit single power law index is $\langle\alpha\rangle=-1.75\,\pm\,0.01$. The mean best-fit double power law indices are $\langle\alpha_1\rangle=-1.40\,\pm\,0.03$ and $\langle\alpha_2\rangle=-2.33\,\pm\,0.04$. We conclude that neither a single nor a double power law is strongly favored over the other. The LFs show some variation when we separate the \hii region sample into subsets by heliocentric distance, physical size, Galactocentric radius, and location relative to the spiral arms, but blending individual \hii regions into larger complexes does not change the value of the power law indices of the best-fit LF models. 
The consistency of the power law indices across multiple wavelengths suggests that the LF is independent of wavelength. This implies that infrared and radio tracers can be employed in place of H$\alpha$.
\end{abstract}



\section{Introduction} \label{sec:intro}

The radiation from a high-mass star (M~$\geq$~8\,M$_{\odot}$) ionizes the surrounding interstellar medium (ISM), creating an \hii region. High-mass stars often form in associations with each other \citep{mot18}. Considered on a broad scale, these associations of high-mass stars exert a significant influence on their host galaxies; in the Milky Way, OB stars are thought to be a primary source of the approximately 20\% by mass of interstellar hydrogen that is ionized \citep{dra11,web19}.  

The stellar initial mass function (IMF) of most galaxies can be modeled by a power law \citep{sal55}, and by extension it is expected that the \hii region luminosity function (LF) has a similar form \citep{mw97}. If, however, solitary high-mass stars form via a different mechanism than those within clusters, then the \hii region LF could be distinct from the form of the IMF \citep{ple00}. \citet{oey04} find that this is not the case and that both the IMF and the clustering distribution of OB stars are power laws, supporting the expectation that the \hii region LF will have the same power law form as the IMF.

\hii regions are both unambiguous tracers of high-mass star formation and, as the brightest objects in the Galaxy at infrared and radio wavelengths, are readily detectable across the entirety of the Galactic disk \citep[][submitted]{che20}. Determining the shape of the LF can provide insight into the history of Galactic star formation as well as the current structure of the Galaxy. The shape of the LF is set by many processes, including bursts of star formation \citep{oey98} and formation mechanisms of OB associations \citep[e.g.,][]{mw97,bra06}. In turn, these processes impact the evolution of the ISM. A galaxy's \hii region LF is a crucial tool in understanding the galaxy's history and star formation characteristics.

The majority of \hii region LF studies examine extragalactic \hii region populations. For these populations, the \hii region LFs are generally well-represented by single power laws of the form \useshortskip\begin{equation} \label{eq:singlaw}
    N(L)\,dL=A\,L^{\alpha}\,dL,   
\end{equation} 
where $N(L)dL$ is the number of \hii regions with luminosities between $L$ and $L+dL$, $A$ is a scaling factor, and $\alpha$ is the power law index \citep[e.g.,][]{ken89}. Some studies find that the shape of the \hii region LF is dependent on a galaxy's Hubble type with early-type spirals having shallower LFs and fewer \hii regions \citep{ken89,you99}, whereas others find no such trend \citep{gon97,thi02}. Nonetheless, the LF power law index is consistently found to be $\alpha \approx -2$\footnote{Our reported power law indices are comparable to those in \citet{ken89}, who} subtract 1 from the indices of the fitted LFs to account for the usage of logarithmic binning in contrast to the usage of linear binning elsewhere in the literature. All reported and cited indices are provided in this way, i.e. an $\alpha$ of $-2$ in this paper is equivalent to an $\alpha$ of $-1$ found with linear binning. \citep[e.g.,][]{ken86,hod89,mw97,thi00,cas00}. Moreover, the form of the \hii region LF generally appears to be independent of wavelength \citep[e.g.,][]{mw97,pal09,liu13}, galactic radius \citep[e.g.,][]{ken80,ken89}, and surface brightness \citep{hel09}. 

Some LFs show evidence of a break or ``knee'' at the so-called Str{\"o}mgren luminosity $L_{\rm{H}\alpha}$ $\approx$ 10$^{39}$ erg s$^{-1}$ \citep{ken89,you99,bra06}. These LFs are better fit by a double or `broken' power law \citep{ken89} of the form 
\begin{equation}
\label{eq:double_law}
    N(L)\,dL=
    \begin{cases}
     A\,(L/L_{\rm knee})^{\alpha_{1}}\,dL, & L < L_{\rm knee} \\
     A\,(L/L_{\rm knee})^{\alpha_{2}}\,dL, & L > L_{\rm knee} \\
    \end{cases},
\end{equation}
where $A$ is a scaling factor, $\alpha_1$ and $\alpha_2$ are the power law indices, and $L_{\rm knee}$ is the value of the knee luminosity. A double power law LF is not, however, found in all galaxies \citep[e.g.,][]{gon97}, and not all galaxies have a knee at the Str{\"o}mgren luminosity \citep[e.g.,][]{lee11}. This break may correspond to a transition from ``giant'' to ``supergiant'' \hii regions \citep{ken89}, the evolution of the \hii region H$\alpha$ luminosities, or the transition between \hii regions ionized by a single star and \hii regions ionized by multiple stars \citep{oey98}. \citet{bec00} suggest that \hii regions above the break are density bounded and those below the break are ionization bounded. There is some evidence that double power law LFs are preferentially found in early-type spiral galaxies \citep{ken89}, though work by \citet{you99} and \citet{bra06} raise the possibility that these double power law LFs represent a more general form of the \hii region LF. If this is the case, single power law LFs should be found only in galaxies that have a lack of nebulae with H$\alpha$ luminosities above the Str{\"o}mgren luminosity relative to the galaxy's total number of nebulae \citep{mw97}. With only $\sim30$ out of thousands of \hii regions with H$\alpha$ luminosities above the Str{\"o}mgren luminosity \citep[see the right panel of Figure 5 in][]{rah11}, the Milky Way is such a galaxy.

The power law index of the Galactic \hii region LF, $\alpha \approx -2$, is similar to those found for other galaxies. This same index is found by studies using multiple wavelengths, Galactic spatial distributions, and \hii region sizes \citep{smi89,com96,mw97,cas00,pal09,mur10,mot11}. Other analyses, however, show local differences in the LF across the Galactic disk. Using far-infrared observations of ultra-compact \hii regions, \citet{cas00} find that the LF within the Solar circle has a peak luminosity nearly an order of magnitude greater than the LF in the outer Galaxy. \citet{pal09} find that the Galactic \hii region LF is dependent on Galactic longitude, with notably different results for the first and fourth quadrants. They also find weak evidence for a break in the fourth quadrant LF corresponding to $L_{\rm{H}\alpha}$ $\approx$ 10$^{37.75}$ erg~s$^{-1}$. The authors hypothesize that these results arise from distinct \hii region populations in each quadrant.

Previous studies of the Galactic \hii region LF are hampered by limited spatial coverage, small sample sizes, and incompleteness at lower luminosities. Here, we improve the determination of the Galactic LF by addressing and investigating each of these issues. We analyze the Galactic LF at multiple infrared ($8, 12, 22, 24, 70,$ and $160 \microns$) and radio ($20$ and $21 \cm$) wavelengths to assess whether the form of the LF depends on the observed wavelength. Unlike previous studies performed in H$\alpha$ and to a lesser extent the study by \citet{liu13} in Pa$\alpha$, observations at these frequencies are unaffected by extinction.

\section{Data}
\label{sec:data}
We use a sample of 797 first Galactic quadrant \hii regions spanning $18\degree<\ell<65\degree$ taken from the \textit{WISE} Catalog of Galactic \hii Regions V2.2\footnote{\url{http://astro.phys.wvu.edu/wise/}} \citep{and14}.  We include only those regions for which the kinematic distance ambiguity (KDA) is resolved or for which we have parallax distances. The Heliocentric distances, $d$, Galactocentric radii, $R$, and their associated uncertainties are determined using the method given by \citet{wen18}. Through population synthesis modeling, Armentrout et al. (2021, in press) conclude that the sample is complete for all sources ionized by single stars of spectral type O9.5V and earlier.  

Although the \textit{WISE} Catalog covers the entire Galactic disk, we restrict the present analysis to the first quadrant because the \hii region sample here is the most complete and fractional distance errors are relatively small \citep{and12, wen18}. The current sample of fourth quadrant \hii regions is notably incomplete compared with the first quadrant sample \citep{and14, wen19}.  \hii region samples in the second and third quadrants have comparatively high fractional distance errors relative to those for the first quadrant sample \citep{and12,wen18}. Consequently, the resulting errors in luminosity are large in those quadrants. Analysis of the entirety of the inner Galaxy will be possible upon publication of the full Southern \hii Region Discovery Survey catalog \citep[][Wenger et al. 2021, submitted]{bro17,wen19}.


\hii regions produce thermal radio continuum emission from free-free interactions between ions and electrons. Here, we use the results of \citet{mak17}, who calculated the flux densities for all known Galactic \hii regions in the Galactic longitude range $17\fd5 < \ell < 65$\degree. Radio flux densities come from the $20 \cm$ Multi-Array Galactic Plane Imaging Survey \citep[MAGPIS, hereafter denoted with the subscript ``M'';][]{hel06} and $21 \cm$ Very Large Array Galactic Plane Survey \citep[VGPS, hereafter denoted with the subscript ``V'';][]{sti06}. In addition, we analyze the combined VGPS and MAGPIS data set (hereafter referred to as the MAGPIS+VGPS $20 \cm$ data set and denoted with the subscript ``M+V'') described in \citet{mak17}.

Infrared flux densities come from the \textit{Spitzer} Galactic Legacy Infrared Mid-Plane Survey Extraordinaire \citep[GLIMPSE;][]{ben03,chu09} $8 \microns$ band, the Wide-field Infrared Survey Explorer \citep[\textit{WISE};][]{wri10} $12$ and $22 \microns$ bands, the \textit{Spitzer} MIPSGAL \citep{car09} $24 \microns$ band, and the Herschel infrared Galactic Plane Survey \citep[Hi-GAL;  ][]{mol10,mol16} $70$ and $160 \microns$ bands.  

Each infrared data set contains a small number of saturated sources ($\leq 26$) for which we can estimate infrared flux densities using their radio flux densities. If more than $0.1\%$ of pixels within the defined region are saturated (`NaN' value), we use the region's $21\,\cm$ MAGPIS+VGPS flux density and the infrared-to-radio relationships found by \citet{mak17} to infer their infrared flux density:
\useshortskip\begin{equation} \label{eq:corr}
        S_{\nu}=B_{\nu}\,S_{\rm{M+V}}^{\beta_{\nu}}\,.
\end{equation}
Here, $S_{\rm{M+V}}$ is the MAGPIS+VGPS 21\,cm flux density, and $B_{\nu}$ and $\beta_{\nu}$ are the scaling factor and power law index at infrared frequency $\nu$, respectively. We reproduce values for $B_{\nu}$ and $\beta_{\nu}$  from \citet{mak17} in Table~\ref{tab:corr}. As noted by \citet{mak17}, the $24\,\microns$ MIPSGAL and $22\,\microns$ \textit{WISE} \hii region flux densities are essentially identical.  We therefore use the MIPSGAL correlation parameters to find both the MIPSGAL and $22 \microns$ \textit{WISE} flux densities.

\begin{deluxetable}{cRR}
\setlength{\tabcolsep}{11pt}
\renewcommand{\arraystretch}{1.1}
\tablecaption{Infrared and Radio Flux Density Correlation Parameters\tablenotemark{a} \label{tab:corr}}
\tablecolumns{3}
\tablehead{
\colhead{Wavelength} &
\colhead{$B_\nu$} &
\colhead{$\beta_\nu$} 
}
\startdata
$8\,\microns$  &  30.6 &   0.809 \\
$12\,\microns$ &  34.4 &  0.815 \\
$22\,\microns$ &  82.2 &  0.822 \\
$24\,\microns$ &  82.2 &   0.822 \\
$70\,\microns$ &  2220 &  0.756 \\
$160\,\microns$ &  3700 &  0.726 
\enddata
\tablenotetext{a}{From \citep{mak17}; see Equation~\ref{eq:corr}.}
\vspace{-4mm}
\end{deluxetable}
  
We calculate \hii region monochromatic luminosities (hereafter `luminosities') from the flux densities using \useshortskip
\begin{equation} \label{eq:nulnu}
    \left(\frac{\nu L_{\nu}}{\rm erg~s^{-1}}\right)=4\pi\times10^{-23}\left(\frac{S_\nu}{\jy}\right)\left(\frac{d}{\rm cm}\right)^{2}\left(\frac{\nu}{\rm Hz}\right),
\end{equation}
where $S_\nu$ is the flux density at observing frequency $\nu$ and $d$ is the heliocentric distance to the \hii region.  For the radio flux densities, we combine Equation~6 from \citet{rub68} with Equation \ref{eq:nulnu} to additionally determine the number of Lyman continuum photons emitted per second
\begin{multline} \label{eq:nly}
    \left(\frac{N_{\rm{ly}}}{\rm s^{-1}}\right)\simeq5.01\,\times\,10^{26}\!\left(\frac{\nu L_{\nu}}{\rm erg~s^{-1}}\right) \\ \times\left(\frac{T_e}{\K}\right)^{-0.45}\,\left(\frac{\nu}{\rm Hz}\right)^{-0.9}\!,
\end{multline}
where $T_e$ is the electron temperature of the \hii region. We assume a typical electron temperature of $\expo4\K$ for all regions.

Since the highest-mass star in a cluster frequently produces the majority of the ionizing photons, we assume that each \hii region has a single ionizing source \citep[e.g.,][]{mw97}, and we can therefore determine completeness limits in spectral type from limits in ionizing photon rates. In the same manner, we can associate the location of the double power law knees with a spectral type. We convert $N_{\rm{ly}}$ to main sequence spectral types using the calibration in Table 1 in \citet{mar05}. If $N_{\rm{ly}}$ lies between two or more of the tabulated spectral types, we assign the less luminous type to the region. 

We calculate \hii region H$\alpha$ luminosities from Lyman photon rates to facilitate comparisons between this work and previous studies of the LF, which have been primarily performed in H$\alpha$. We use the method outlined in \citet{pal09} for these calculations. Beginning by substituting Equation~\ref{eq:nly} for the Lyman photon rate term in Equation~16 in \citet{pal09}, we calculate the H$\alpha$ luminosities
\begin{multline}
    \left(\frac{L_{\rm{H}\alpha}}{\rm erg~s^{-1}}\right)\simeq 8.03\times10^{14}\!\left(\frac{\nu L_{\nu}}{\rm erg~s^{-1}}\right)\!\left(\frac{T_e}{\K}\right)^{-0.45} \\ \times\left(\frac{\nu}{\rm Hz}\right)^{-0.9}\left(\frac{h\nu_{\rm{H}\beta}}{\ev}\right)\frac{\alpha^{\rm{eff}}_{\rm{H}\beta}}{\alpha_{\rm{B}}}\frac{j_{\rm{H}\alpha}}{j_{\rm{H}\beta}},
\end{multline}
where $h\nu_{\rm{H}\beta}=2.55$ \ev is the energy of an H$\beta$ photon, $\alpha^{\rm{eff}}_{\rm{H}\beta}=\SI{3.03e-14}{\cm\cubed\per\second}$ and $\alpha_{\rm{B}}=\SI{2.59e-13}{\cm\cubed\per\second}$ are, respectively, the effective H$\beta$ and Case B recombination coefficients, and $\frac{j_{\rm{H}\alpha}}{j_{\rm{H}\beta}}=2.87$ is the line intensity ratio of H$\alpha$ to H$\beta$ \citep{ost06}.

\subsection{Monte Carlo-Generated Luminosity Distributions}
\label{subsec:mcdist}
To better account for uncertainties in the \hii\ region luminosities as a result of flux density and distance uncertainties, we generate \hii\ region luminosity distributions using a Monte Carlo routine. For each \textit{WISE} Catalog \hii region in our sample, we randomly draw a new flux density value at each wavelength from a Gaussian distribution whose mean is that \hii\ region's \textit{WISE} Catalog flux density. The associated standard deviation is the quadrature sum of the error in the flux density and $20\%$ of the flux density. We include the factor of $20\%$ to account for a possible systematic uncertainty in the flux density measurements. For the sources with only kinematic distances, we generate new distances by randomly sampling the radio recombination line-derived distance probability density functions using the Monte Carlo kinematic distance code of \citet{wen17,wen18}. In the case of the 44 regions for which we have parallax distances, we generate new distance values by randomly sampling a Gaussian distribution whose mean and standard deviation are those given in the \textit{WISE} Catalog. In turn, we calculate new luminosities and so build new luminosity distributions.

Using this methodology, we generate \num{15000} simulated \hii region luminosity distributions at each wavelength. We fit and analyze these distributions using the method described in Section \ref{subsec:fit}. Here, unless otherwise specified, the quoted values and errors of the power law fits refer to the median values and median absolute deviations (MAD) of these \num{15000} Monte Carlo-generated distributions. 

\subsection{Does the \hii Region Luminosity Function Depend on the Data Sample?}
In order to investigate whether the \hii region LF changes as a function of various physical properties of the nebulae, we define and analyze subsets of our \hii region sample. We  divide each full wavelength data set into roughly equal subsets by heliocentric distance, Galactocentric radius \rgal, physical size, and region location relative to spiral arms. With the exception of the arm/interarm subset, there is no \textit{a priori} physically meaningful point of division for each subset. We include the dividing values for each subset in the appropriate subsections of Section \ref{sec:lf_form}.

We additionally investigate if source blending (or confusion) has an impact on the \hii region LF. We combine sources into larger complexes and examine the subsequent effects on the form of the LF pattern.  This change simulates the effect of lower spatial resolution, mimicking what an observer might see in an extragalactic sample. We simulate the five complexes in Table \ref{tab:blend}, which probe a range of longitudes, flux densities, size scales, and number of constituent regions. We define the complexes as circular masks with centroid positions and radii given in Table~\ref{tab:blend}.  We remove any \textit{WISE} Catalog region that falls within these zones and also has a radial velocity within $10~\kms$ of the central local standard of rest (LSR) velocity listed in Table~\ref{tab:blend}.  We then add the total flux density and subtract the background flux density within the circular zones using the method outlined in Section~3 of \citet{mak17}. We diverge from their method by using the median flux density per pixel of a single background zone to determine the background flux density, whereas \citet{mak17} use the mean flux density per pixel of four background zones.

\begin{deluxetable}{cCCCCC}
\setlength{\tabcolsep}{5pt}
\tablecaption{Parameters of Blended \hii Region Complexes\label{tab:blend}}
\tablecolumns{6}
\tablehead{
\colhead{Complex} &
\colhead{$\ell$} &
\colhead{\textit{b}} &
\colhead{Radius} &
\colhead{V$_{\rm LSR}$} &
\colhead{Number}\\[-6pt]
\colhead{} &
\colhead{deg.} &
\colhead{deg.} &
\colhead{deg.} &
\colhead{$\kms$} &
\colhead{of Regions}
}
\startdata
G19.6$-$0.2  &  19.606 &  -0.190 & 0.160 & 45 & 6 \\
G33.1$-$0.1 &  33.080 &  -0.067 & 0.153 & 96 & 6 \\
W47 &  37.357 &  -0.170 & 0.180 & 40 & 3 \\
W49 &  43.169 &   +0.002 & 0.092 & 10 & 6 \\
W51 &  49.110 &  -0.366 & 0.520 & 58 & 15
\enddata
\end{deluxetable}

Since the MAGPIS+VPGS data set is comprised of a mix of two distinct data sets, a blended \hii region complex may contain regions from both of these data sets. As a result, we cannot simply define a combined representative background flux density for such a complex. We therefore use only the VGPS data to find the background and \hii region flux density values for the blended MAGPIS+VGPS complexes.\\

\section{Fitting the \hii Region LF
\label{sec:met}}

We characterize the \hii region LF using power laws and examine whether the form of the LF is dependent on the location or size of the \hii regions. We also investigate whether the LF is better described by a single power law of the form given by Equation~\ref{eq:singlaw} or by a double power law of the form given by Equation~\ref{eq:double_law}.

\subsection{Single and Double Power Law Fits}
\label{subsec:fit}

In order to compare the results of fitting a single power law to those of a double power law, we fit both functional forms to the \hii region data using the method of maximum likelihood estimation (MLE). We discuss the model fitting in detail in Section~\ref{appsubsec:limit} of the Appendix. In addition to the power law indices and -- in the case of the double power law model -- the knee luminosity, we fit for the completeness limit of each LF. We refer to the sections of the LF with luminosities lower and higher than the completeness limit as the `incomplete' and `complete' sections, respectively.

\begin{figure*}[h!]
\centering
 \subfloat{\label{fig:glimpse_complete}%
    \includegraphics[scale=0.6,trim={3.75cm 8.5cm 3.5cm 8.5cm}, clip]{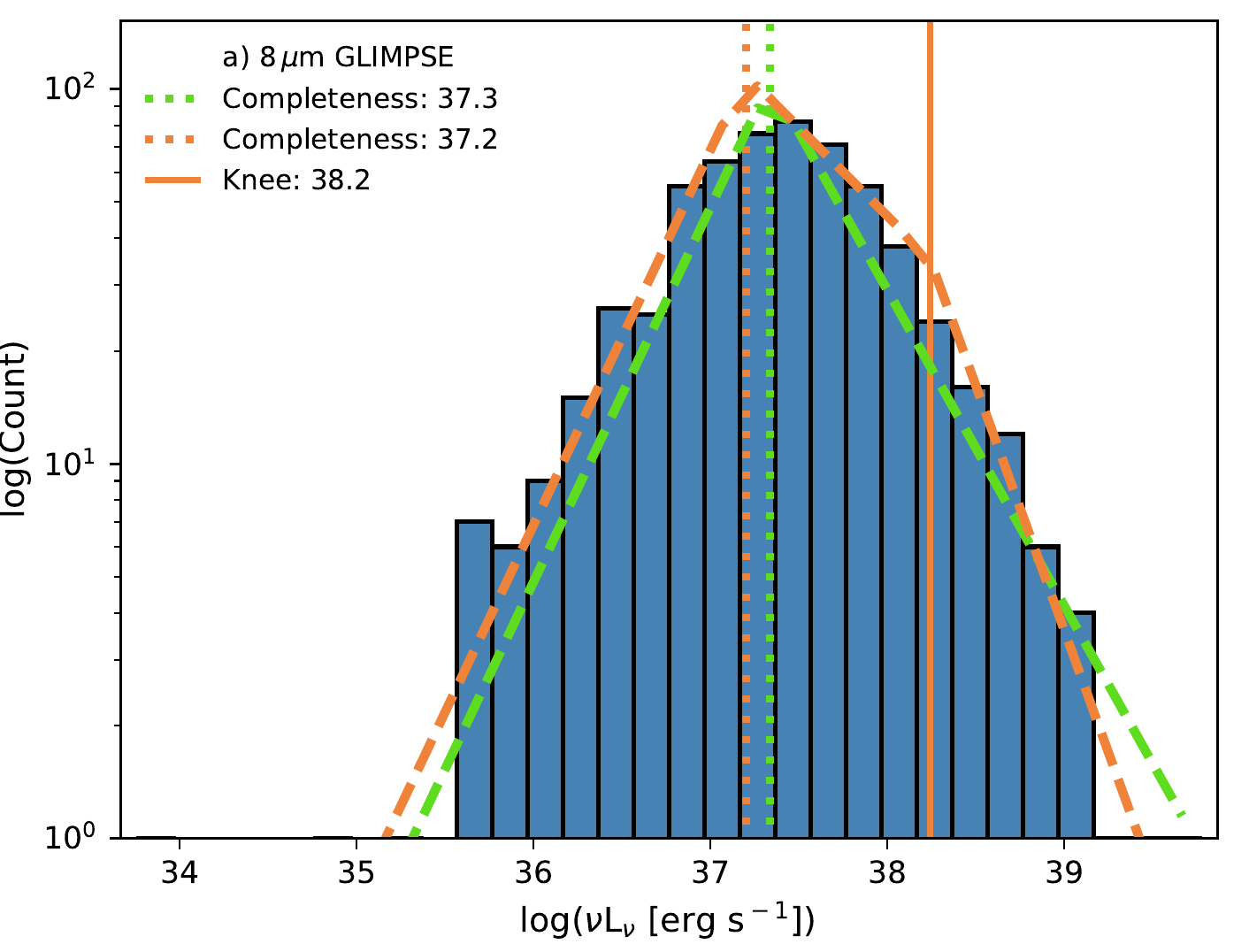}}\qquad
  \subfloat{\label{fig:wise3_complete}%
    \includegraphics[scale=0.6,trim={3.75cm 8.5cm 3.5cm 8.5cm}, clip]{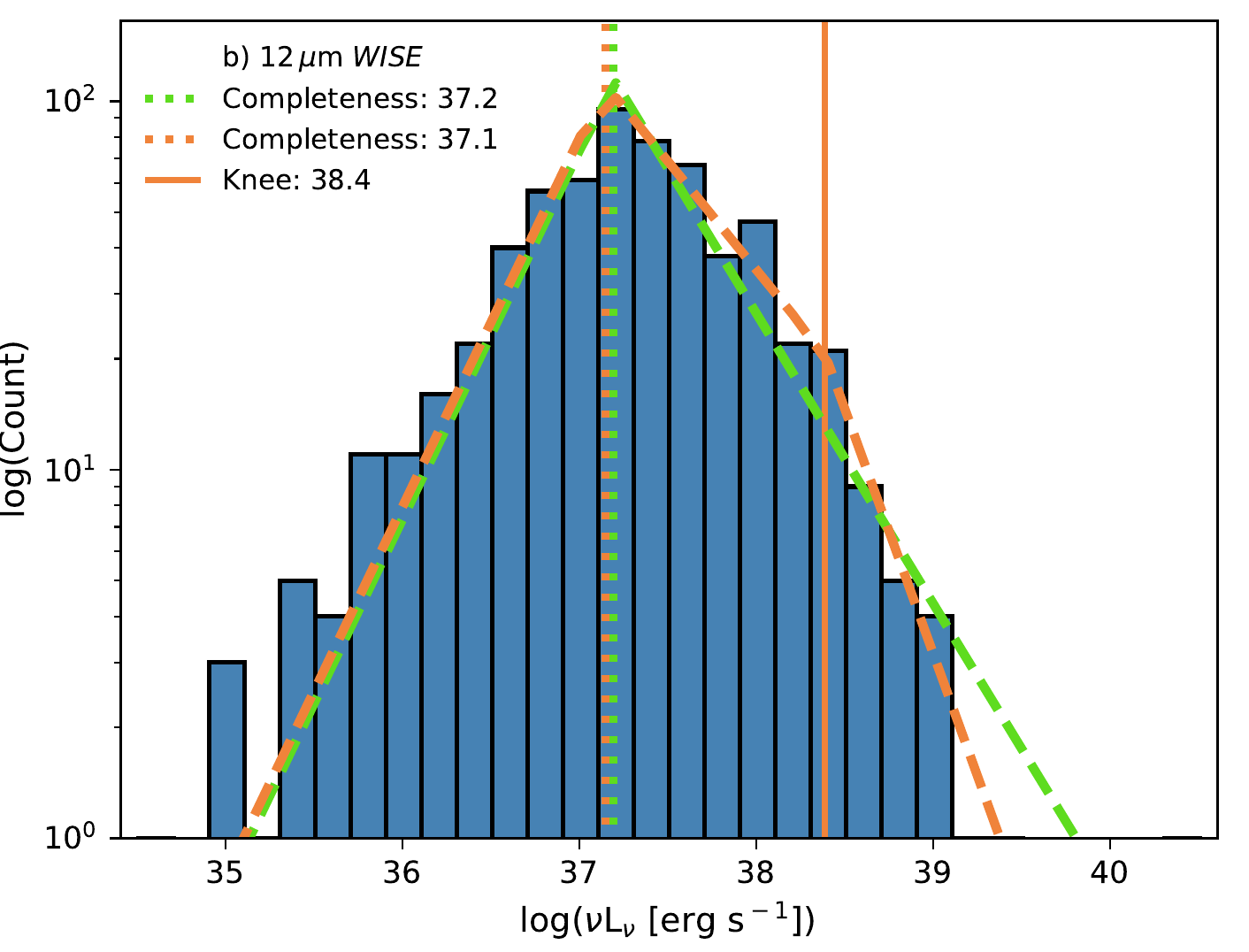}}\\
  \subfloat{\label{fig:wise4_complete}%
    \includegraphics[scale=0.6,trim={3.75cm 8.5cm 3.5cm 8.5cm}, clip]{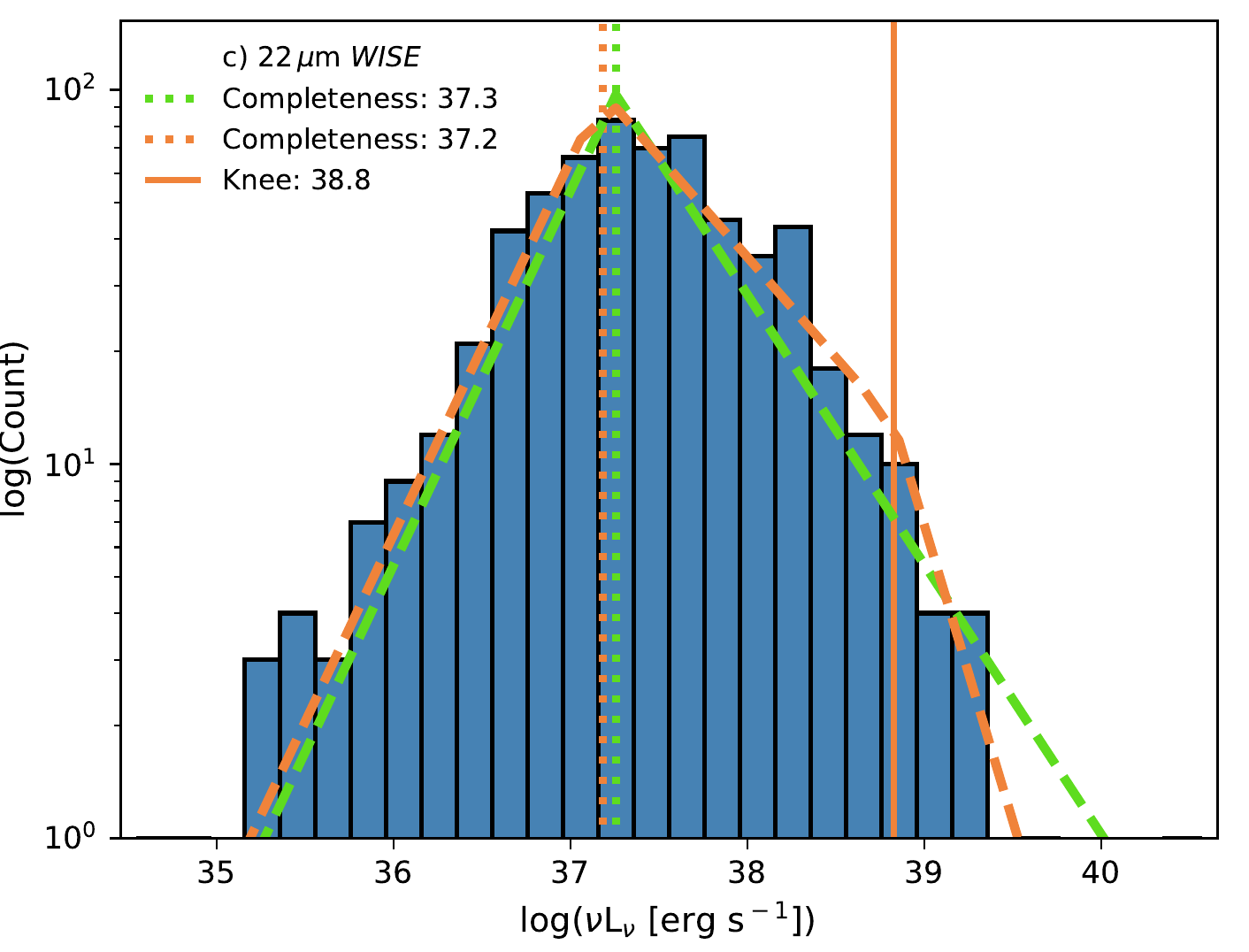}}\qquad
  \subfloat{\label{fig:mipsgal_complete}%
    \includegraphics[scale=0.6,trim={3.75cm 8.5cm 3.5cm 8.5cm}, clip]{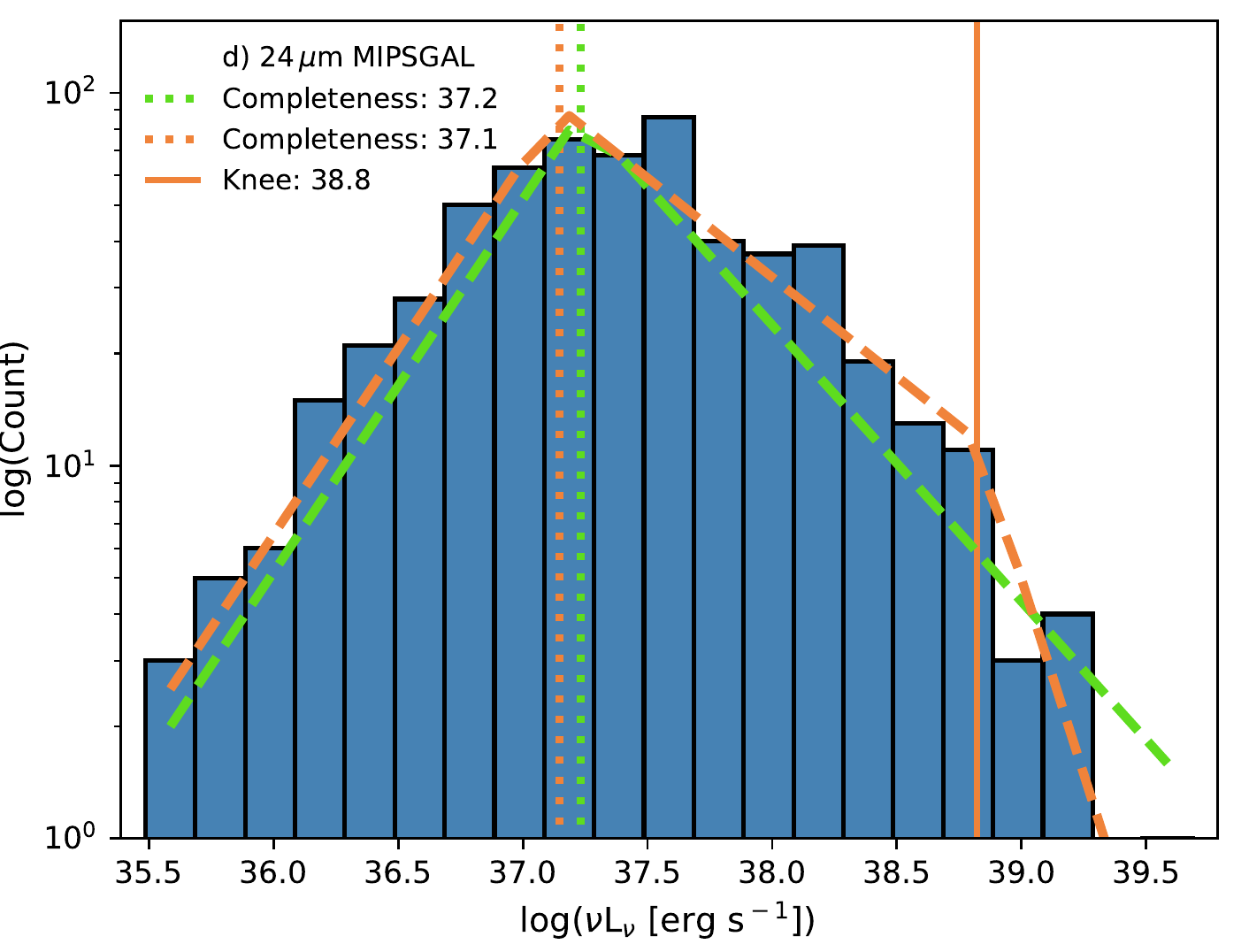}}\\
  \subfloat{\label{fig:higal70_complete}%
    \includegraphics[scale=0.6,trim={3.75cm 8.5cm 3.5cm 8.5cm}, clip]{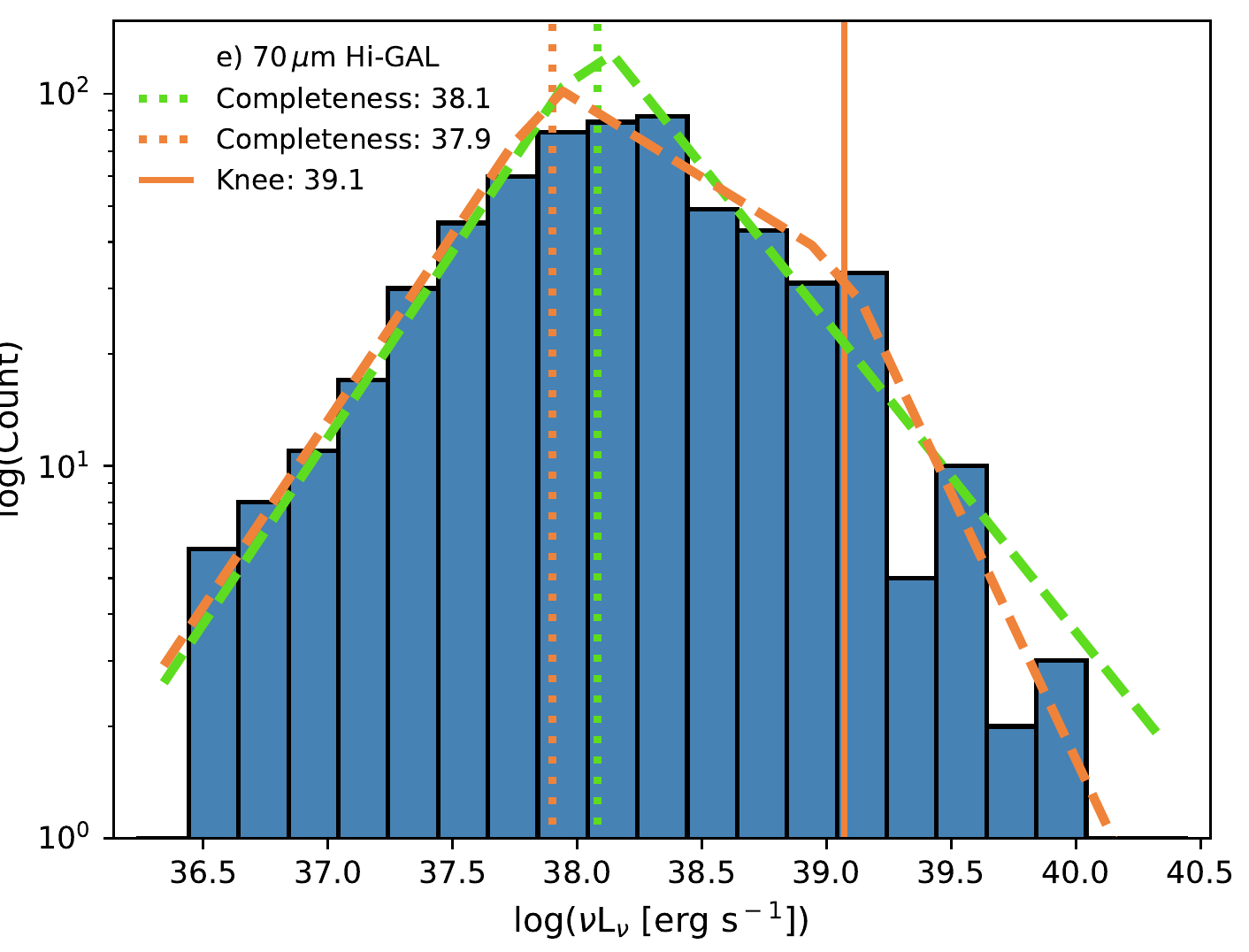}}\qquad
  \subfloat{\label{fig:higal160_complete}%
    \includegraphics[scale=0.6,trim={3.75cm 8.5cm 3.5cm 8.5cm}, clip]{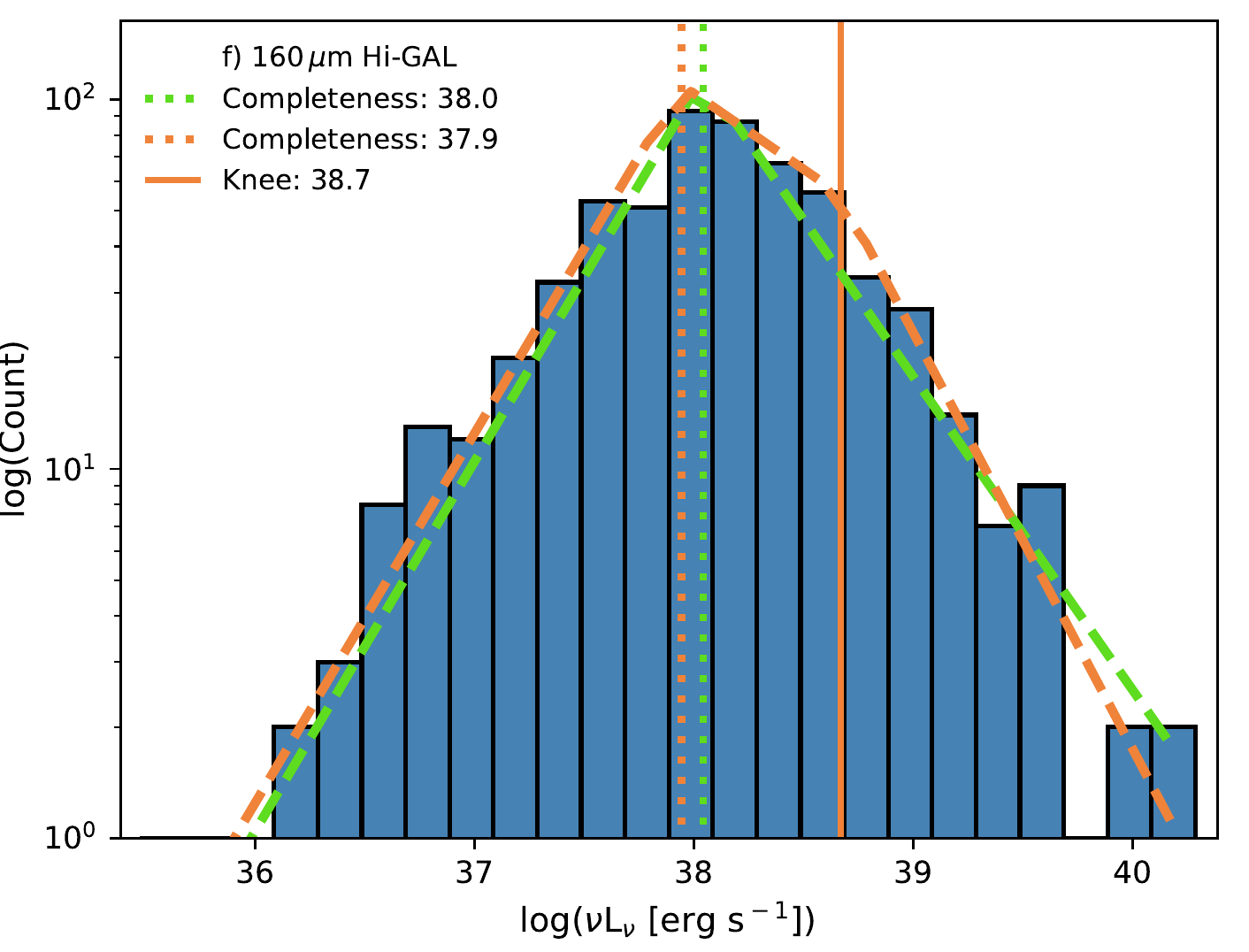}}
\caption{\hii region LFs of example infrared data sets: $8\,\microns$ GLIMPSE (panel \subref*{fig:glimpse_complete}), $12\,\microns$ \textit{WISE} (panel \subref*{fig:wise3_complete}), $22\,\microns$ \textit{WISE} (panel \subref*{fig:wise4_complete}), $24\,\microns$ MIPSGAL (panel \subref*{fig:mipsgal_complete}), $70\,\microns$ Hi-GAL (panel \subref*{fig:higal70_complete}), and $160\,\microns$ Hi-GAL (panel \subref*{fig:higal160_complete}). Completeness limits and knee luminosities are respectively denoted by dotted and solid vertical lines. The model fits are represented by dashed lines. Fits and values associated with the single power law model are in green and those associated with the double power law model are in orange. Note that the fit parameters and distributions in this figure are from a single Monte Carlo iteration and not the full analysis, and that they are presented only as examples of the fitting methods and results.}
\label{fig:ir_full_lfs}
\end{figure*}

\begin{figure*}[ht!]
\centering
  \subfloat{\label{fig:magpis_complete}%
    \includegraphics[scale=0.6,trim={3.75cm 6.25cm 3.5cm 6.5cm}, clip]{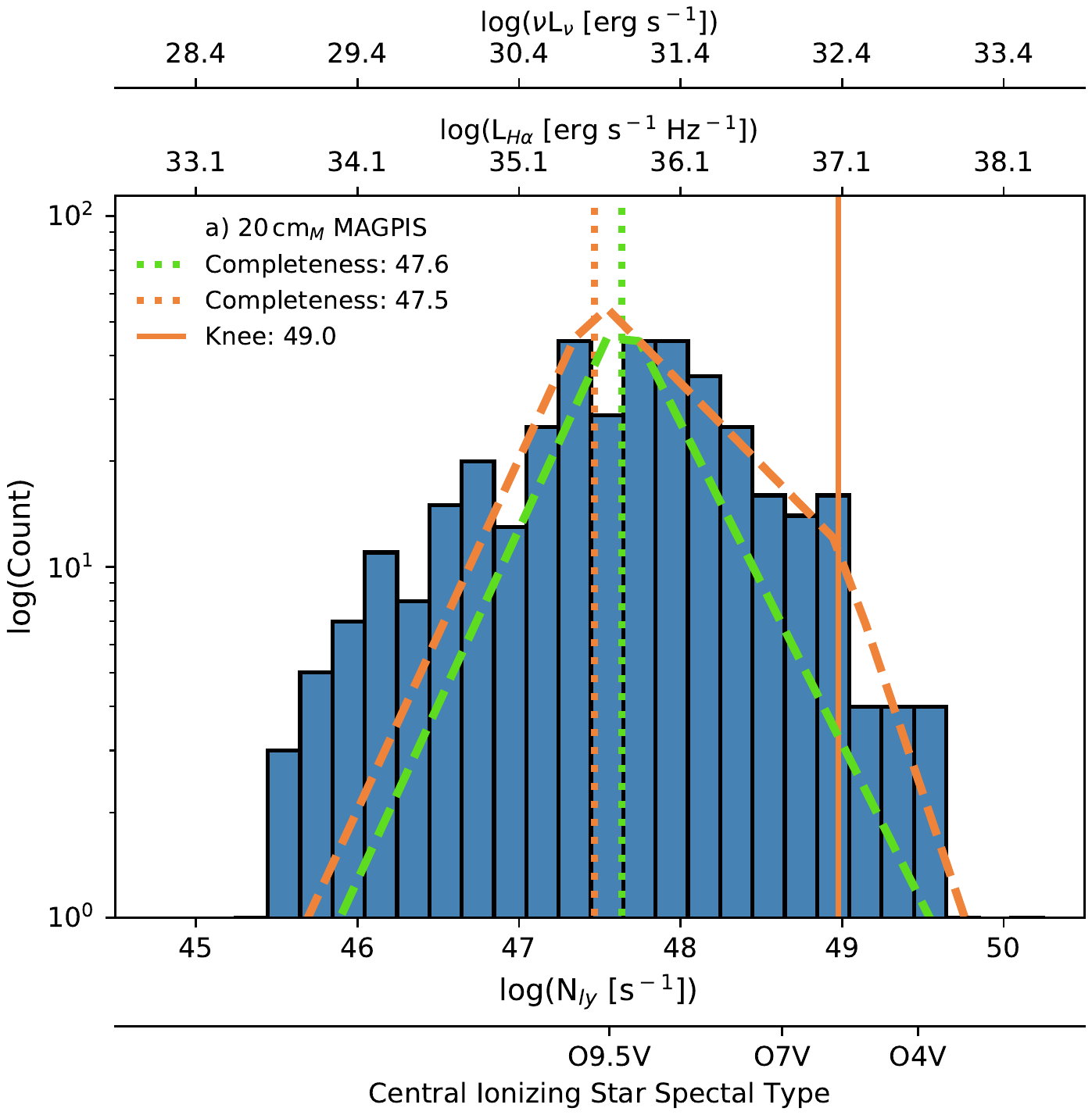}}\qquad
  \subfloat{\label{fig:magpis_vgps_complete}%
    \includegraphics[scale=0.6,trim={3.75cm 6.25cm 3.5cm 6.5cm}, clip]{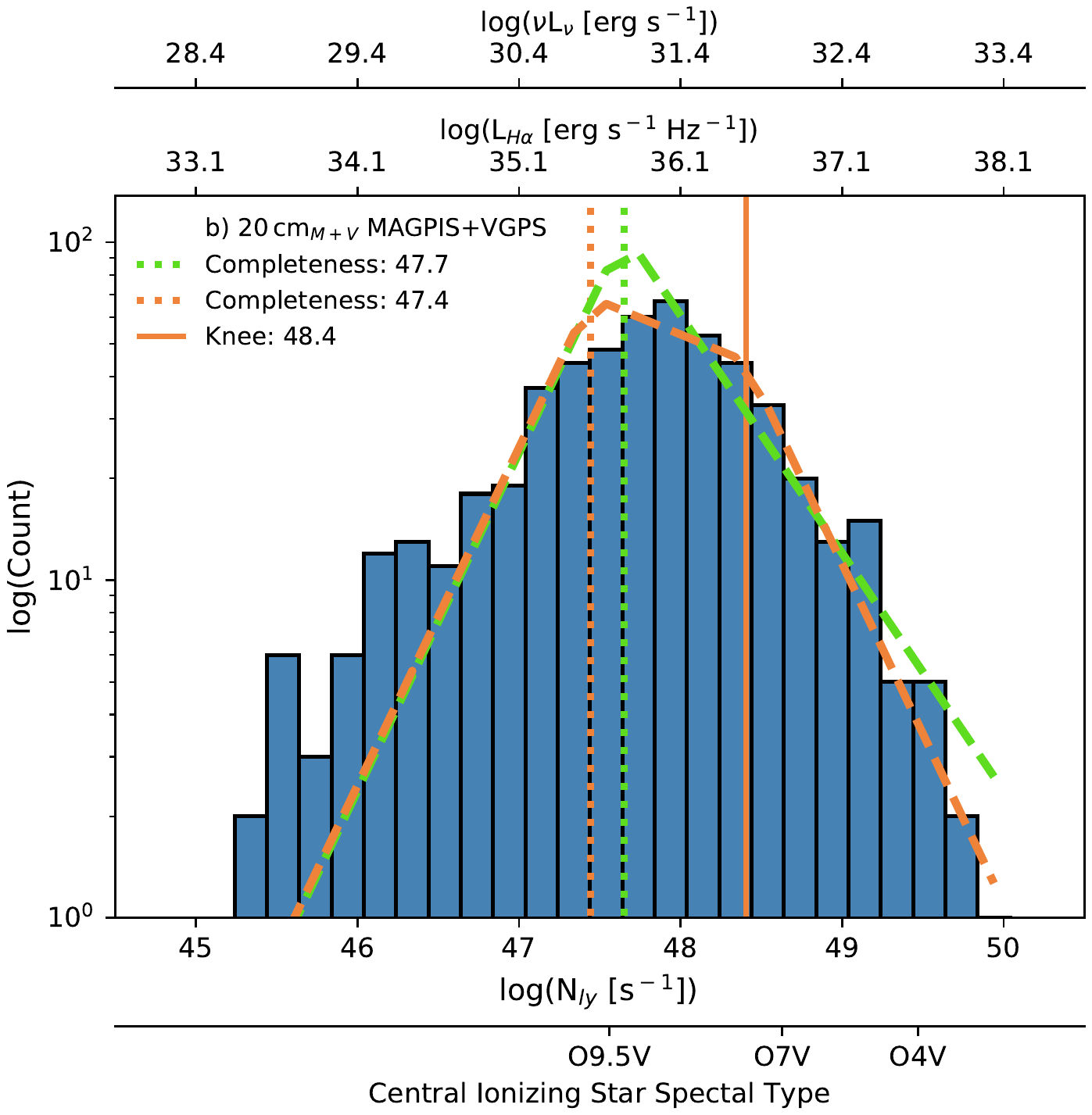}}\qquad
  \subfloat{\label{fig:vgps_complete}%
    \includegraphics[scale=0.6,trim={3.75cm 6.25cm 3.5cm 6.5cm}, clip]{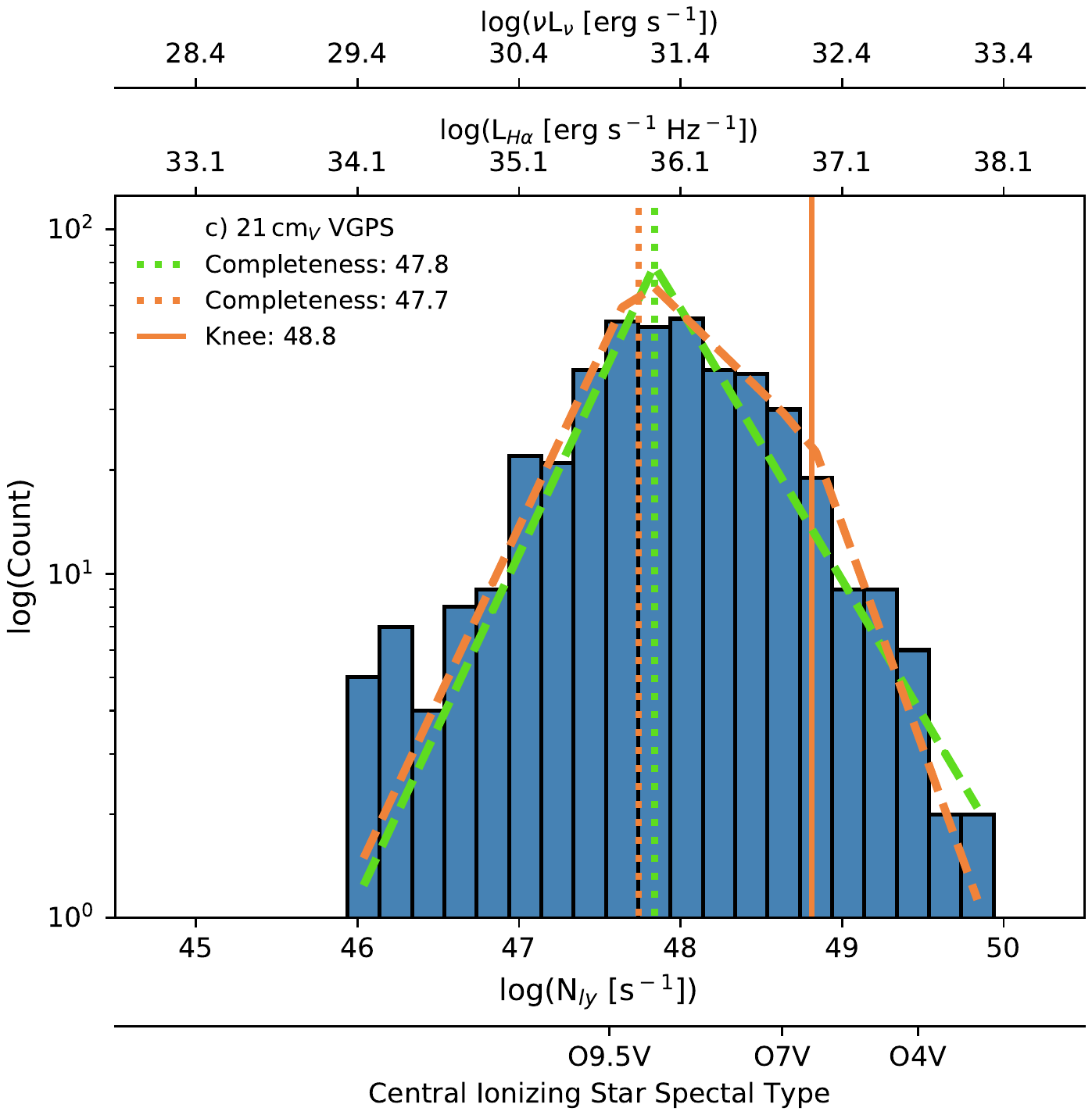}}
\caption{\hii region LFs of example radio data sets: $20\,\cm$ MAGPIS (panel \subref*{fig:magpis_complete}), $21\,\cm$ MAGPIS+VGPS (panel \subref*{fig:magpis_vgps_complete}), and $21\,\cm$ VGPS (panel \subref*{fig:vgps_complete}). Completeness limits and knee luminosities are respectively denoted by dotted and solid vertical lines. The model fits are represented by dashed lines. Fits and values associated with the single power law model are in green and those associated with the double power law model are in orange. We include separate x-axes for ease of comparison: the monochromatic luminosity $\nu L_{\nu}$, H$\alpha$ luminosity $L_{\rm{H}\alpha}$, Lyman continuum photon rate $N_{\rm{ly}}$, and the spectral type of the central ionizing star derived from \citet{mar05}. Note that the fit parameters and distributions in this figure are from a single Monte Carlo iteration and not the full analysis, and that they are presented only as examples of the fitting methods and results.}
\label{fig:radio_full_lfs}
\end{figure*}

When maximizing the likelihood functions of our models, we impose minimally restrictive bounds on the power law indices, completeness limits, and the knee luminosity in order to limit incorrect fits.  We restrict the indices to a minimum of $0.5$ and leave the maximum unbounded as no previous work has found an index much less than one. Likewise, we constrain the completeness limits and knee luminosity to the central $80\%$ of the full range of data being fit. A completeness limit outside these bounds would only be the result of a meaningless fit, as we \textit{a priori} assume that our data are complete neither to less-luminous B stars nor only to the most luminous O stars. As an illustration of our methods, example fits to infrared and radio LFs are shown in Figure \ref{fig:ir_full_lfs} and Figure \ref{fig:radio_full_lfs}, respectively.



\subsection{Modeled LF Comparison}
\label{subsec:comp}
We use the Bayesian Information Criterion (BIC) to determine whether a double power law model is preferred over a single power law model as outlined in Section ~\ref{appsubsec:bic}. We compare the median single and double power law model BICs for each wavelength in each data subset. If the double power law model BIC is lower than the single power law model BIC by at least ten \citep{kas95}, then we conclude that the former is favored over the latter. If the double power law BIC is not lower than this threshold, we favor the single power law as the best-fit model. There are no cases in which the median single and double power law model BICs are exactly equal.

\subsubsection{Power Law Fitting Method Comparison}
\label{subsubsec:fitting_comp}
In contrast to the traditional method of determining the form of the LF by fitting linear functions to logarithmically binned data in log-log space (i.e. bins of constant separation and width in log-log space), we use MLE to determine the model parameters. If we are to compare our results with those in the literature that use the traditional method, we must compare the two methods.

In order to validate the direct comparison of the results of our fitting method to those from previous work, we fit the non-Monte Carlo data from the \textit{WISE} Catalog using both techniques. We calculate the ratio of the single power law indices obtained using the two methods for each wavelength in every data subset. We find that the median and MAD of these ratios is $1.01\,\pm\,0.03$ with minimum and maximum ratios of $0.70$ and $1.10$. As a ratio of unity means that both fitting methods return the same power law index, we conclude that the two methods return the same results. We therefore can directly compare the model parameters we obtain using MLE with those reported in the literature.


\section{The Form of the First Galactic Quadrant \hii Region LF}
\label{sec:lf_form}

We show the full list of median power law indices from the fits to the Monte Carlo-generated \hii region luminosity distributions in Table~\ref{tab:fits} and all of the median completeness limit and knee luminosities -- both with their associated spectral types -- in Tables~\ref{tab:lim} and \ref{tab:knee}, respectively. Table~\ref{tab:stats} in the Appendix summarizes the BIC model choice results. In order to better compare our results with the literature, the values given in the text are the mean and standard deviations of the parameters unless otherwise specified. We collectively refer to the $8, 12, 22, 24, 70,$ and $160 \microns$ data using the term ``infrared'' and the $20$ and $21 \cm$ data using the term ``radio.''

We include example LF fits for each subset-wavelength combination as well as graphical summaries of the power law indices presented in Appendices~\ref{appsubsec:lf} and \ref{appsubsec:sim_summary}. We report knee luminosities and completeness limits in units of erg~s$^{-1}$ for both the $\nu L_{\nu}$ and $L_{\rm{H}\alpha}$ values, and Lyman continuum photon rates $N_{\rm{ly}}$ of the radio data in units of s$^{-1}$.

\begin{figure}[h]
   \centering
   \includegraphics[width=0.95\columnwidth,trim={3cm 7.75cm 3cm 8cm}, clip]{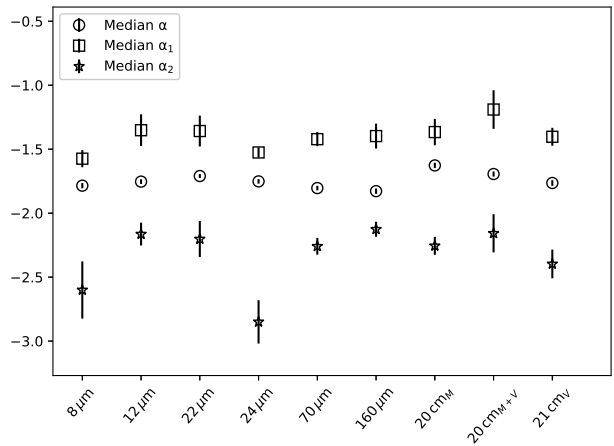}
   \caption{Median single and double power law indices of all first Galactic quadrant \hii regions. The median absolute deviations are represented by the error bars. The indices for the complete range of data subsets are shown in Figure \ref{fig:alphacomp2}.}
   \label{fig:constalpha_complete}
\end{figure}

\vspace{2mm}
\begin{figure}[h]
   \centering
   \includegraphics[width=0.95\columnwidth,trim={3cm 7.75cm 3cm 8cm}, clip]{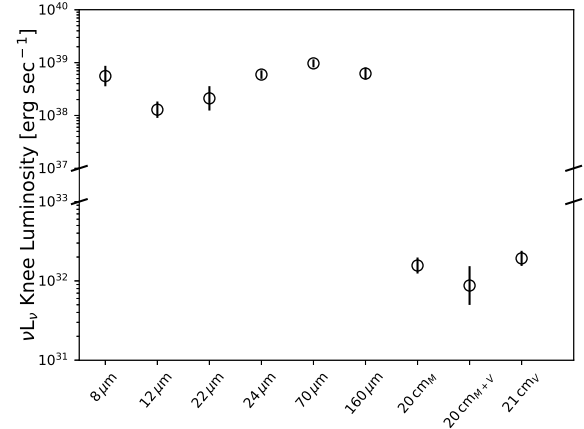}
   \caption{Median knee luminosities of all first Galactic quadrant \hii regions. The median absolute deviations are represented by the error bars. The knee luminosities for the complete range of data subsets are shown in Figure \ref{fig:kneecomp2}.}
   \label{fig:knee_complete}
\end{figure}

\subsection{Full Data Set}
\label{subsec:all}

Averaged across all wavelengths, the full \hii region LF best-fit single power law model has a mean power law index and standard deviation of $\langle\alpha\rangle=-1.75\,\pm\,0.01$. The best-fit double power law model has mean power law indices and standard deviations of $\langle\alpha_1\rangle=-1.40\,\pm\,0.03$ and $\langle\alpha_2\rangle=-2.33\,\pm\,0.04$. A double power law model is favored in five of the nine wavelengths.  All wavelengths have similar power law indices (see Figure~\ref{fig:constalpha_complete}).  From the radio continuum data we find that the \textit{WISE} Catalog is statistically complete for all first Galactic quadrant \hii regions ionized by single O9.5V stars (see Table~\ref{tab:lim}), a value in agreement with that found by Armentrout et al. (2021, in press).



\begin{figure}[h]
   \centering
   \includegraphics[width=0.95\columnwidth,trim={3cm 7.75cm 3.25cm 8cm}, clip]{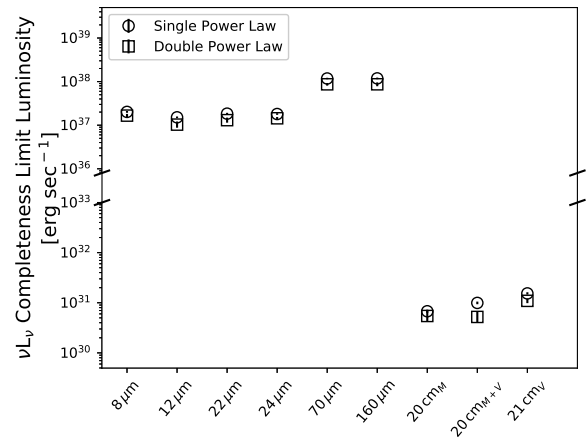}
   \caption{Median completeness limit luminosities of all first Galactic quadrant \hii regions. The median absolute deviations (MADs) are represented by the error bars. The completeness limit luminosities for the complete range of data subsets are shown in Figure \ref{fig:limitcomp2}. Note that the MADs are smaller than the size of the circular markers.}
   \label{fig:lim_complete}
\end{figure}



The knee (Figure~\ref{fig:knee_complete}) and completeness (Figure~\ref{fig:lim_complete}) luminosities are consistent across the infrared wavelengths and, separately, between all the radio data sets. We note that the best-fit double power laws to the Galactic LF listed in Table~\ref{tab:knee} have $\rm{H} \alpha$ knee values nearly two orders of magnitude below the canonical Str{\"o}mgren luminosity of $L_{\rm{H} \alpha} \approx 10^{39}$~erg~s$^{-1}$. This is likely a consequence of the lack of \hii regions with luminosities above the Str{\"o}mgren luminosity in the first Galactic quadrant and not an indication of an inherently unusual \hii region population.



\startlongtable
\begin{deluxetable*}{ccCCCC}
\tablecaption{Median Single and Double Power Law Fit Parameters \label{tab:fits}}
\tablecolumns{6}
\tablehead{
\colhead{} &
\colhead{} &
\multicolumn{1}{c}{Single} &
\colhead{} &
\multicolumn{2}{c}{Double} 
\\ \cline{3-3} \cline{5-6} 
\colhead{Subset} &
\colhead{Data} &
\colhead{$\alpha$} &
\colhead{} &
\colhead{$\alpha_{1}$} &
\colhead{$\alpha_{2}$} 
}
\startdata
All & $8\,\microns$ & -1.79 \pm 0.02 & & -1.57 \pm 0.07 & -2.60 \pm 0.22 \\
& $12\,\microns$ & -1.75 \pm 0.02 & & -1.35 \pm 0.12 & -2.16 \pm 0.09 \\
& $22\,\microns$ & -1.71 \pm 0.02 & & -1.36 \pm 0.12 & -2.20 \pm 0.14 \\
& $24\,\microns$ & -1.75 \pm 0.02 & & -1.53 \pm 0.04 & -2.85 \pm 0.17 \\
& $70\,\microns$ & -1.80 \pm 0.02 & & -1.42 \pm 0.05 & -2.26 \pm 0.07 \\
& $160\,\microns$ & -1.83 \pm 0.02 & & -1.40 \pm 0.10 & -2.13 \pm 0.06 \\
& $20\,\cm_{\textrm{{\tiny M}}}$ & -1.63 \pm 0.02 & & -1.37 \pm 0.10 & -2.26 \pm 0.07 \\
& $20\,\cm_{\textrm{{\tiny M+V}}}$ & -1.69 \pm 0.02 & & -1.19 \pm 0.15 & -2.16 \pm 0.15 \\
& $21\,\cm_{\textrm{{\tiny V}}}$ & -1.76 \pm 0.02 & & -1.40 \pm 0.07 & -2.40 \pm 0.11 \\\hline
$d_\sun \leq 7.75$ \kpc & $8\,\microns$ & -1.71 \pm 0.03 & & -1.36 \pm 0.11 & -2.60 \pm 0.25 \\
& $12\,\microns$ & -1.67 \pm 0.03 & & -1.28 \pm 0.07 & -2.58 \pm 0.18 \\
& $22\,\microns$ & -1.63 \pm 0.02 & & -1.32 \pm 0.05 & -2.60 \pm 0.17 \\
& $24\,\microns$ & -1.65 \pm 0.02 & & -1.36 \pm 0.05 & -2.61 \pm 0.18 \\
& $70\,\microns$ & -1.71 \pm 0.02 & & -1.38 \pm 0.04 & -2.50 \pm 0.11 \\
& $160\,\microns$ & -1.69 \pm 0.03 & & -1.20 \pm 0.08 & -2.32 \pm 0.11 \\
& $20\,\cm_{\textrm{{\tiny M}}}$ & -1.77 \pm 0.06 & & -1.70 \pm 0.08 & -2.17 \pm 0.09 \\
& $20\,\cm_{\textrm{{\tiny M+V}}}$ & -1.65 \pm 0.06 & & -1.52 \pm 0.10 & -2.23 \pm 0.08 \\
& $21\,\cm_{\textrm{{\tiny V}}}$ & -1.70 \pm 0.05 & & -1.43 \pm 0.13 & -2.24 \pm 0.07 \\\hline
$d_\sun > 7.75$ \kpc & $8\,\microns$ & -1.83 \pm 0.03 & & -1.67 \pm 0.05 & -2.56 \pm 0.15 \\
& $12\,\microns$ & -1.79 \pm 0.02 & & -1.63 \pm 0.07 & -2.15 \pm 0.08 \\
& $22\,\microns$ & -1.75 \pm 0.02 & & -1.55 \pm 0.05 & -2.30 \pm 0.09 \\
& $24\,\microns$ & -1.81 \pm 0.02 & & -1.60 \pm 0.04 & -2.83 \pm 0.13 \\
& $70\,\microns$ & -1.85 \pm 0.02 & & -1.51 \pm 0.07 & -2.23 \pm 0.09 \\
& $160\,\microns$ & -1.87 \pm 0.03 & & -1.65 \pm 0.07 & -2.17 \pm 0.13 \\
& $20\,\cm_{\textrm{{\tiny M}}}$ & -1.72 \pm 0.03 & & -1.37 \pm 0.06 & -2.37 \pm 0.10 \\
& $20\,\cm_{\textrm{{\tiny M+V}}}$ & -1.79 \pm 0.02 & & -1.51 \pm 0.05 & -2.48 \pm 0.14 \\
& $21\,\cm_{\textrm{{\tiny V}}}$ & -1.84 \pm 0.03 & & -1.52 \pm 0.07 & -2.69 \pm 0.25 \\\hline
$\rgal \leq 5$ \kpc & $8\,\microns$ & -1.70 \pm 0.04 & & -1.46 \pm 0.07 & -2.46 \pm 0.17 \\
& $12\,\microns$ & -1.65 \pm 0.04 & & -1.36 \pm 0.10 & -2.42 \pm 0.17 \\
& $22\,\microns$ & -1.66 \pm 0.03 & & -1.44 \pm 0.05 & -2.52 \pm 0.15 \\
& $24\,\microns$ & -1.67 \pm 0.03 & & -1.47 \pm 0.04 & -2.60 \pm 0.16 \\
& $70\,\microns$ & -1.77 \pm 0.03 & & -1.44 \pm 0.09 & -2.12 \pm 0.08 \\
& $160\,\microns$ & -1.76 \pm 0.03 & & -1.36 \pm 0.14 & -2.01 \pm 0.07 \\
& $20\,\cm_{\textrm{{\tiny M}}}$ & -1.62 \pm 0.04 & & -1.43 \pm 0.09 & -2.06 \pm 0.05 \\
& $20\,\cm_{\textrm{{\tiny M+V}}}$ & -1.63 \pm 0.03 & & -1.37 \pm 0.09 & -2.16 \pm 0.06 \\
& $21\,\cm_{\textrm{{\tiny V}}}$ & -1.73 \pm 0.06 & & -1.35 \pm 0.09 & -2.23 \pm 0.09 \\\hline
$\rgal > 5$ \kpc & $8\,\microns$ & -1.82 \pm 0.03 & & -1.58 \pm 0.07 & -2.76 \pm 0.21 \\
& $12\,\microns$ & -1.79 \pm 0.02 & & -1.39 \pm 0.09 & -2.24 \pm 0.08 \\
& $22\,\microns$ & -1.74 \pm 0.02 & & -1.35 \pm 0.09 & -2.30 \pm 0.11 \\
& $24\,\microns$ & -1.80 \pm 0.02 & & -1.51 \pm 0.05 & -3.06 \pm 0.25 \\
& $70\,\microns$ & -1.82 \pm 0.02 & & -1.40 \pm 0.06 & -2.41 \pm 0.09 \\
& $160\,\microns$ & -1.86 \pm 0.03 & & -1.43 \pm 0.09 & -2.29 \pm 0.11 \\
& $20\,\cm_{\textrm{{\tiny M}}}$ & -1.66 \pm 0.02 & & -1.32 \pm 0.07 & -2.53 \pm 0.18 \\
& $20\,\cm_{\textrm{{\tiny M+V}}}$ & -1.75 \pm 0.03 & & -1.34 \pm 0.12 & -2.40 \pm 0.23 \\
& $21\,\cm_{\textrm{{\tiny V}}}$ & -1.82 \pm 0.03 & & -1.50 \pm 0.07 & -2.68 \pm 0.25 \\\hline
$r \leq 2.4 \pc$ & $8\,\microns$ & -2.23 \pm 0.06 & & -1.26 \pm 0.22 & -3.13 \pm 0.23 \\
& $12\,\microns$ & -2.10 \pm 0.06 & & -1.18 \pm 0.22 & -2.80 \pm 0.21 \\
& $22\,\microns$ & -2.03 \pm 0.05 & & -1.26 \pm 0.17 & -3.23 \pm 0.38 \\
& $24\,\microns$ & -2.16 \pm 0.06 & & -1.63 \pm 0.17 & -3.46 \pm 0.37 \\
& $70\,\microns$ & -2.05 \pm 0.04 & & -1.40 \pm 0.12 & -2.68 \pm 0.15 \\
& $160\,\microns$ & -2.21 \pm 0.06 & & -1.33 \pm 0.14 & -3.08 \pm 0.21 \\
& $20\,\cm_{\textrm{{\tiny M}}}$ & -1.75 \pm 0.03 & & -1.18 \pm 0.08 & -2.96 \pm 0.21 \\
& $20\,\cm_{\textrm{{\tiny M+V}}}$ & -1.76 \pm 0.04 & & -1.04 \pm 0.10 & -2.76 \pm 0.21 \\
& $21\,\cm_{\textrm{{\tiny V}}}$ & -2.08 \pm 0.07 & & -1.39 \pm 0.18 & -3.40 \pm 0.45 \\\hline
$r > 2.4 \pc$ & $8\,\microns$ & -1.82 \pm 0.03 & & -1.53 \pm 0.07 & -2.60 \pm 0.18 \\
& $12\,\microns$ & -1.77 \pm 0.03 & & -1.42 \pm 0.15 & -2.23 \pm 0.12 \\
& $22\,\microns$ & -1.65 \pm 0.02 & & -1.28 \pm 0.08 & -2.36 \pm 0.16 \\
& $24\,\microns$ & -1.71 \pm 0.03 & & -1.33 \pm 0.05 & -2.90 \pm 0.19 \\
& $70\,\microns$ & -1.73 \pm 0.02 & & -1.31 \pm 0.08 & -2.26 \pm 0.12 \\
& $160\,\microns$ & -1.76 \pm 0.03 & & -1.21 \pm 0.11 & -2.02 \pm 0.09 \\
& $20\,\cm_{\textrm{{\tiny M}}}$ & -1.72 \pm 0.03 & & -1.41 \pm 0.08 & -2.26 \pm 0.07 \\
& $20\,\cm_{\textrm{{\tiny M+V}}}$ & -1.75 \pm 0.02 & & -1.45 \pm 0.05 & -2.40 \pm 0.11 \\
& $21\,\cm_{\textrm{{\tiny V}}}$ & -1.76 \pm 0.02 & & -1.40 \pm 0.06 & -2.50 \pm 0.15 \\\hline
Arm & $8\,\microns$ & -1.75 \pm 0.02 & & -1.53 \pm 0.05 & -2.87 \pm 0.19 \\
& $12\,\microns$ & -1.74 \pm 0.02 & & -1.44 \pm 0.07 & -2.71 \pm 0.18 \\
& $22\,\microns$ & -1.69 \pm 0.02 & & -1.44 \pm 0.04 & -2.90 \pm 0.15 \\
& $24\,\microns$ & -1.74 \pm 0.02 & & -1.51 \pm 0.04 & -2.94 \pm 0.16 \\
& $70\,\microns$ & -1.77 \pm 0.02 & & -1.45 \pm 0.06 & -2.26 \pm 0.10 \\
& $160\,\microns$ & -1.78 \pm 0.02 & & -1.42 \pm 0.09 & -2.14 \pm 0.09 \\
& $20\,\cm_{\textrm{{\tiny M}}}$ & -1.60 \pm 0.02 & & -1.32 \pm 0.09 & -2.40 \pm 0.11 \\
& $20\,\cm_{\textrm{{\tiny M+V}}}$ & -1.66 \pm 0.02 & & -1.26 \pm 0.07 & -2.30 \pm 0.09 \\
& $21\,\cm_{\textrm{{\tiny V}}}$ & -1.75 \pm 0.03 & & -1.42 \pm 0.07 & -2.41 \pm 0.11 \\\hline
Interarm & $8\,\microns$ & -1.93 \pm 0.05 & & -1.85 \pm 0.09 & -2.26 \pm 0.16 \\
& $12\,\microns$ & -1.83 \pm 0.04 & & -1.85 \pm 0.05 & -1.77 \pm 0.08 \\
& $22\,\microns$ & -1.77 \pm 0.03 & & -1.62 \pm 0.17 & -2.05 \pm 0.07 \\
& $24\,\microns$ & -1.83 \pm 0.04 & & -1.65 \pm 0.09 & -2.63 \pm 0.19 \\
& $70\,\microns$ & -1.92 \pm 0.04 & & -1.31 \pm 0.11 & -2.31 \pm 0.11 \\
& $160\,\microns$ & -2.00 \pm 0.05 & & -1.22 \pm 0.17 & -2.21 \pm 0.09 \\
& $20\,\cm_{\textrm{{\tiny M}}}$ & -1.73 \pm 0.04 & & -1.55 \pm 0.13 & -2.13 \pm 0.08 \\
& $20\,\cm_{\textrm{{\tiny M+V}}}$ & -1.83 \pm 0.04 & & -1.48 \pm 0.20 & -2.33 \pm 0.12 \\
& $21\,\cm_{\textrm{{\tiny V}}}$ & -1.88 \pm 0.04 & & -1.55 \pm 0.13 & -2.51 \pm 0.17 \\
\enddata
\end{deluxetable*}

\onecolumngrid

\startlongtable
\begin{deluxetable*}{ccccccccccc}
\tabletypesize{\scriptsize}
\setlength{\tabcolsep}{5pt}
\renewcommand{\arraystretch}{1.25}
\tablecaption{Median Completeness Limits \label{tab:lim}}
\tablecolumns{11}
\tablehead{
\colhead{} &
\colhead{} &
\multicolumn{4}{c}{Single Power Law} &
\colhead{} &
\multicolumn{4}{c}{Double Power Law} 
\\ \cline{3-6} \cline{8-11}
\colhead{Subset} &
\colhead{Data} &
\colhead{log($\nu L_{\nu}$)} &
\colhead{log($N_{\rm{ly}}$)} &
\colhead{log($L_{\rm{H}\alpha}$)} &
\colhead{Spectral} &
\colhead{} &
\colhead{log($\nu L_{\nu}$)} &
\colhead{log($N_{\rm{ly}}$)} &
\colhead{log($L_{\rm{H}\alpha}$)} &
\colhead{Spectral} \\[-6pt]
\colhead{} &
\colhead{} &
\colhead{log(erg~s$^{-1}$)} &
\colhead{log(s$^{-1}$)} &
\colhead{log(erg~s$^{-1}$)} &
\colhead{Type} &
\colhead{} &
\colhead{log(erg~s$^{-1}$)} &
\colhead{log(s$^{-1}$)} &
\colhead{log(erg~s$^{-1}$)} &
\colhead{Type} 
}
\startdata
All	& $8\,\microns$ &	37.31 $\pm$ 0.02	&	\nodata	&	\nodata	&	\nodata	& &	37.22 $\pm$ 0.04	&	\nodata	&	\nodata	&	\nodata	\\
	& $12\,\microns$ &	37.18 $\pm$ 0.02	&	\nodata	&	\nodata	&	\nodata	& &	37.01 $\pm$ 0.07	&	\nodata	&	\nodata	&	\nodata	\\
	& $22\,\microns$ &	37.26 $\pm$ 0.02	&	\nodata	&	\nodata	&	\nodata	& &	37.12 $\pm$ 0.06	&	\nodata	&	\nodata	&	\nodata	\\
	& $24\,\microns$ &	37.26 $\pm$ 0.02	&	\nodata	&	\nodata	&	\nodata	& &	37.16 $\pm$ 0.03	&	\nodata	&	\nodata	&	\nodata	\\
	& $70\,\microns$ &	38.07 $\pm$ 0.02	&	\nodata	&	\nodata	&	\nodata	& &	37.94 $\pm$ 0.03	&	\nodata	&	\nodata	&	\nodata	\\
	& $160\,\microns$ &	38.07 $\pm$ 0.02	&	\nodata	&	\nodata	&	\nodata	& &	37.94 $\pm$ 0.04	&	\nodata	&	\nodata	&	\nodata	\\
	& $20\,\cm_{\textrm{{\tiny M}}}$ &	30.80 $\pm$ 0.03	&	47.47 $\pm$ 0.03	&	35.61 $\pm$ 0.03	&	$<$ O9.5V	& &	30.71 $\pm$ 0.11	&	47.38 $\pm$ 0.11	&	35.52 $\pm$ 0.11	&	$<$ O9.5V	\\
	& $20\,\cm_{\textrm{{\tiny M+V}}}$ &	30.97 $\pm$ 0.03	&	47.64 $\pm$ 0.03	&	35.78 $\pm$ 0.03	&	O9.5V	& &	30.69 $\pm$ 0.11	&	47.36 $\pm$ 0.11	&	35.50 $\pm$ 0.11	&	$<$ O9.5V	\\
	& $21\,\cm_{\textrm{{\tiny V}}}$ &	31.16 $\pm$ 0.02	&	47.83 $\pm$ 0.02	&	35.97 $\pm$ 0.02	&	O9.5V	& &	31.01 $\pm$ 0.05	&	47.68 $\pm$ 0.05	&	35.82 $\pm$ 0.05	&	O9.5V	\\\hline
$d_\sun \leq 7.75$ \kpc	& $8\,\microns$ &	37.06 $\pm$ 0.04	&	\nodata	&	\nodata	&	\nodata	& &	36.88 $\pm$ 0.10	&	\nodata	&	\nodata	&	\nodata	\\
	& $12\,\microns$ &	36.89 $\pm$ 0.04	&	\nodata	&	\nodata	&	\nodata	& &	36.68 $\pm$ 0.06	&	\nodata	&	\nodata	&	\nodata	\\
	& $22\,\microns$ &	36.98 $\pm$ 0.03	&	\nodata	&	\nodata	&	\nodata	& &	36.82 $\pm$ 0.05	&	\nodata	&	\nodata	&	\nodata	\\
	& $24\,\microns$ &	36.93 $\pm$ 0.03	&	\nodata	&	\nodata	&	\nodata	& &	36.80 $\pm$ 0.04	&	\nodata	&	\nodata	&	\nodata	\\
	& $70\,\microns$ &	37.77 $\pm$ 0.03	&	\nodata	&	\nodata	&	\nodata	& &	37.63 $\pm$ 0.03	&	\nodata	&	\nodata	&	\nodata	\\
	& $160\,\microns$ &	37.74 $\pm$ 0.03	&	\nodata	&	\nodata	&	\nodata	& &	37.52 $\pm$ 0.06	&	\nodata	&	\nodata	&	\nodata	\\
	& $20\,\cm_{\textrm{{\tiny M}}}$ &	30.74 $\pm$ 0.07	&	47.41 $\pm$ 0.07	&	35.54 $\pm$ 0.07	&	$<$ O9.5V	& &	30.73 $\pm$ 0.07	&	47.40 $\pm$ 0.07	&	35.54 $\pm$ 0.07	&	$<$ O9.5V	\\
	& $20\,\cm_{\textrm{{\tiny M+V}}}$ &	30.72 $\pm$ 0.09	&	47.39 $\pm$ 0.09	&	35.53 $\pm$ 0.09	&	$<$ O9.5V	& &	30.72 $\pm$ 0.09	&	47.39 $\pm$ 0.09	&	35.53 $\pm$ 0.09	&	$<$ O9.5V	\\
	& $21\,\cm_{\textrm{{\tiny V}}}$ &	30.92 $\pm$ 0.06	&	47.59 $\pm$ 0.06	&	35.73 $\pm$ 0.06	&	O9.5V	& &	30.80 $\pm$ 0.12	&	47.47 $\pm$ 0.12	&	35.61 $\pm$ 0.12	&	$<$ O9.5V	\\\hline
$d_\sun > 7.75$ \kpc	& $8\,\microns$ &	37.45 $\pm$ 0.02	&	\nodata	&	\nodata	&	\nodata	& &	37.40 $\pm$ 0.03	&	\nodata	&	\nodata	&	\nodata	\\
	& $12\,\microns$ &	37.34 $\pm$ 0.02	&	\nodata	&	\nodata	&	\nodata	& &	37.29 $\pm$ 0.03	&	\nodata	&	\nodata	&	\nodata	\\
	& $22\,\microns$ &	37.42 $\pm$ 0.02	&	\nodata	&	\nodata	&	\nodata	& &	37.34 $\pm$ 0.03	&	\nodata	&	\nodata	&	\nodata	\\
	& $24\,\microns$ &	37.43 $\pm$ 0.02	&	\nodata	&	\nodata	&	\nodata	& &	37.36 $\pm$ 0.03	&	\nodata	&	\nodata	&	\nodata	\\
	& $70\,\microns$ &	38.23 $\pm$ 0.02	&	\nodata	&	\nodata	&	\nodata	& &	38.15 $\pm$ 0.03	&	\nodata	&	\nodata	&	\nodata	\\
	& $160\,\microns$ &	38.22 $\pm$ 0.02	&	\nodata	&	\nodata	&	\nodata	& &	38.17 $\pm$ 0.03	&	\nodata	&	\nodata	&	\nodata	\\
	& $20\,\cm_{\textrm{{\tiny M}}}$ &	31.05 $\pm$ 0.03	&	47.72 $\pm$ 0.03	&	35.86 $\pm$ 0.03	&	O9.5V	& &	30.88 $\pm$ 0.05	&	47.55 $\pm$ 0.05	&	35.69 $\pm$ 0.05	&	$<$ O9.5V	\\
	& $20\,\cm_{\textrm{{\tiny M+V}}}$ &	31.19 $\pm$ 0.02	&	47.85 $\pm$ 0.02	&	35.99 $\pm$ 0.02	&	O9.5V	& &	31.09 $\pm$ 0.04	&	47.76 $\pm$ 0.04	&	35.89 $\pm$ 0.04	&	O9.5V	\\
	& $21\,\cm_{\textrm{{\tiny V}}}$ &	31.32 $\pm$ 0.03	&	47.99 $\pm$ 0.03	&	36.13 $\pm$ 0.03	&	O9V	& &	31.23 $\pm$ 0.04	&	47.90 $\pm$ 0.04	&	36.03 $\pm$ 0.04	&	O9.5V	\\\hline
$\rgal \leq 5$ \kpc	& $8\,\microns$ &	37.26 $\pm$ 0.05	&	\nodata	&	\nodata	&	\nodata	& &	37.11 $\pm$ 0.07	&	\nodata	&	\nodata	&	\nodata	\\
	& $12\,\microns$ &	37.08 $\pm$ 0.06	&	\nodata	&	\nodata	&	\nodata	& &	36.89 $\pm$ 0.10	&	\nodata	&	\nodata	&	\nodata	\\
	& $22\,\microns$ &	37.30 $\pm$ 0.04	&	\nodata	&	\nodata	&	\nodata	& &	37.18 $\pm$ 0.05	&	\nodata	&	\nodata	&	\nodata	\\
	& $24\,\microns$ &	37.25 $\pm$ 0.04	&	\nodata	&	\nodata	&	\nodata	& &	37.15 $\pm$ 0.04	&	\nodata	&	\nodata	&	\nodata	\\
	& $70\,\microns$ &	38.09 $\pm$ 0.03	&	\nodata	&	\nodata	&	\nodata	& &	37.99 $\pm$ 0.04	&	\nodata	&	\nodata	&	\nodata	\\
	& $160\,\microns$ &	38.08 $\pm$ 0.03	&	\nodata	&	\nodata	&	\nodata	& &	37.97 $\pm$ 0.06	&	\nodata	&	\nodata	&	\nodata	\\
	& $20\,\cm_{\textrm{{\tiny M}}}$ &	30.88 $\pm$ 0.06	&	47.55 $\pm$ 0.06	&	35.69 $\pm$ 0.06	&	$<$ O9.5V	& &	30.83 $\pm$ 0.10	&	47.49 $\pm$ 0.10	&	35.63 $\pm$ 0.10	&	$<$ O9.5V	\\
	& $20\,\cm_{\textrm{{\tiny M+V}}}$ &	30.89 $\pm$ 0.05	&	47.56 $\pm$ 0.05	&	35.70 $\pm$ 0.05	&	O9.5V	& &	30.79 $\pm$ 0.10	&	47.46 $\pm$ 0.10	&	35.60 $\pm$ 0.10	&	$<$ O9.5V	\\
	& $21\,\cm_{\textrm{{\tiny V}}}$ &	31.17 $\pm$ 0.08	&	47.84 $\pm$ 0.08	&	35.97 $\pm$ 0.08	&	O9.5V	& &	30.99 $\pm$ 0.08	&	47.66 $\pm$ 0.08	&	35.79 $\pm$ 0.08	&	O9.5V	\\\hline
$\rgal > 5$ \kpc	& $8\,\microns$ &	37.32 $\pm$ 0.02	&	\nodata	&	\nodata	&	\nodata	& &	37.23 $\pm$ 0.04	&	\nodata	&	\nodata	&	\nodata	\\
	& $12\,\microns$ &	37.21 $\pm$ 0.02	&	\nodata	&	\nodata	&	\nodata	& &	37.07 $\pm$ 0.05	&	\nodata	&	\nodata	&	\nodata	\\
	& $22\,\microns$ &	37.26 $\pm$ 0.02	&	\nodata	&	\nodata	&	\nodata	& &	37.10 $\pm$ 0.05	&	\nodata	&	\nodata	&	\nodata	\\
	& $24\,\microns$ &	37.26 $\pm$ 0.02	&	\nodata	&	\nodata	&	\nodata	& &	37.15 $\pm$ 0.04	&	\nodata	&	\nodata	&	\nodata	\\
	& $70\,\microns$ &	38.06 $\pm$ 0.02	&	\nodata	&	\nodata	&	\nodata	& &	37.90 $\pm$ 0.04	&	\nodata	&	\nodata	&	\nodata	\\
	& $160\,\microns$ &	38.07 $\pm$ 0.02	&	\nodata	&	\nodata	&	\nodata	& &	37.93 $\pm$ 0.05	&	\nodata	&	\nodata	&	\nodata	\\
	& $20\,\cm_{\textrm{{\tiny M}}}$ &	30.79 $\pm$ 0.03	&	47.46 $\pm$ 0.03	&	35.60 $\pm$ 0.03	&	$<$ O9.5V	& &	30.63 $\pm$ 0.07	&	47.29 $\pm$ 0.07	&	35.43 $\pm$ 0.07	&	$<$ O9.5V	\\
	& $20\,\cm_{\textrm{{\tiny M+V}}}$ &	31.03 $\pm$ 0.03	&	47.70 $\pm$ 0.03	&	35.84 $\pm$ 0.03	&	O9.5V	& &	30.81 $\pm$ 0.09	&	47.48 $\pm$ 0.09	&	35.61 $\pm$ 0.09	&	$<$ O9.5V	\\
	& $21\,\cm_{\textrm{{\tiny V}}}$ &	31.19 $\pm$ 0.03	&	47.86 $\pm$ 0.03	&	36.00 $\pm$ 0.03	&	O9.5V	& &	31.07 $\pm$ 0.05	&	47.74 $\pm$ 0.05	&	35.88 $\pm$ 0.05	&	O9.5V	\\\hline
$r \leq 2.4 \pc$	& $8\,\microns$ &	37.22 $\pm$ 0.03	&	\nodata	&	\nodata	&	\nodata	& &	36.97 $\pm$ 0.08	&	\nodata	&	\nodata	&	\nodata	\\
	& $12\,\microns$ &	37.05 $\pm$ 0.04	&	\nodata	&	\nodata	&	\nodata	& &	36.77 $\pm$ 0.08	&	\nodata	&	\nodata	&	\nodata	\\
	& $22\,\microns$ &	37.23 $\pm$ 0.03	&	\nodata	&	\nodata	&	\nodata	& &	36.97 $\pm$ 0.08	&	\nodata	&	\nodata	&	\nodata	\\
	& $24\,\microns$ &	37.25 $\pm$ 0.03	&	\nodata	&	\nodata	&	\nodata	& &	37.11 $\pm$ 0.07	&	\nodata	&	\nodata	&	\nodata	\\
	& $70\,\microns$ &	38.06 $\pm$ 0.02	&	\nodata	&	\nodata	&	\nodata	& &	37.89 $\pm$ 0.05	&	\nodata	&	\nodata	&	\nodata	\\
	& $160\,\microns$ &	38.06 $\pm$ 0.03	&	\nodata	&	\nodata	&	\nodata	& &	37.83 $\pm$ 0.07	&	\nodata	&	\nodata	&	\nodata	\\
	& $20\,\cm_{\textrm{{\tiny M}}}$ &	30.58 $\pm$ 0.03	&	47.25 $\pm$ 0.03	&	35.39 $\pm$ 0.03	&	$<$ O9.5V	& &	30.26 $\pm$ 0.10	&	46.93 $\pm$ 0.10	&	35.07 $\pm$ 0.10	&	$<$ O9.5V	\\
	& $20\,\cm_{\textrm{{\tiny M+V}}}$ &	30.65 $\pm$ 0.04	&	47.32 $\pm$ 0.04	&	35.46 $\pm$ 0.04	&	$<$ O9.5V	& &	30.23 $\pm$ 0.11	&	46.90 $\pm$ 0.11	&	35.03 $\pm$ 0.11	&	$<$ O9.5V	\\
	& $21\,\cm_{\textrm{{\tiny V}}}$ &	31.02 $\pm$ 0.04	&	47.69 $\pm$ 0.04	&	35.82 $\pm$ 0.04	&	O9.5V	& &	30.81 $\pm$ 0.09	&	47.48 $\pm$ 0.09	&	35.62 $\pm$ 0.09	&	$<$ O9.5V	\\\hline
$r > 2.4 \pc$	& $8\,\microns$ &	37.65 $\pm$ 0.03	&	\nodata	&	\nodata	&	\nodata	& &	37.55 $\pm$ 0.04	&	\nodata	&	\nodata	&	\nodata	\\
	& $12\,\microns$ &	37.51 $\pm$ 0.03	&	\nodata	&	\nodata	&	\nodata	& &	37.39 $\pm$ 0.06	&	\nodata	&	\nodata	&	\nodata	\\
	& $22\,\microns$ &	37.45 $\pm$ 0.03	&	\nodata	&	\nodata	&	\nodata	& &	37.24 $\pm$ 0.06	&	\nodata	&	\nodata	&	\nodata	\\
	& $24\,\microns$ &	37.47 $\pm$ 0.03	&	\nodata	&	\nodata	&	\nodata	& &	37.27 $\pm$ 0.05	&	\nodata	&	\nodata	&	\nodata	\\
	& $70\,\microns$ &	38.18 $\pm$ 0.02	&	\nodata	&	\nodata	&	\nodata	& &	38.02 $\pm$ 0.05	&	\nodata	&	\nodata	&	\nodata	\\
	& $160\,\microns$ &	38.25 $\pm$ 0.03	&	\nodata	&	\nodata	&	\nodata	& &	38.07 $\pm$ 0.05	&	\nodata	&	\nodata	&	\nodata	\\
	& $20\,\cm_{\textrm{{\tiny M}}}$ &	31.36 $\pm$ 0.03	&	48.03 $\pm$ 0.03	&	36.17 $\pm$ 0.03	&	O9V	& &	31.31 $\pm$ 0.07	&	47.98 $\pm$ 0.07	&	36.11 $\pm$ 0.07	&	O9V	\\
	& $20\,\cm_{\textrm{{\tiny M+V}}}$ &	31.30 $\pm$ 0.02	&	47.97 $\pm$ 0.02	&	36.10 $\pm$ 0.02	&	O9V	& &	31.20 $\pm$ 0.04	&	47.87 $\pm$ 0.04	&	36.01 $\pm$ 0.04	&	O9.5V	\\
	& $21\,\cm_{\textrm{{\tiny V}}}$ &	31.32 $\pm$ 0.03	&	47.99 $\pm$ 0.03	&	36.13 $\pm$ 0.03	&	O9V	& &	31.20 $\pm$ 0.04	&	47.87 $\pm$ 0.04	&	36.01 $\pm$ 0.04	&	O9.5V	\\\hline
Arm	& $8\,\microns$ &	37.29 $\pm$ 0.02	&	\nodata	&	\nodata	&	\nodata	& &	37.20 $\pm$ 0.04	&	\nodata	&	\nodata	&	\nodata	\\
	& $12\,\microns$ &	37.18 $\pm$ 0.02	&	\nodata	&	\nodata	&	\nodata	& &	37.05 $\pm$ 0.05	&	\nodata	&	\nodata	&	\nodata	\\
	& $22\,\microns$ &	37.27 $\pm$ 0.02	&	\nodata	&	\nodata	&	\nodata	& &	37.16 $\pm$ 0.03	&	\nodata	&	\nodata	&	\nodata	\\
	& $24\,\microns$ &	37.27 $\pm$ 0.02	&	\nodata	&	\nodata	&	\nodata	& &	37.18 $\pm$ 0.03	&	\nodata	&	\nodata	&	\nodata	\\
	& $70\,\microns$ &	38.09 $\pm$ 0.02	&	\nodata	&	\nodata	&	\nodata	& &	37.97 $\pm$ 0.04	&	\nodata	&	\nodata	&	\nodata	\\
	& $160\,\microns$ &	38.07 $\pm$ 0.02	&	\nodata	&	\nodata	&	\nodata	& &	37.96 $\pm$ 0.04	&	\nodata	&	\nodata	&	\nodata	\\
	& $20\,\cm_{\textrm{{\tiny M}}}$ &	30.79 $\pm$ 0.04	&	47.46 $\pm$ 0.04	&	35.59 $\pm$ 0.04	&	$<$ O9.5V	& &	30.67 $\pm$ 0.11	&	47.34 $\pm$ 0.11	&	35.48 $\pm$ 0.11	&	$<$ O9.5V	\\
	& $20\,\cm_{\textrm{{\tiny M+V}}}$ &	30.95 $\pm$ 0.03	&	47.62 $\pm$ 0.03	&	35.76 $\pm$ 0.03	&	O9.5V	& &	30.73 $\pm$ 0.07	&	47.40 $\pm$ 0.07	&	35.53 $\pm$ 0.07	&	$<$ O9.5V	\\
	& $21\,\cm_{\textrm{{\tiny V}}}$ &	31.18 $\pm$ 0.04	&	47.85 $\pm$ 0.04	&	35.99 $\pm$ 0.04	&	O9.5V	& &	31.05 $\pm$ 0.06	&	47.72 $\pm$ 0.06	&	35.85 $\pm$ 0.06	&	O9.5V	\\\hline
Interarm	& $8\,\microns$ &	37.41 $\pm$ 0.04	&	\nodata	&	\nodata	&	\nodata	& &	37.38 $\pm$ 0.05	&	\nodata	&	\nodata	&	\nodata	\\
	& $12\,\microns$ &	37.24 $\pm$ 0.03	&	\nodata	&	\nodata	&	\nodata	& &	37.25 $\pm$ 0.04	&	\nodata	&	\nodata	&	\nodata	\\
	& $22\,\microns$ &	37.31 $\pm$ 0.03	&	\nodata	&	\nodata	&	\nodata	& &	37.23 $\pm$ 0.09	&	\nodata	&	\nodata	&	\nodata	\\
	& $24\,\microns$ &	37.30 $\pm$ 0.04	&	\nodata	&	\nodata	&	\nodata	& &	37.23 $\pm$ 0.06	&	\nodata	&	\nodata	&	\nodata	\\
	& $70\,\microns$ &	38.12 $\pm$ 0.03	&	\nodata	&	\nodata	&	\nodata	& &	37.94 $\pm$ 0.05	&	\nodata	&	\nodata	&	\nodata	\\
	& $160\,\microns$ &	38.15 $\pm$ 0.03	&	\nodata	&	\nodata	&	\nodata	& &	37.96 $\pm$ 0.08	&	\nodata	&	\nodata	&	\nodata	\\
	& $20\,\cm_{\textrm{{\tiny M}}}$ &	30.88 $\pm$ 0.04	&	47.55 $\pm$ 0.04	&	35.69 $\pm$ 0.04	&	$<$ O9.5V	& &	30.83 $\pm$ 0.08	&	47.50 $\pm$ 0.08	&	35.64 $\pm$ 0.08	&	$<$ O9.5V	\\
	& $20\,\cm_{\textrm{{\tiny M+V}}}$ &	31.04 $\pm$ 0.04	&	47.71 $\pm$ 0.04	&	35.85 $\pm$ 0.04	&	O9.5V	& &	30.91 $\pm$ 0.10	&	47.58 $\pm$ 0.10	&	35.71 $\pm$ 0.10	&	O9.5V	\\
	& $21\,\cm_{\textrm{{\tiny V}}}$ &	31.20 $\pm$ 0.03	&	47.86 $\pm$ 0.03	&	36.00 $\pm$ 0.03	&	O9.5V	& &	31.10 $\pm$ 0.06	&	47.76 $\pm$ 0.06	&	35.90 $\pm$ 0.06	&	O9.5V	\\
\enddata
\end{deluxetable*}

\startlongtable
\begin{deluxetable*}{cccccc}
\tabletypesize{\scriptsize}
\setlength{\tabcolsep}{5pt}
\renewcommand{\arraystretch}{1.25}
\tablecaption{Median Knee Luminosities \label{tab:knee}}
\tablecolumns{6}
\tablehead{
\colhead{Subset} &
\colhead{Data} &
\colhead{log($\nu L_{\nu}$)} &
\colhead{log($N_{\rm{ly}}$)} &
\colhead{log($L_{\rm{H}\alpha}$)} &
\colhead{Spectral} \\[-6pt]
\colhead{} &
\colhead{} &
\colhead{log(erg~s$^{-1}$)} &
\colhead{log(s$^{-1}$)} &
\colhead{log(erg~s$^{-1}$)} &
\colhead{Type}
}
\startdata
All	& $8\,\microns$ &	38.75 $\pm$ 0.19	&	\nodata	&	\nodata	&	\nodata	\\
	& $12\,\microns$ &	38.11 $\pm$ 0.16	&	\nodata	&	\nodata	&	\nodata	\\
	& $22\,\microns$ &	38.32 $\pm$ 0.23	&	\nodata	&	\nodata	&	\nodata	\\
	& $24\,\microns$ &	38.77 $\pm$ 0.08	&	\nodata	&	\nodata	&	\nodata	\\
	& $70\,\microns$ &	38.99 $\pm$ 0.07	&	\nodata	&	\nodata	&	\nodata	\\
	& $160\,\microns$ &	38.79 $\pm$ 0.11	&	\nodata	&	\nodata	&	\nodata	\\
	& $20\,\cm_{\textrm{{\tiny M}}}$ &	32.17 $\pm$ 0.10	&	48.84 $\pm$ 0.10	&	36.97 $\pm$ 0.10	&	O6.5V	\\
	& $20\,\cm_{\textrm{{\tiny M+V}}}$ &	31.91 $\pm$ 0.24	&	48.58 $\pm$ 0.24	&	36.72 $\pm$ 0.24	&	O7.5V	\\
	& $21\,\cm_{\textrm{{\tiny V}}}$ &	32.26 $\pm$ 0.10	&	48.93 $\pm$ 0.10	&	37.06 $\pm$ 0.10	&	O6.5V	\\\hline
$d_\sun \leq 7.75$ \kpc	& $8\,\microns$ &	38.40 $\pm$ 0.17	&	\nodata	&	\nodata	&	\nodata	\\
	& $12\,\microns$ &	38.20 $\pm$ 0.12	&	\nodata	&	\nodata	&	\nodata	\\
	& $22\,\microns$ &	38.47 $\pm$ 0.11	&	\nodata	&	\nodata	&	\nodata	\\
	& $24\,\microns$ &	38.39 $\pm$ 0.13	&	\nodata	&	\nodata	&	\nodata	\\
	& $70\,\microns$ &	39.05 $\pm$ 0.06	&	\nodata	&	\nodata	&	\nodata	\\
	& $160\,\microns$ &	38.74 $\pm$ 0.11	&	\nodata	&	\nodata	&	\nodata	\\
	& $20\,\cm_{\textrm{{\tiny M}}}$ &	32.28 $\pm$ 0.05	&	48.95 $\pm$ 0.05	&	37.09 $\pm$ 0.05	&	O6.5V	\\
	& $20\,\cm_{\textrm{{\tiny M+V}}}$ &	32.22 $\pm$ 0.09	&	48.89 $\pm$ 0.09	&	37.03 $\pm$ 0.09	&	O6.5V	\\
	& $21\,\cm_{\textrm{{\tiny V}}}$ &	32.18 $\pm$ 0.10	&	48.85 $\pm$ 0.10	&	36.99 $\pm$ 0.10	&	O6.5V	\\\hline
$d_\sun > 7.75$ \kpc	& $8\,\microns$ &	38.89 $\pm$ 0.09	&	\nodata	&	\nodata	&	\nodata	\\
	& $12\,\microns$ &	38.58 $\pm$ 0.23	&	\nodata	&	\nodata	&	\nodata	\\
	& $22\,\microns$ &	38.82 $\pm$ 0.09	&	\nodata	&	\nodata	&	\nodata	\\
	& $24\,\microns$ &	38.86 $\pm$ 0.05	&	\nodata	&	\nodata	&	\nodata	\\
	& $70\,\microns$ &	39.03 $\pm$ 0.13	&	\nodata	&	\nodata	&	\nodata	\\
	& $160\,\microns$ &	39.10 $\pm$ 0.26	&	\nodata	&	\nodata	&	\nodata	\\
	& $20\,\cm_{\textrm{{\tiny M}}}$ &	32.27 $\pm$ 0.08	&	48.93 $\pm$ 0.08	&	37.07 $\pm$ 0.08	&	O6.5V	\\
	& $20\,\cm_{\textrm{{\tiny M+V}}}$ &	32.37 $\pm$ 0.09	&	49.04 $\pm$ 0.09	&	37.17 $\pm$ 0.09	&	O6V	\\
	& $21\,\cm_{\textrm{{\tiny V}}}$ &	32.41 $\pm$ 0.11	&	49.08 $\pm$ 0.11	&	37.21 $\pm$ 0.11	&	O6V	\\\hline
$\rgal \leq 5$ \kpc	& $8\,\microns$ &	38.90 $\pm$ 0.12	&	\nodata	&	\nodata	&	\nodata	\\
	& $12\,\microns$ &	38.63 $\pm$ 0.21	&	\nodata	&	\nodata	&	\nodata	\\
	& $22\,\microns$ &	38.95 $\pm$ 0.07	&	\nodata	&	\nodata	&	\nodata	\\
	& $24\,\microns$ &	38.94 $\pm$ 0.07	&	\nodata	&	\nodata	&	\nodata	\\
	& $70\,\microns$ &	38.97 $\pm$ 0.15	&	\nodata	&	\nodata	&	\nodata	\\
	& $160\,\microns$ &	38.75 $\pm$ 0.12	&	\nodata	&	\nodata	&	\nodata	\\
	& $20\,\cm_{\textrm{{\tiny M}}}$ &	32.25 $\pm$ 0.08	&	48.91 $\pm$ 0.08	&	37.05 $\pm$ 0.08	&	O6.5V	\\
	& $20\,\cm_{\textrm{{\tiny M+V}}}$ &	32.20 $\pm$ 0.08	&	48.87 $\pm$ 0.08	&	37.00 $\pm$ 0.08	&	O6.5V	\\
	& $21\,\cm_{\textrm{{\tiny V}}}$ &	32.23 $\pm$ 0.09	&	48.90 $\pm$ 0.09	&	37.04 $\pm$ 0.09	&	O6.5V	\\\hline
$\rgal > 5$ \kpc	& $8\,\microns$ &	38.63 $\pm$ 0.15	&	\nodata	&	\nodata	&	\nodata	\\
	& $12\,\microns$ &	38.10 $\pm$ 0.09	&	\nodata	&	\nodata	&	\nodata	\\
	& $22\,\microns$ &	38.28 $\pm$ 0.12	&	\nodata	&	\nodata	&	\nodata	\\
	& $24\,\microns$ &	38.60 $\pm$ 0.09	&	\nodata	&	\nodata	&	\nodata	\\
	& $70\,\microns$ &	39.02 $\pm$ 0.07	&	\nodata	&	\nodata	&	\nodata	\\
	& $160\,\microns$ &	38.90 $\pm$ 0.11	&	\nodata	&	\nodata	&	\nodata	\\
	& $20\,\cm_{\textrm{{\tiny M}}}$ &	32.09 $\pm$ 0.13	&	48.76 $\pm$ 0.13	&	36.90 $\pm$ 0.13	&	O7V	\\
	& $20\,\cm_{\textrm{{\tiny M+V}}}$ &	32.18 $\pm$ 0.20	&	48.85 $\pm$ 0.20	&	36.99 $\pm$ 0.20	&	O6.5V	\\
	& $21\,\cm_{\textrm{{\tiny V}}}$ &	32.36 $\pm$ 0.13	&	49.03 $\pm$ 0.13	&	37.16 $\pm$ 0.13	&	O6V	\\\hline
$r \leq 2.4 \pc$	& $8\,\microns$ &	37.73 $\pm$ 0.11	&	\nodata	&	\nodata	&	\nodata	\\
	& $12\,\microns$ &	37.59 $\pm$ 0.13	&	\nodata	&	\nodata	&	\nodata	\\
	& $22\,\microns$ &	37.96 $\pm$ 0.14	&	\nodata	&	\nodata	&	\nodata	\\
	& $24\,\microns$ &	38.09 $\pm$ 0.10	&	\nodata	&	\nodata	&	\nodata	\\
	& $70\,\microns$ &	38.71 $\pm$ 0.10	&	\nodata	&	\nodata	&	\nodata	\\
	& $160\,\microns$ &	38.63 $\pm$ 0.07	&	\nodata	&	\nodata	&	\nodata	\\
	& $20\,\cm_{\textrm{{\tiny M}}}$ &	31.67 $\pm$ 0.07	&	48.34 $\pm$ 0.07	&	36.47 $\pm$ 0.07	&	O8V	\\
	& $20\,\cm_{\textrm{{\tiny M+V}}}$ &	31.56 $\pm$ 0.10	&	48.22 $\pm$ 0.10	&	36.36 $\pm$ 0.10	&	O8.5V	\\
	& $21\,\cm_{\textrm{{\tiny V}}}$ &	31.79 $\pm$ 0.11	&	48.46 $\pm$ 0.11	&	36.60 $\pm$ 0.11	&	O7.5V	\\\hline
$r > 2.4 \pc$	& $8\,\microns$ &	38.83 $\pm$ 0.13	&	\nodata	&	\nodata	&	\nodata	\\
	& $12\,\microns$ &	38.46 $\pm$ 0.29	&	\nodata	&	\nodata	&	\nodata	\\
	& $22\,\microns$ &	38.72 $\pm$ 0.16	&	\nodata	&	\nodata	&	\nodata	\\
	& $24\,\microns$ &	38.82 $\pm$ 0.07	&	\nodata	&	\nodata	&	\nodata	\\
	& $70\,\microns$ &	39.18 $\pm$ 0.12	&	\nodata	&	\nodata	&	\nodata	\\
	& $160\,\microns$ &	38.88 $\pm$ 0.14	&	\nodata	&	\nodata	&	\nodata	\\
	& $20\,\cm_{\textrm{{\tiny M}}}$ &	32.37 $\pm$ 0.04	&	49.04 $\pm$ 0.04	&	37.18 $\pm$ 0.04	&	O6V	\\
	& $20\,\cm_{\textrm{{\tiny M+V}}}$ &	32.42 $\pm$ 0.08	&	49.09 $\pm$ 0.08	&	37.23 $\pm$ 0.08	&	O6V	\\
	& $21\,\cm_{\textrm{{\tiny V}}}$ &	32.39 $\pm$ 0.08	&	49.06 $\pm$ 0.08	&	37.19 $\pm$ 0.08	&	O6V	\\\hline
Arm	& $8\,\microns$ &	38.80 $\pm$ 0.10	&	\nodata	&	\nodata	&	\nodata	\\
	& $12\,\microns$ &	38.54 $\pm$ 0.11	&	\nodata	&	\nodata	&	\nodata	\\
	& $22\,\microns$ &	38.82 $\pm$ 0.06	&	\nodata	&	\nodata	&	\nodata	\\
	& $24\,\microns$ &	38.79 $\pm$ 0.07	&	\nodata	&	\nodata	&	\nodata	\\
	& $70\,\microns$ &	39.11 $\pm$ 0.12	&	\nodata	&	\nodata	&	\nodata	\\
	& $160\,\microns$ &	38.93 $\pm$ 0.15	&	\nodata	&	\nodata	&	\nodata	\\
	& $20\,\cm_{\textrm{{\tiny M}}}$ &	32.25 $\pm$ 0.08	&	48.92 $\pm$ 0.08	&	37.06 $\pm$ 0.08	&	O6.5V	\\
	& $20\,\cm_{\textrm{{\tiny M+V}}}$ &	32.17 $\pm$ 0.09	&	48.84 $\pm$ 0.09	&	36.97 $\pm$ 0.09	&	O6.5V	\\
	& $21\,\cm_{\textrm{{\tiny V}}}$ &	32.32 $\pm$ 0.08	&	48.98 $\pm$ 0.08	&	37.12 $\pm$ 0.08	&	O6V	\\\hline
Interarm	& $8\,\microns$ &	38.95 $\pm$ 0.10	&	\nodata	&	\nodata	&	\nodata	\\
	& $12\,\microns$ &	39.02 $\pm$ 0.02	&	\nodata	&	\nodata	&	\nodata	\\
	& $22\,\microns$ &	38.81 $\pm$ 0.27	&	\nodata	&	\nodata	&	\nodata	\\
	& $24\,\microns$ &	38.81 $\pm$ 0.10	&	\nodata	&	\nodata	&	\nodata	\\
	& $70\,\microns$ &	38.73 $\pm$ 0.10	&	\nodata	&	\nodata	&	\nodata	\\
	& $160\,\microns$ &	38.54 $\pm$ 0.11	&	\nodata	&	\nodata	&	\nodata	\\
	& $20\,\cm_{\textrm{{\tiny M}}}$ &	32.22 $\pm$ 0.12	&	48.89 $\pm$ 0.12	&	37.03 $\pm$ 0.12	&	O6.5V	\\
	& $20\,\cm_{\textrm{{\tiny M+V}}}$ &	32.12 $\pm$ 0.27	&	48.78 $\pm$ 0.27	&	36.92 $\pm$ 0.27	&	O7V	\\
	& $21\,\cm_{\textrm{{\tiny V}}}$ &	32.23 $\pm$ 0.16	&	48.89 $\pm$ 0.16	&	37.03 $\pm$ 0.16	&	O6.5V	\\
\enddata
\end{deluxetable*}

\twocolumngrid
\clearpage

\subsection{Heliocentric Distance \label{subsec:heliocomp}}

Slight variations in sample completeness as a function of heliocentric distance will likely appear in the LF. As a test, we divide the heliocentric distance subset into two groups: sources nearer than $7.75~\kpc$ and sources further than $7.75~\kpc$. We choose this distance in order to obtain groups containing roughly equal numbers of regions; the same reasoning underlies the choice of dividing value for the other subsets with the exception of the arm/interarm subset. We find that the mean power law indices of both the single and double power law fits are lower for the near subset than the far subset.

Since the completeness limits of the far subset are slightly higher than those of the near subset, the far subset contains proportionally fewer low-luminosity \hii regions than the near subset. Power law fits to the far subset are therefore more strongly influenced by the relatively few high-luminosity regions in comparison to the near subset and would have a steeper LF. The mean $\langle\alpha_{2}\rangle$ values of the near and far subsets are within $1\sigma$ of each other while the near and far subset $\langle\alpha_{1}\rangle$ values are substantially more distinct. This is expected under our hypothesis as both subsets are equally complete at the high-luminosity end represented by $\langle\alpha_{2}\rangle$. 

If the far subset is comprised of proportionally more high-luminosity regions than the near subset, its knee luminosity should be higher than the knee luminosity of the near subset as well. Indeed, we find exactly this relation.


\subsection{Galactocentric Radius}
\label{subsec:rgal}

It is possible that the Galactic \hii region population is not uniform across the disk. Variation of the LF as a function of Galactocentric radius may reflect such a population inhomogeneity. We therefore divide the \rgal subset at $5~\kpc$ and find that the \hii region LF in the inner Galaxy is flatter than in the outer Galaxy by a more than $3\sigma$ difference in both the single and double power law models. As Galactocentric radius increases, the number of \hii regions decreases. If the number of high-luminosity regions decreases at a faster rate than the number of low-luminosity regions, then the \hii region LF at small Galactocentric radii will be flatter than the LF at large Galactocentric radii, thus explaining our results.

\subsection{Physical Size}
\label{subsec:angsize}

We divide the \hii region population by physical size and find that the LF of the subset containing regions with physical sizes smaller than $2.4 \pc$ has steeper $\langle\alpha\rangle$ and $\langle\alpha_2\rangle$ values than that of the subset with regions larger than $2.4 \pc$.

Most notably, $\langle\alpha_2\rangle=-3.06\,\pm\,0.10$ of the small subset is especially steep in comparison to that of the other subsets. This is perhaps not unexpected since more luminous regions expand more rapidly and are less likely to be physically small \citep{kru09}. As a consequence, we expect few high-luminosity sources in the small physical size subset, and the high-luminosity end of such a distribution is likely to have a sharp decrease in number of regions (see Figure \ref{fig:magpis_vgps_small}).

The subset of larger regions also has higher knee luminosities than every subset other than the subset of regions at large heliocentric distances, which is unsurprising. Just as the subset of regions with smaller physical sizes will preferentially contain low-luminosity sources, the large subset is expected to consist of proportionally more high-luminosity \hii regions and so be more likely to have a knee at a relatively high luminosity.

\subsection{Arm/Interarm Membership}
\label{subsec:armcomp}



Since \hii regions trace high-mass star formation and are consequently found primarily in spiral arms, it is possible that arm and interarm \hii region populations have distinct formation histories and characteristics and, consequently, dissimilar LFs. \citet{kre16} find that the \hii populations in arm and interarm regions in NGC 628 have very similar properties, however, which would imply an LF independent of arm/interarm membership. Moreover, \citet{azi11} find that the LFs of the arm and interarm \hii region populations in M31 have identical power law indices even though the LF of the former population peaks at a higher luminosity than the LF of the latter. Some studies have found discrepancies between the \hii region arm and interarm LFs in other galaxies \citep[e.g.,][]{thi00,sco01} though a large number have not \citep[e.g.,][]{kna98, azi11, kre16}.

The spiral structure map of the Milky Way has yet to be definitively established so here we adopt one model parameterization out of many as a fiducial pattern to use to establish an \hii region's arm or interarm location. For the arm/interarm subset, we assign each region to a spiral arm or interarm region based on the recommended best-fit parameters from \citet{hh14} for a four-arm logarithmic spiral arm model. This model is of the form 
\begin{equation}\label{eq:arm_model}
    r_i=R_ie^{(\theta-\theta_i)\tan\Psi_i}
\end{equation}
where $r_i$ is the Galactocentric radius of the \textit{i}th spiral arm at an azimuth $\theta$ around the Galactic Center defined such that $\theta=0\degree$ lies on the positive x-axis and increases counterclockwise as viewed from the Galactic north pole\footnote{Note that \citet{hh14} define Galactic azimuth differently from the standard definition used in the \textit{WISE} Catalog where $\theta=0\degree$ lies on the Galactic Center-Sun line and increases clockwise as viewed from the Galactic north pole. Therefore, the azimuths in the \textit{WISE} Catalog, $\theta_{\rm WISE}$, can be transformed to the azimuths in Equation~\ref{eq:arm_model} using the relation ${\theta=(90\degree-\theta_{\rm WISE})\bmod{360\degree}}$.}. The pitch angle, initial Galactocentric radius, and initial azimuth of arm $i$ are $\Psi_i$, $R_i$, and $\theta_i$, respectively. Though there is evidence that spiral arm widths are not constant with increasing Galactic radii \citep{reid14}, we set arm half-widths to a constant value of $0.5~\kpc$ for all Galactic longitudes since the variation in arm width is small relative to the uncertainties in the distances to most \hii regions. Since the location of spiral arms is still not fully certain, our analysis seeks only to broadly examine if there is a difference in the \hii region LF within and between spiral arms. Consequently, we do not further refine the location or width of the arms. 

We find that there is some evidence that the Galactic \hii region LF has disparate forms for the arm and interarm populations defined using the \citet{hh14} model; the interarm subset has a slightly steeper $\langle\alpha_{1}\rangle$, but a shallower $\langle\alpha_{2}\rangle$. While these results may have an explanation similar to that of the variation of the LF with Galactocentric radius in that they possibly reflect distinct populations, the location and width of spiral arms is highly model-dependent.  As a preliminary test of potential model dependence, we change the pitch angles to those in \citet{reid14} and, separately, the arm widths to $0.4~\kpc$ and perform the same analysis. We find no variations in the results when adjusting the pitch angle or arm width, but because of the inherent uncertainty in the location of the arms we cannot make significant conclusions about the arm and interarm LFs. As such, we encourage further study of this topic.

\subsection{Effects of Spatial Resolution and Blending}
\label{subsec:blending}


Many studies of \hii region LFs, whether within the Galaxy or in other galaxies, are potentially hampered by low spatial resolution. While previous work by \citet{mur10} has shown the effects of observational blending to be small, our high-spatial resolution data set provides us with the ability to better examine the potential effect of spatial resolution on such studies. 

\begin{figure}[ht]
   \centering
   \includegraphics[width=0.95\columnwidth,trim={3cm 7.75cm 3cm 8cm}, clip]{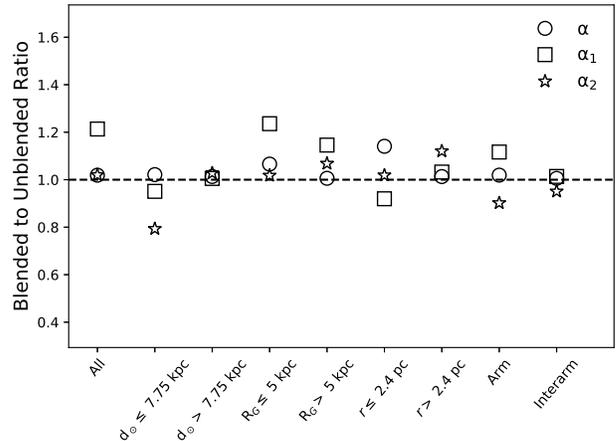}
   \caption{Ratio of blended fit $\alpha$ values divided by unblended fit $\alpha$ values for each $12\,\microns$ \textit{WISE} subset. The complete range of wavelengths is shown in Figure \ref{fig:alphacomp_blend2}.}
   \label{fig:wise3_alpha_blend}
\end{figure}

 The majority of our LF model parameters for the blended data --- including all single power law indices --- are in good agreement with the parameters of the fits to the unblended Monte Carlo-generated LFs as in the case of the $12\,\microns$ \textit{WISE} data presented in Figure \ref{fig:wise3_alpha_blend}. Since only a small minority of power law indices have notably different unblended and blended values, we conclude that potential source blending has a small-to-insignificant effect on the shape of the LF.

As with the power law indices, the knee values are largely consistent between the blended and unblended data. Similarly, we find that the completeness limits are also consistent in all subsets at all wavelengths with no systematic trends in how they differ between the blended and unblended data. We include plots of all power law index and knee comparisons in Section \ref{appsubsec:blend} of the Appendix. 

\subsection{Comparison of Single and Double Power Law Fits}
\label{subsec:statcomp}

Within each subset, the BIC-favored model is generally consistent across all wavelengths, as shown in Table~\ref{tab:stats}. We call the model favored by the majority of the wavelengths of a data subset the ``majority model'' of that subset. Similarly, we call the model favored by a minority of wavelengths the ``minority model'' of the subset. The power law indices are largely similar across all wavelengths with only small differences in a small number of data sets, as shown in Figure~\ref{fig:alphacomp2}. This is true even across wavelengths that are best-fit by different models.

In eight of the nine subsets, the majority model is a single power law, and in three of these it is favored at every wavelength. The one exception is the best-fit model to the LF of the undivided data, which is a double power law. In this case, a double power law is favored by five of the nine wavelengths. In conjunction with the consistency of the power law indices across the studied wavelengths, this suggests that the LF of the undivided data may not be best described solely by a single or double power law.

We therefore conclude that, although a single power law is the best-fit model for the LF of a majority of the sub-populations in the Milky Way, it is not the favored model for the first Galactic quadrant \hii region LF in all sub-samples at all wavelengths.

\section{Discussion and Conclusions}
\label{sec:conc}

Using a sample of first Galactic quadrant \hii regions from the \textit{WISE} Catalog of Galactic \hii Regions V2.2, we investigate the form of the Galactic \hii region luminosity function (LF) at multiple infrared and radio wavelengths. In light of previous work that suggests differences in the LFs of \hii region sub-populations, we examine the effects of separating our data into subsets by the regions' heliocentric distance, Galactocentric radius, physical size, and region location relative to the spiral arms.

Averaged across all wavelengths, the mean and standard deviation of the indices from the best-fit double power law function to the full data set are $\langle\alpha_1\rangle=-1.40\,\pm\,0.03$ and $\langle\alpha_2\rangle=-2.33\,\pm\,0.04$ with a mean knee H$\alpha$ luminosity of $10^{36.92\,\pm\,0.09}$~erg~s$^{-1}$. This is a lower knee luminosity than the Str{\"o}mgren luminosity of $L_{\rm{H} \alpha} \approx 10^{39}$~erg~s$^{-1}$ found in most galaxies with double power law LFs. The corresponding mean single power law index is $\langle\alpha\rangle=-1.75\,\pm\,0.01$, which is less steep than the value of $\approx-2$ found by the few previous studies of the Galactic LF. These studies found that while the average LF index of other galaxies is also $\approx-2$, the index of an individual galaxy's LF can be significantly higher or lower \citep[e.g.,][]{you99,liu13}. Our result is consistent with this established broad variation in the \hii region LF power law index. The best-fit power law indices of both models are nearly independent of wavelength. This suggests that future observations can be made at radio and infrared wavelengths that are less affected by absorption without the uncertainty of the viability of inter-wavelength comparisons.

The knee luminosities do not vary across the infrared wavelengths or, separately, the radio data sets. The completeness limit luminosities, which are nearly identical between the two models, are similarly consistent across both wavelength categories and are generally lower than the knee luminosities by approximately an order of magnitude. 
 

We simulate the effects of spatial resolution and find that our results are not affected by such potential confusion.  The LF model parameters from fits to blended data are largely consistent with those of the unblended data. This consistency suggests that studies of extragalactic LFs, which often have lower angular resolutions than this work, can provide results equivalent to those with higher angular resolution.


The form of the Galactic \hii region LF varies when we divide the data into subsets by heliocentric distance, Galactocentric radius, physical size, and location relative to the spiral arms. We think that the variation in the LF with heliocentric distance is a product of observational bias and does not reflect an underlying physical difference. Spiral arms are often, but not universally, observed to have shallower LFs than interarm regions \citep{ken80,ran92,ban93,thi00,kna93,roz96,kna98}. \citet{oey98} suggest that this result stems from two distinct \hii region populations: spiral arm nebulae are on average younger whereas interarm \hii regions are generally older. In light of this, we suggest that a difference in population characteristics may explain the variation of the LF with Galactocentric radius and region location relative to the spiral arms, though both cases are in need of further study since the literature is mixed. We propose that the variation found between physical size subsets is a reflection of the faster expansion rates of high-luminosity \hii regions. Simulations of the Galactic \hii region population would likely play a major role in clarifying these results.

This analysis places the Milky Way in the broader context of extragalactic \hii region LF research at an unprecedented level of detail. Previous research has found that some galaxies have LFs best modeled by single power laws while others are best described by double power laws. We conclude that neither a single nor a double power law is strongly favored over the other as a best-fit model for the full first Galactic quadrant \hii region LF, and that most sub-population LFs are better modeled with a single power law. 




\section{Acknowledgements}

JLM would like to thank Nate Garver-Daniels for his vital assistance with unexpected computer problems and Evan Kelner-Levine for his support in the final stages of preparing this manuscript. The authors would also like to thank the referee for their valuable input and comments. This work is supported by NSF grant AST-1516021 to LDA. TMB was supported by NSF grant AST-1714688. The Green Bank Observatory and National Radio Astronomy Observatory are facilities of the National Science Foundation operated under cooperative agreement by Associated Universities, Inc.

\newpage
\bibliographystyle{aasjournal}
\bibliography{lfpaper}

\onecolumngrid
\newpage
\appendix
\section{Fitting Power Law Distributions and Power Law Model Selection}

\subsection{Single and Double Power Law Models Incorporating a Free Completeness Limit}
\label{appsubsec:limit}

Consider the probability density function of a single power law model:
\begin{equation}
  p(x;\alpha, x_m) = \begin{cases}
    0 & x < x_s \\
    A & x_s \leq x < x_m \\
    A\left(\frac{x}{x_m}\right)^{-\alpha} & x \geq x_m \,,
  \end{cases}
\end{equation}
where the completeness limit \(x_m > 0\), the power law index \(\alpha > 1\), and \(A\) is a normalizing constant.
This distribution follows a single power law above \(x_m\), but is
constant between \(x_s = \text{min}(\vect{X})\) and \(x_m\). 

The likelihood that data \(\vect{X} = (x_0, x_1, ..., x_N)\) are drawn
from a model is given by the likelihood function
\begin{equation}
  L(\vect{\theta} | \vect{X}) = \prod_i^N p(x_i ; \vect{\theta})\,,
\end{equation}
where \(\vect{\theta}\) are the model parameters (i.e.,
\(\vect{\theta} = (\alpha, x_m)\) for a single power law distribution, and
\(\vect{\theta} = (\alpha_1, \alpha_2, x_m, x_{\rm knee})\) for a double power law
distribution).
We identify the parameters that best represent the underlying data
distribution by maximizing the likelihood function, or, equivalently,
the logarithm of the likelihood function,
\begin{equation}
  \log L(\vect{\theta} | \vect{X}) = \sum_i^N \log p(x_i ; \vect{\theta})\,.
\end{equation}

The logarithm of the likelihood function for a single power law is therefore
\begin{equation}
  \log L(\alpha, x_m|\vect{X}) = N\log A - \alpha \sum_{i; x_i \geq x_m}^N \left(\log x_i - \log x_m\right)\,,
\end{equation}
where 
\begin{equation}
  \log A = \log(\alpha - 1) - \log[\alpha(x_m - x_s) + x_s]\,.
\end{equation}

This function is maximized for \(\vect{\theta} = \hat{\vect{\theta}} = (\hat{\alpha}, \hat{x}_m)\) such that
\begin{align}
  0 & = \frac{d\log L(\hat{\vect{\theta}} | \vect{X})}{d\hat{\vect{\theta}}} \nonumber\,.
\end{align}  
We maximize the likelihood function numerically.

  


Similarly, the probability density function of a double power law is given by
\begin{equation}
  p(x;\alpha_1, \alpha_2, x_m, x_{\rm knee}) = \begin{cases}
    0 & x < x_s \\
    A\left(\frac{x_m}{x_{\rm knee}}\right)^{-\alpha_1} & x_s \leq x < x_m \\
    A\left(\frac{x}{x_{\rm knee}}\right)^{-\alpha_1} & x_m \leq x < x_{\rm knee} \\
    A\left(\frac{x}{x_{\rm knee}}\right)^{-\alpha_2} & x \geq x_{\rm knee}\,,
  \end{cases}
\end{equation}
where the completeness limit \(x_m > 0\), the location of the power law discontinuity \(x_{\rm knee} > x_m\), the power law indices \(\alpha_1 > 1\) and \(\alpha_2 > 1\), and \(A\) is a normalizing constant.

The logarithm of the likelihood function for a double power law is
\begin{align}
  \log L(\alpha_1, \alpha_2, x_m, x_{\rm knee}|\vect{X}) & = N\log A - N_s \alpha_1 \left(\log x_m - \log x_{\rm knee}\right) \nonumber \\
  &  - \alpha_1 \sum_{i; x_m \leq x_i < x_{\rm knee}}^N \left(\log x_i - \log x_{\rm knee}\right) - \alpha_2 \sum_{i; x_i \geq x_{\rm knee}}^N \left(\log x_i - \log x_{\rm knee}\right),
\end{align}
where \(N_s\) is the number of data satisfying \(x_s \leq x_i < x_m\)
and
\begin{equation}
  \log A = -\log\left[\left(\frac{x_m}{x_{\rm knee}}\right)^{-\alpha_1}(x_m - x_s)  + \frac{x_{\rm knee}}{1 - \alpha_1}\left[1 - \left(\frac{x_m}{x_{\rm knee}}\right)^{1 - \alpha_1}\right] + \frac{x_{\rm knee}}{\alpha_2 - 1}\right].
\end{equation}

This function is maximized for \(\vect{\theta} = \hat{\vect{\theta}} = (\hat{\alpha_1}, \hat{\alpha_2}, \hat{x}_{\rm knee}, \hat{x}_m)\). Although there may be an analytic solution to maximizing the likelihood functions, we approach this problem numerically in the same manner as the single power law.




\subsection{Model Selection}
\label{appsubsec:bic}

We use the Bayesian Information Criterion (BIC) to select between
the single and double power law models. BIC is defined as
\begin{equation}
  \text{BIC} = k\log N - 2 \log \hat{L}
\end{equation}
where \(k\) is the number of free parameters in the model, \(N\) is
the number of data elements, and \(\hat{L}\) is the maximized likelihood. If the double power law model BIC is lower than the single power law model BIC by at least ten \citep{kas95}, then the double power law model is preferred.  Otherwise, the single power law model is preferred.

\section{Model Comparison Results}
\label{app:stat}

Here we show the results of the comparison of the LFs that are fit to the Monte Carlo-generated luminosity distributions. If \(\Delta{\rm BIC} = {\rm BIC_{Single}} - {\rm BIC_{Double}}\) is greater than ten, then the double power law model is preferred. Otherwise, the single power law model is preferred. We outline the analytical and comparison methods in Section~\ref{subsec:comp} and discuss the results in Section~\ref{subsec:statcomp}.

\startlongtable
\begin{deluxetable*}{cccR}
\tabletypesize{\small}
\renewcommand{\arraystretch}{1.1}
\tablecaption{Favored Model as Determined by the Bayesian Information Criterion (BIC)\label{tab:stats}}
\tablecolumns{4}
\tablehead{
\colhead{Subset} &
\colhead{Data} &
\colhead{Favored Model} &
\colhead{$\Delta${\rm BIC}}
}
\startdata
All	& $8\,\microns$ &	Single	&	0.46	\\
	& $12\,\microns$ &	Single	&	9.05	\\
	& $22\,\microns$ &	Double	&	14.35	\\
	& $24\,\microns$ &	Double	&	10.78	\\
	& $70\,\microns$ &	Single	&	9.97	\\
	& $160\,\microns$ &	Single	&	4.85	\\
	& $20\,\cm_{\textrm{{\tiny M}}}$ &	Double	&	15.68	\\
	& $20\,\cm_{\textrm{{\tiny M+V}}}$ &	Double	&	33.33	\\
	& $21\,\cm_{\textrm{{\tiny V}}}$ &	Double	&	10.84	\\\hline
$d_\sun \leq 7.75$ \kpc	& $8\,\microns$ &	Single	&	9.11	\\
	& $12\,\microns$ &	Double	&	13.17	\\
	& $22\,\microns$ &	Double	&	12.46	\\
	& $24\,\microns$ &	Double	&	10.08	\\
	& $70\,\microns$ &	Single	&	6.96	\\
	& $160\,\microns$ &	Double	&	13.34	\\
	& $20\,\cm_{\textrm{{\tiny M}}}$ &	Single	&	-7.56	\\
	& $20\,\cm_{\textrm{{\tiny M+V}}}$ &	Single	&	-1.29	\\
	& $21\,\cm_{\textrm{{\tiny V}}}$ &	Single	&	-0.98	\\\hline
$d_\sun > 7.75$ \kpc	& $8\,\microns$ &	Single	&	-5.61	\\
	& $12\,\microns$ &	Single	&	-6.56	\\
	& $22\,\microns$ &	Single	&	-2.05	\\
	& $24\,\microns$ &	Single	&	-2.32	\\
	& $70\,\microns$ &	Single	&	-2.73	\\
	& $160\,\microns$ &	Single	&	-55.25	\\
	& $20\,\cm_{\textrm{{\tiny M}}}$ &	Single	&	2.04	\\
	& $20\,\cm_{\textrm{{\tiny M+V}}}$ &	Single	&	0.07	\\
	& $21\,\cm_{\textrm{{\tiny V}}}$ &	Single	&	0.84	\\\hline
$\rgal \leq 5$ \kpc	& $8\,\microns$ &	Single	&	-3.23	\\
	& $12\,\microns$ &	Single	&	-2.42	\\
	& $22\,\microns$ &	Single	&	-1.36	\\
	& $24\,\microns$ &	Single	&	-3.01	\\
	& $70\,\microns$ &	Single	&	-4.49	\\
	& $160\,\microns$ &	Single	&	-4.81	\\
	& $20\,\cm_{\textrm{{\tiny M}}}$ &	Single	&	-3.46	\\
	& $20\,\cm_{\textrm{{\tiny M+V}}}$ &	Single	&	-0.29	\\
	& $21\,\cm_{\textrm{{\tiny V}}}$ &	Single	&	-0.68	\\\hline
$\rgal > 5$ \kpc	& $8\,\microns$ &	Single	&	1.55	\\
	& $12\,\microns$ &	Single	&	2.17	\\
	& $22\,\microns$ &	Single	&	9.45	\\
	& $24\,\microns$ &	Single	&	9.22	\\
	& $70\,\microns$ &	Single	&	6.55	\\
	& $160\,\microns$ &	Single	&	2.83	\\
	& $20\,\cm_{\textrm{{\tiny M}}}$ &	Single	&	9.7	\\
	& $20\,\cm_{\textrm{{\tiny M+V}}}$ &	Double	&	17.54	\\
	& $21\,\cm_{\textrm{{\tiny V}}}$ &	Single	&	2.31	\\\hline
$r \leq 2.4 \pc$	& $8\,\microns$ &	Single	&	9.42	\\
	& $12\,\microns$ &	Double	&	11.13	\\
	& $22\,\microns$ &	Double	&	18.05	\\
	& $24\,\microns$ &	Single	&	3.78	\\
	& $70\,\microns$ &	Single	&	2.24	\\
	& $160\,\microns$ &	Single	&	3.83	\\
	& $20\,\cm_{\textrm{{\tiny M}}}$ &	Double	&	23.62	\\
	& $20\,\cm_{\textrm{{\tiny M+V}}}$ &	Double	&	37.32	\\
	& $21\,\cm_{\textrm{{\tiny V}}}$ &	Single	&	4.14	\\\hline
$r > 2.4 \pc$	& $8\,\microns$ &	Single	&	-1.26	\\
	& $12\,\microns$ &	Single	&	0.13	\\
	& $22\,\microns$ &	Double	&	12.64	\\
	& $24\,\microns$ &	Double	&	12.21	\\
	& $70\,\microns$ &	Single	&	5.67	\\
	& $160\,\microns$ &	Single	&	0.86	\\
	& $20\,\cm_{\textrm{{\tiny M}}}$ &	Single	&	-0.65	\\
	& $20\,\cm_{\textrm{{\tiny M+V}}}$ &	Single	&	2.86	\\
	& $21\,\cm_{\textrm{{\tiny V}}}$ &	Single	&	6.69	\\\hline
Arm	& $8\,\microns$ &	Single	&	3.09	\\
	& $12\,\microns$ &	Single	&	7.54	\\
	& $22\,\microns$ &	Single	&	9.58	\\
	& $24\,\microns$ &	Single	&	5.13	\\
	& $70\,\microns$ &	Single	&	2.8	\\
	& $160\,\microns$ &	Single	&	-0.18	\\
	& $20\,\cm_{\textrm{{\tiny M}}}$ &	Double	&	10.36	\\
	& $20\,\cm_{\textrm{{\tiny M+V}}}$ &	Double	&	16.7	\\
	& $21\,\cm_{\textrm{{\tiny V}}}$ &	Single	&	3.84	\\\hline
Interarm	& $8\,\microns$ &	Single	&	-8.57	\\
	& $12\,\microns$ &	Single	&	-9.64	\\
	& $22\,\microns$ &	Single	&	-4.22	\\
	& $24\,\microns$ &	Single	&	-5.02	\\
	& $70\,\microns$ &	Single	&	-2.04	\\
	& $160\,\microns$ &	Single	&	-3.69	\\
	& $20\,\cm_{\textrm{{\tiny M}}}$ &	Single	&	-5.31	\\
	& $20\,\cm_{\textrm{{\tiny M+V}}}$ &	Single	&	-2.21	\\
	& $21\,\cm_{\textrm{{\tiny V}}}$ &	Single	&	-4.09	\\
\enddata
\end{deluxetable*}

\onecolumngrid
\newpage
\section{Graphical Luminosity Functions, Power Law Indices, and Blending Analyses}
\label{app:graphs}
\subsection{Example Luminosity Functions}
\label{appsubsec:lf}

Here we show the \hii region luminosity distributions and example power law fits. Each distribution is fit with a single (dashed green line) and double (dashed orange line) power law over the complete section of the distribution (completeness limits are denoted by dotted vertical lines in the same colors as the fits). The location of the knee luminosity is denoted by a solid vertical orange line.

\begin{figure*}[h]
\centering
  \subfloat{\label{fig:glimpse_neardist}%
    \includegraphics[scale=0.5,trim={3.75cm 8.5cm 3.5cm 8.5cm}, clip]{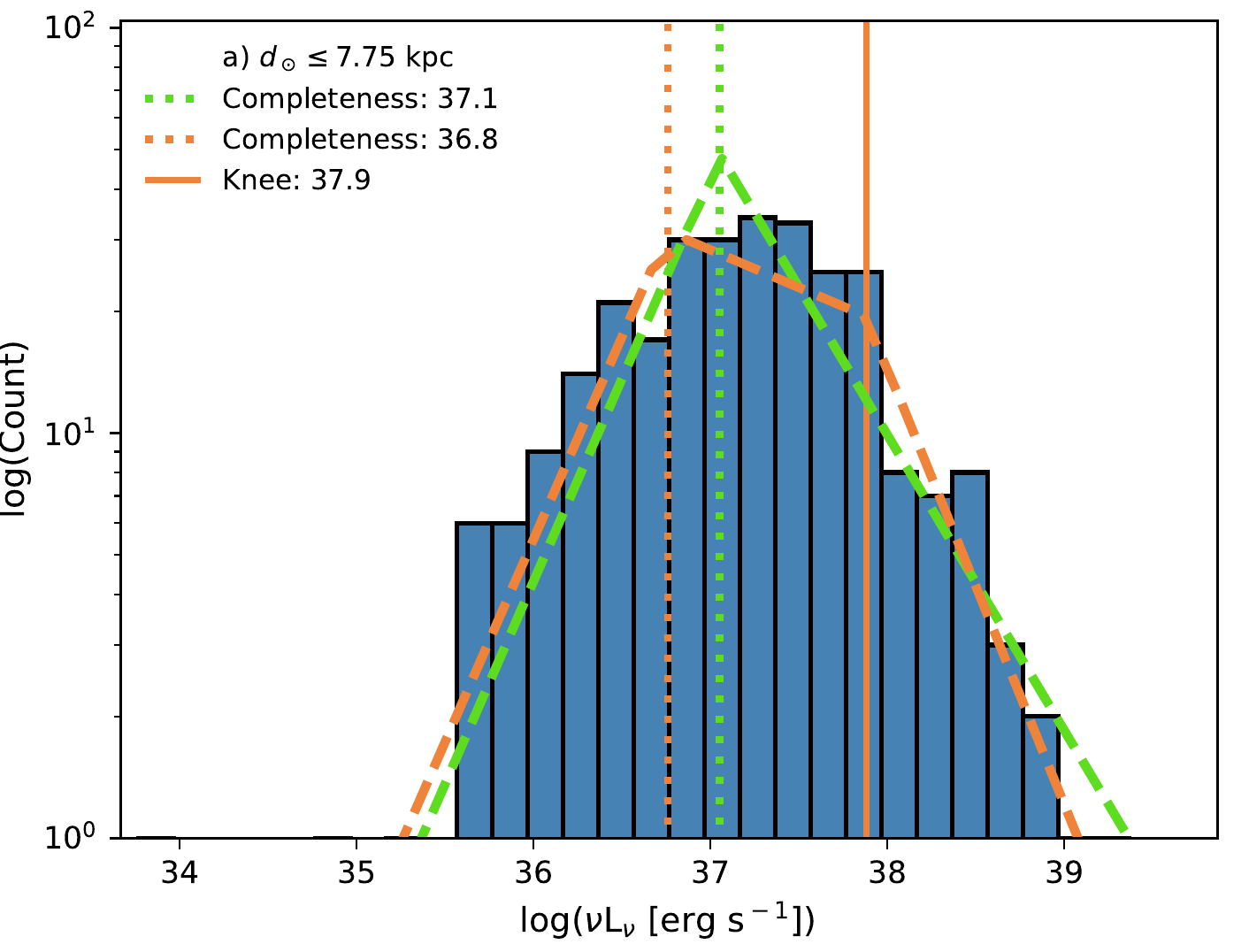}}\qquad
  \subfloat{\label{fig:glimpse_fardist}%
    \includegraphics[scale=0.5,trim={3.75cm 8.5cm 3.5cm 8.5cm}, clip]{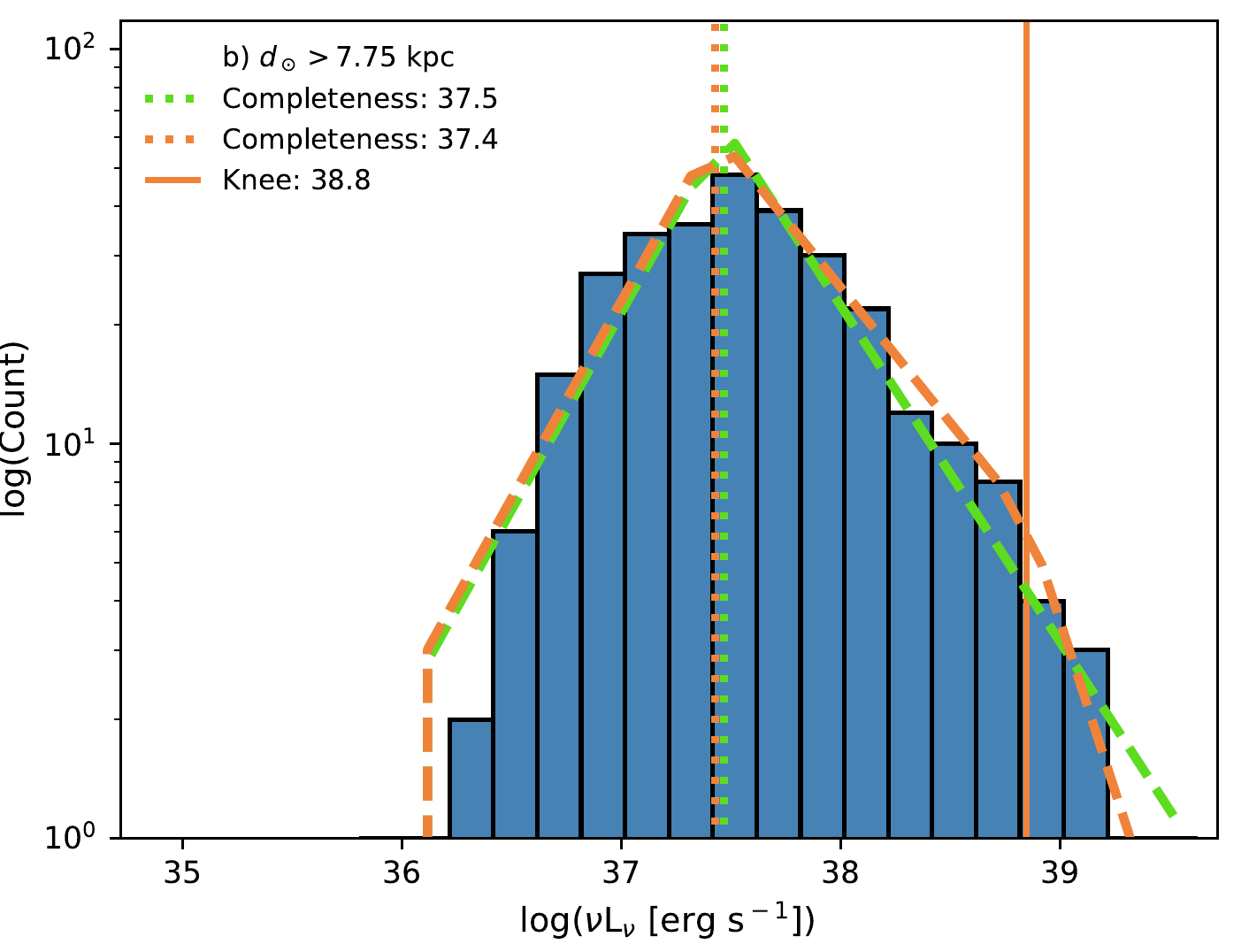}}\\
  \subfloat{\label{fig:glimpse_nearrgal}%
    \includegraphics[scale=0.5,trim={3.75cm 8.5cm 3.5cm 8.5cm}, clip]{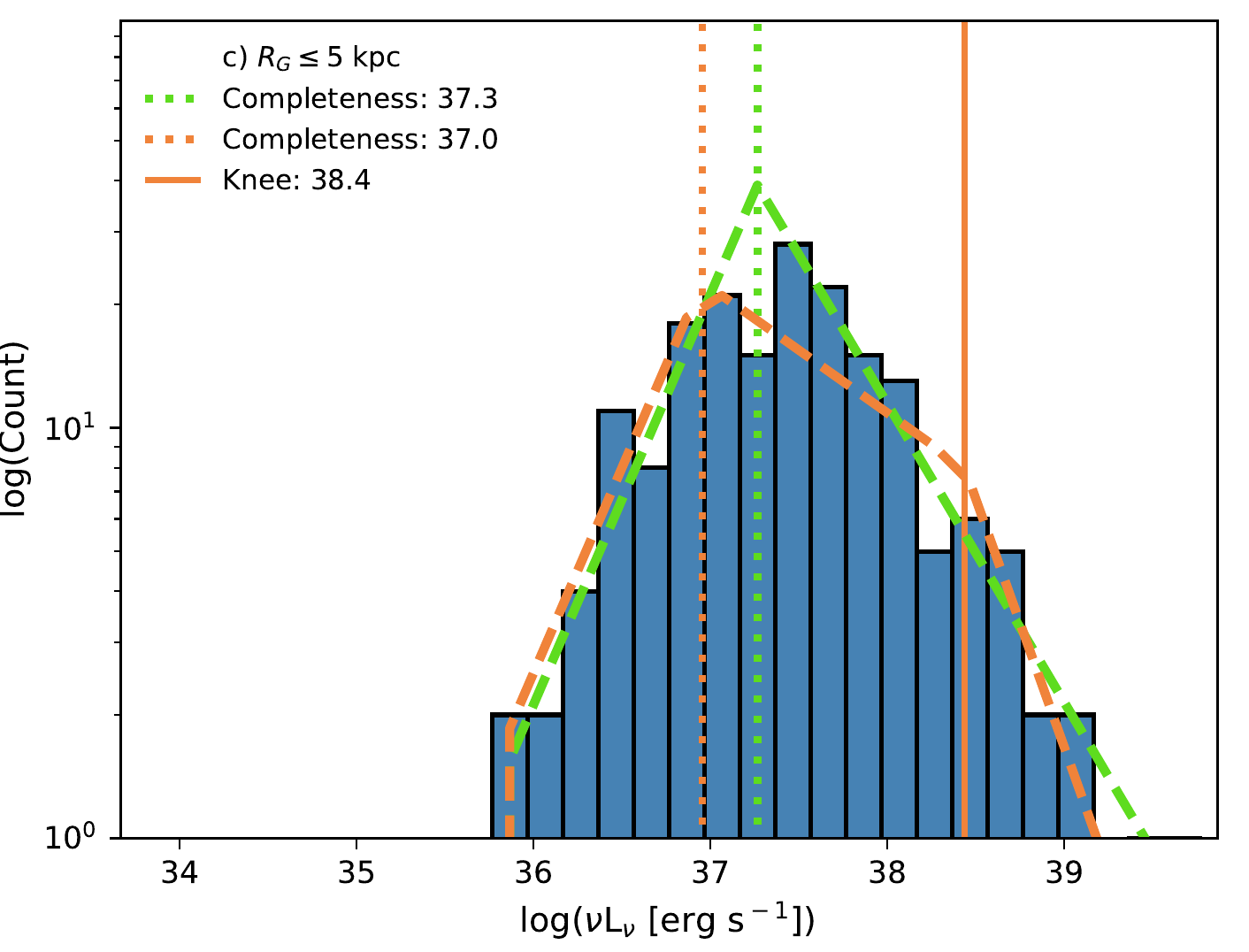}}\qquad
  \subfloat{\label{fig:glimpse_farrgal}%
    \includegraphics[scale=0.5,trim={3.75cm 8.5cm 3.5cm 8.5cm}, clip]{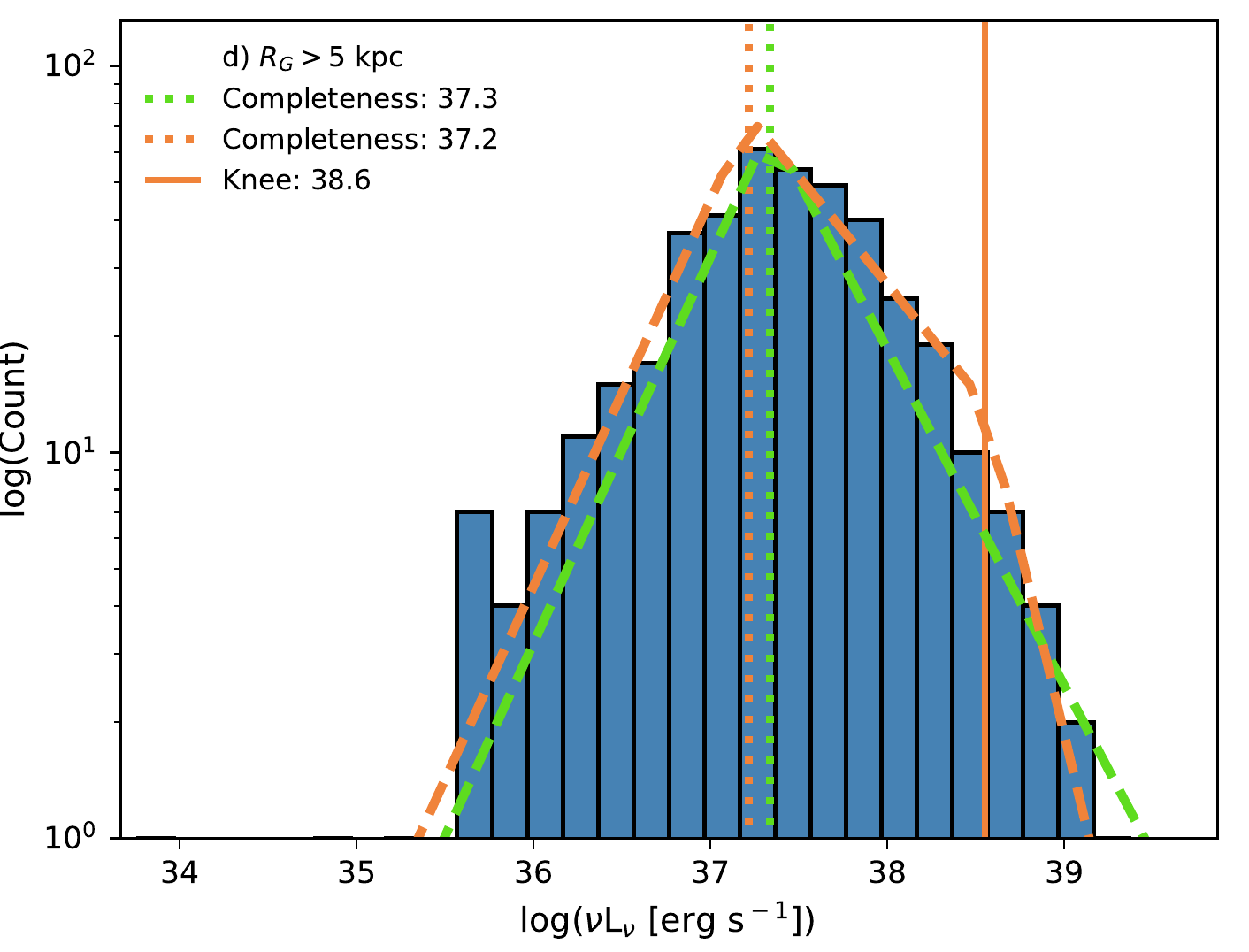}}\\
  \subfloat{\label{fig:glimpse_small}%
    \includegraphics[scale=0.5,trim={3.75cm 8.5cm 3.5cm 8.5cm}, clip]{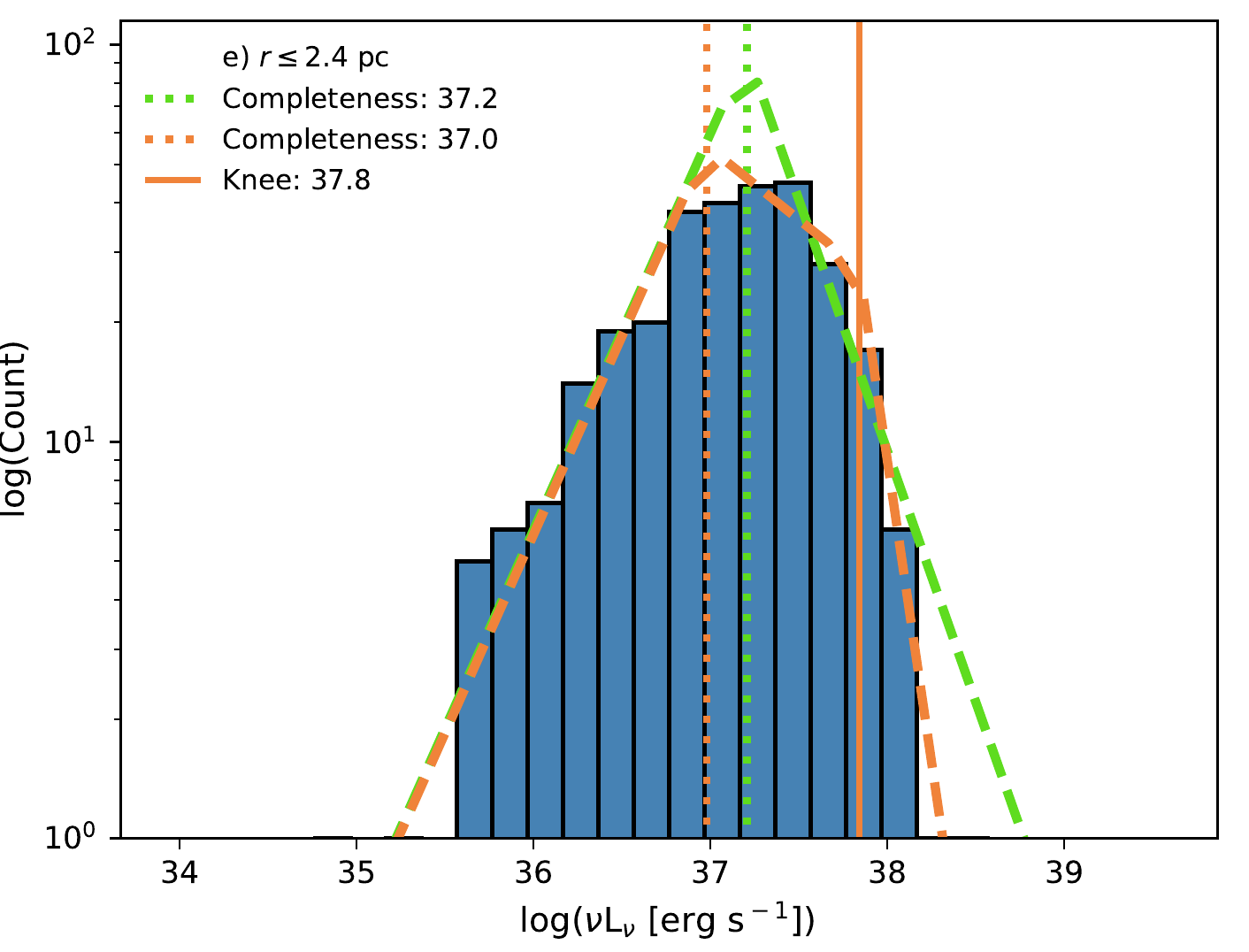}}\qquad
  \subfloat{\label{fig:glimpse_large}%
    \includegraphics[scale=0.5,trim={3.75cm 8.5cm 3.5cm 8.5cm}, clip]{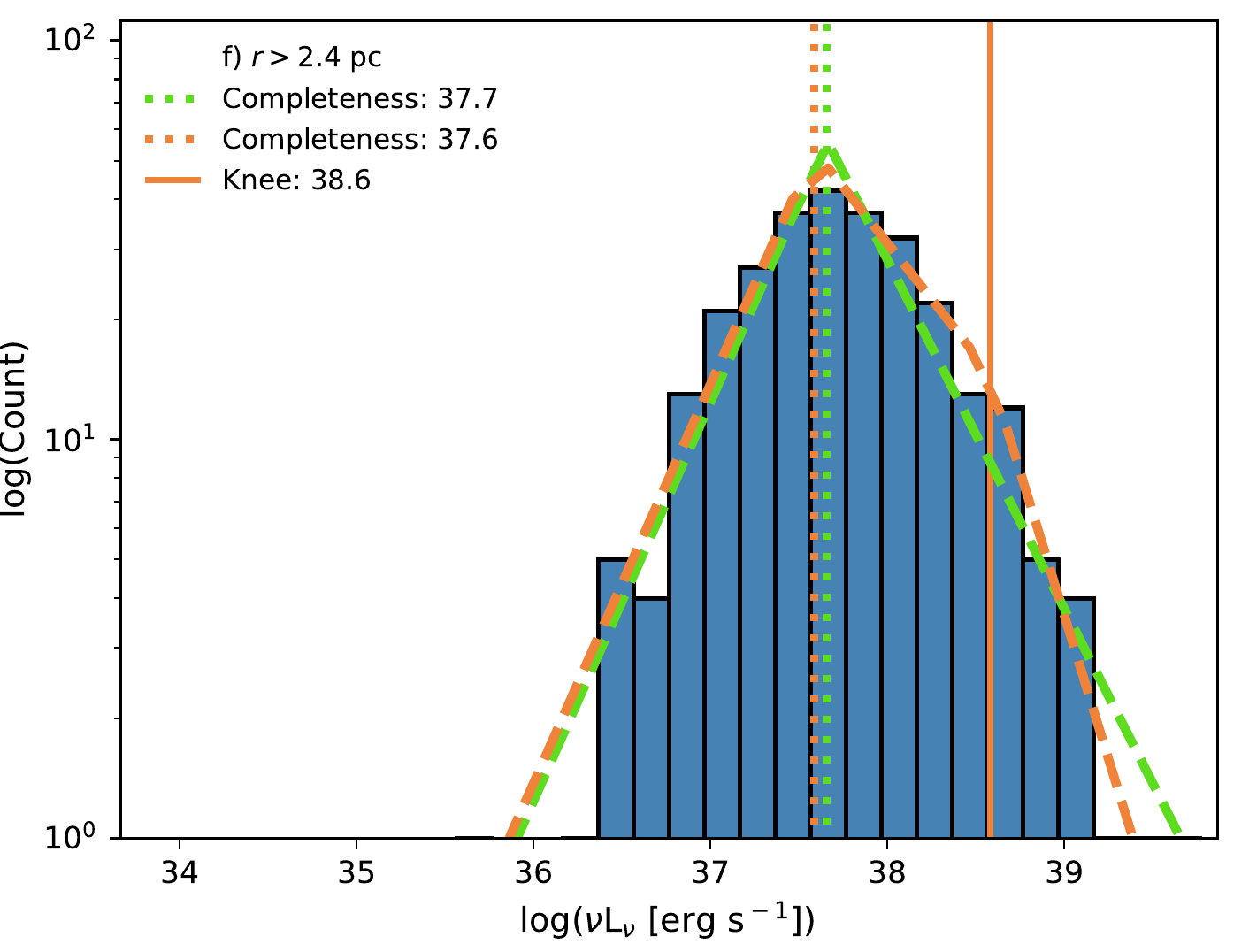}}\\
  \subfloat{\label{fig:glimpse_arm}%
    \includegraphics[scale=0.5,trim={3.75cm 8.5cm 3.5cm 8.5cm}, clip]{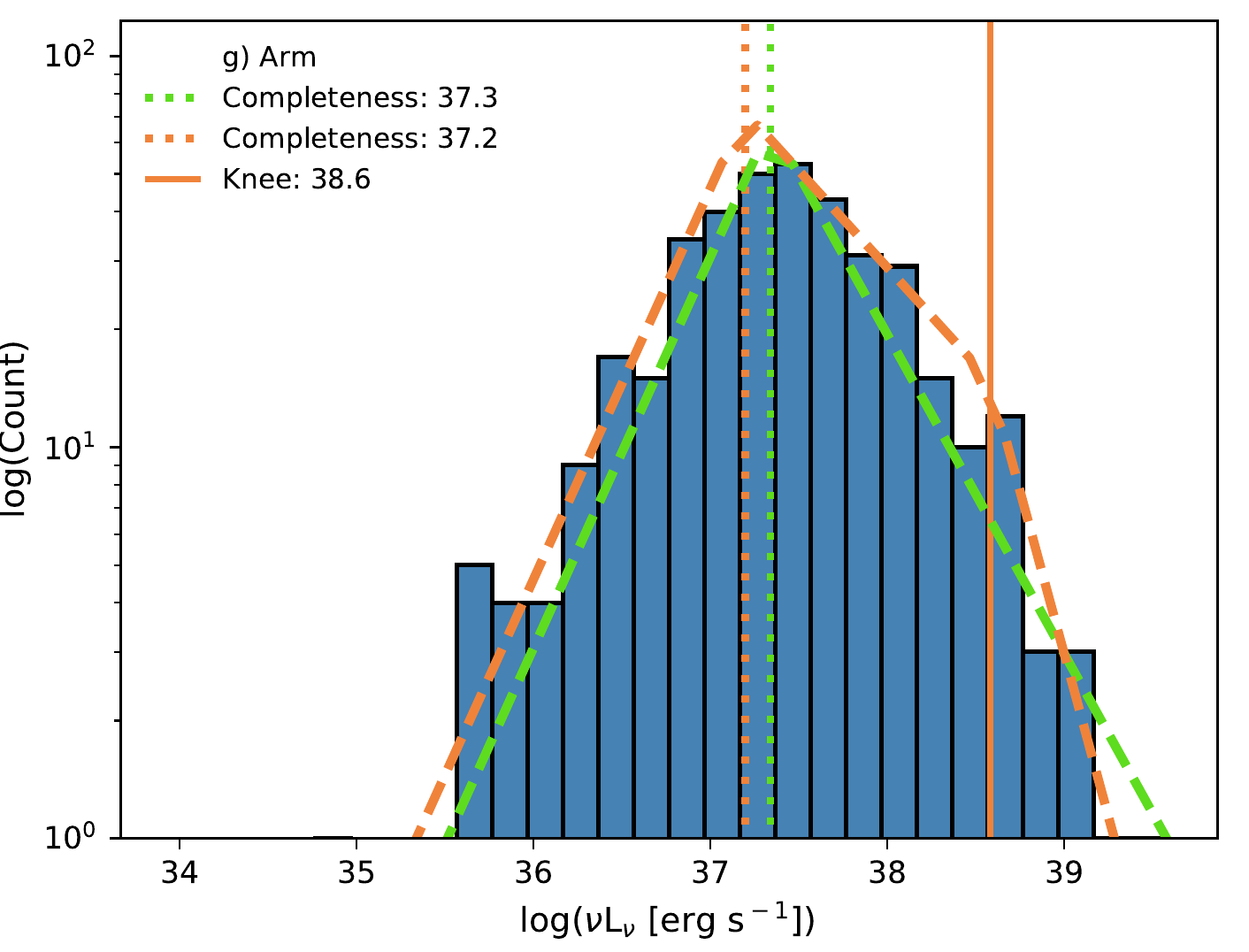}}\qquad
  \subfloat{\label{fig:glimpse_interarm}%
    \includegraphics[scale=0.5,trim={3.75cm 8.5cm 3.5cm 8.5cm}, clip]{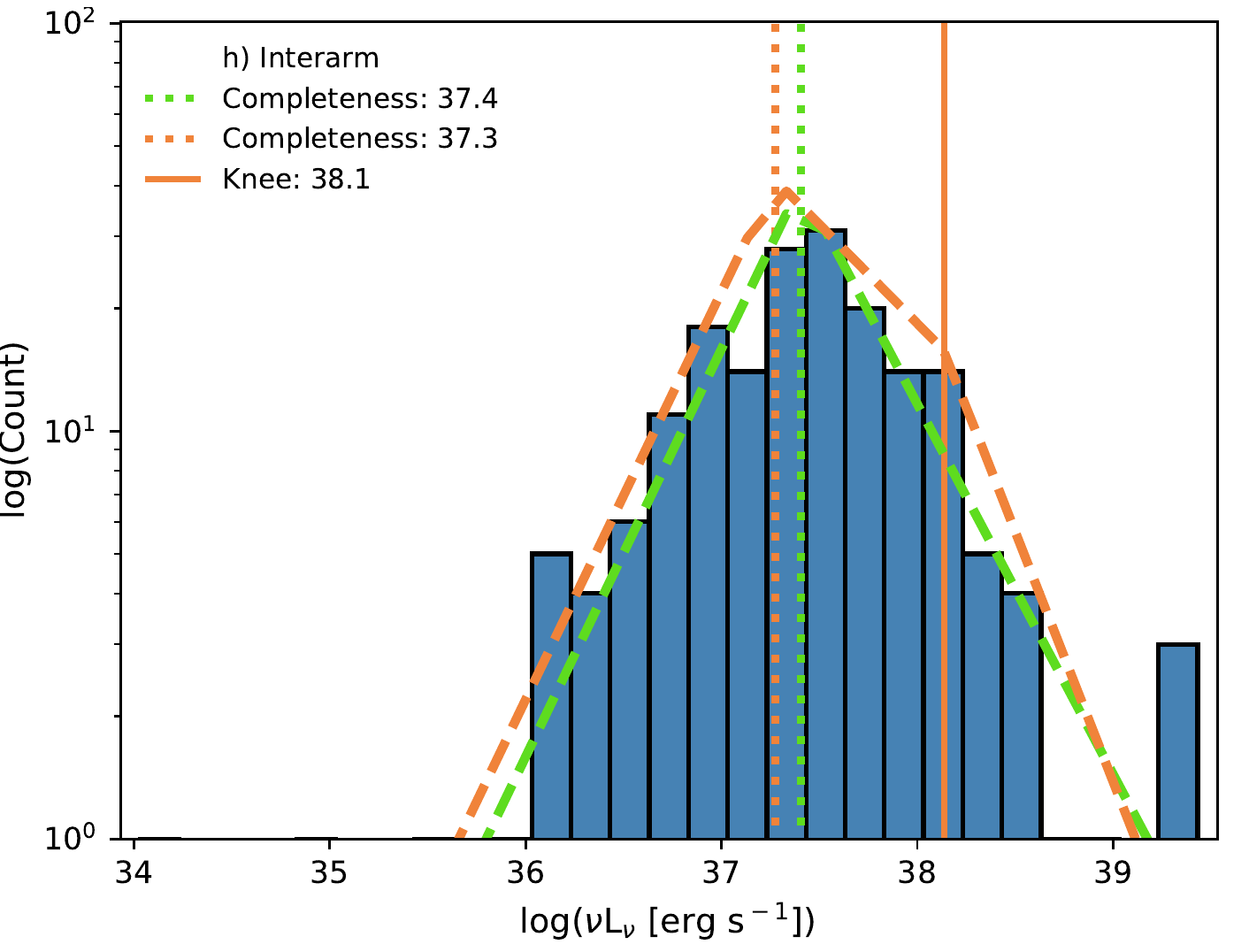}}
\caption{Single and double power law fits to the catalog-only $8\,\microns$ GLIMPSE data subsets: $d_\sun \leq 7.75$ \kpc (panel \subref*{fig:glimpse_neardist}), $d_\sun > 7.75$ \kpc (panel \subref*{fig:glimpse_fardist}), $\rgal \leq 5$ \kpc (panel \subref*{fig:glimpse_nearrgal}), $\rgal > 5$ \kpc (panel \subref*{fig:glimpse_farrgal}), $r \leq 2.4 \pc$ (panel \subref*{fig:glimpse_small}), $r > 2.4 \pc$ (panel \subref*{fig:glimpse_large}), arm (panel \subref*{fig:glimpse_arm}), and interarm (panel \subref*{fig:glimpse_interarm}).}
\label{fig:glimpse}
\end{figure*}

\begin{figure*}[h]
\centering
\subfloat{\label{fig:wise3_neardist}%
    \includegraphics[scale=0.5,trim={3.75cm 8.5cm 3.5cm 8.5cm}, clip]{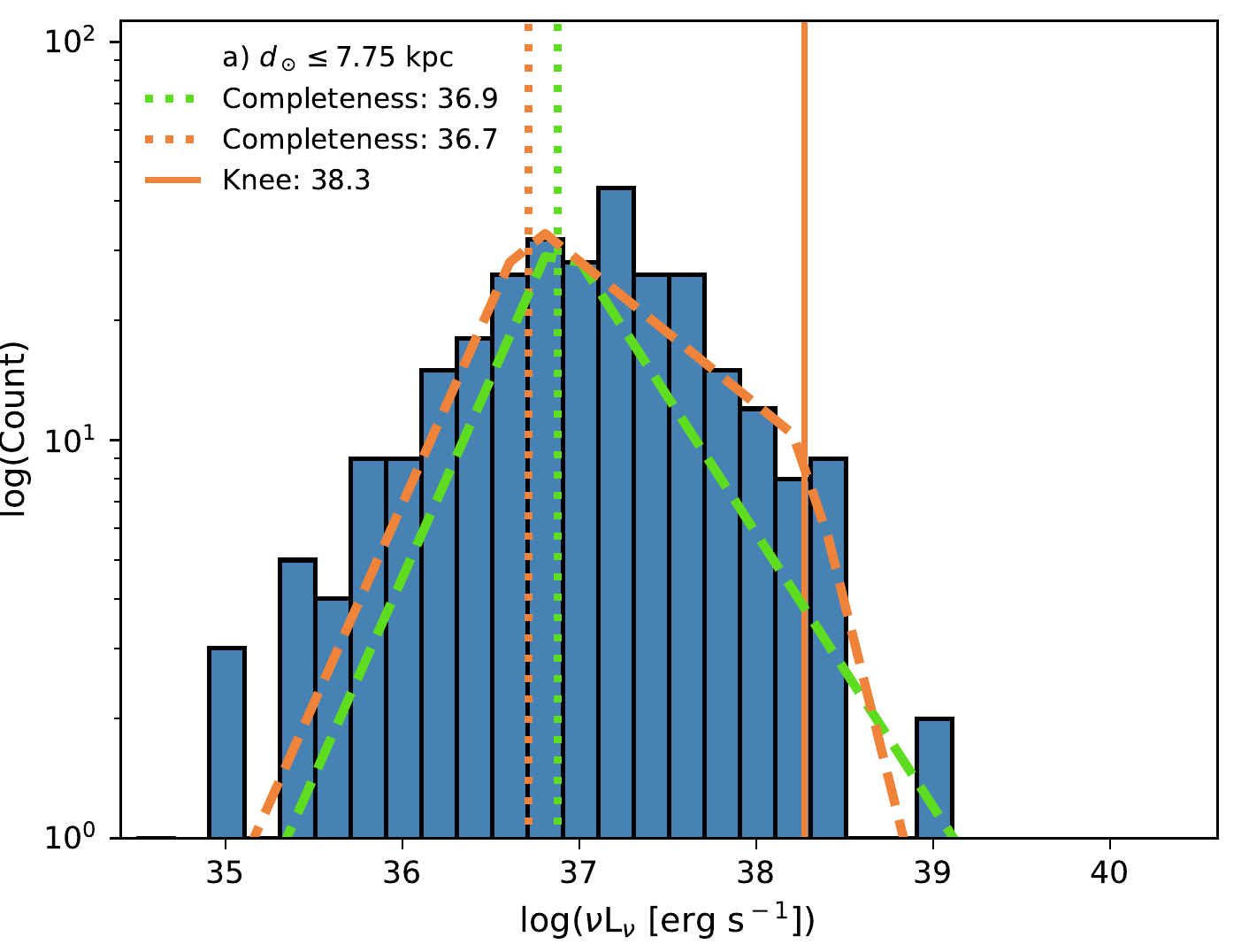}}\qquad
\subfloat{\label{fig:wise3_fardist}%
    \includegraphics[scale=0.5,trim={3.75cm 8.5cm 3.5cm 8.5cm}, clip]{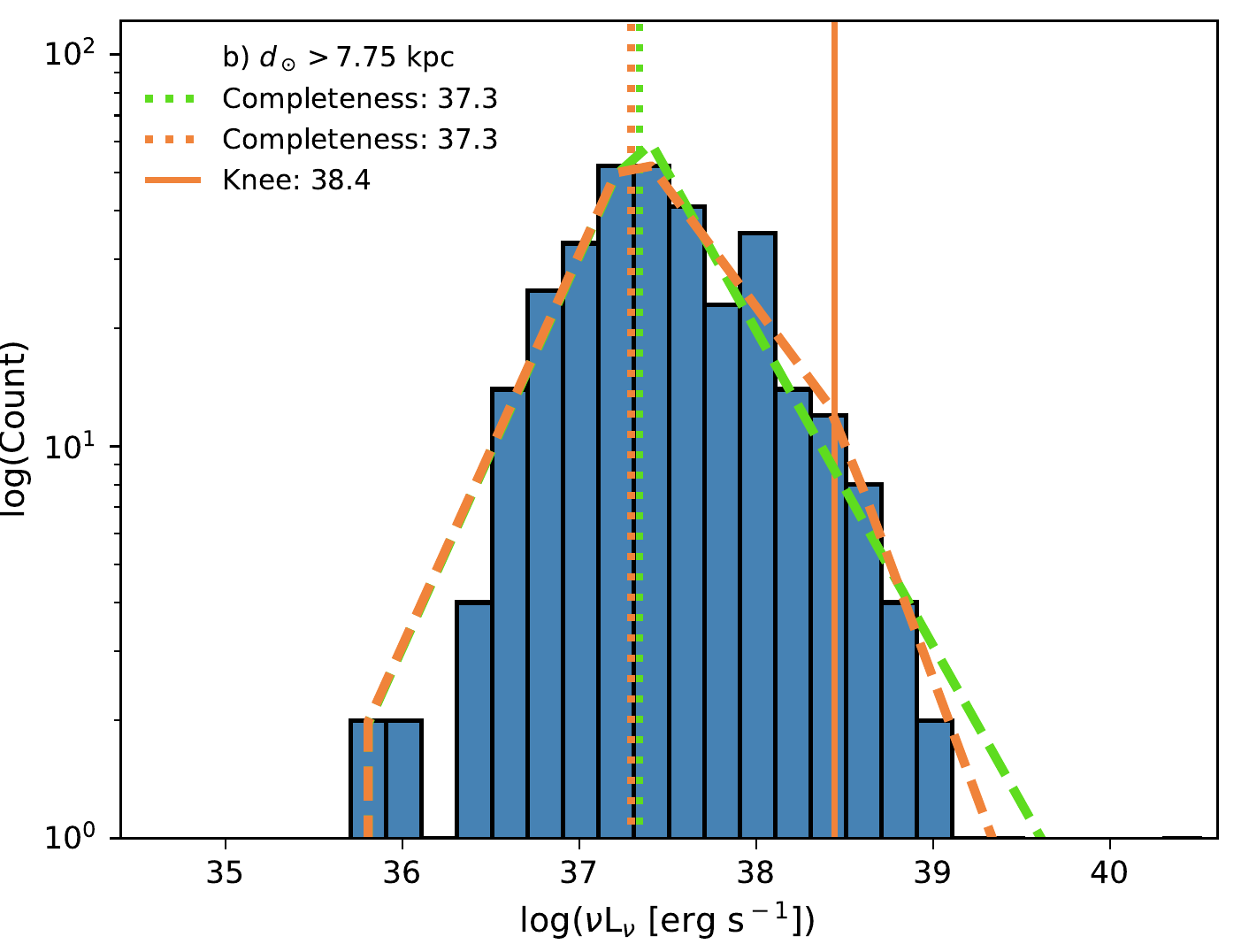}}\\
\subfloat{\label{fig:wise3_nearrgal}%
    \includegraphics[scale=0.5,trim={3.75cm 8.5cm 3.5cm 8.5cm}, clip]{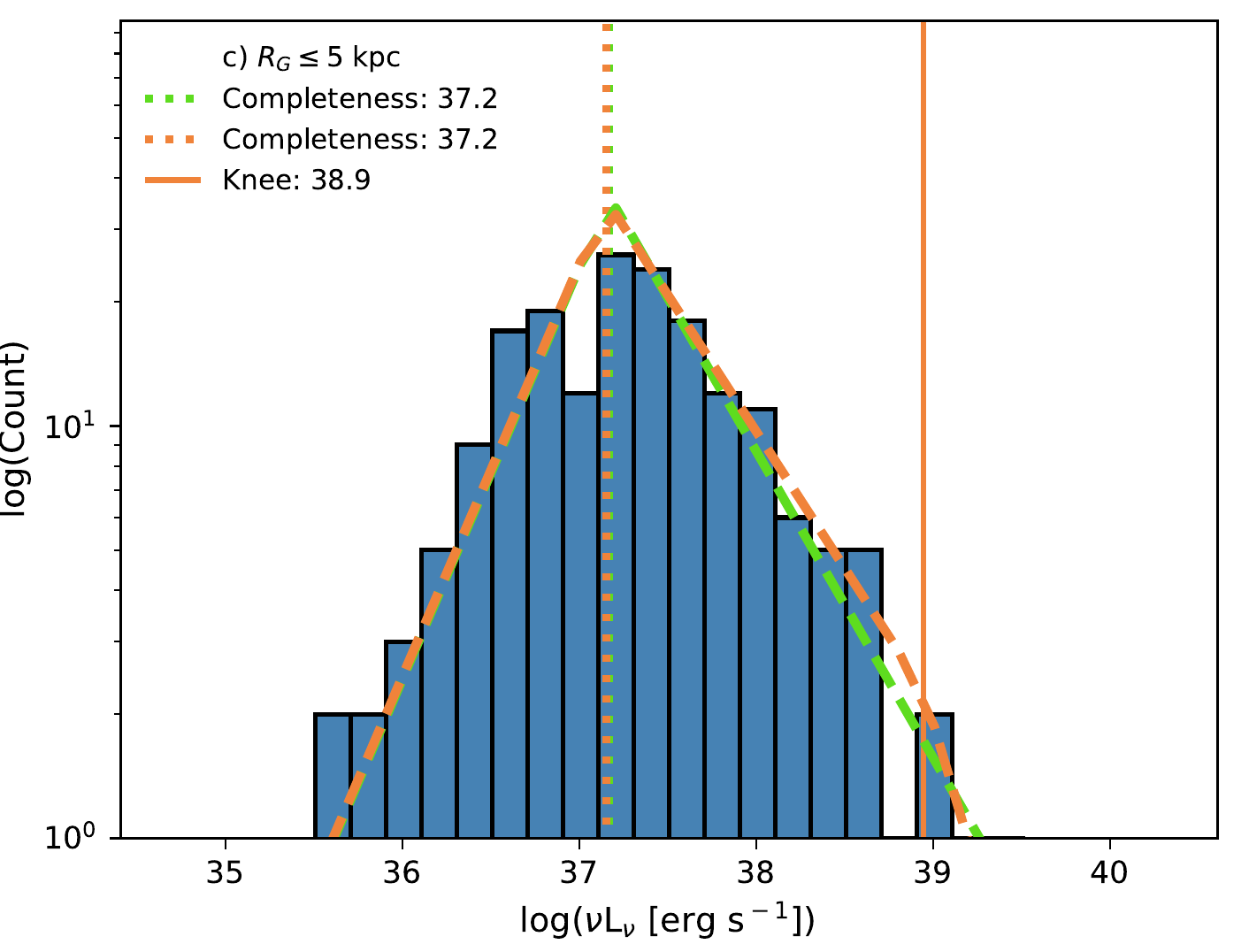}}\qquad
\subfloat{\label{fig:wise3_farrgal}%
    \includegraphics[scale=0.5,trim={3.75cm 8.5cm 3.5cm 8.5cm}, clip]{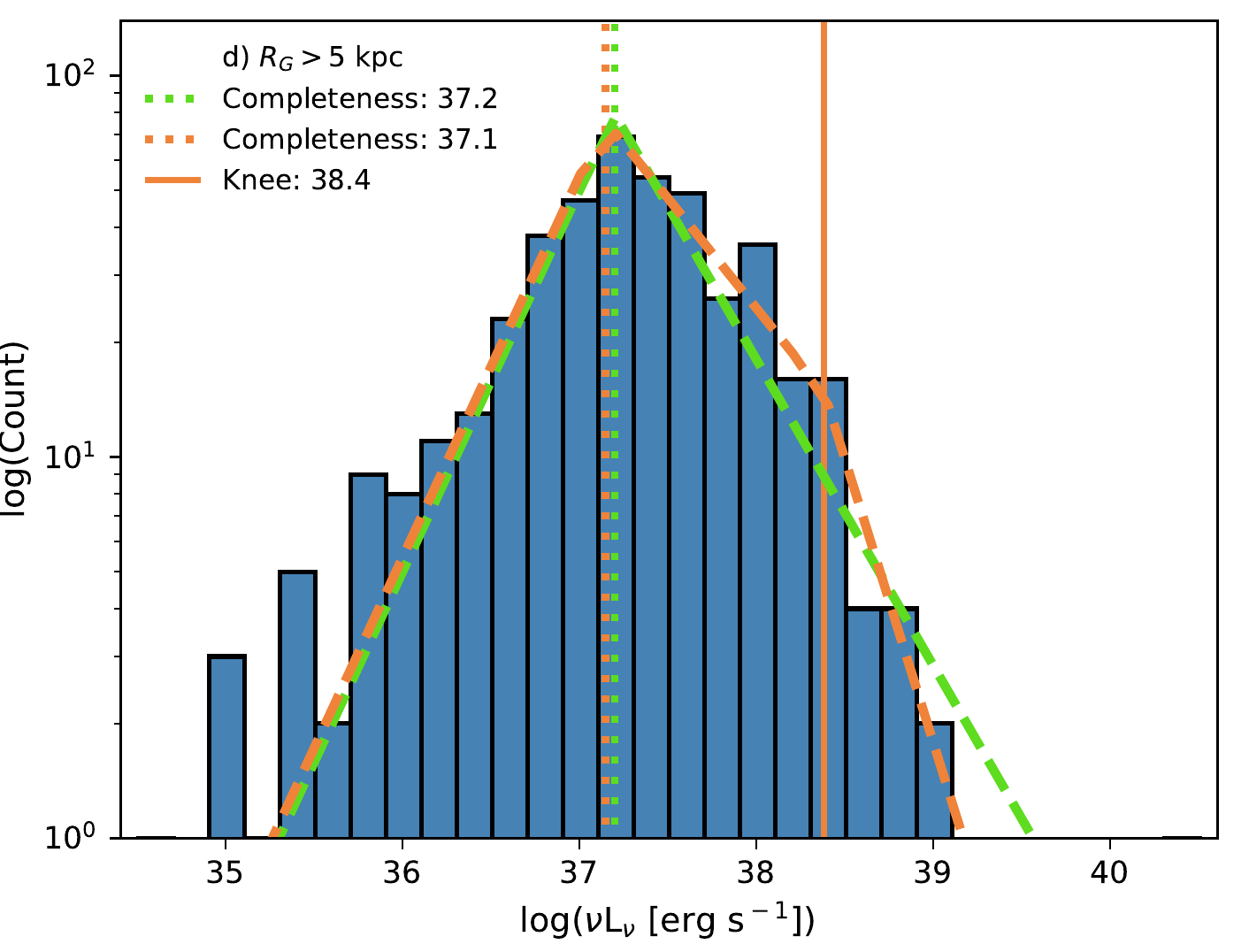}}\\
\subfloat{\label{fig:wise3_small}%
    \includegraphics[scale=0.5,trim={3.75cm 8.5cm 3.5cm 8.5cm}, clip]{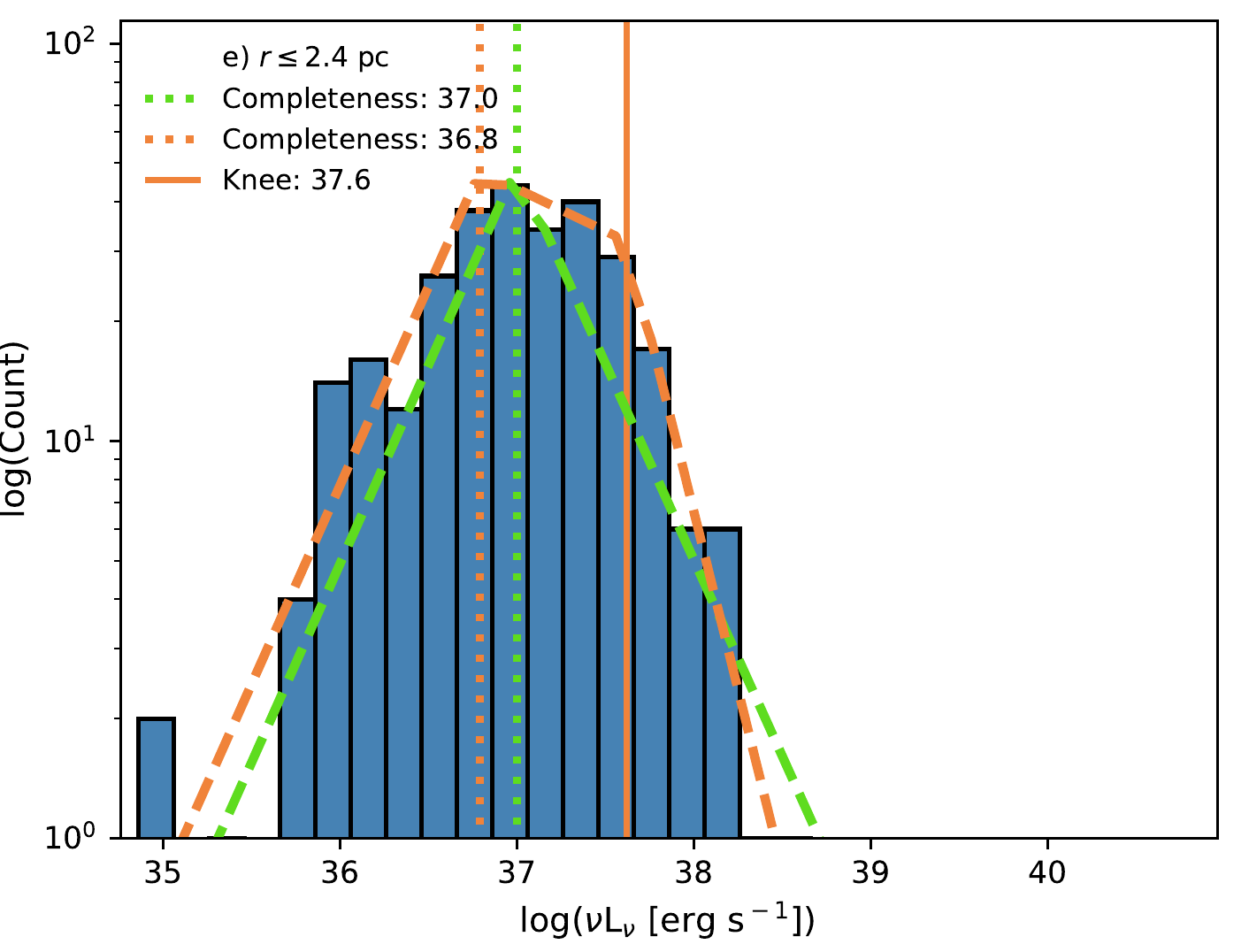}}\qquad
\subfloat{\label{fig:wise3_large}%
    \includegraphics[scale=0.5,trim={3.75cm 8.5cm 3.5cm 8.5cm}, clip]{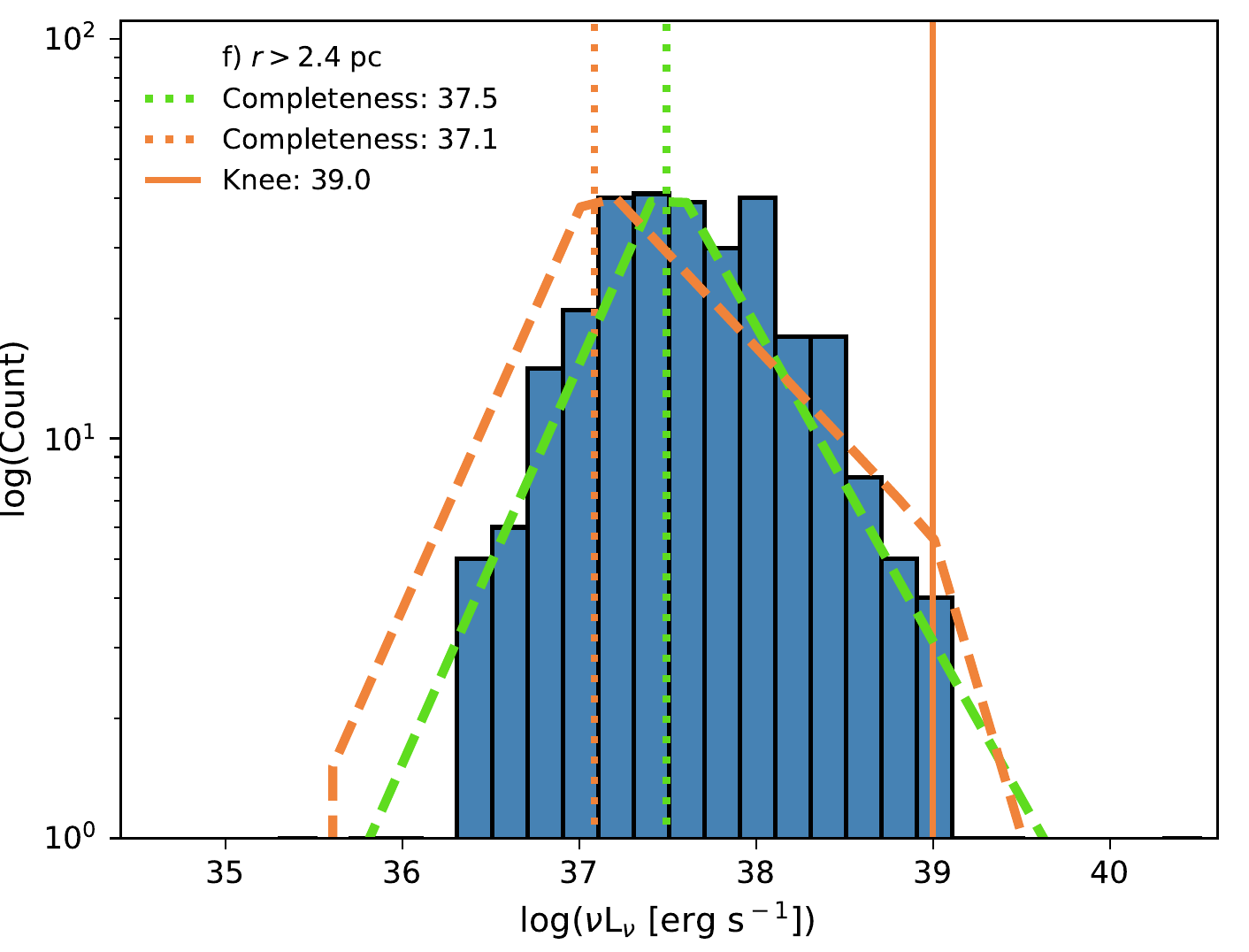}}\\
\subfloat{\label{fig:wise3_arm}%
    \includegraphics[scale=0.5,trim={3.75cm 8.5cm 3.5cm 8.5cm}, clip]{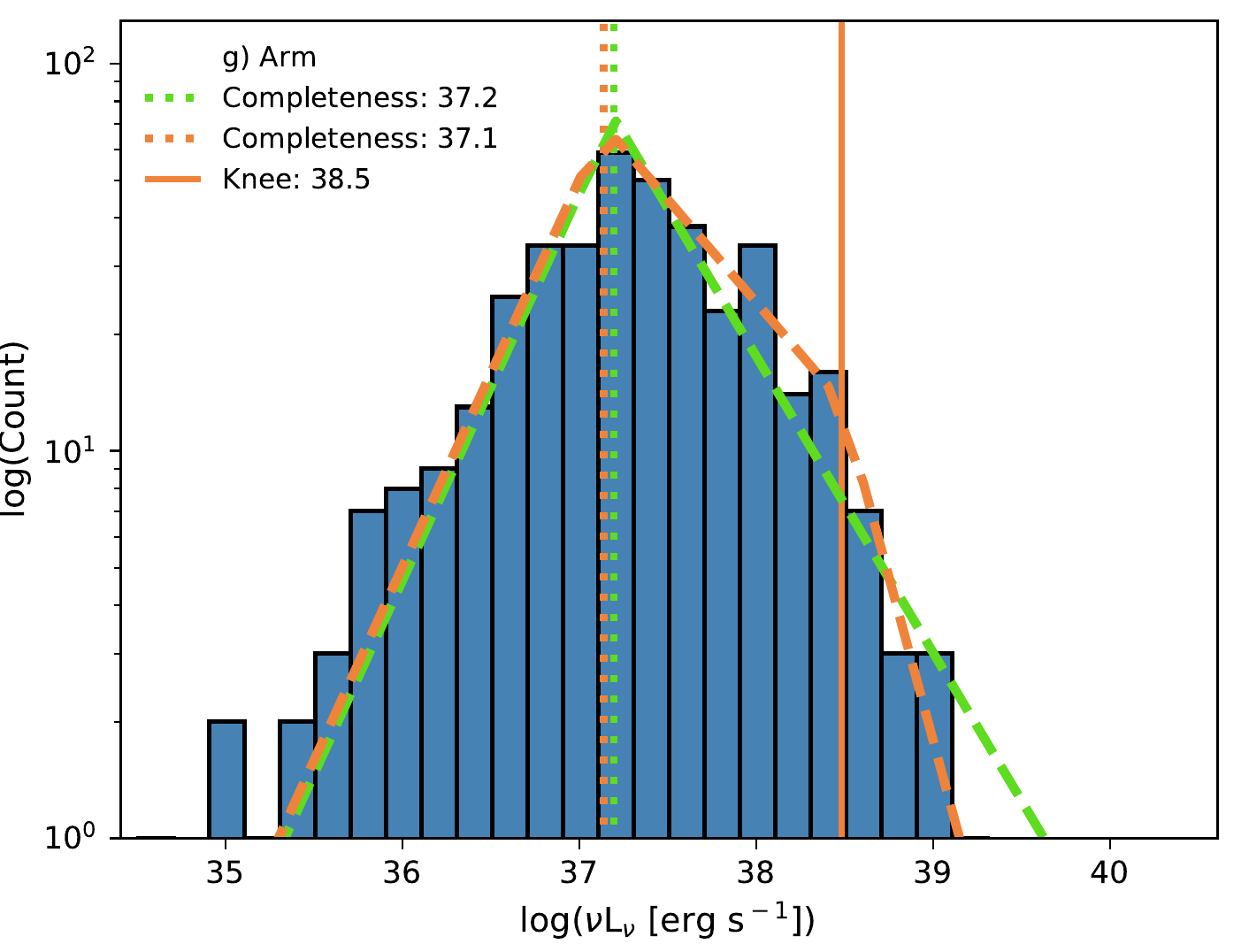}}\qquad
\subfloat{\label{fig:wise3_interarm}%
    \includegraphics[scale=0.5,trim={3.75cm 8.5cm 3.5cm 8.5cm}, clip]{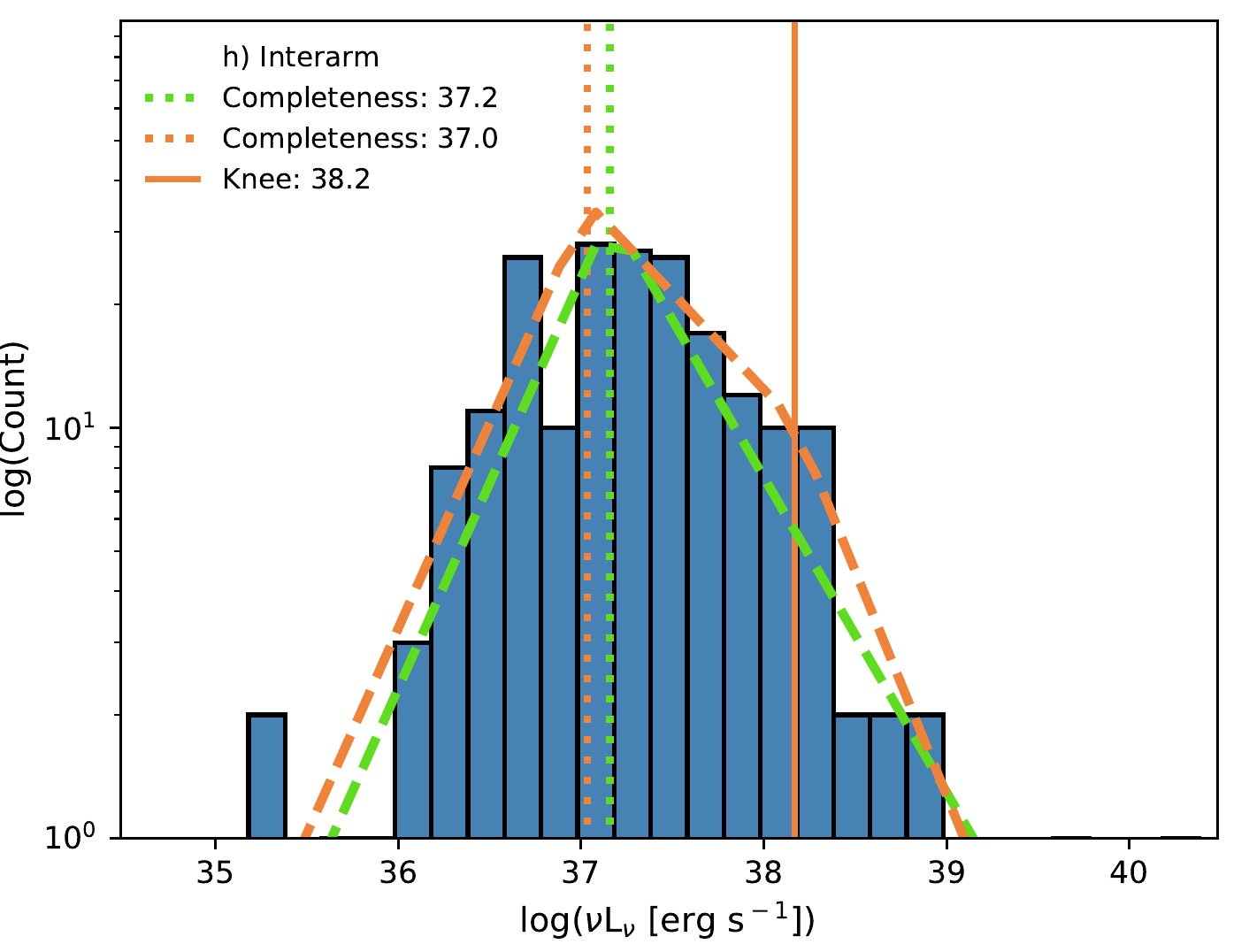}}
\caption{Single and double power law fits to the catalog-only $12\,\microns$ \textit{WISE} data subsets: $d_\sun \leq 7.75$ \kpc (panel \subref*{fig:wise3_neardist}), $d_\sun > 7.75$ \kpc (panel \subref*{fig:wise3_fardist}), $\rgal \leq 5$ \kpc (panel \subref*{fig:wise3_nearrgal}), $\rgal > 5$ \kpc (panel \subref*{fig:wise3_farrgal}), $r \leq 2.4 \pc$ (panel \subref*{fig:wise3_small}), $r > 2.4 \pc$ (panel \subref*{fig:wise3_large}), arm (panel \subref*{fig:wise3_arm}), and interarm (panel \subref*{fig:wise3_interarm}).}
\label{fig:wise32}
\end{figure*}

\begin{figure*}[h]
\centering
\subfloat{\label{fig:wise4_neardist}%
    \includegraphics[scale=0.5,trim={3.75cm 8.5cm 3.5cm 8.5cm}, clip]{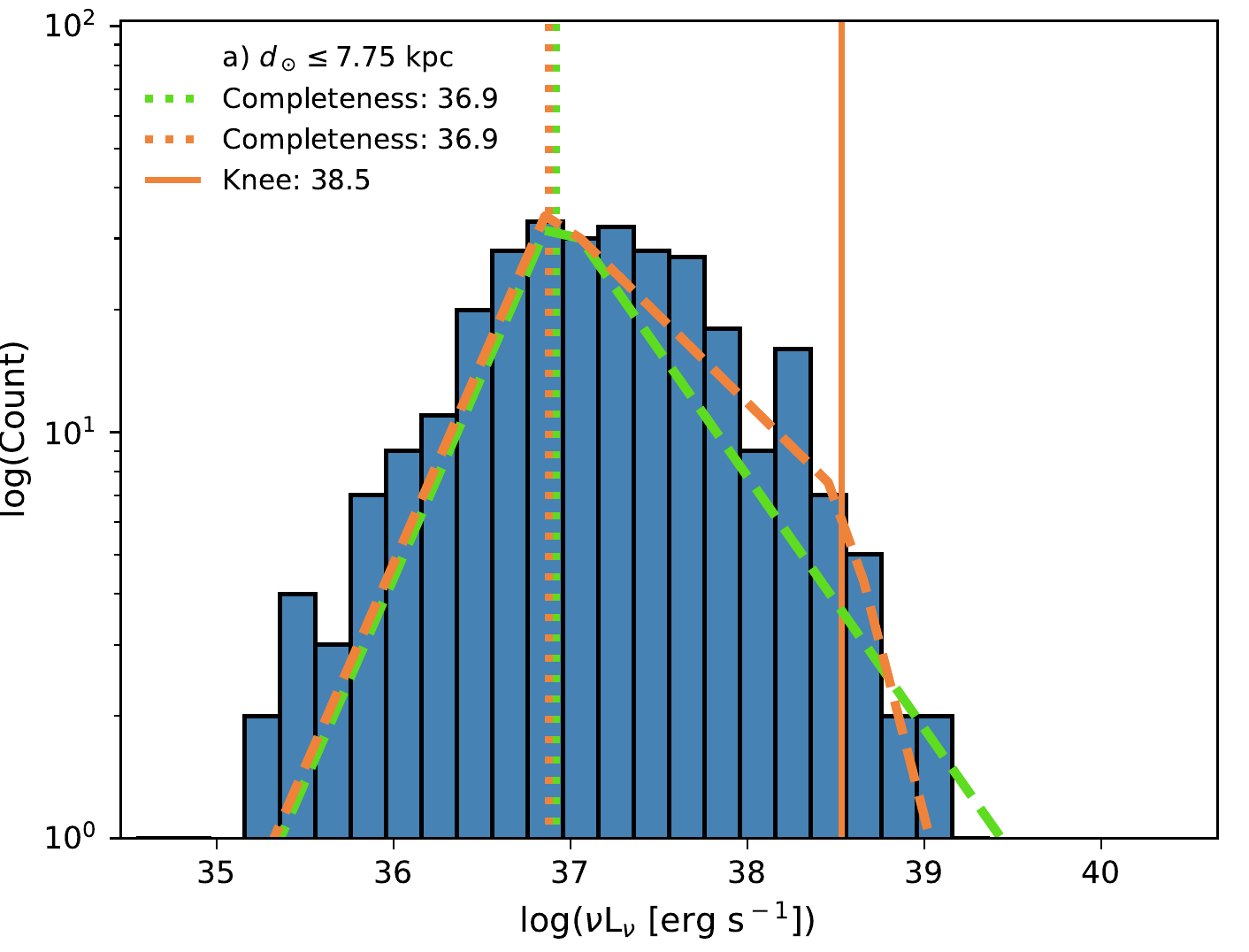}}\qquad
\subfloat{\label{fig:wise4_fardist}%
    \includegraphics[scale=0.5,trim={3.75cm 8.5cm 3.5cm 8.5cm}, clip]{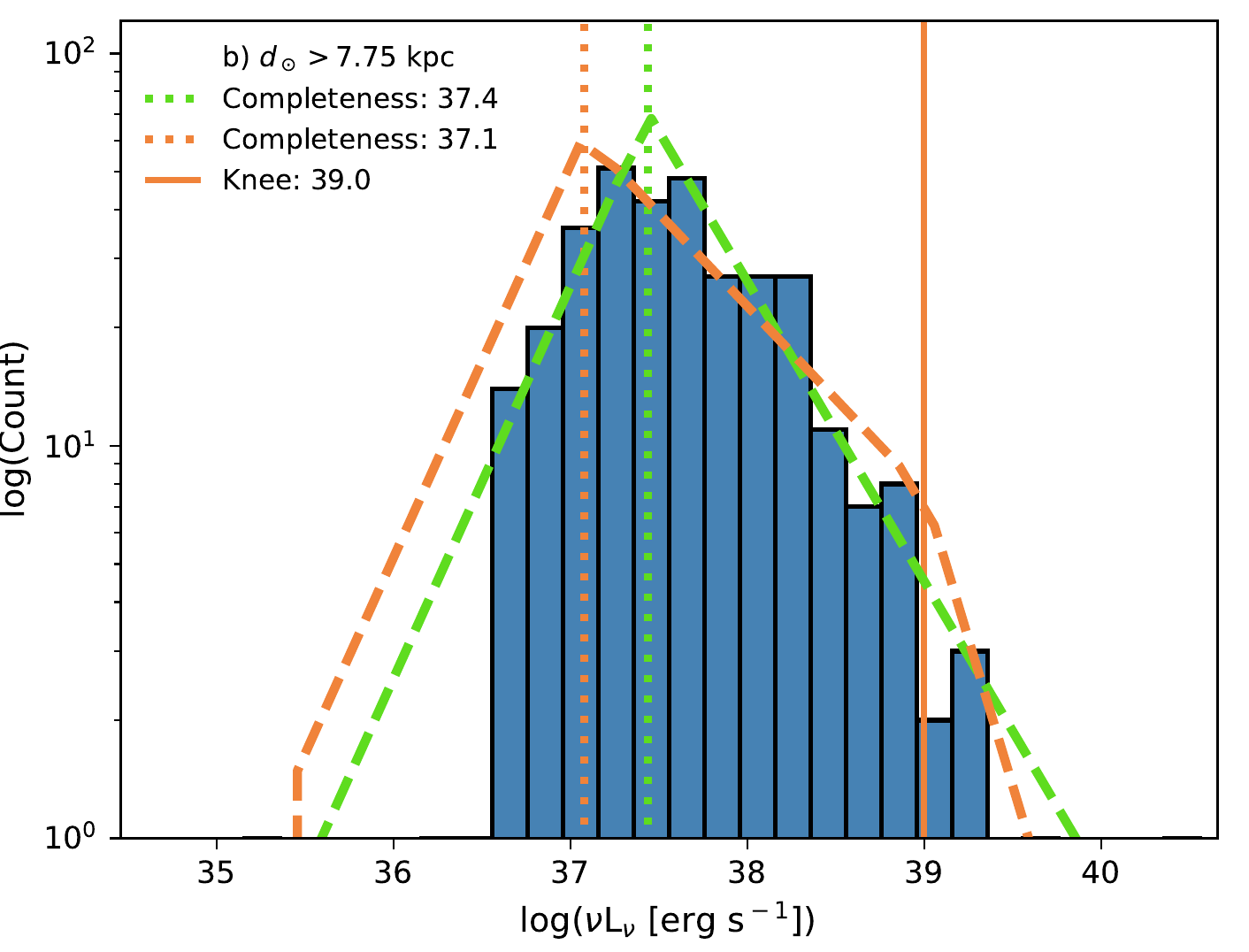}}\\
\subfloat{\label{fig:wise4_nearrgal}%
    \includegraphics[scale=0.5,trim={3.75cm 8.5cm 3.5cm 8.5cm}, clip]{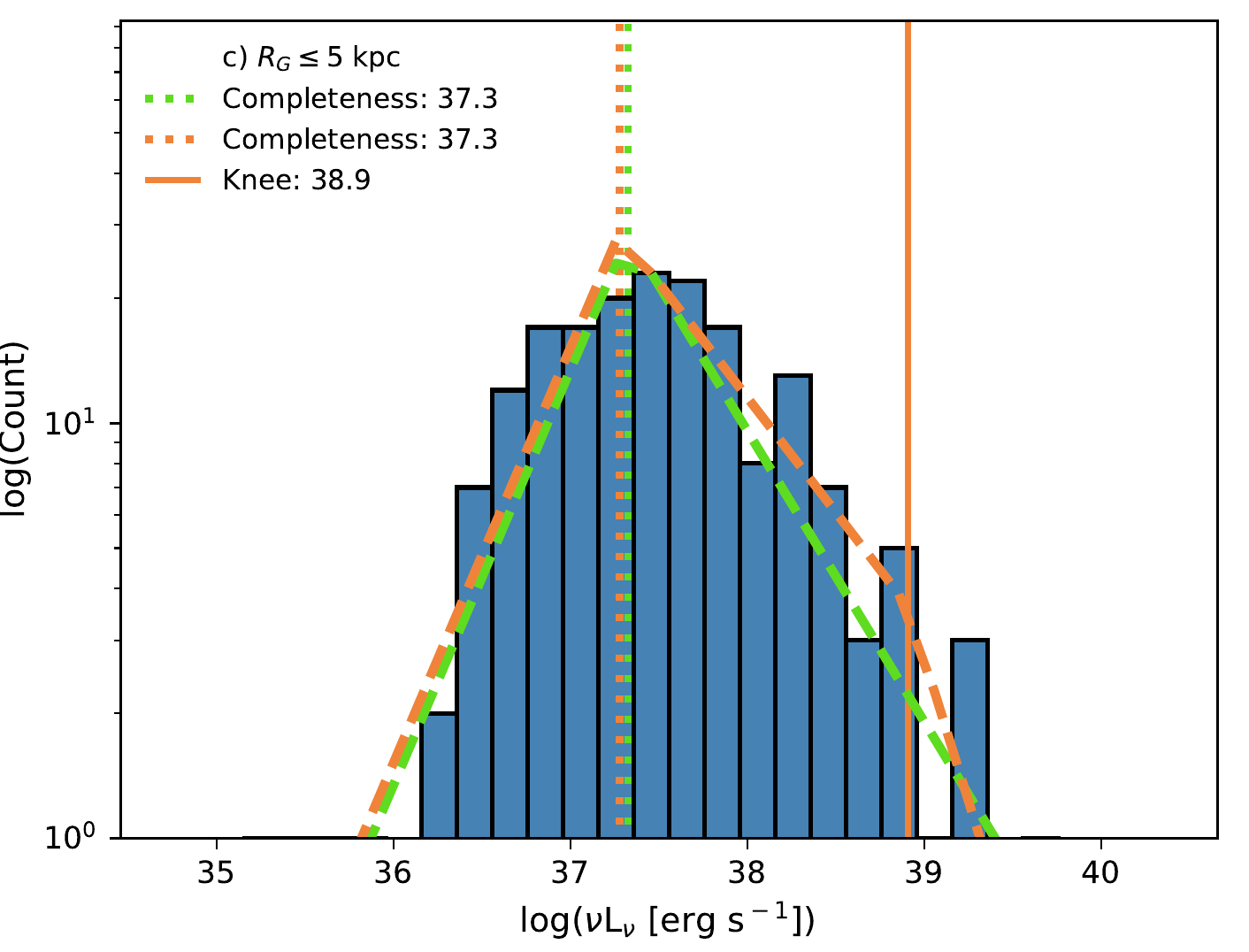}}\qquad
\subfloat{\label{fig:wise4_farrgal}%
    \includegraphics[scale=0.5,trim={3.75cm 8.5cm 3.5cm 8.5cm}, clip]{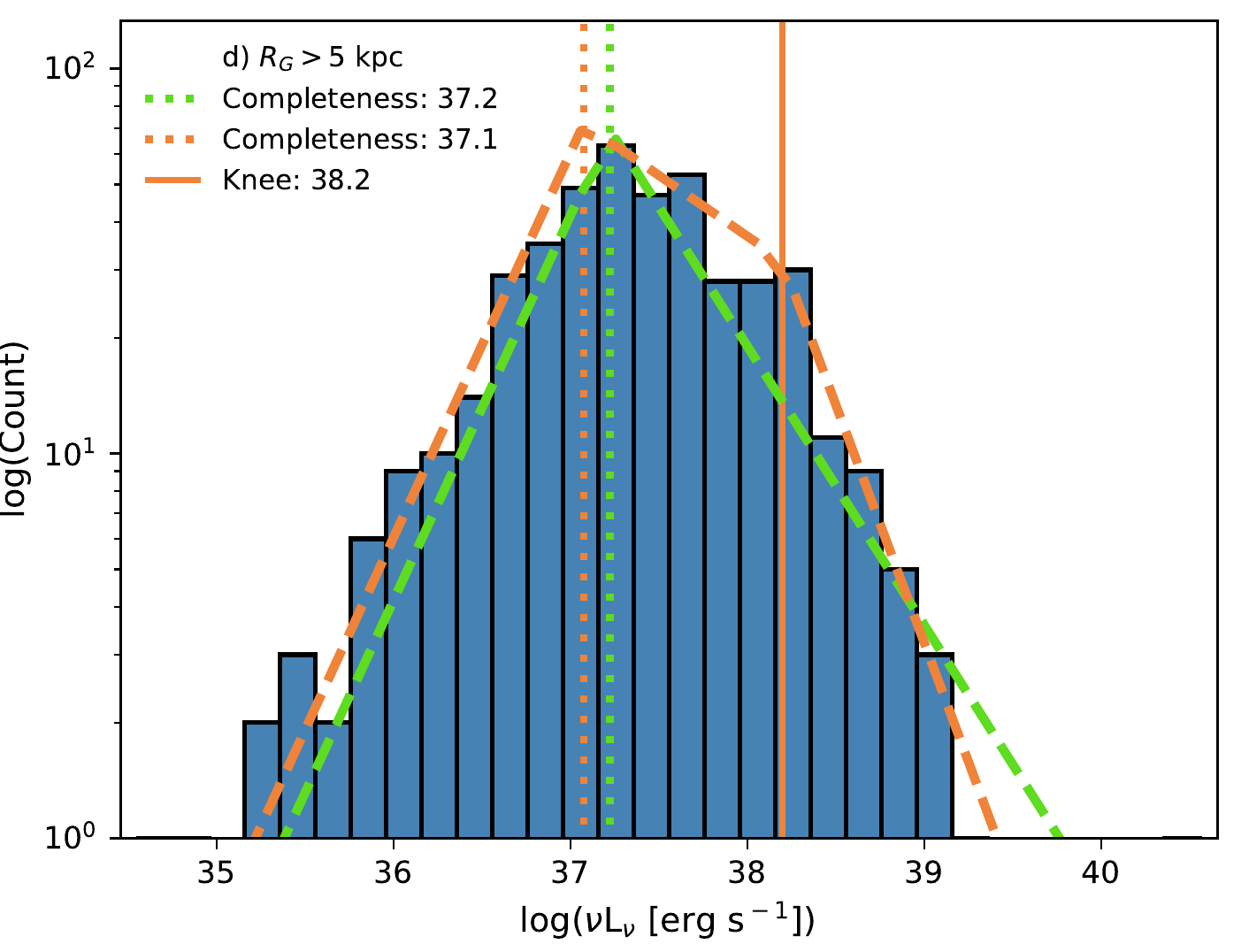}}\\
\subfloat{\label{fig:wise4_small}%
    \includegraphics[scale=0.5,trim={3.75cm 8.5cm 3.5cm 8.5cm}, clip]{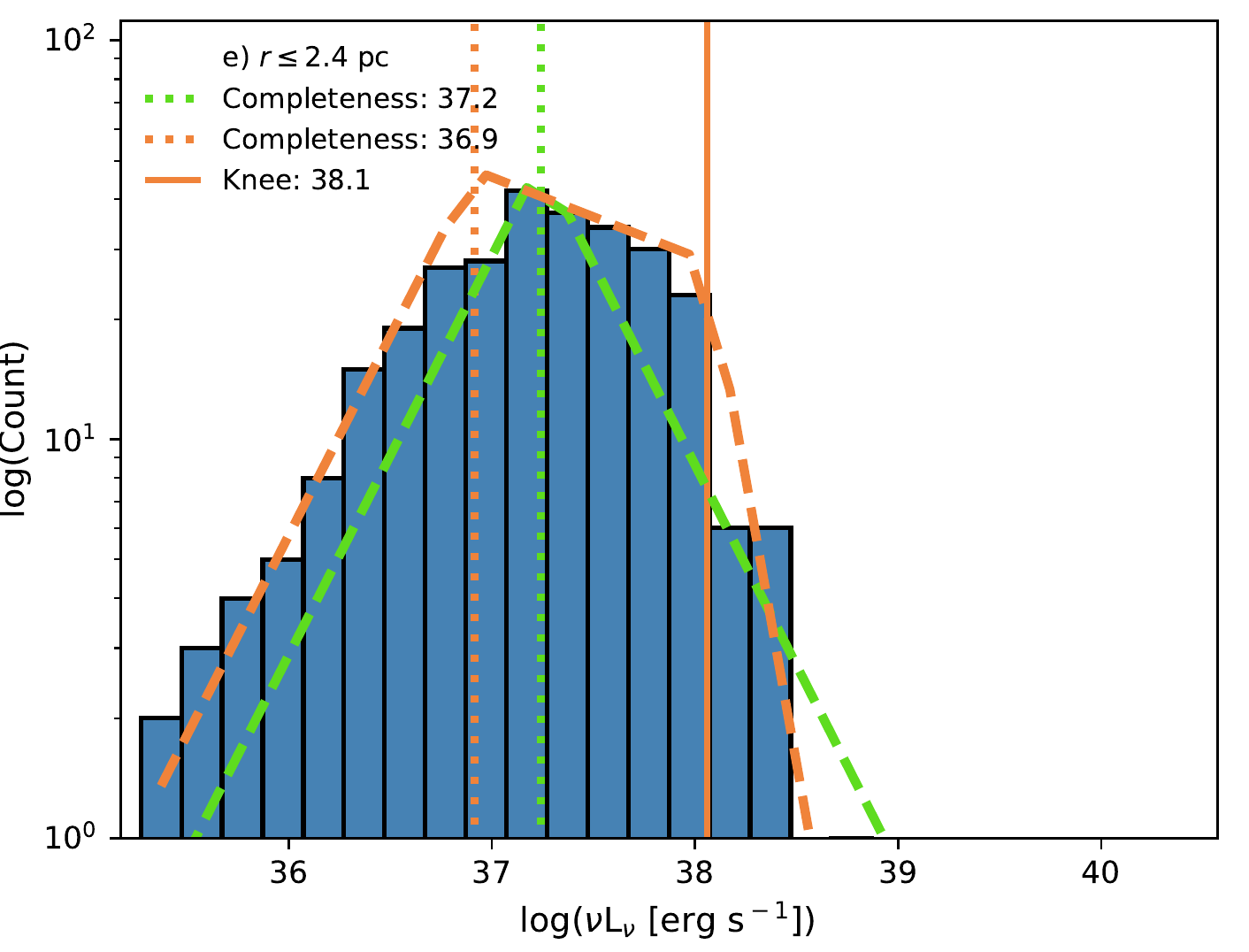}}\qquad
\subfloat{\label{fig:wise4_large}%
    \includegraphics[scale=0.5,trim={3.75cm 8.5cm 3.5cm 8.5cm}, clip]{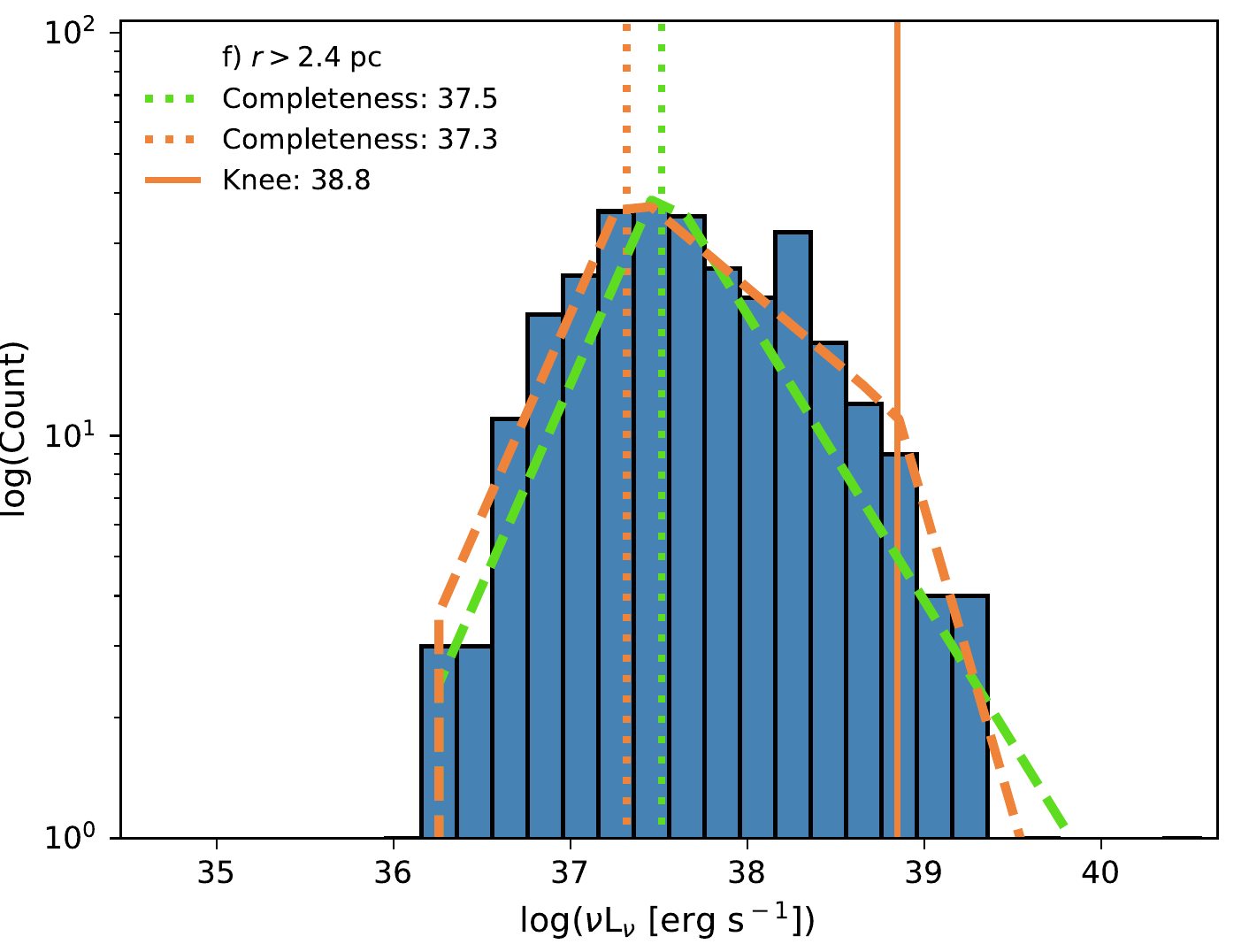}}\\
\subfloat{\label{fig:wise4_arm}%
    \includegraphics[scale=0.5,trim={3.75cm 8.5cm 3.5cm 8.5cm}, clip]{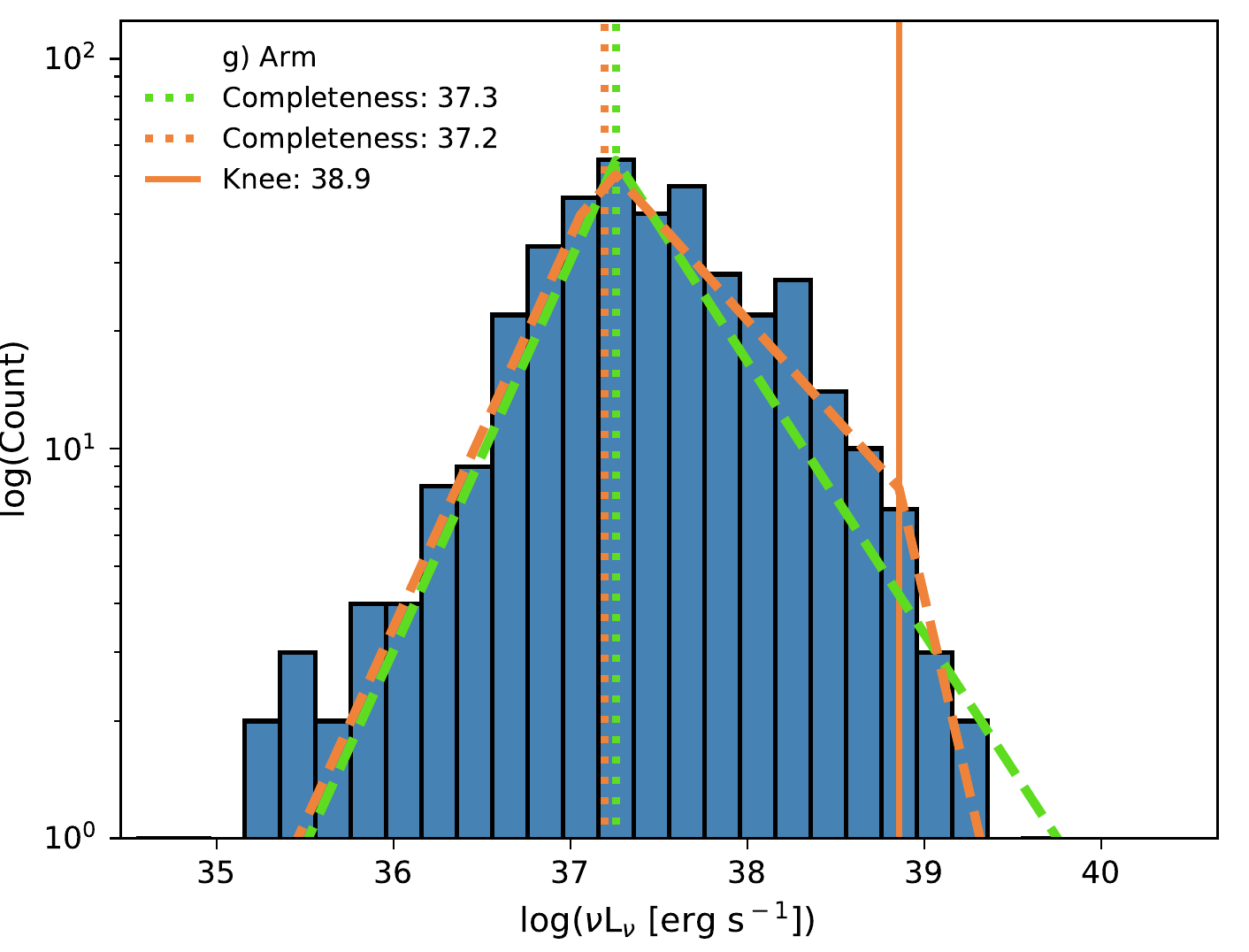}}\qquad
\subfloat{\label{fig:wise4_interarm}%
    \includegraphics[scale=0.5,trim={3.75cm 8.5cm 3.5cm 8.5cm}, clip]{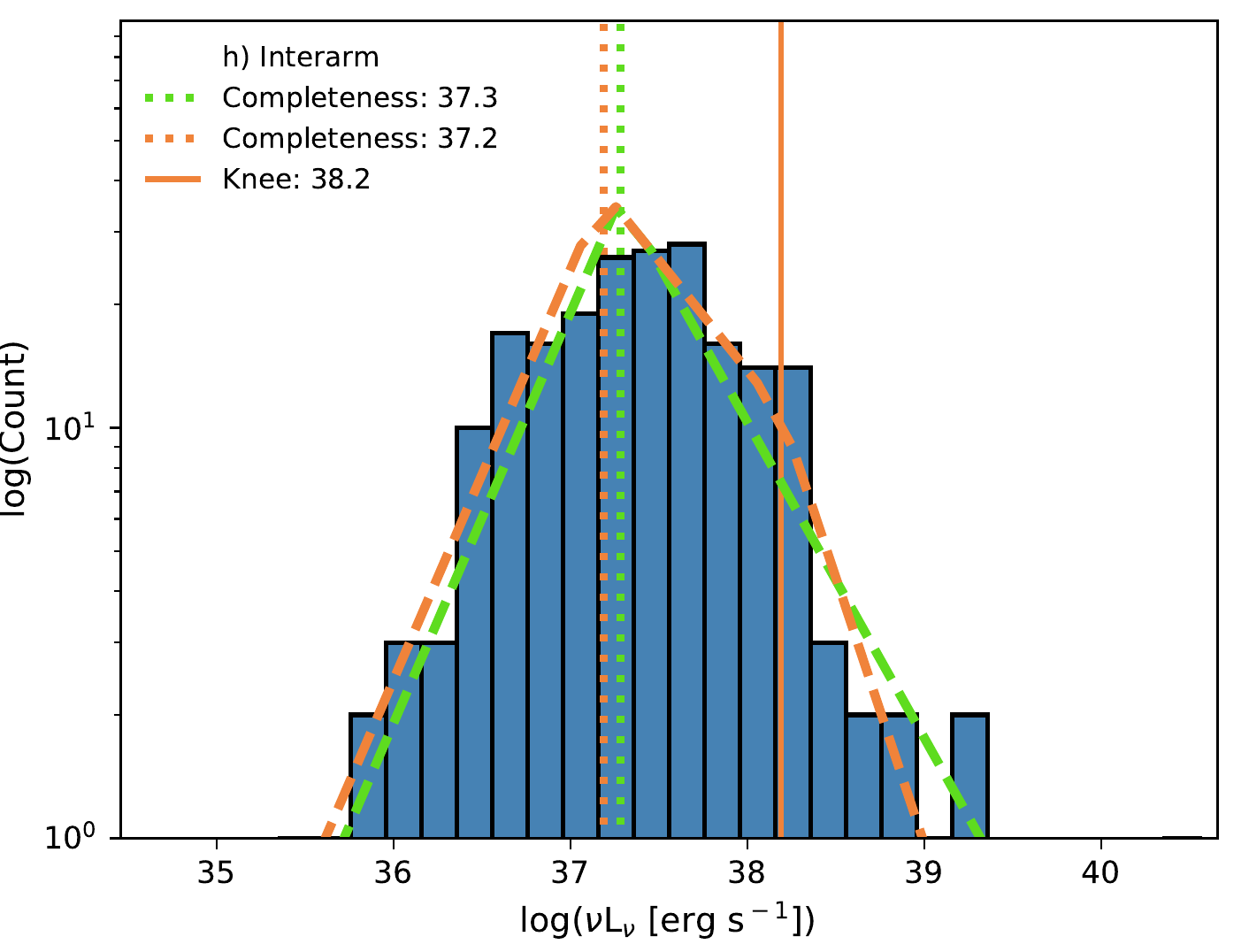}}
\caption{Single and double power law fits to the catalog-only $22\,\microns$ \textit{WISE} data subsets: $d_\sun \leq 7.75$ \kpc (panel \subref*{fig:wise4_neardist}), $d_\sun > 7.75$ \kpc (panel \subref*{fig:wise4_fardist}), $\rgal \leq 5$ \kpc (panel \subref*{fig:wise4_nearrgal}), $\rgal > 5$ \kpc (panel \subref*{fig:wise4_farrgal}), $r \leq 2.4 \pc$ (panel \subref*{fig:wise4_small}), $r > 2.4 \pc$ (panel \subref*{fig:wise4_large}), arm (panel \subref*{fig:wise4_arm}), and interarm (panel \subref*{fig:wise4_interarm}).}
\label{fig:wise42}
\end{figure*}

\begin{figure*}[h]
\centering
  \subfloat{\label{fig:mipsgal_neardist}%
    \includegraphics[scale=0.5,trim={3.75cm 8.5cm 3.5cm 8.5cm}, clip]{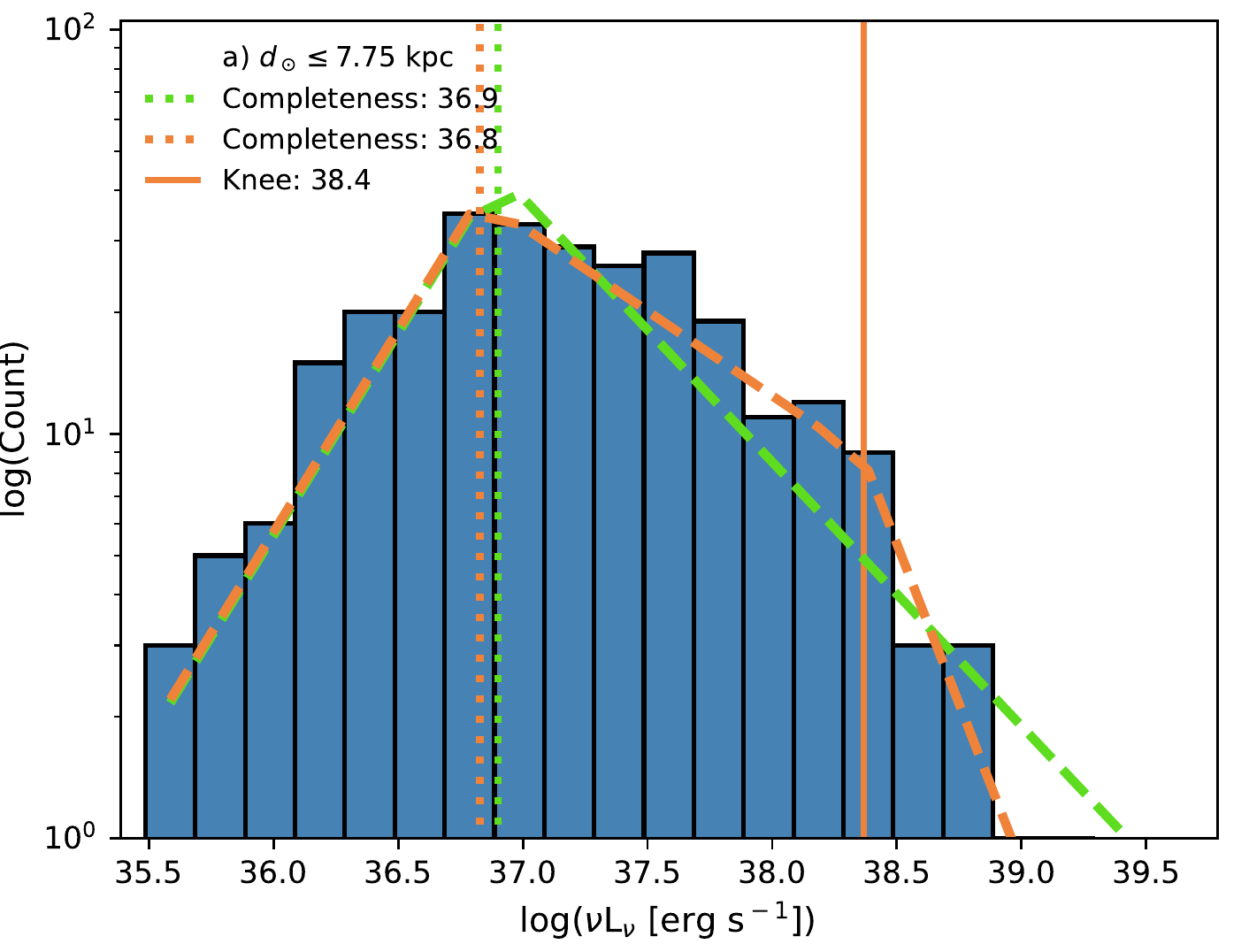}}\qquad
  \subfloat{\label{fig:mipsgal_fardist}%
    \includegraphics[scale=0.5,trim={3.75cm 8.5cm 3.5cm 8.5cm}, clip]{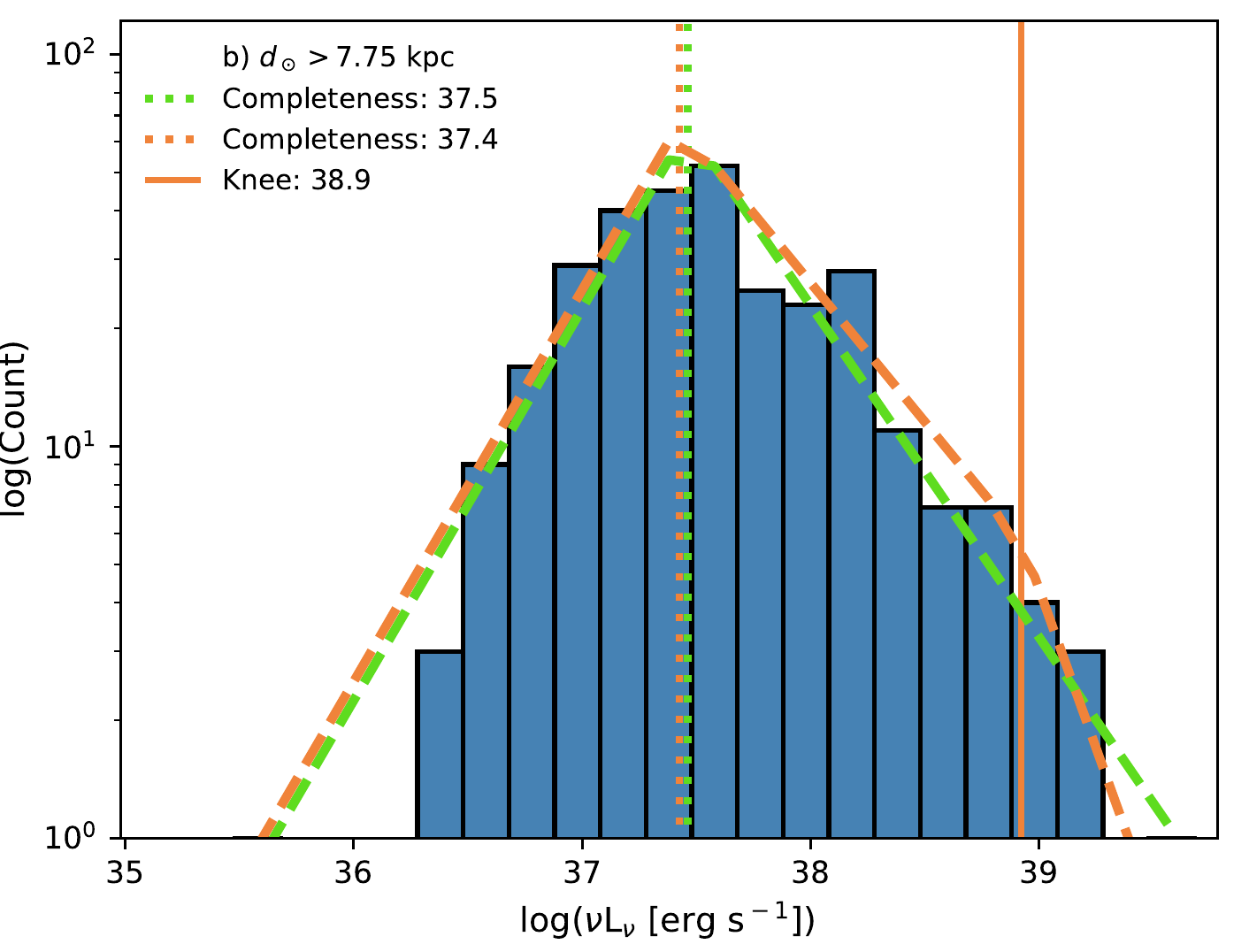}}\\
  \subfloat{\label{fig:mipsgal_nearrgal}%
    \includegraphics[scale=0.5,trim={3.75cm 8.5cm 3.5cm 8.5cm}, clip]{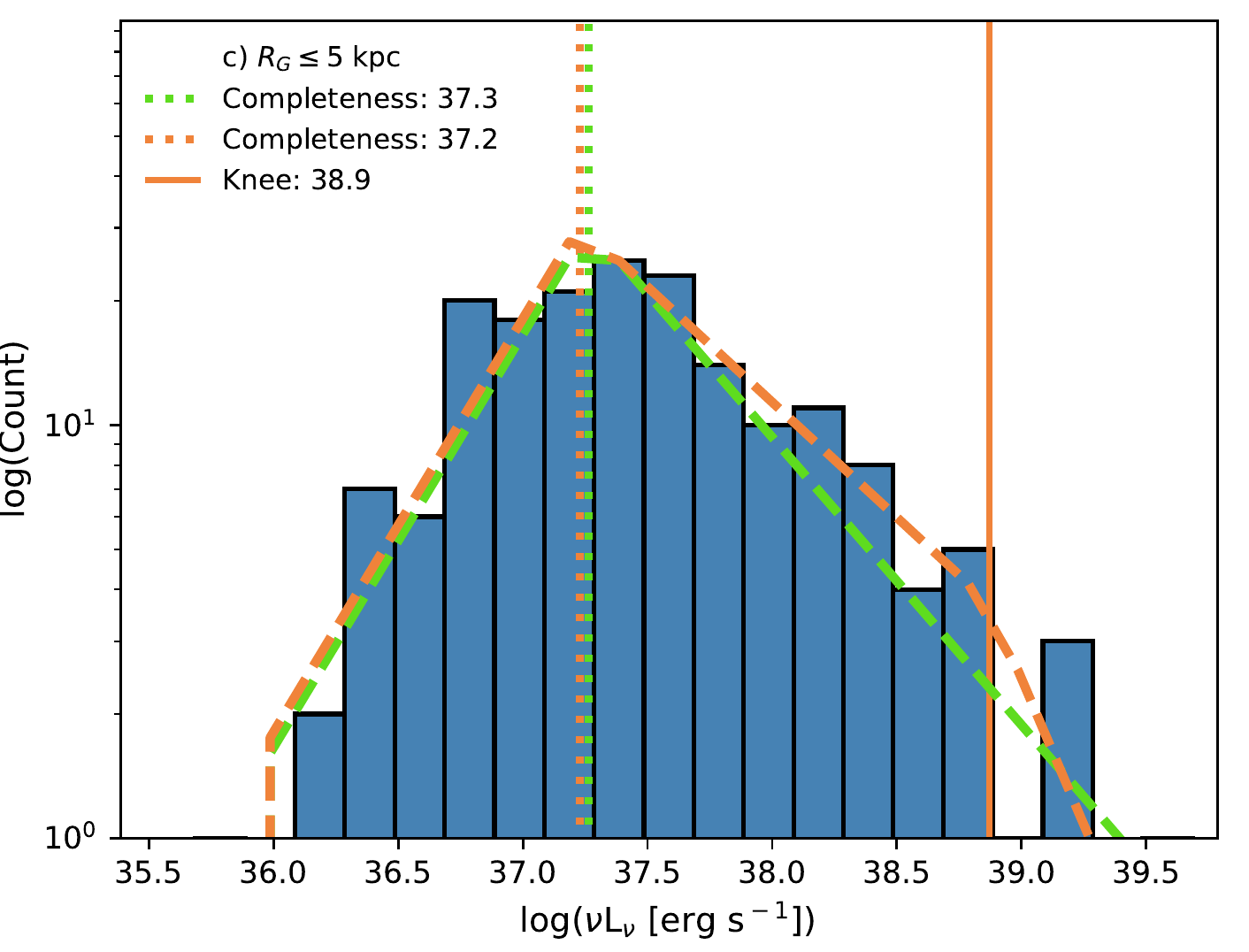}}\qquad
  \subfloat{\label{fig:mipsgal_farrgal}%
    \includegraphics[scale=0.5,trim={3.75cm 8.5cm 3.5cm 8.5cm}, clip]{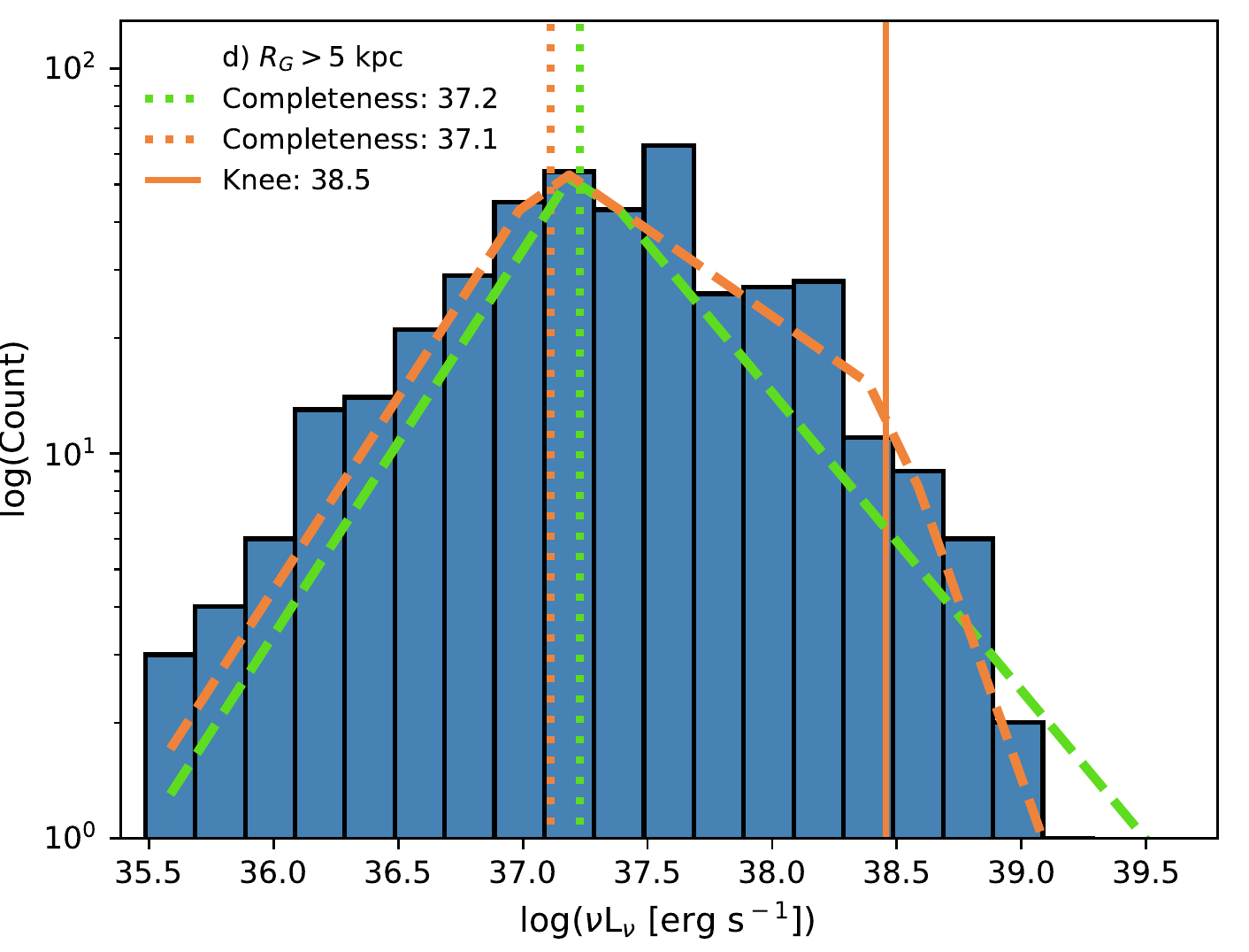}}\\
  \subfloat{\label{fig:mipsgal_small}%
    \includegraphics[scale=0.5,trim={3.75cm 8.5cm 3.5cm 8.5cm}, clip]{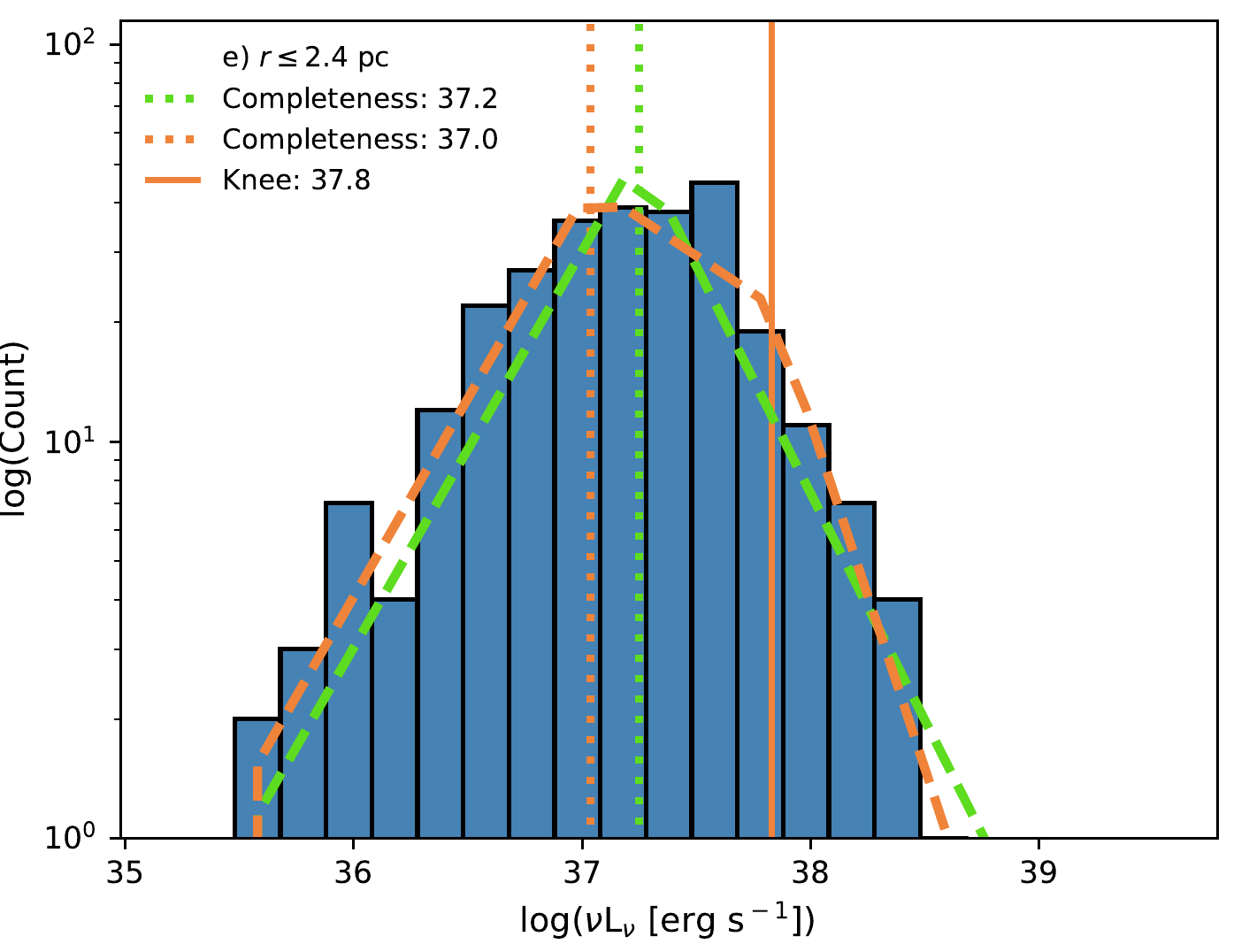}}\qquad
  \subfloat{\label{fig:mipsgal_large}%
    \includegraphics[scale=0.5,trim={3.75cm 8.5cm 3.5cm 8.5cm}, clip]{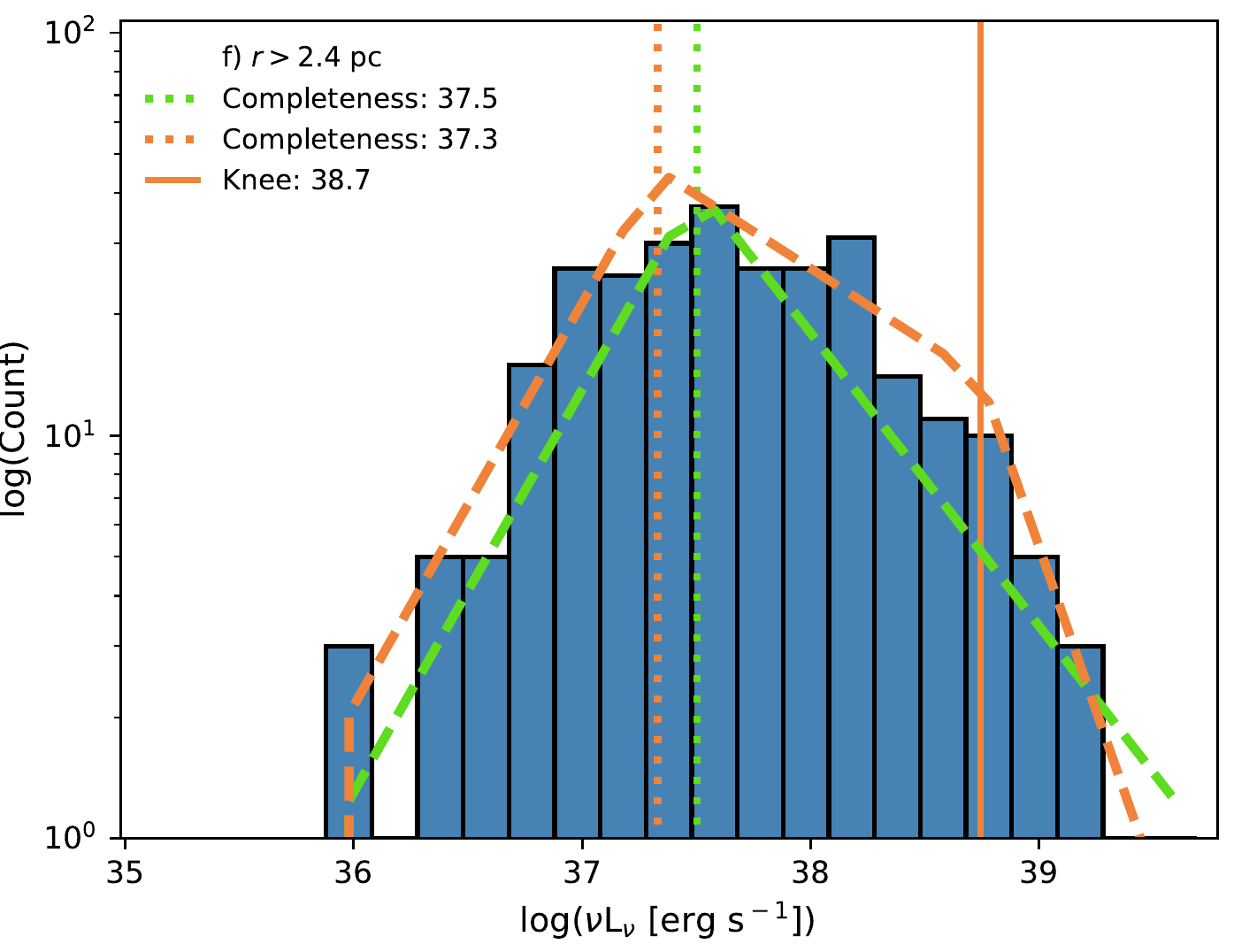}}\\
  \subfloat{\label{fig:mipsgal_arm}%
    \includegraphics[scale=0.5,trim={3.75cm 8.5cm 3.5cm 8.5cm}, clip]{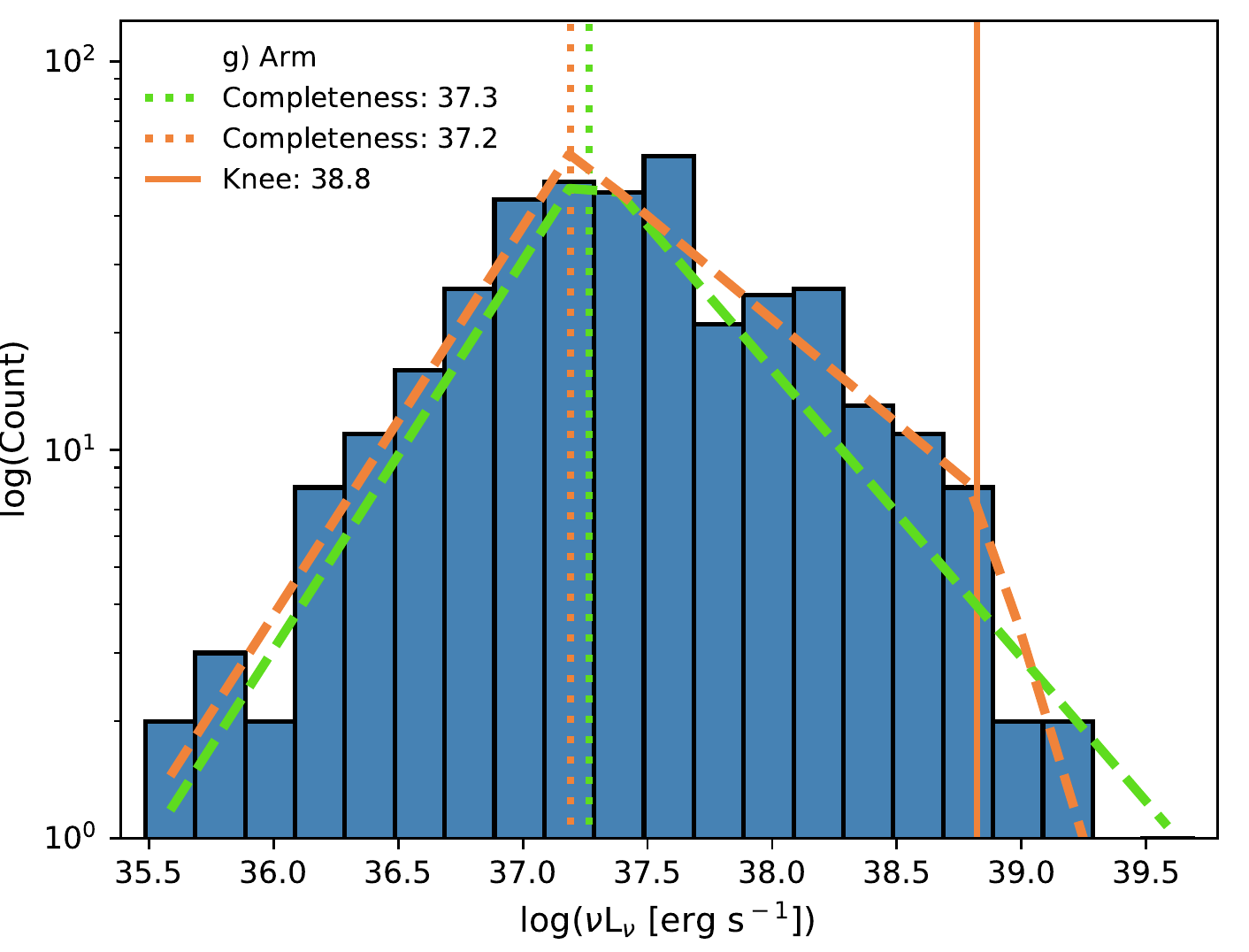}}\qquad
  \subfloat{\label{fig:mipsgal_interarm}%
    \includegraphics[scale=0.5,trim={3.75cm 8.5cm 3.5cm 8.5cm}, clip]{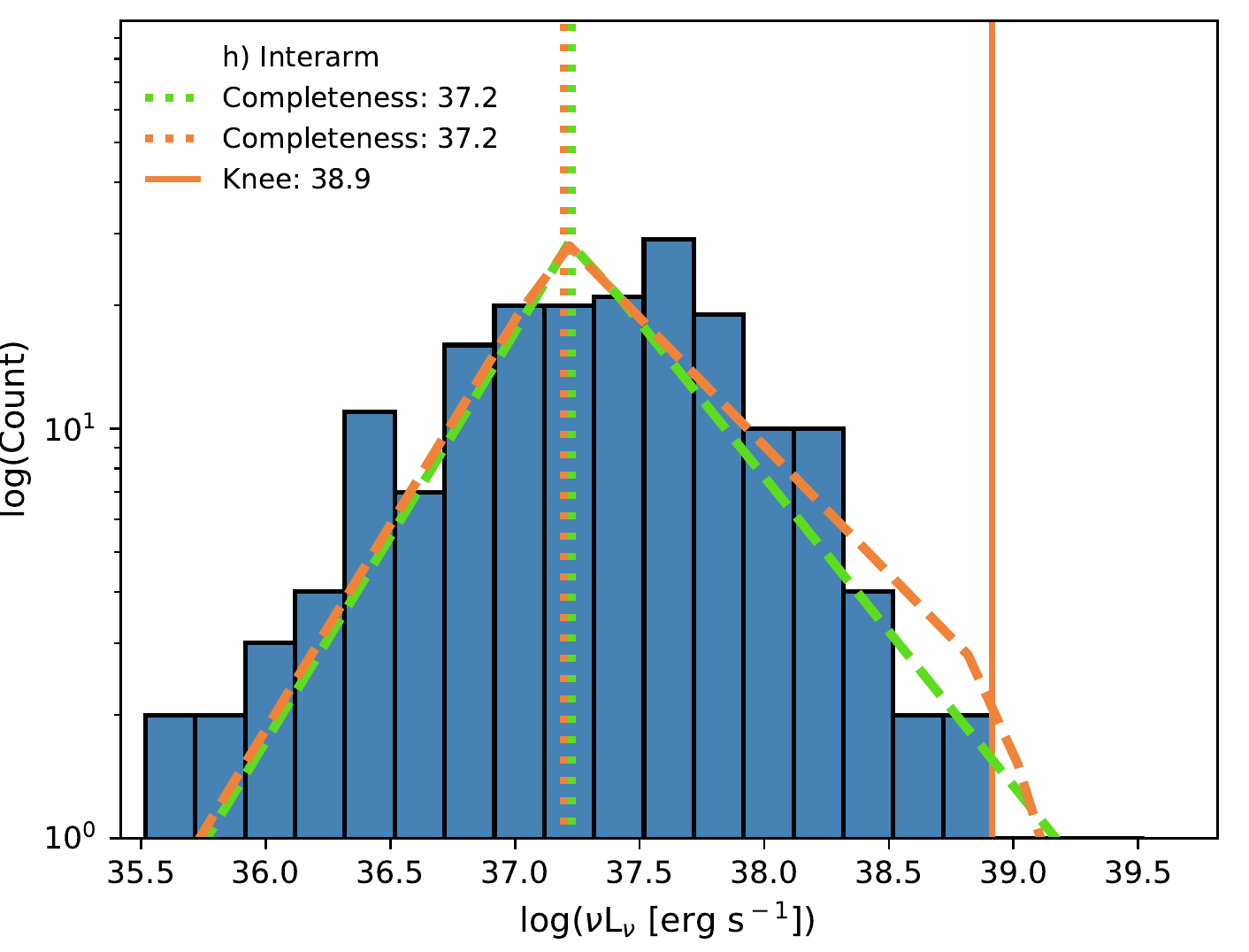}}
\caption{Single and double power law fits to the catalog-only $24\,\microns$ MIPSGAL data subsets: $d_\sun \leq 7.75$ \kpc (panel \subref*{fig:mipsgal_neardist}), $d_\sun > 7.75$ \kpc (panel \subref*{fig:mipsgal_fardist}), $\rgal \leq 5$ \kpc (panel \subref*{fig:mipsgal_nearrgal}), $\rgal > 5$ \kpc (panel \subref*{fig:mipsgal_farrgal}), $r \leq 2.4 \pc$ (panel \subref*{fig:mipsgal_small}), $r > 2.4 \pc$ (panel \subref*{fig:mipsgal_large}), arm (panel \subref*{fig:mipsgal_arm}), and interarm (panel \subref*{fig:mipsgal_interarm}).}
\label{fig:mipsgal2}
\end{figure*}

\begin{figure*}[h]
\centering
  \subfloat{\label{fig:higal70_neardist}%
    \includegraphics[scale=0.5,trim={3.75cm 8.5cm 3.5cm 8.5cm}, clip]{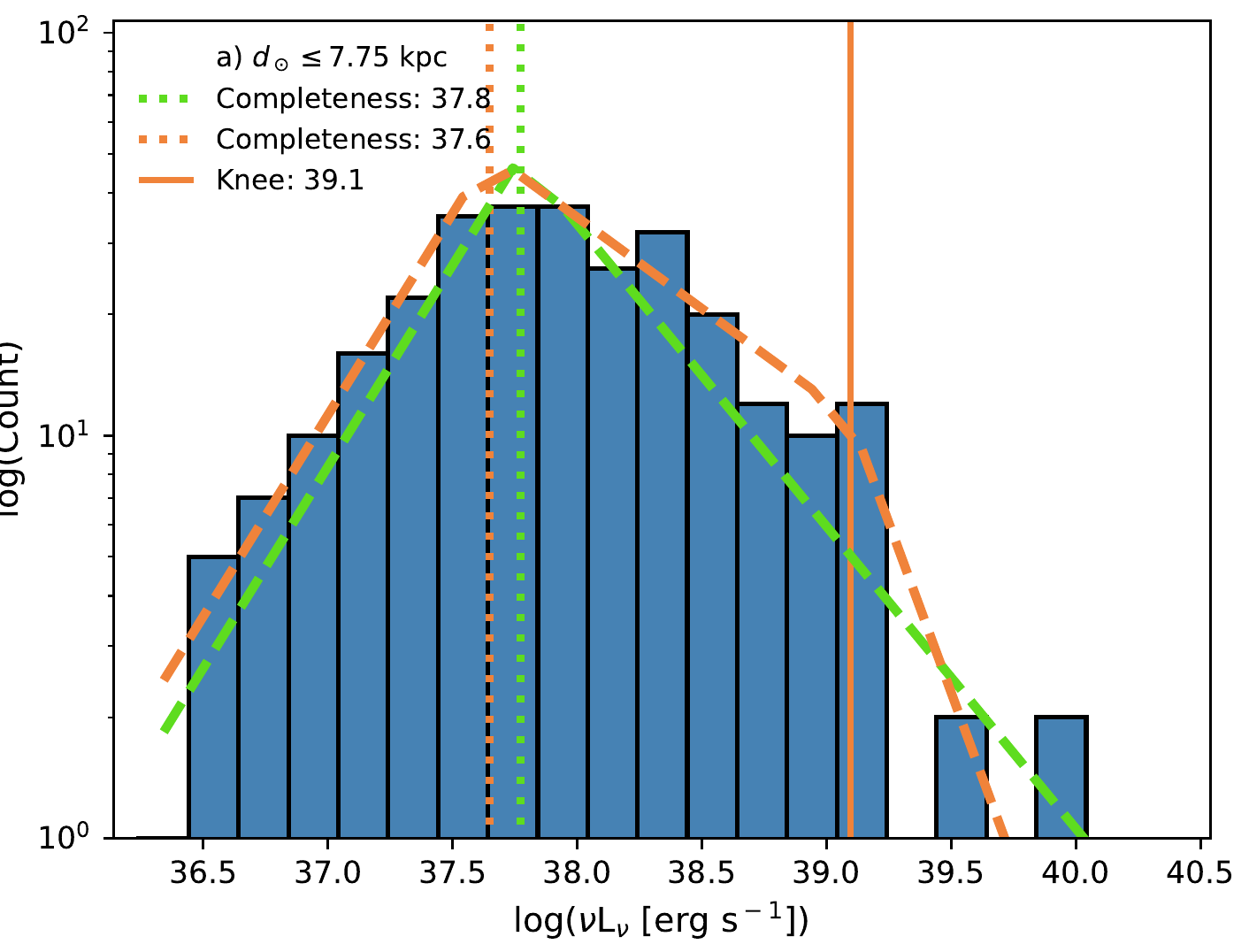}}\qquad
  \subfloat{\label{fig:higal70_fardist}%
    \includegraphics[scale=0.5,trim={3.75cm 8.5cm 3.5cm 8.5cm}, clip]{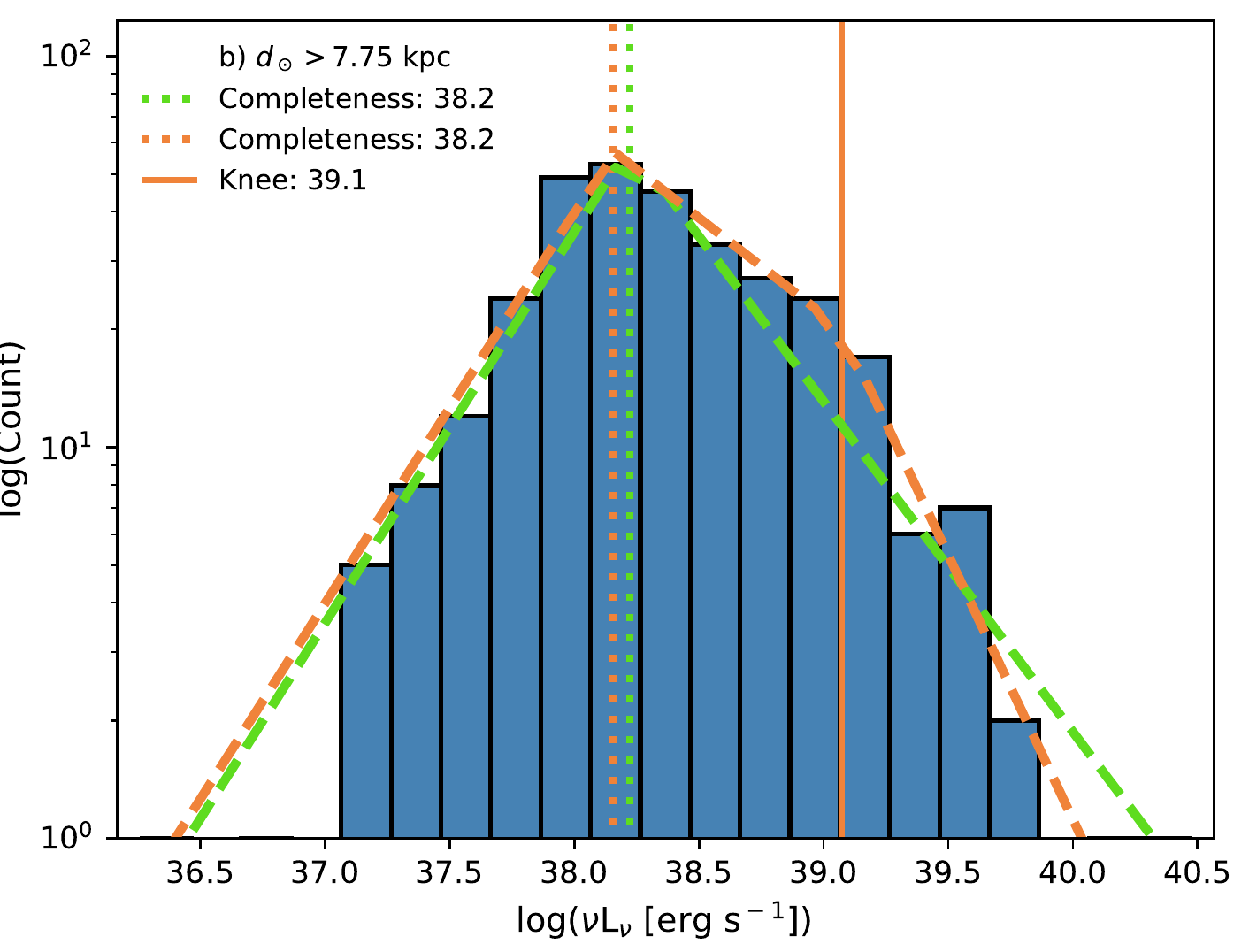}}\\
  \subfloat{\label{fig:higal70_nearrgal}%
    \includegraphics[scale=0.5,trim={3.75cm 8.5cm 3.5cm 8.5cm}, clip]{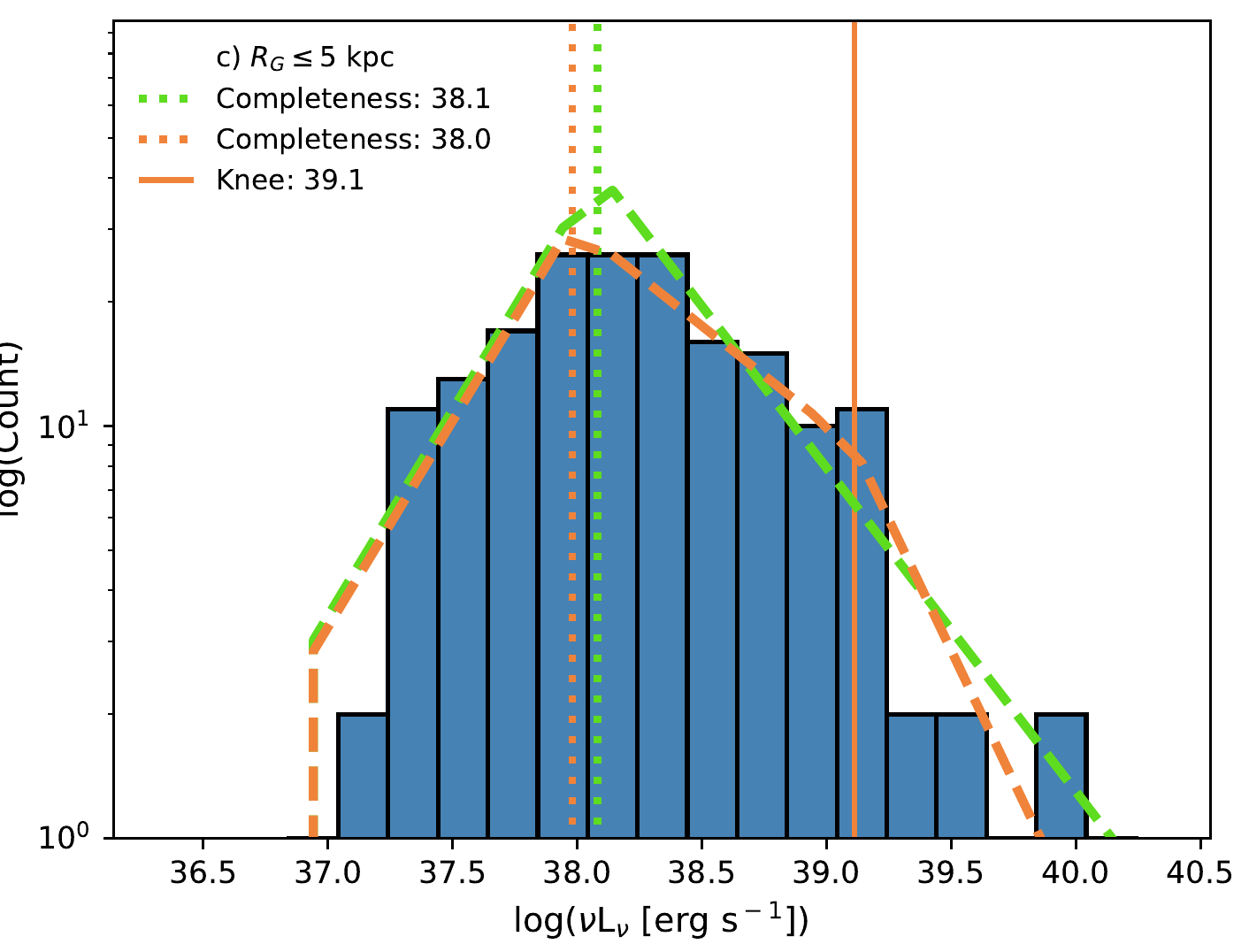}}\qquad
  \subfloat{\label{fig:higal70_farrgal}%
    \includegraphics[scale=0.5,trim={3.75cm 8.5cm 3.5cm 8.5cm}, clip]{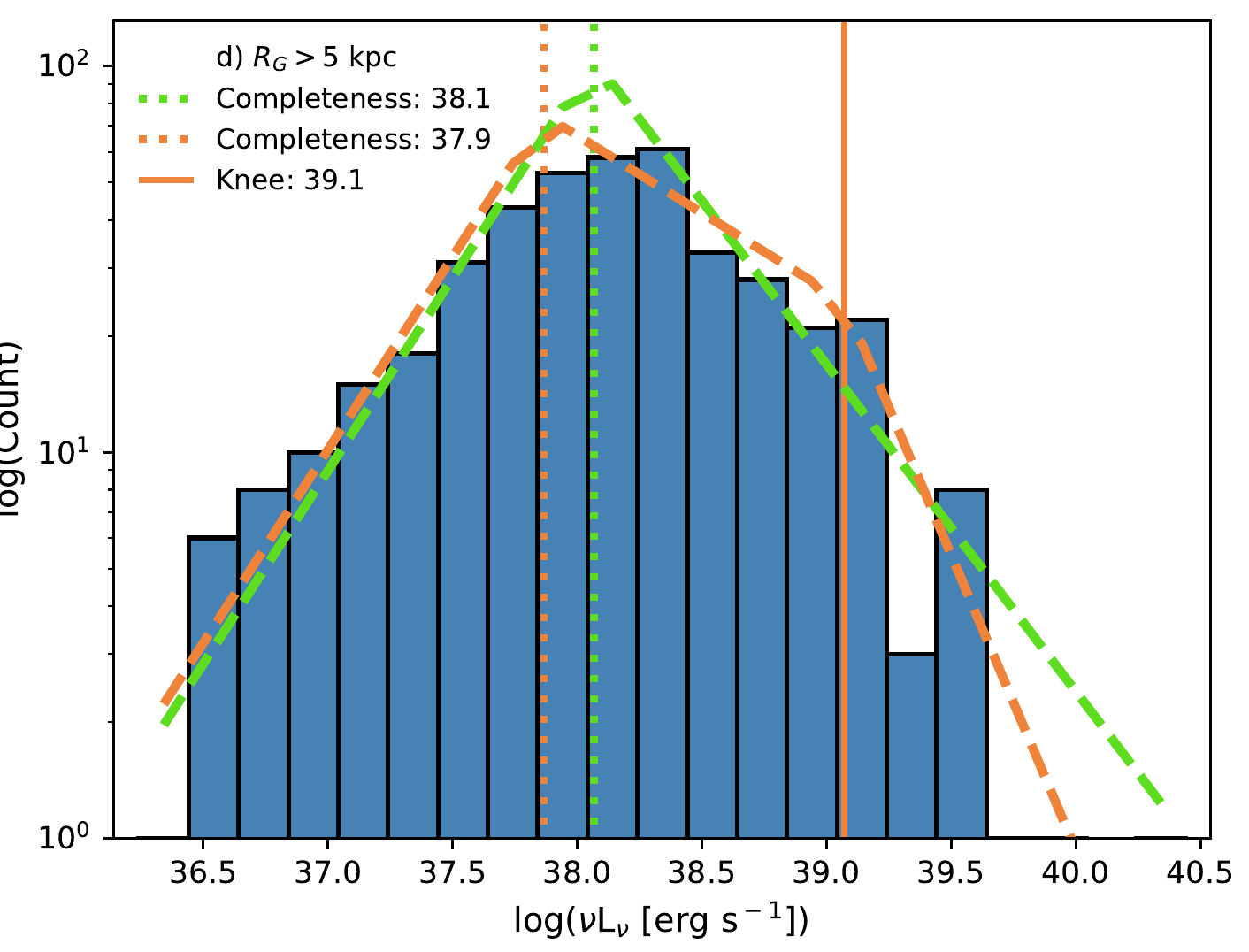}}\\
  \subfloat{\label{fig:higal70_small}%
    \includegraphics[scale=0.5,trim={3.75cm 8.5cm 3.5cm 8.5cm}, clip]{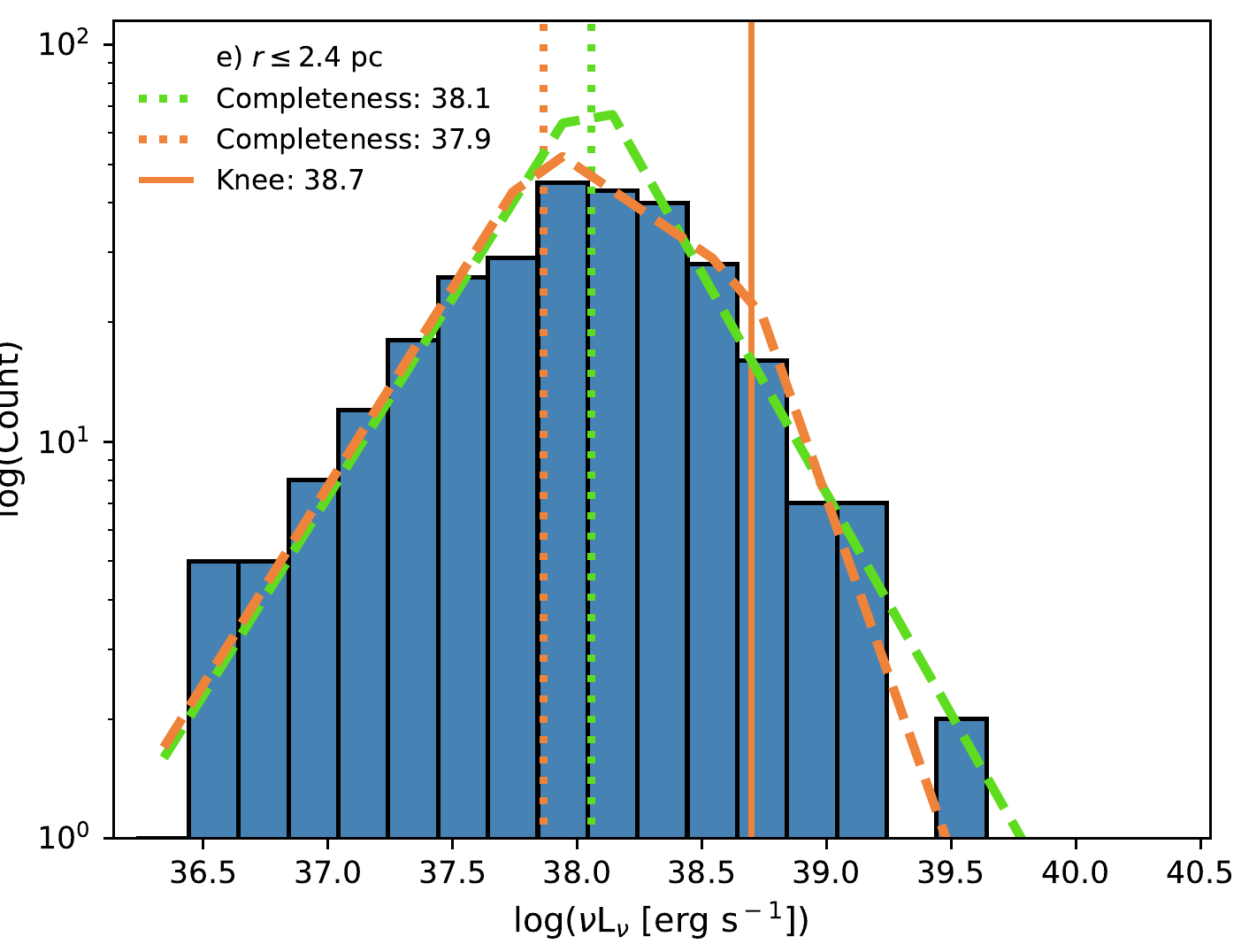}}\qquad
  \subfloat{\label{fig:higal70_large}%
    \includegraphics[scale=0.5,trim={3.75cm 8.5cm 3.5cm 8.5cm}, clip]{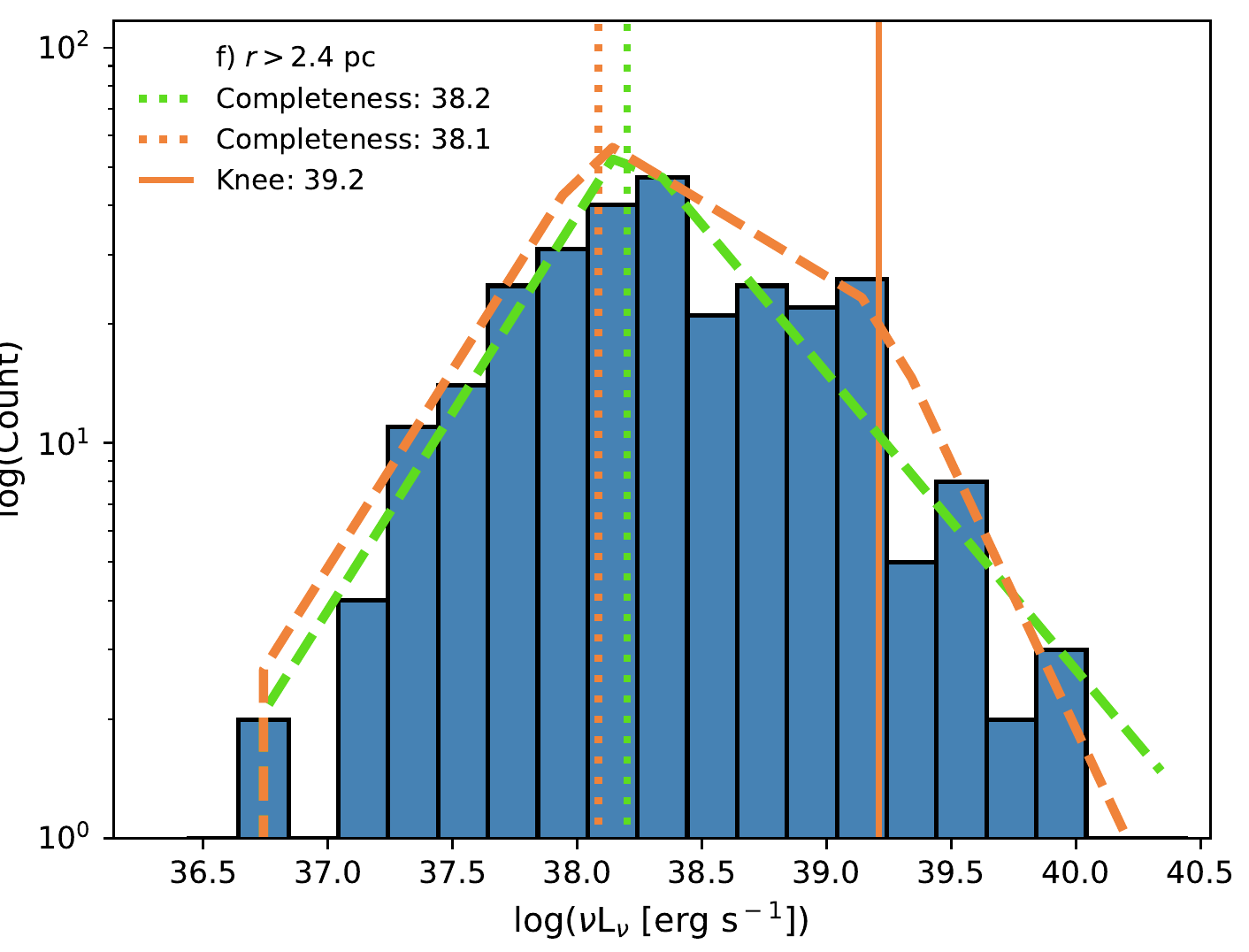}}\\
  \subfloat{\label{fig:higal70_arm}%
    \includegraphics[scale=0.5,trim={3.75cm 8.5cm 3.5cm 8.5cm}, clip]{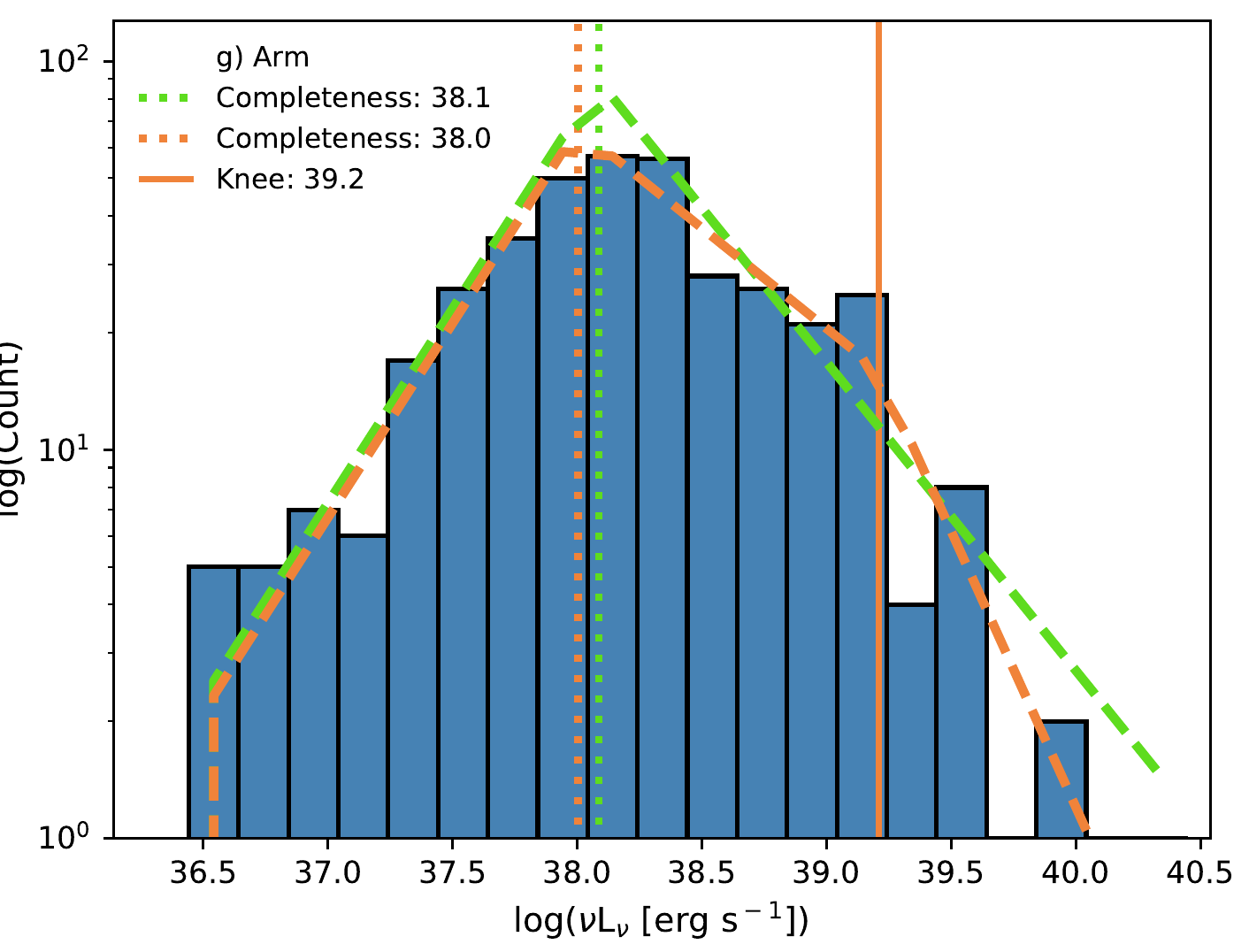}}\qquad
  \subfloat{\label{fig:higal70_interarm}%
    \includegraphics[scale=0.5,trim={3.75cm 8.5cm 3.5cm 8.5cm}, clip]{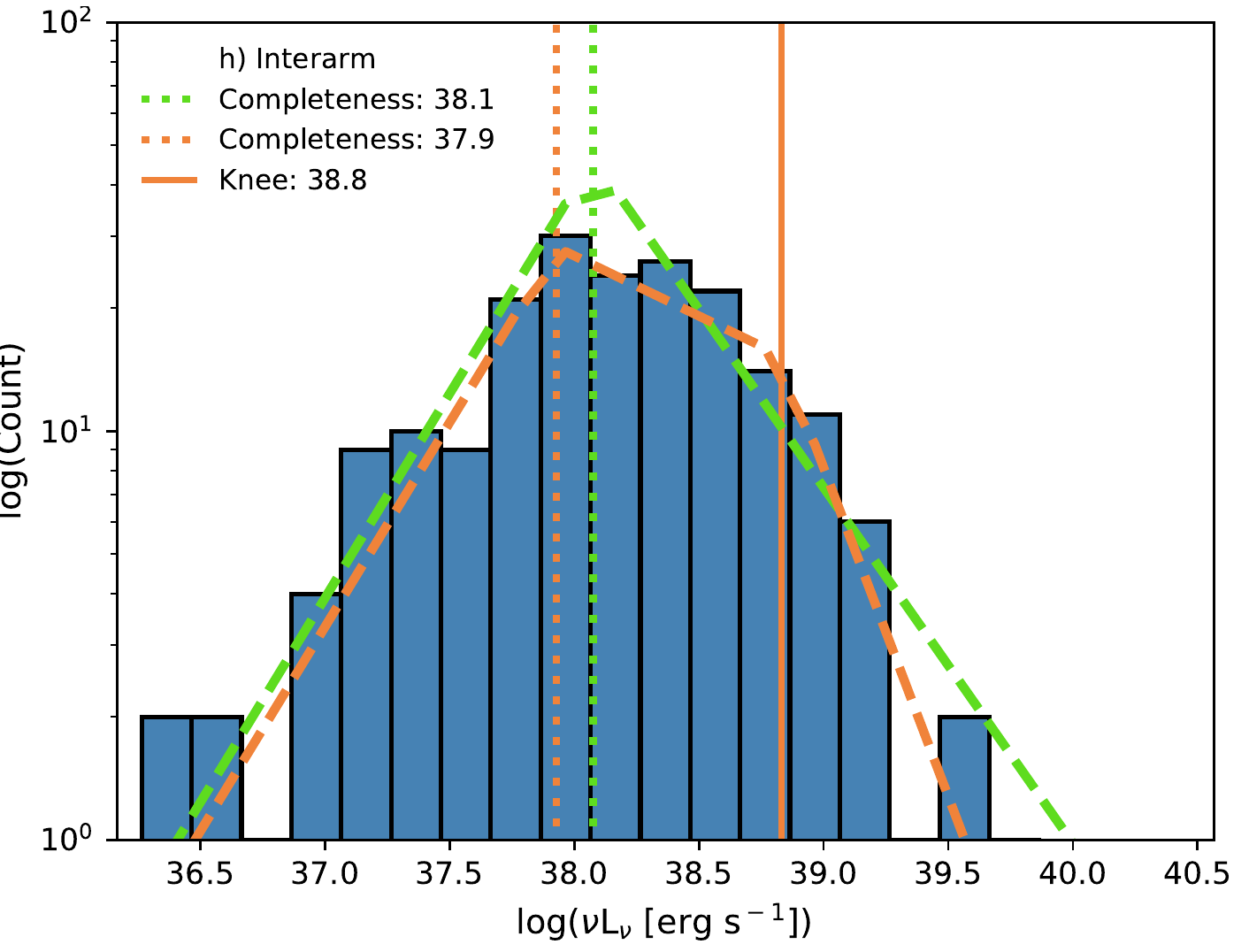}}
\caption{Single and double power law fits to the catalog-only $70\,\microns$ Hi-GAL data subsets: $d_\sun \leq 7.75$ \kpc (panel \subref*{fig:higal70_neardist}), $d_\sun > 7.75$ \kpc (panel \subref*{fig:higal70_fardist}), $\rgal \leq 5$ \kpc (panel \subref*{fig:higal70_nearrgal}), $\rgal > 5$ \kpc (panel \subref*{fig:higal70_farrgal}), $r \leq 2.4 \pc$ (panel \subref*{fig:higal70_small}), $r > 2.4 \pc$ (panel \subref*{fig:higal70_large}), arm (panel \subref*{fig:higal70_arm}), and interarm (panel \subref*{fig:higal70_interarm}). }
\label{fig:higal702}
\end{figure*}

\begin{figure*}[h]
\centering
  \subfloat{\label{fig:higal160_neardist}%
    \includegraphics[scale=0.5,trim={3.75cm 8.5cm 3.5cm 8.5cm}, clip]{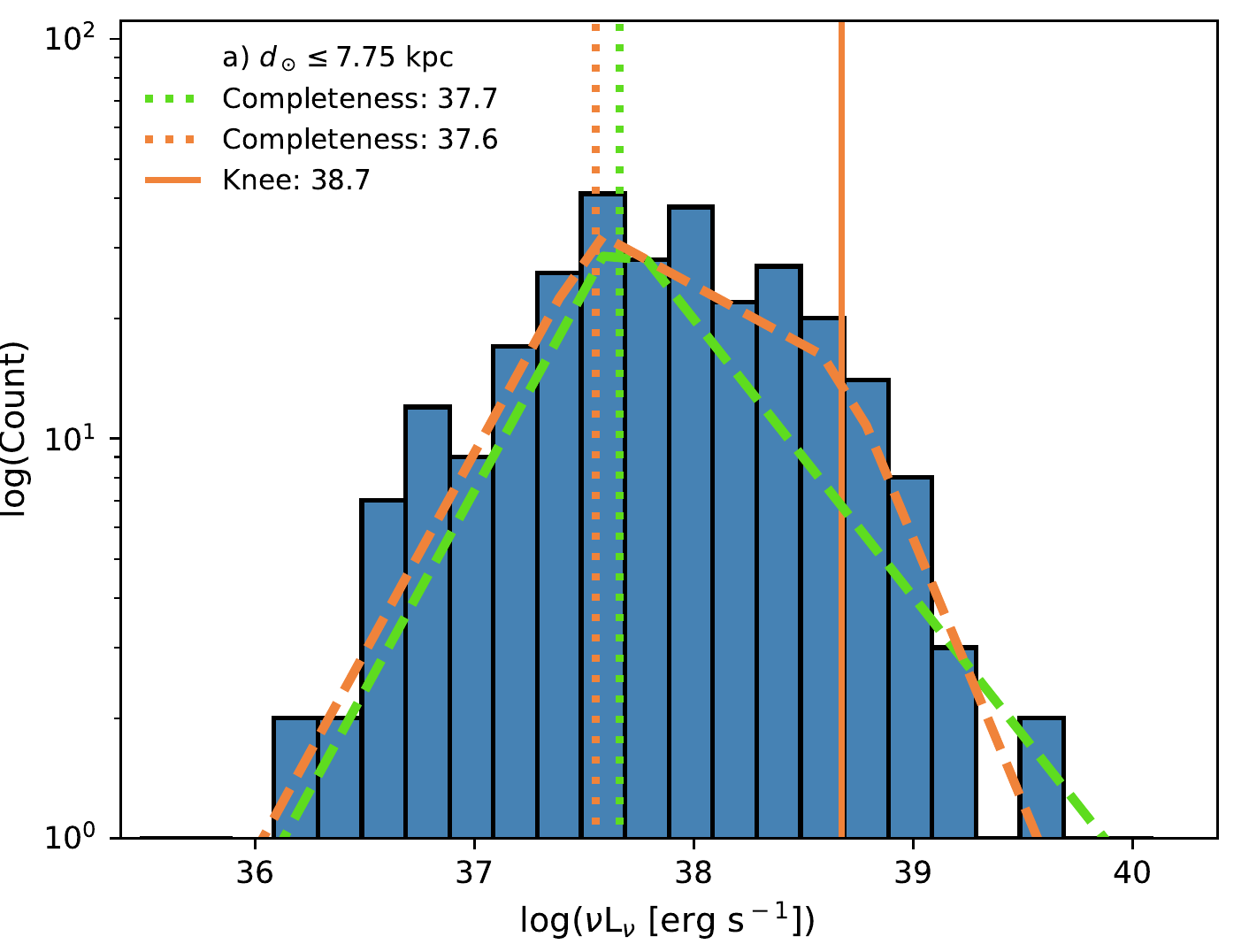}}\qquad
  \subfloat{\label{fig:higal160_fardist}%
    \includegraphics[scale=0.5,trim={3.75cm 8.5cm 3.5cm 8.5cm}, clip]{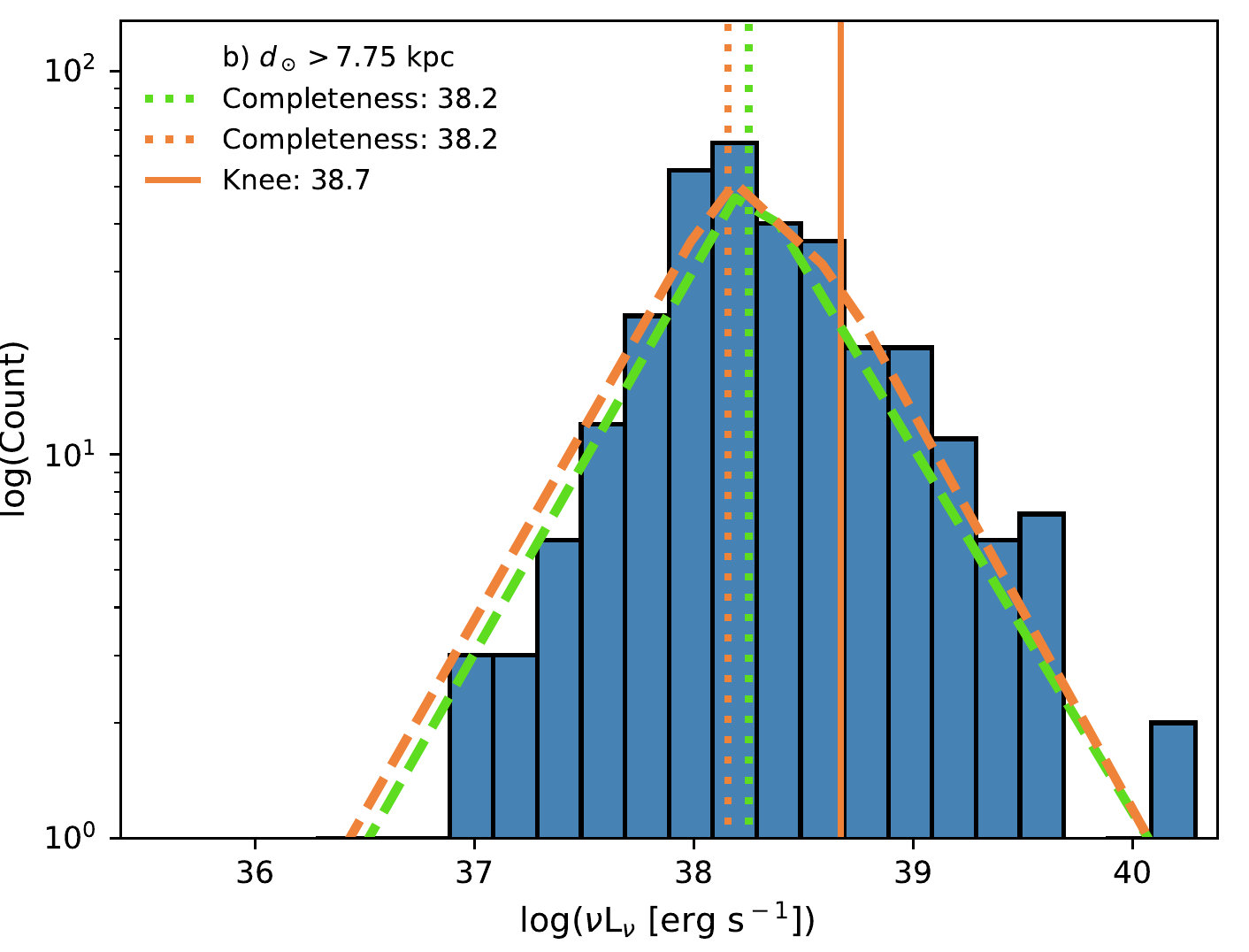}}\\
  \subfloat{\label{fig:higal160_nearrgal}%
    \includegraphics[scale=0.5,trim={3.75cm 8.5cm 3.5cm 8.5cm}, clip]{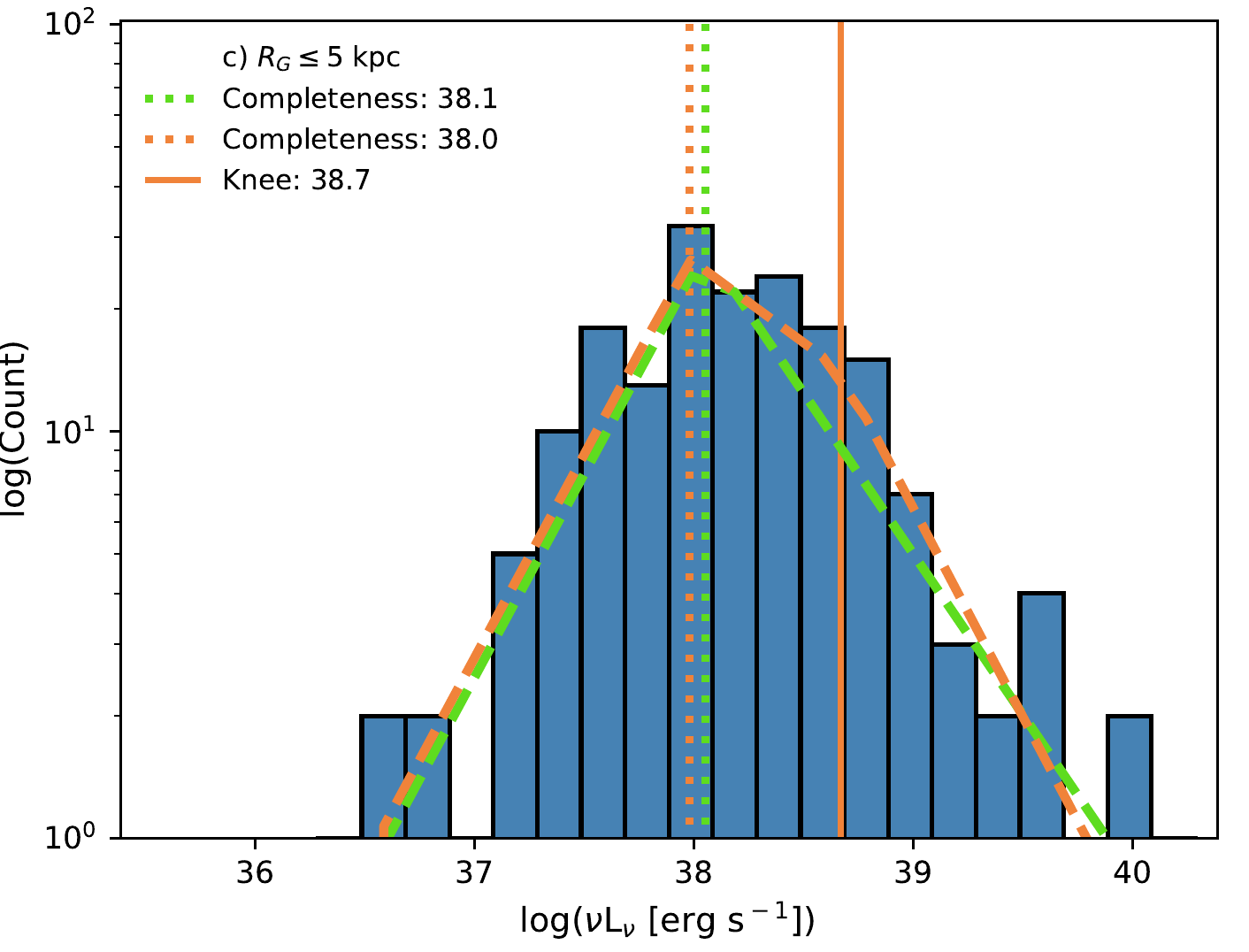}}\qquad
  \subfloat{\label{fig:higal160_farrgal}%
    \includegraphics[scale=0.5,trim={3.75cm 8.5cm 3.5cm 8.5cm}, clip]{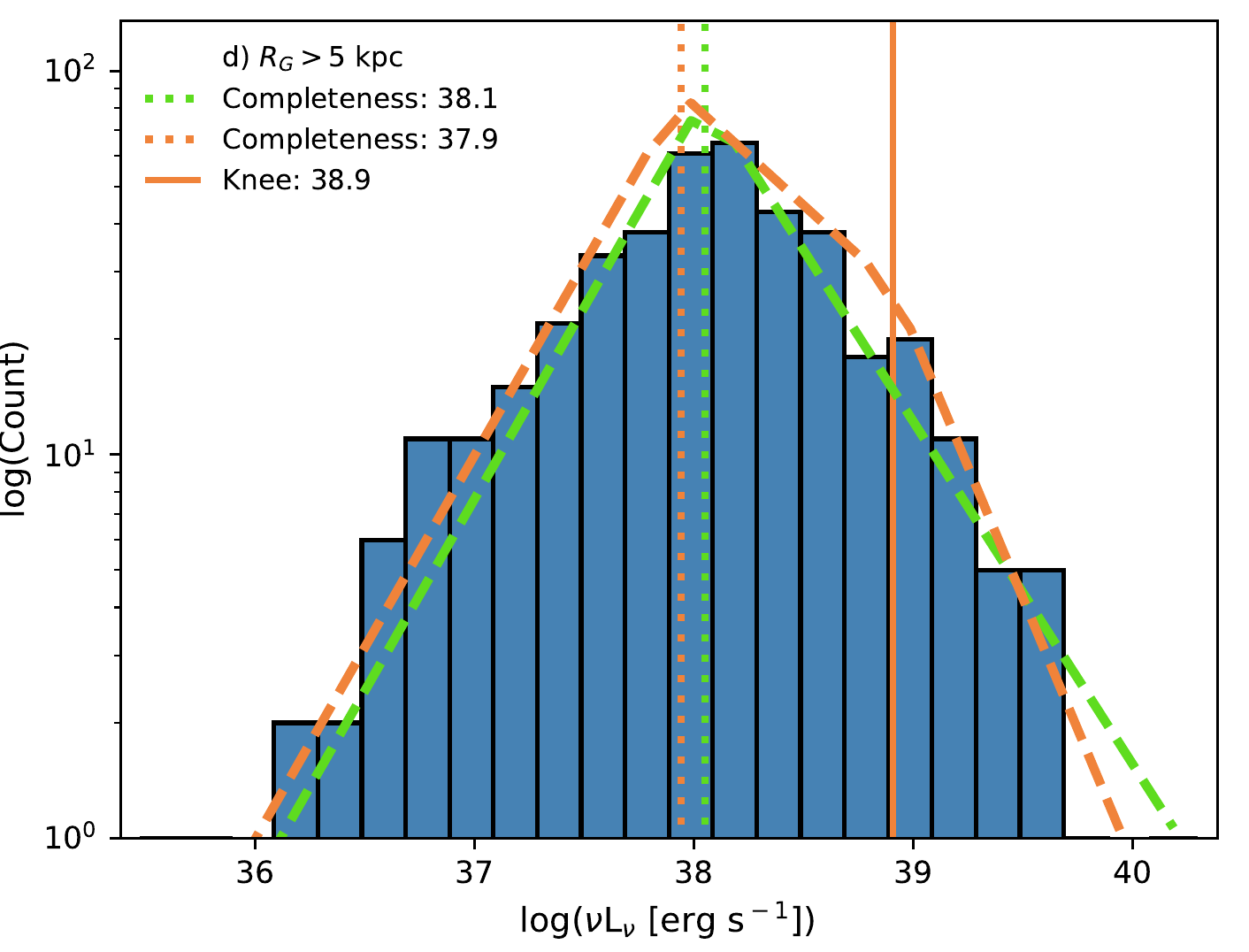}}\\
  \subfloat{\label{fig:higal160_small}%
    \includegraphics[scale=0.5,trim={3.75cm 8.5cm 3.5cm 8.5cm}, clip]{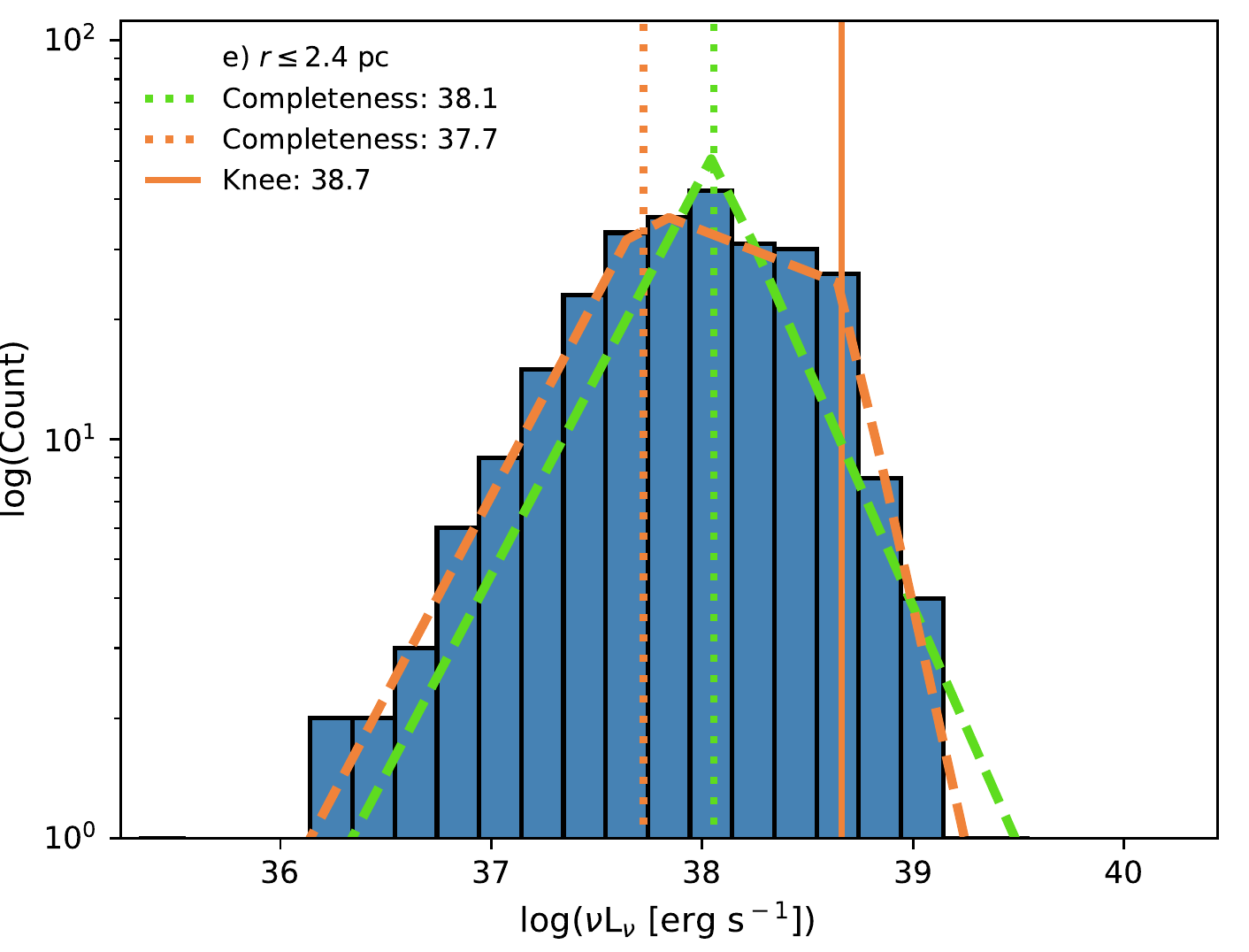}}\qquad
  \subfloat{\label{fig:higal160_large}%
    \includegraphics[scale=0.5,trim={3.75cm 8.5cm 3.5cm 8.5cm}, clip]{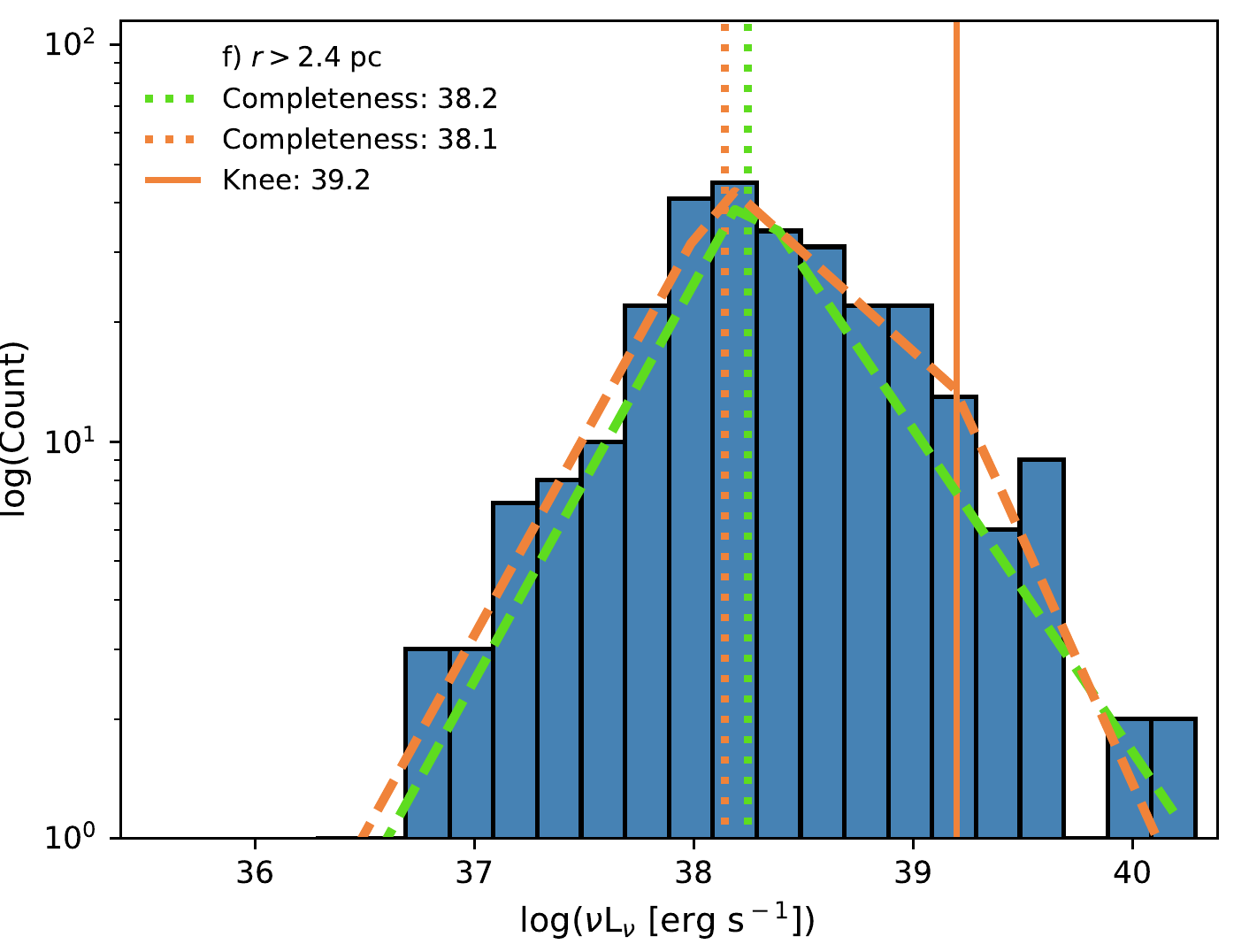}}\\
  \subfloat{\label{fig:higal160_arm}%
    \includegraphics[scale=0.5,trim={3.75cm 8.5cm 3.5cm 8.5cm}, clip]{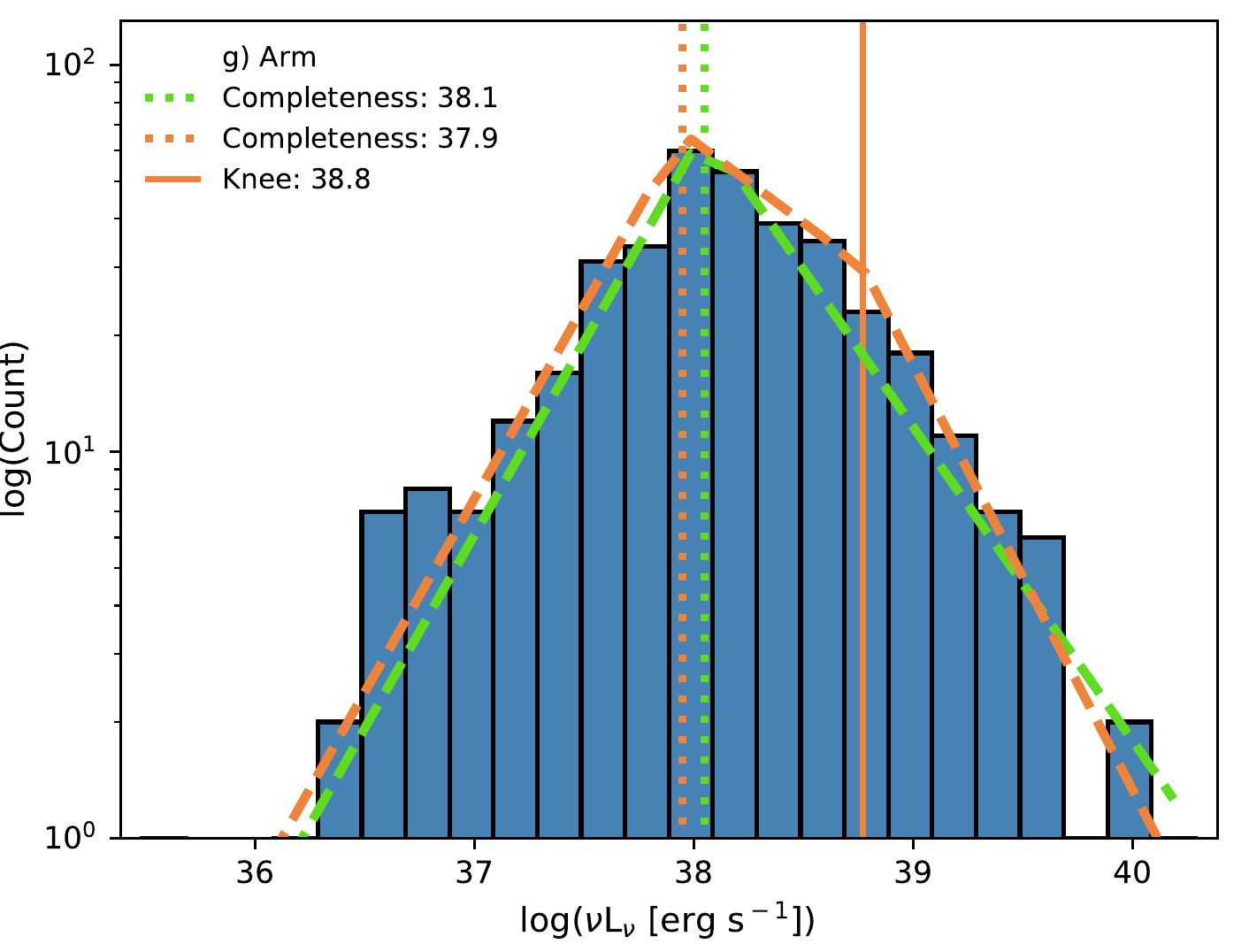}}\qquad
  \subfloat{\label{fig:higal160_interarm}%
    \includegraphics[scale=0.5,trim={3.75cm 8.5cm 3.5cm 8.5cm}, clip]{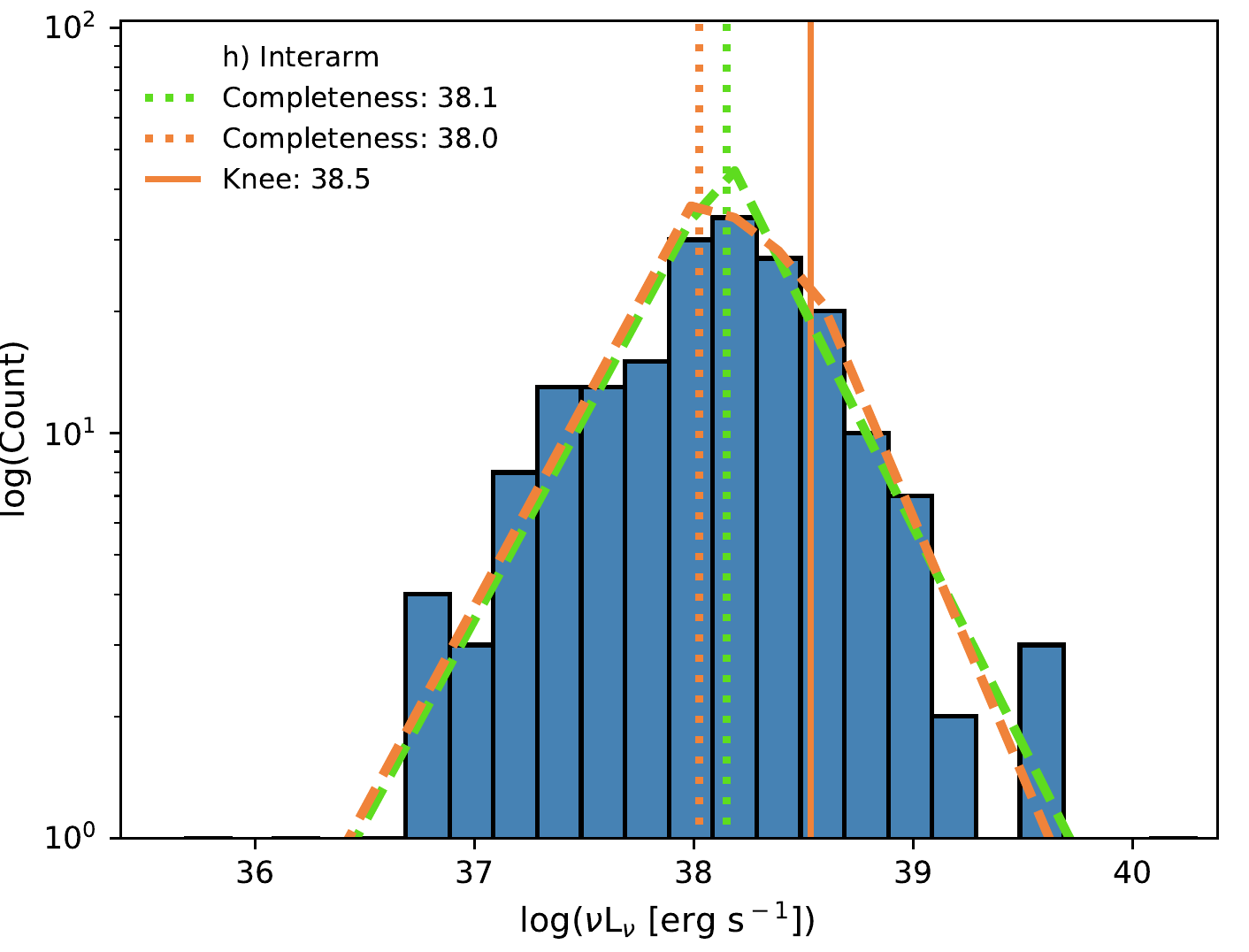}}
\caption{Single and double power law fits to the catalog-only $160\,\microns$ Hi-GAL data subsets: $d_\sun \leq 7.75$ \kpc (panel \subref*{fig:higal160_neardist}), $d_\sun > 7.75$ \kpc (panel \subref*{fig:higal160_fardist}), $\rgal \leq 5$ \kpc (panel \subref*{fig:higal160_nearrgal}), $\rgal > 5$ \kpc (panel \subref*{fig:higal160_farrgal}), $r \leq 2.4 \pc$ (panel \subref*{fig:higal160_small}), $r > 2.4 \pc$ (panel \subref*{fig:higal160_large}), arm (panel \subref*{fig:higal160_arm}), and interarm (panel \subref*{fig:higal160_interarm}).}
\label{fig:higal1602}
\end{figure*}

\begin{sidewaysfigure*}[h]
\centering
  \subfloat{\label{fig:magpis_neardist}%
    \includegraphics[scale=0.385,trim={3.75cm 6.25cm 3.5cm 6.5cm}, clip]{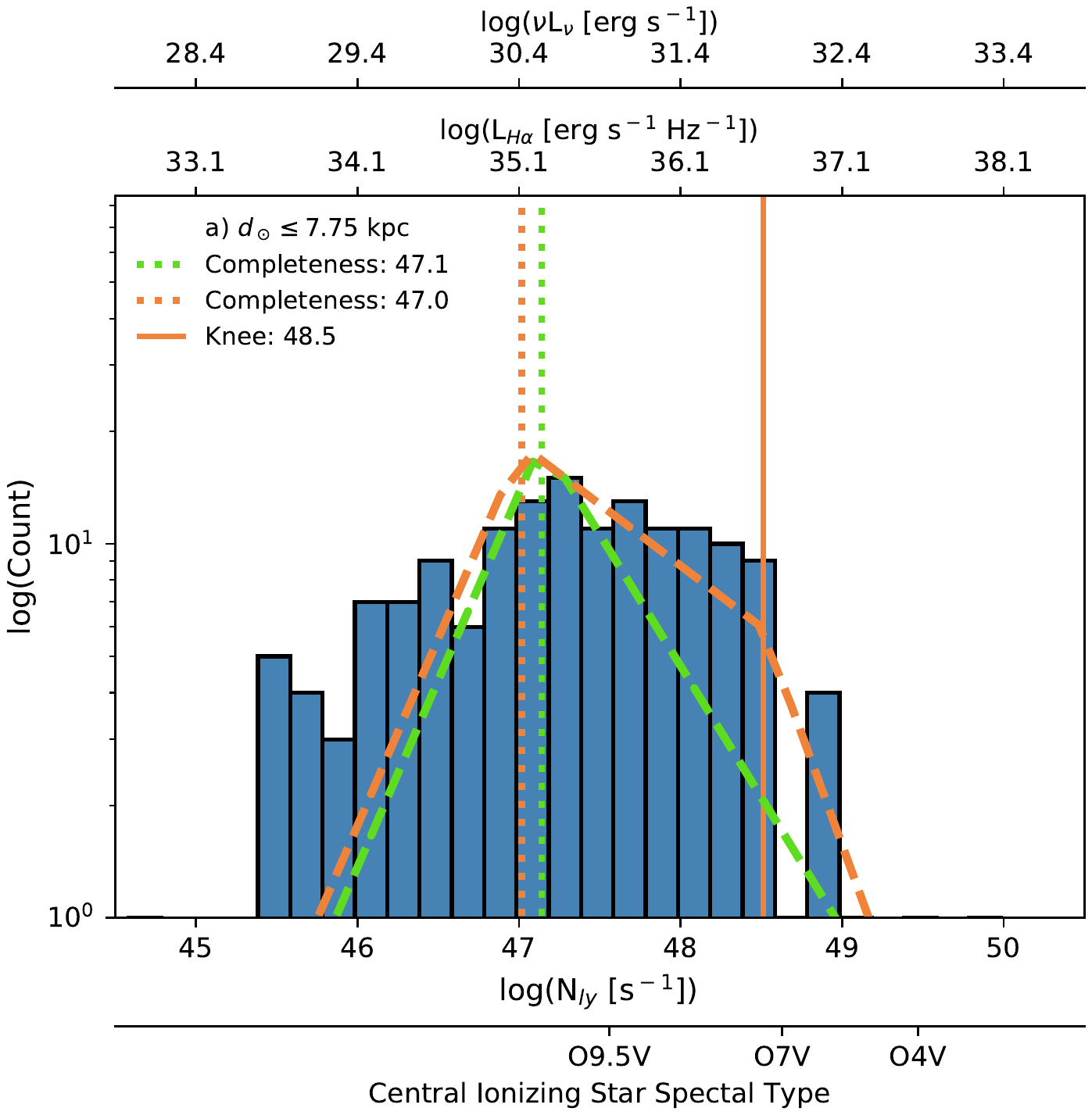}}\qquad
  \subfloat{\label{fig:magpis_fardist}%
    \includegraphics[scale=0.385,trim={3.75cm 6.25cm 3.5cm 6.5cm}, clip]{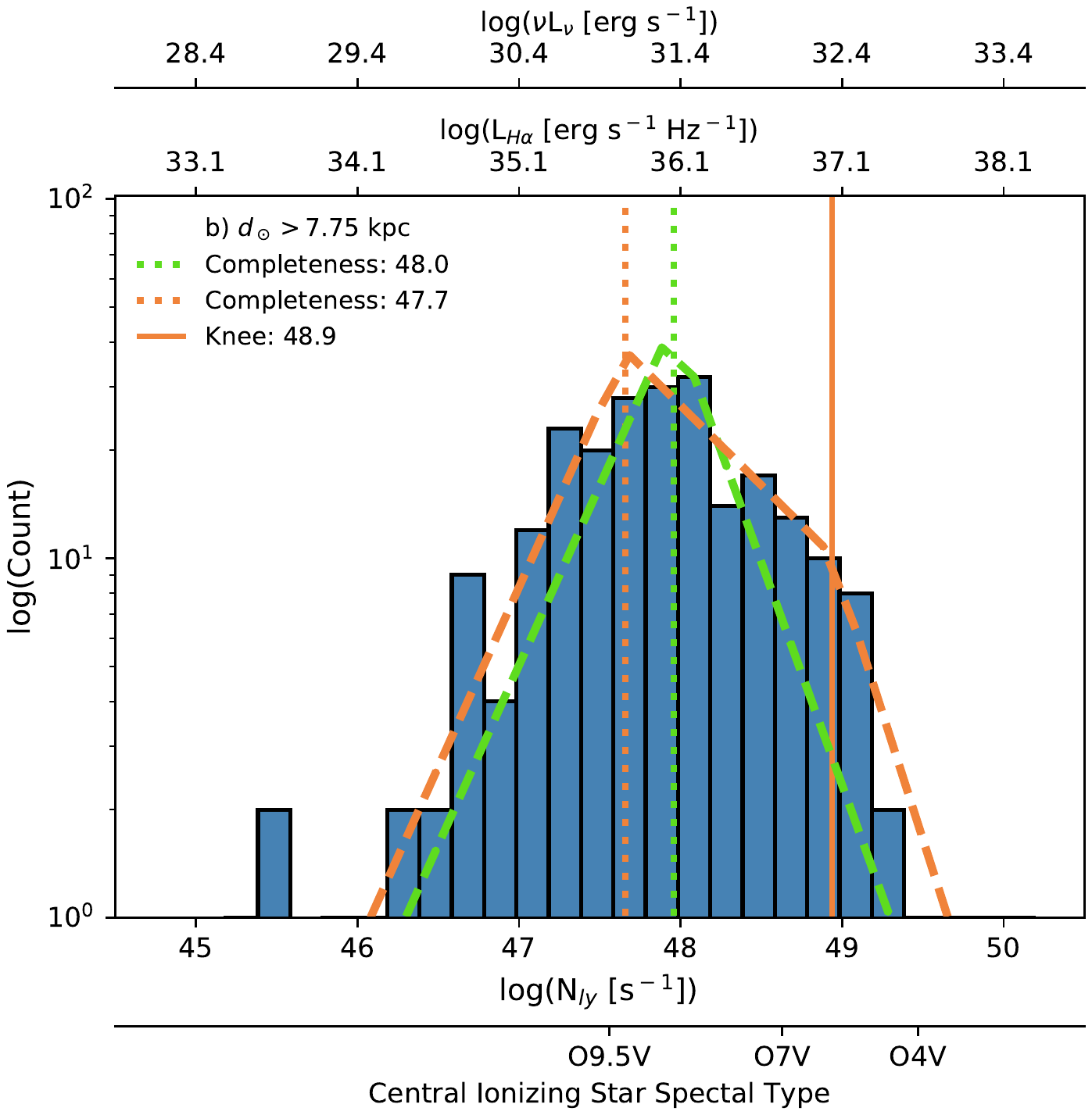}}\qquad
  \subfloat{\label{fig:magpis_nearrgal}%
    \includegraphics[scale=0.385,trim={3.75cm 6.25cm 3.5cm 6.5cm}, clip]{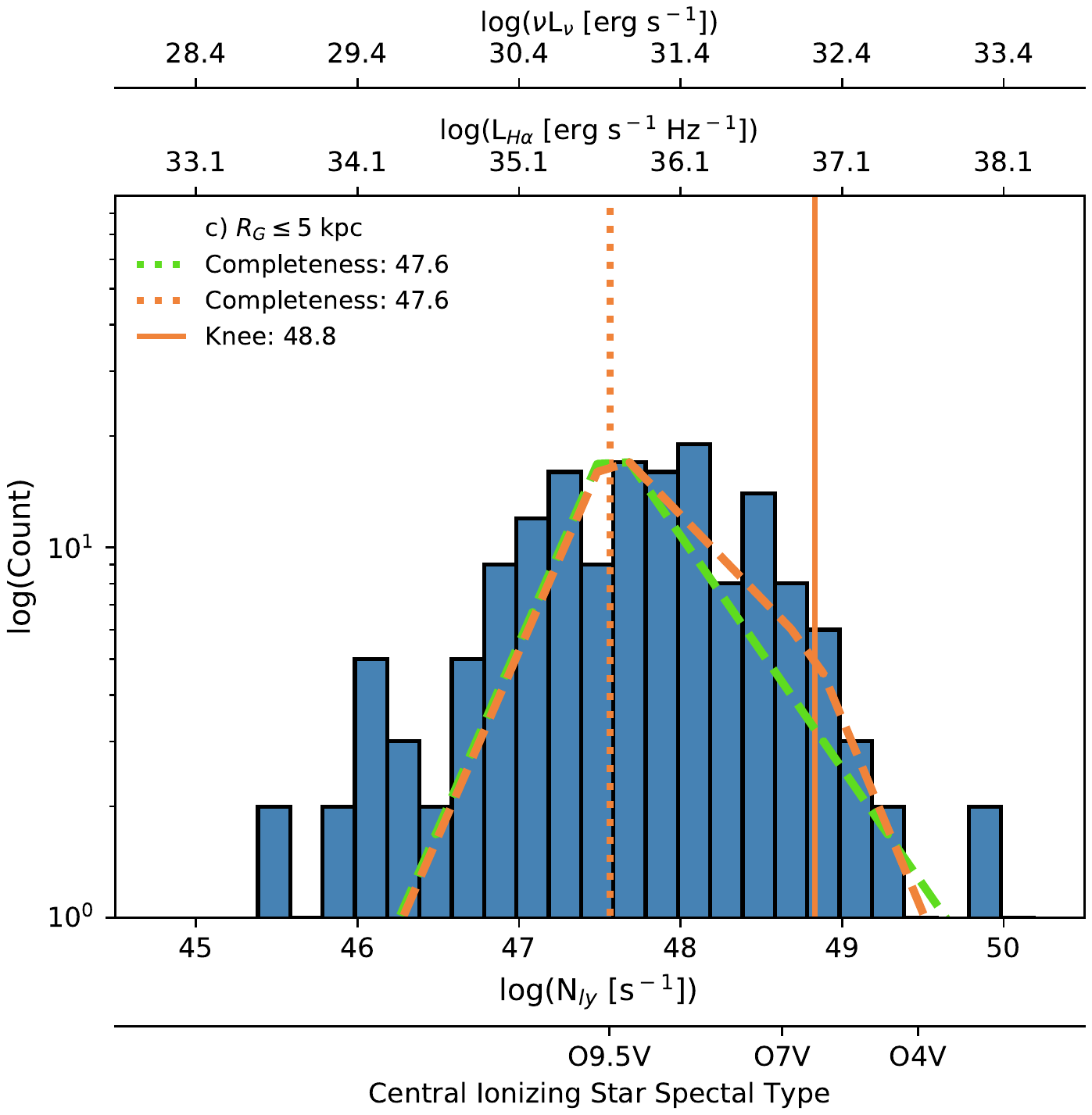}}\\
  \subfloat{\label{fig:magpis_farrgal}%
    \includegraphics[scale=0.385,trim={3.75cm 6.25cm 3.5cm 6.5cm}, clip]{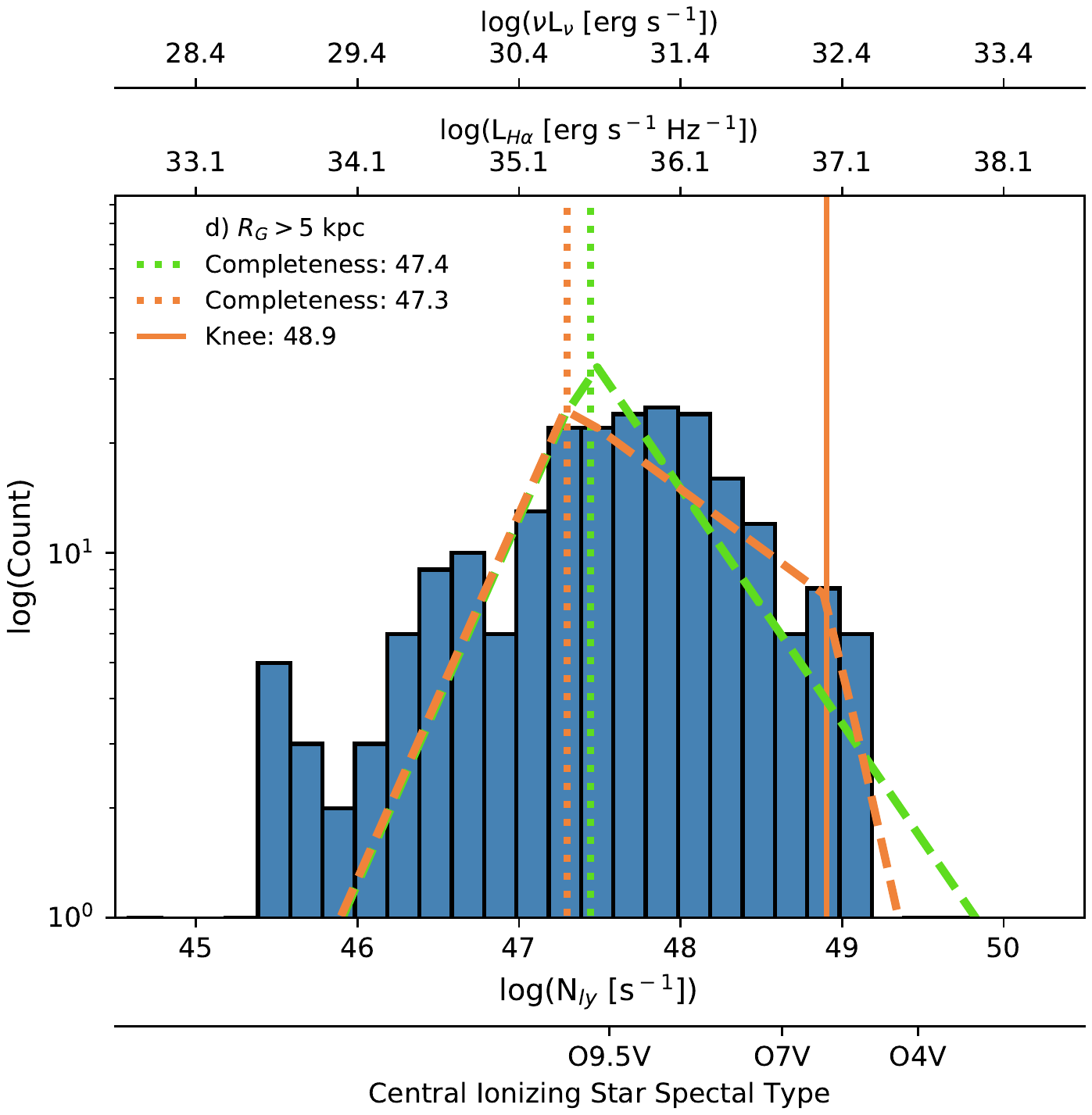}}\qquad
  \subfloat{\label{fig:magpis_small}%
    \includegraphics[scale=0.385,trim={3.75cm 6.25cm 3.5cm 6.5cm}, clip]{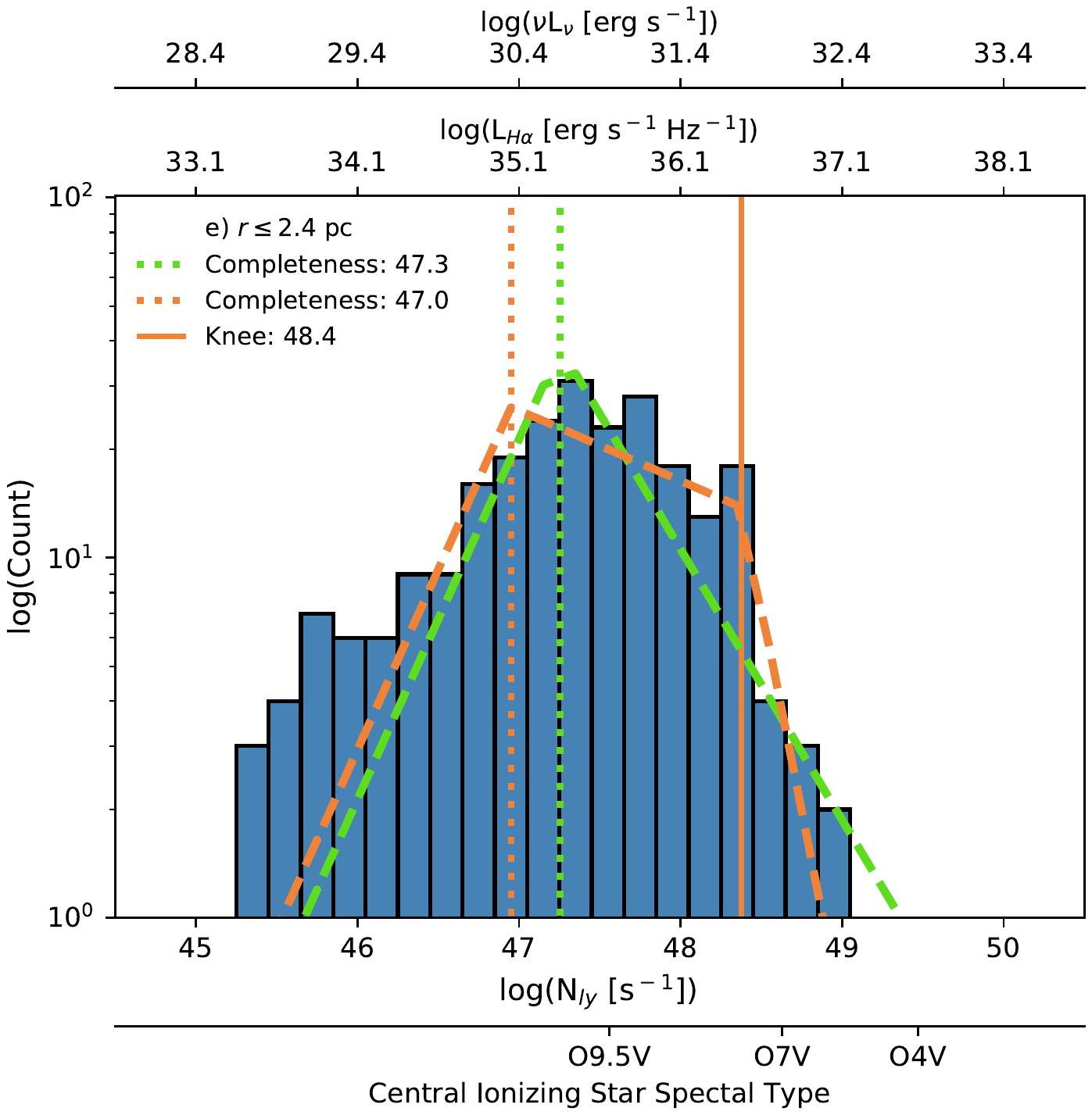}}\qquad
  \subfloat{\label{fig:magpis_large}%
    \includegraphics[scale=0.385,trim={3.75cm 6.25cm 3.5cm 6.5cm}, clip]{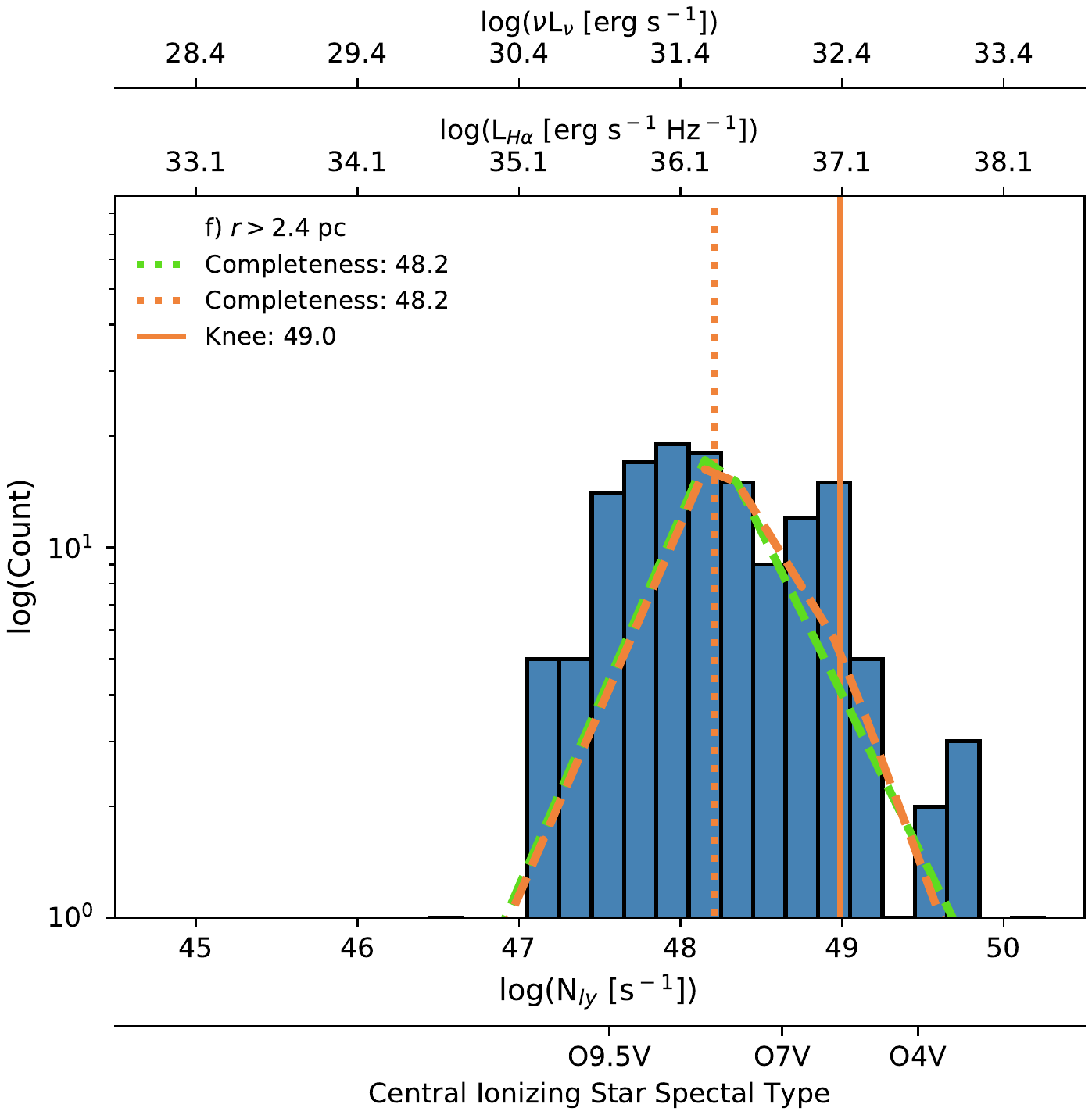}}\\
  \subfloat{\label{fig:magpis_arm}%
    \includegraphics[scale=0.385,trim={3.75cm 6.25cm 3.5cm 6.5cm}, clip]{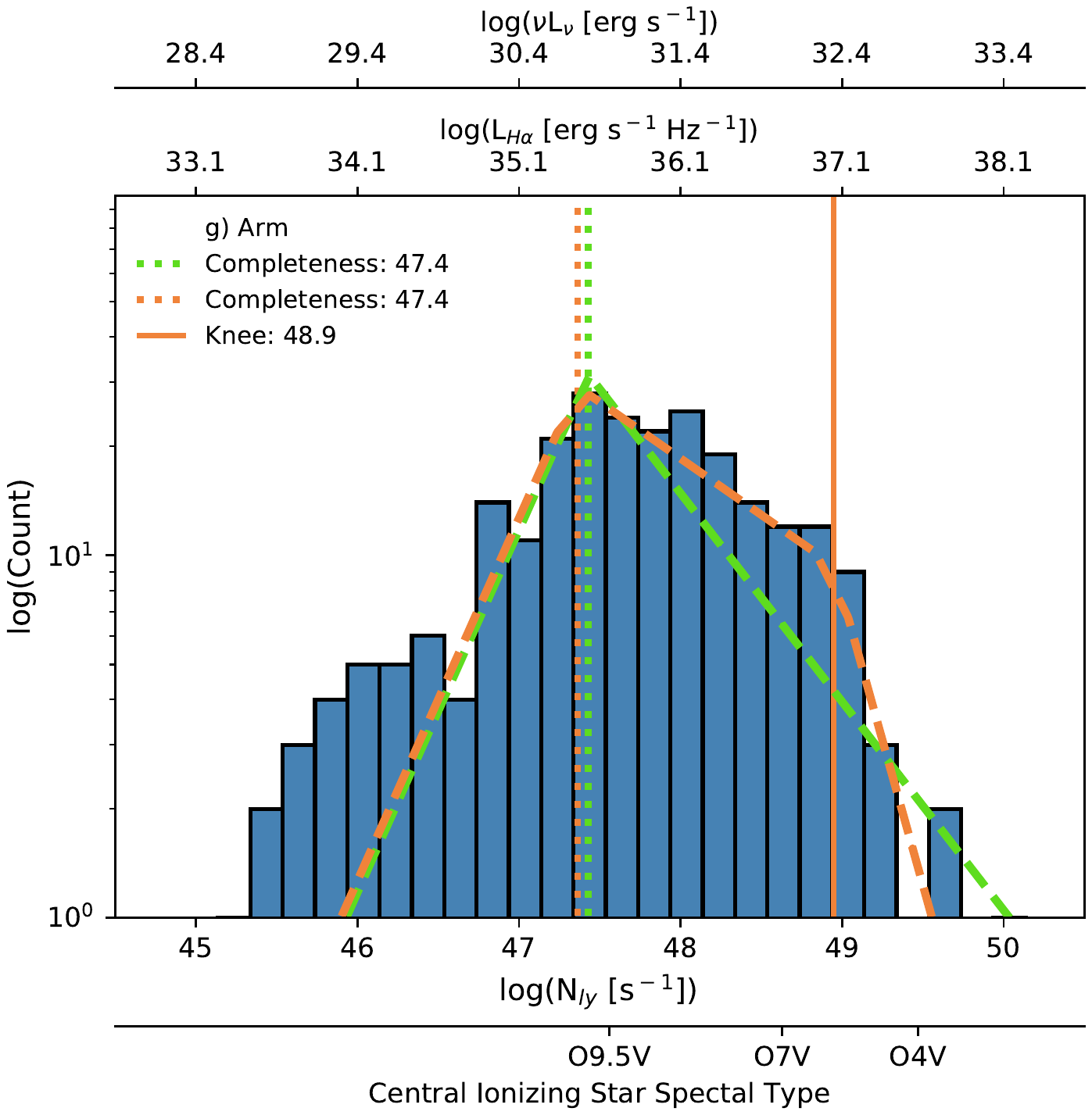}}\qquad
  \subfloat{\label{fig:magpis_interarm}%
    \includegraphics[scale=0.385,trim={3.75cm 6.25cm 3.5cm 6.5cm}, clip]{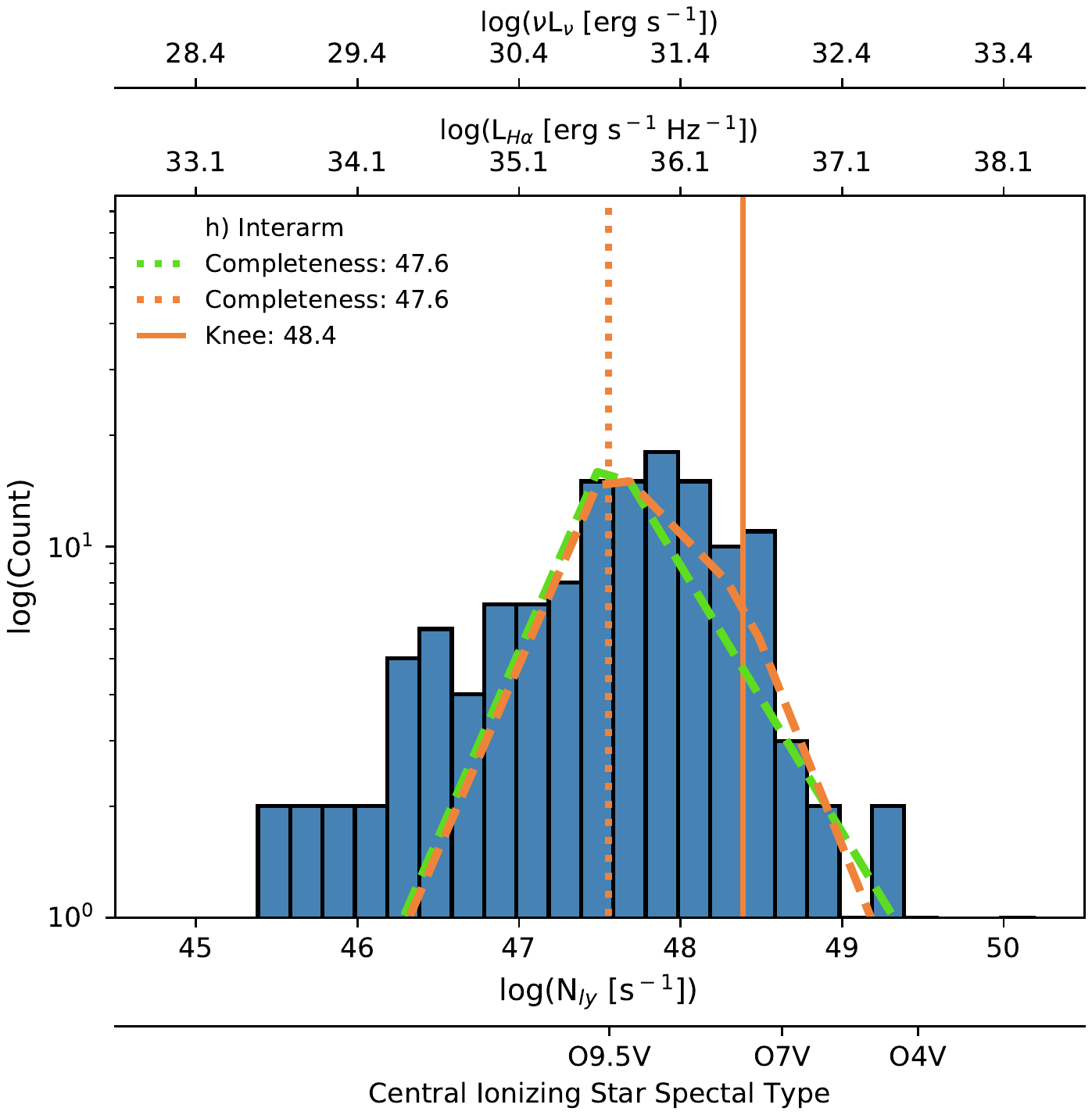}}
\caption{Single and double power law fits to the catalog-only $20\,\cm$ MAGPIS data subsets: $d_\sun \leq 7.75$ \kpc (panel \subref*{fig:magpis_neardist}), $d_\sun > 7.75$ \kpc (panel \subref*{fig:magpis_fardist}), $\rgal \leq 5$ \kpc (panel \subref*{fig:magpis_nearrgal}), $\rgal > 5$ \kpc (panel \subref*{fig:magpis_farrgal}), $r \leq 2.4 \pc$ (panel \subref*{fig:magpis_small}), $r > 2.4 \pc$ (panel \subref*{fig:magpis_large}), arm (panel \subref*{fig:magpis_arm}), and interarm (panel \subref*{fig:magpis_interarm}).}
\label{fig:magpis2}
\end{sidewaysfigure*}

\begin{sidewaysfigure*}[h]
\centering
  \subfloat{\label{fig:magpis_vgps_neardist}%
    \includegraphics[scale=0.385,trim={3.75cm 6.25cm 3.5cm 6.5cm}, clip]{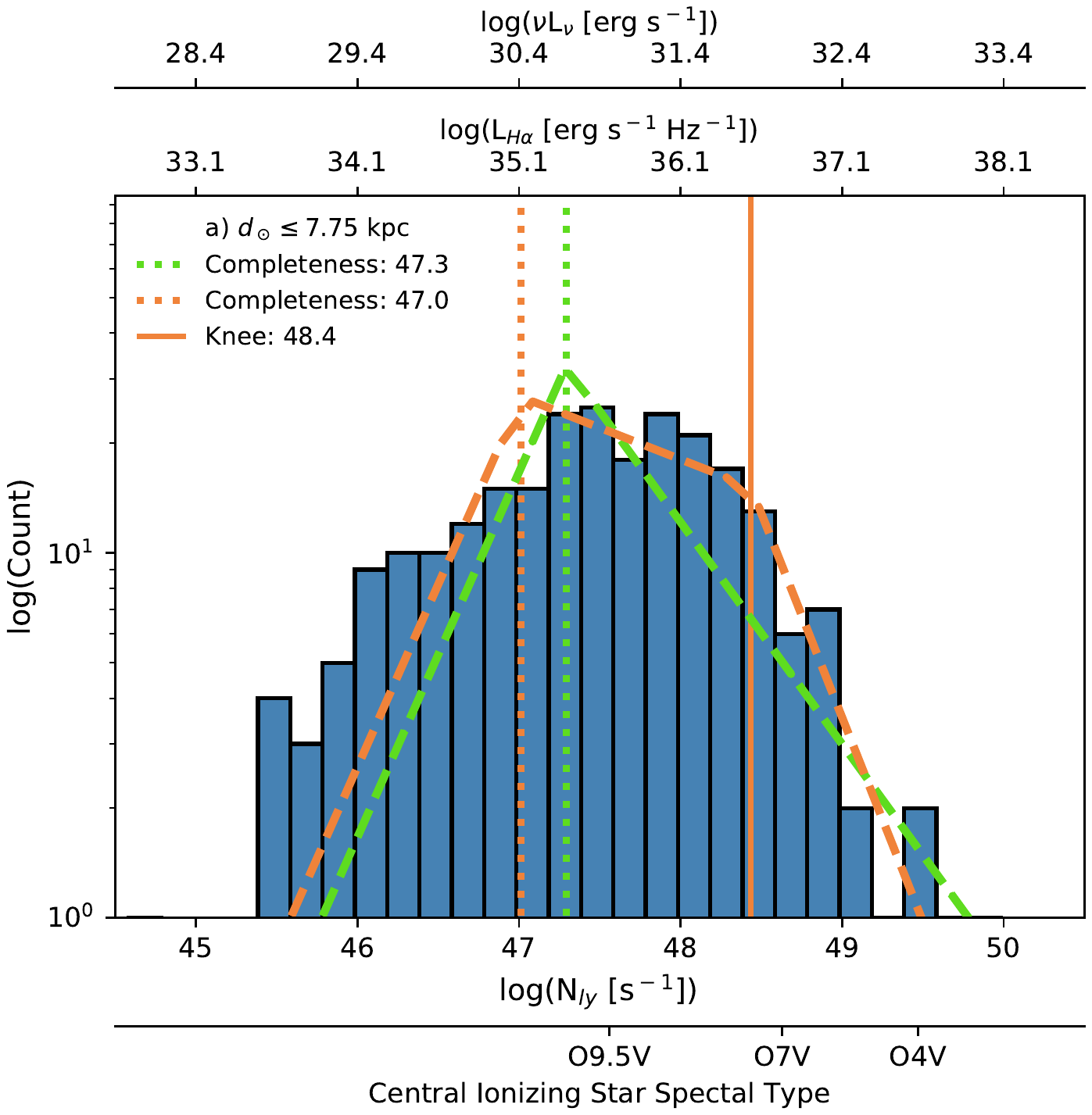}}\qquad
  \subfloat{\label{fig:magpis_vgps_fardist}%
    \includegraphics[scale=0.385,trim={3.75cm 6.25cm 3.5cm 6.5cm}, clip]{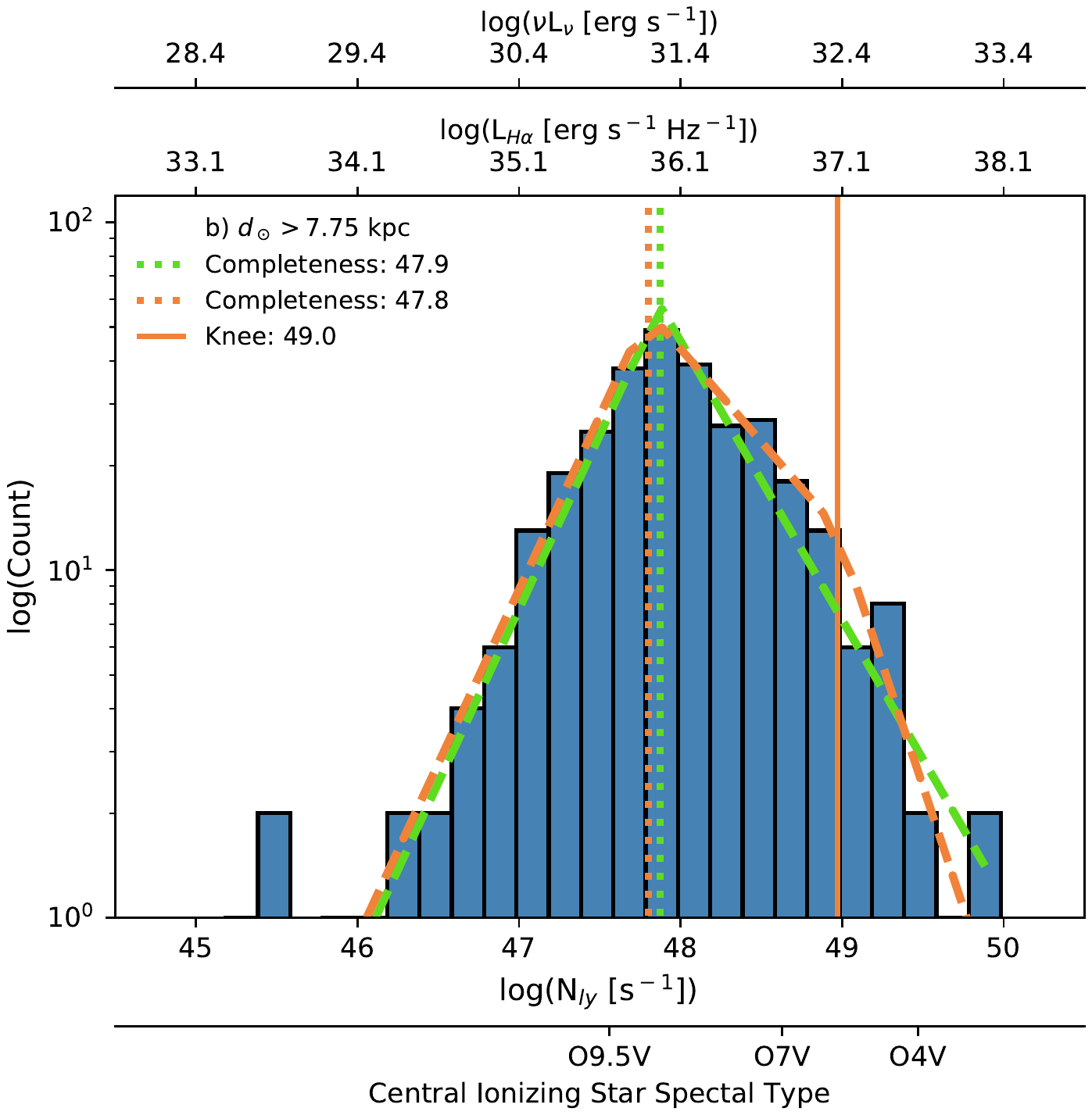}}\qquad
  \subfloat{\label{fig:magpis_vgps_nearrgal}%
    \includegraphics[scale=0.385,trim={3.75cm 6.25cm 3.5cm 6.5cm}, clip]{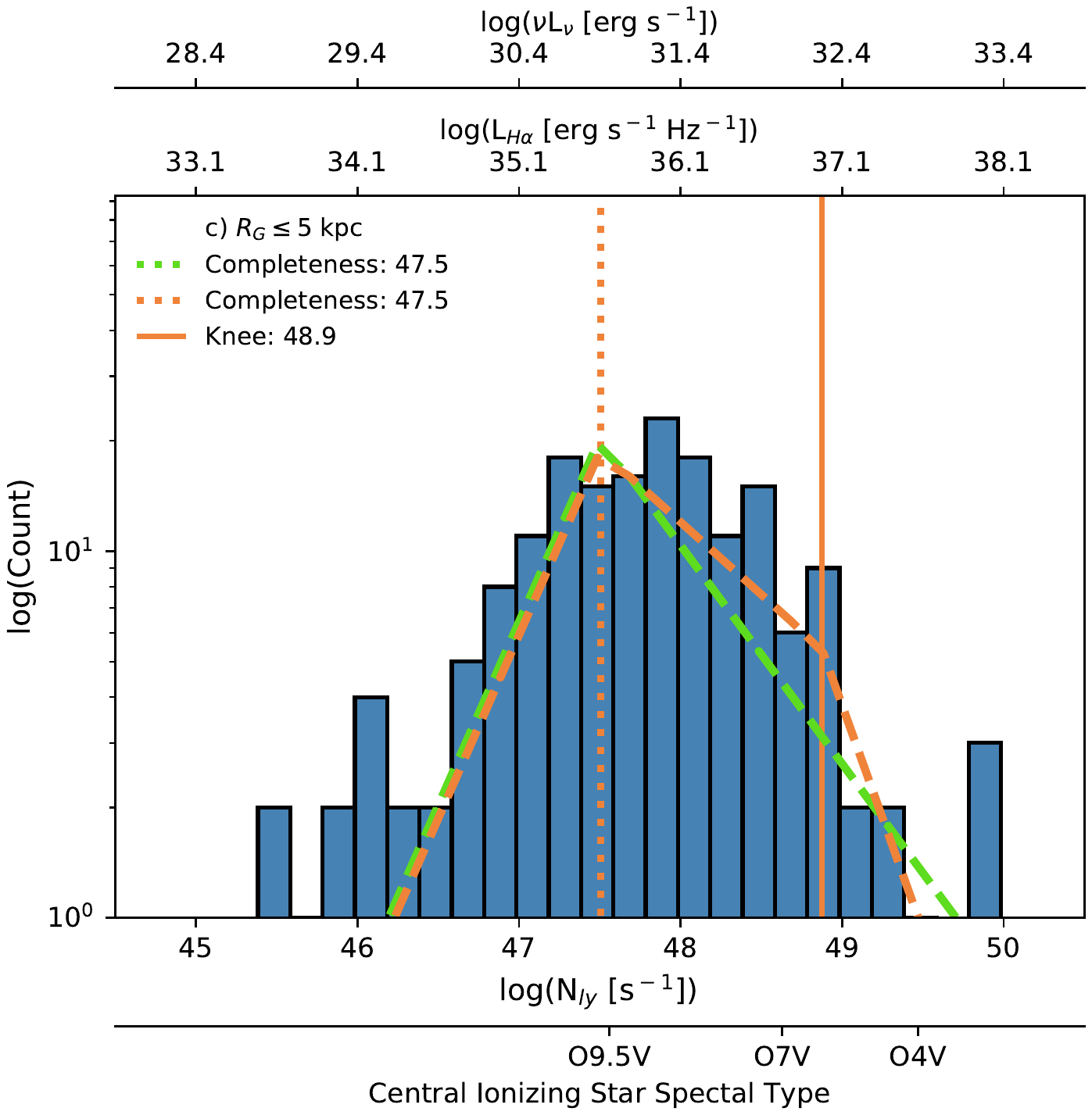}}\\
  \subfloat{\label{fig:magpis_vgps_farrgal}%
    \includegraphics[scale=0.385,trim={3.75cm 6.25cm 3.5cm 6.5cm}, clip]{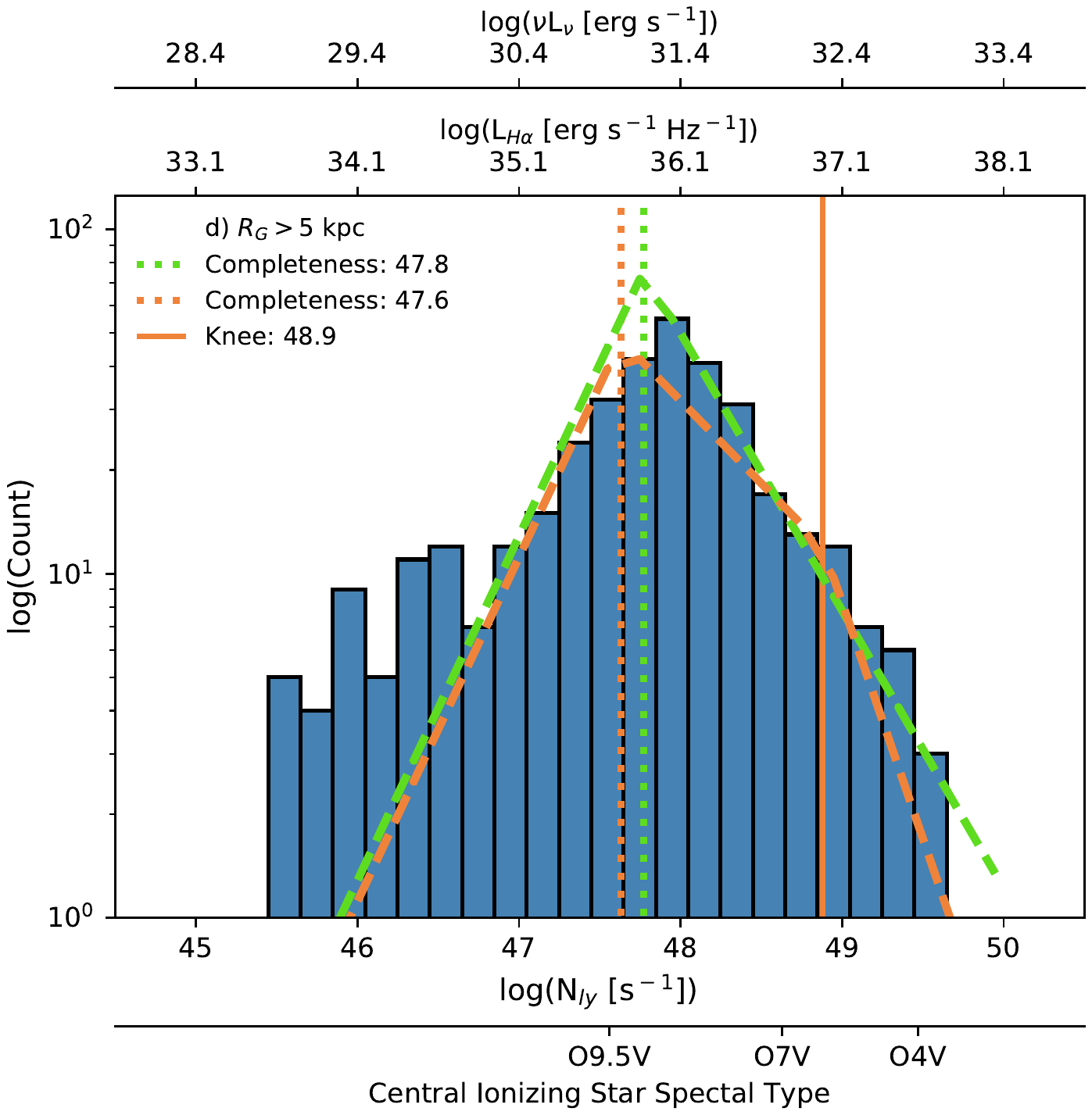}}\qquad
  \subfloat{\label{fig:magpis_vgps_small}%
    \includegraphics[scale=0.385,trim={3.75cm 6.25cm 3.5cm 6.5cm}, clip]{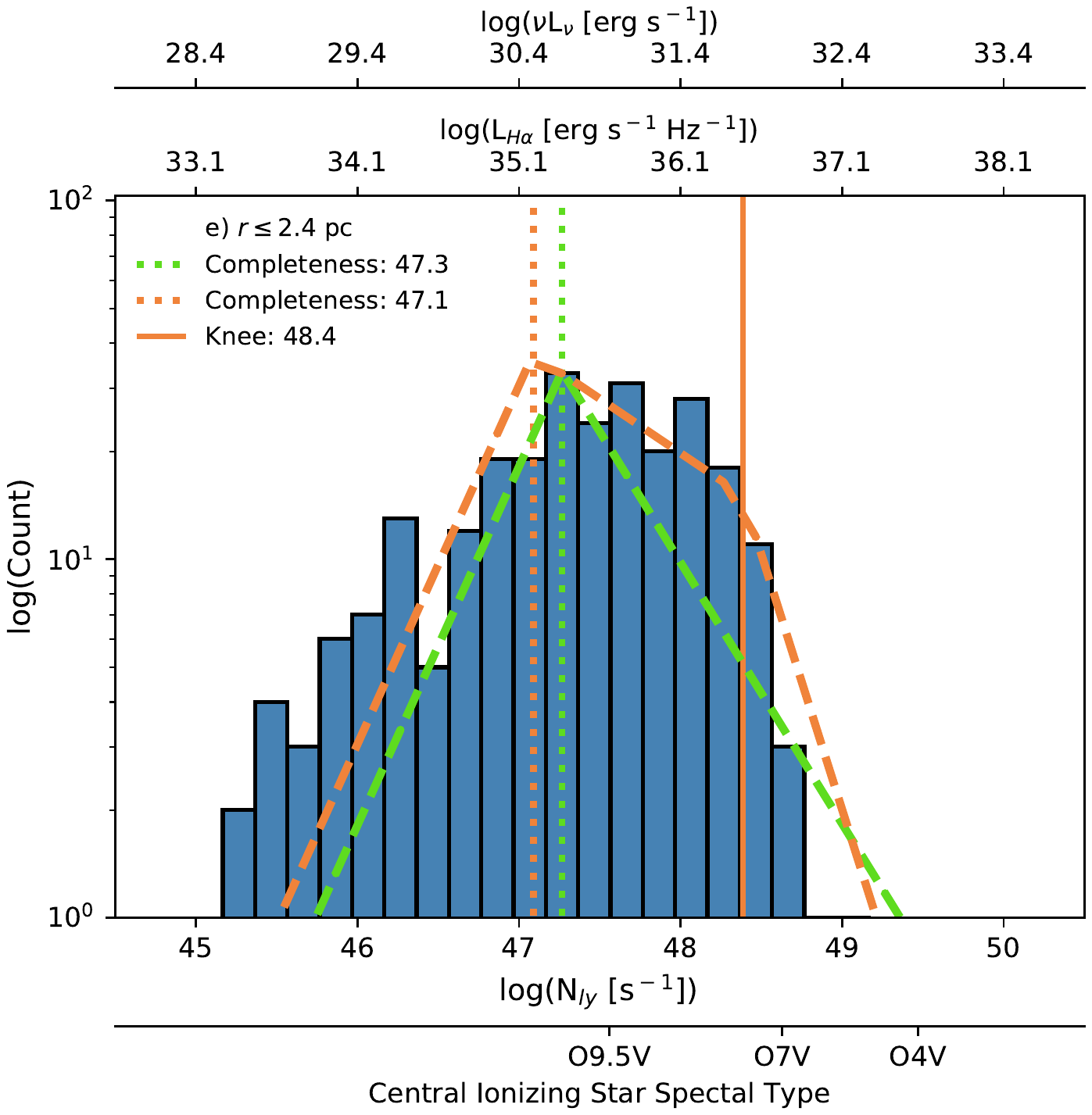}}\qquad
  \subfloat{\label{fig:magpis_vgps_large}%
    \includegraphics[scale=0.385,trim={3.75cm 6.25cm 3.5cm 6.5cm}, clip]{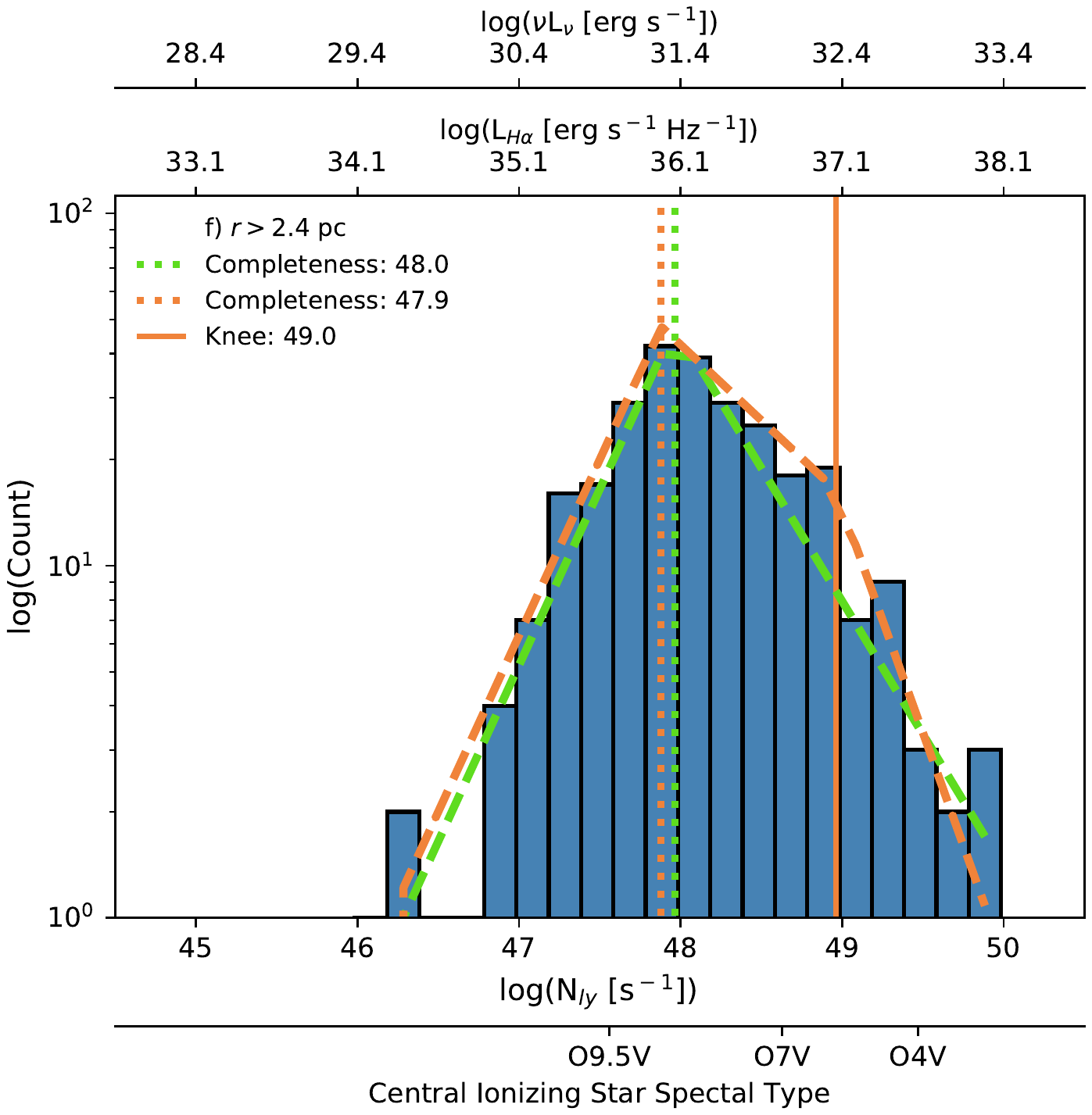}}\\
  \subfloat{\label{fig:magpis_vgps_arm}%
    \includegraphics[scale=0.385,trim={3.75cm 6.25cm 3.5cm 6.5cm}, clip]{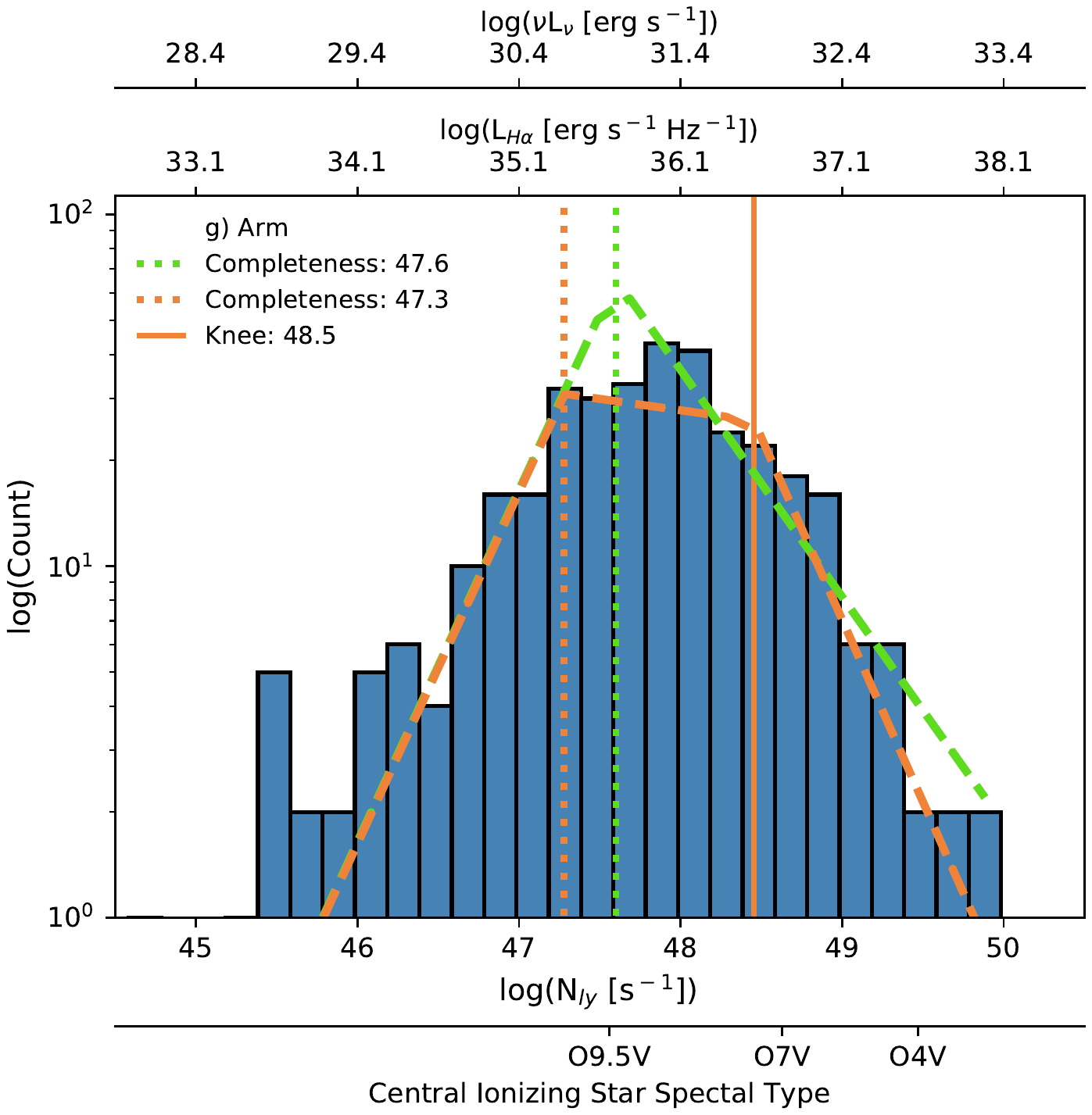}}\qquad
  \subfloat{\label{fig:magpis_vgps_interarm}%
    \includegraphics[scale=0.385,trim={3.75cm 6.25cm 3.5cm 6.5cm}, clip]{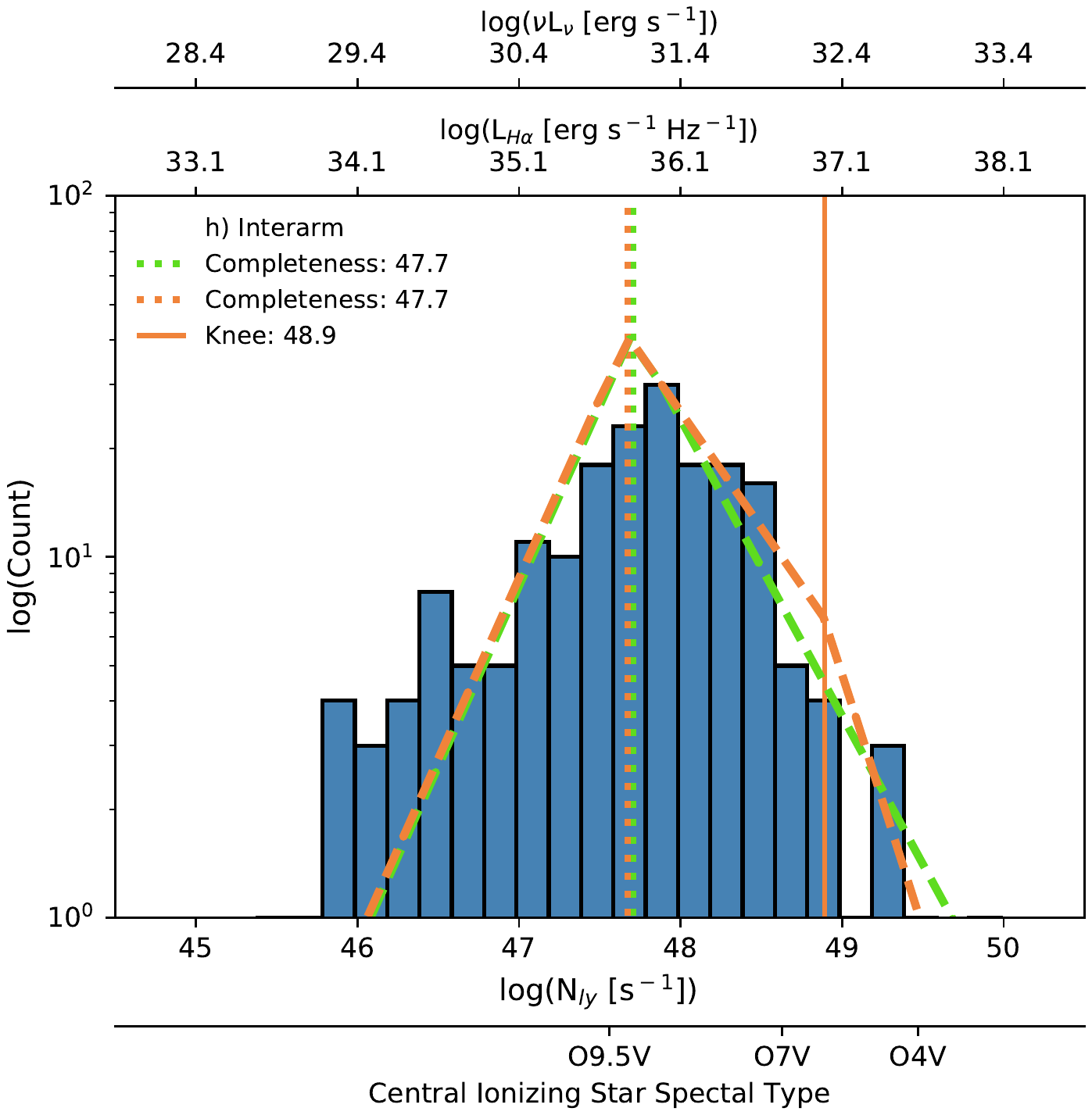}}
\caption{Single and double power law fits to the catalog-only $20\,\cm$ MAGPIS+VGPS data subsets: $d_\sun \leq 7.75$ \kpc (panel \subref*{fig:magpis_vgps_neardist}), $d_\sun > 7.75$ \kpc (panel \subref*{fig:magpis_vgps_fardist}), $\rgal \leq 5$ \kpc (panel \subref*{fig:magpis_vgps_nearrgal}), $\rgal > 5$ \kpc (panel \subref*{fig:magpis_vgps_farrgal}), $r \leq 2.4 \pc$ (panel \subref*{fig:magpis_vgps_small}), $r > 2.4 \pc$ (panel \subref*{fig:magpis_vgps_large}), arm (panel \subref*{fig:magpis_vgps_arm}), and interarm (panel \subref*{fig:magpis_vgps_interarm}).}
\label{fig:magpisvgps2}
\end{sidewaysfigure*}

\begin{sidewaysfigure*}[h]
\centering
  \subfloat{\label{fig:vgps_neardist}%
    \includegraphics[scale=0.385,trim={3.75cm 6.25cm 3.5cm 6.5cm}, clip]{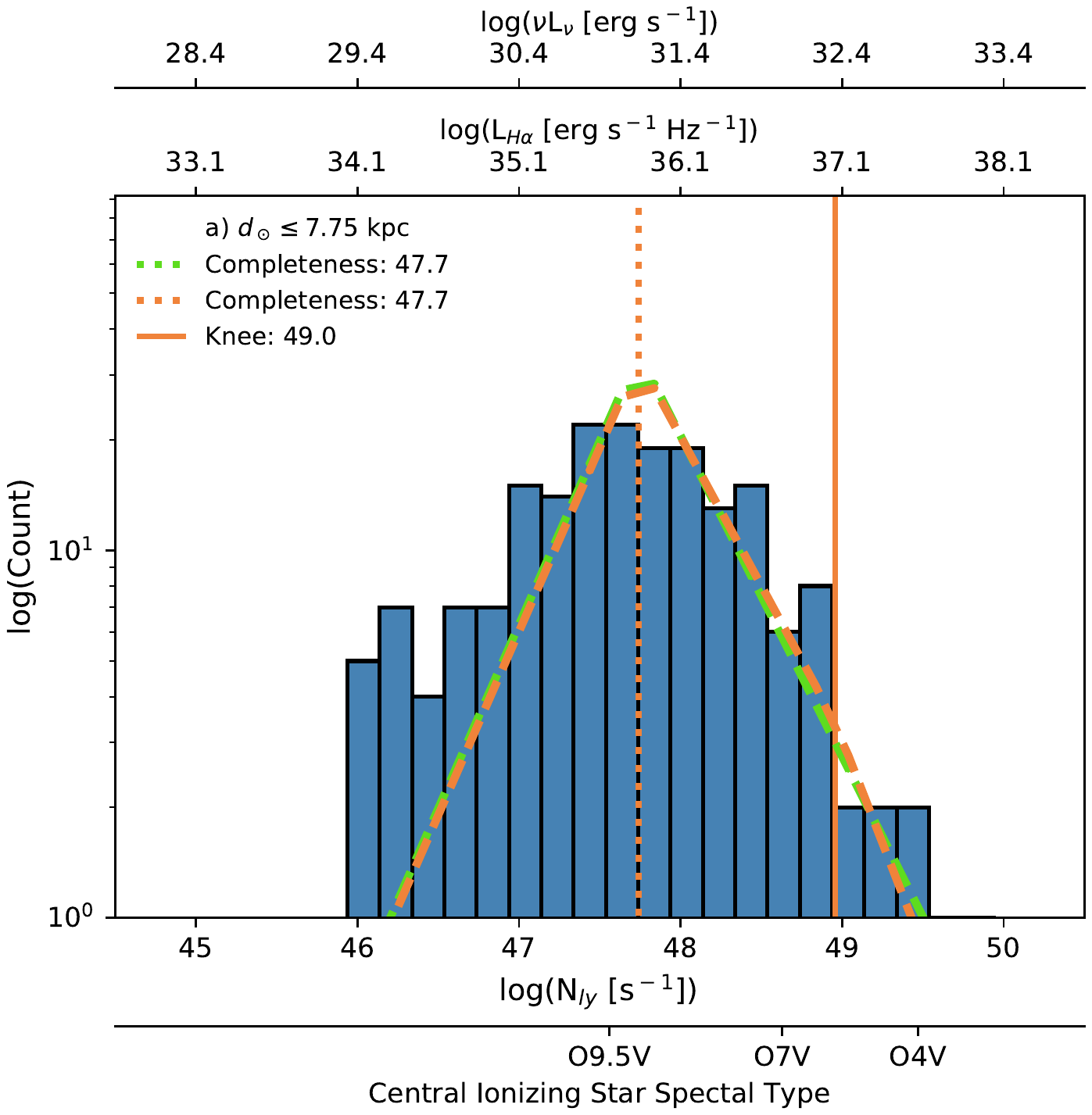}}\qquad
  \subfloat{\label{fig:vgps_fardist}%
    \includegraphics[scale=0.385,trim={3.75cm 6.25cm 3.5cm 6.5cm}, clip]{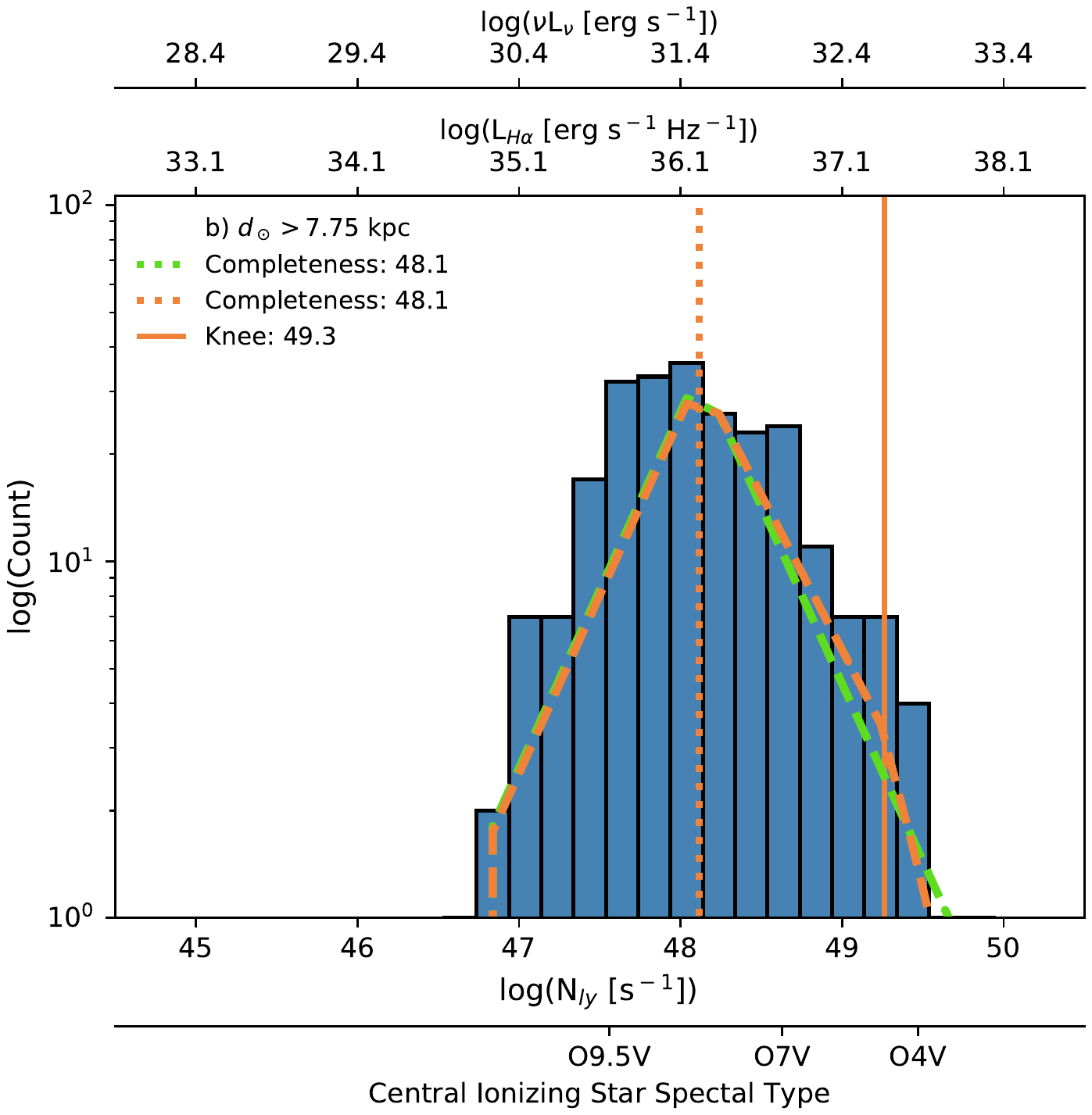}}\qquad
  \subfloat{\label{fig:vgps_nearrgal}%
    \includegraphics[scale=0.385,trim={3.75cm 6.25cm 3.5cm 6.5cm}, clip]{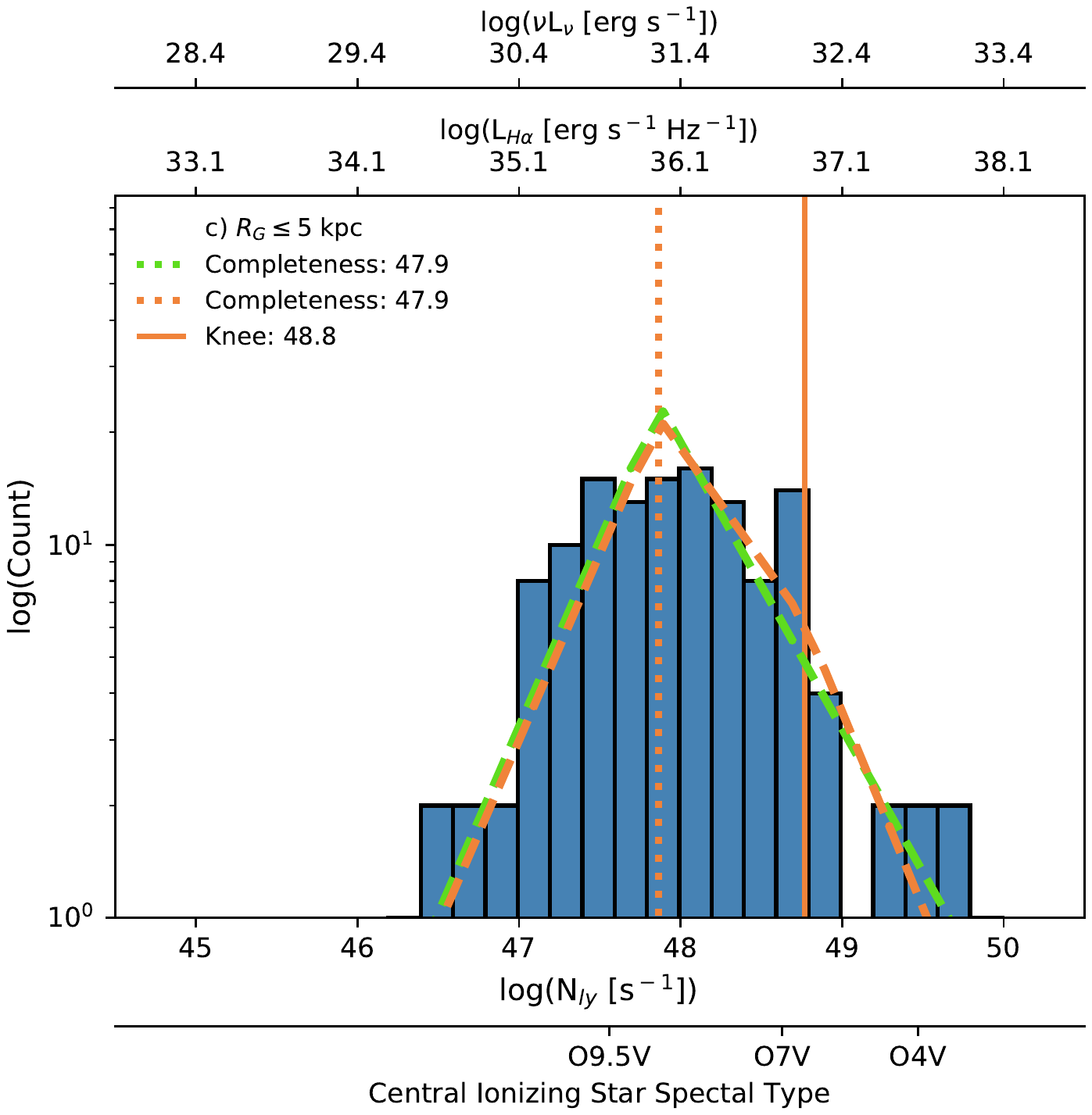}}\\
  \subfloat{\label{fig:vgps_farrgal}%
    \includegraphics[scale=0.385,trim={3.75cm 6.25cm 3.5cm 6.5cm}, clip]{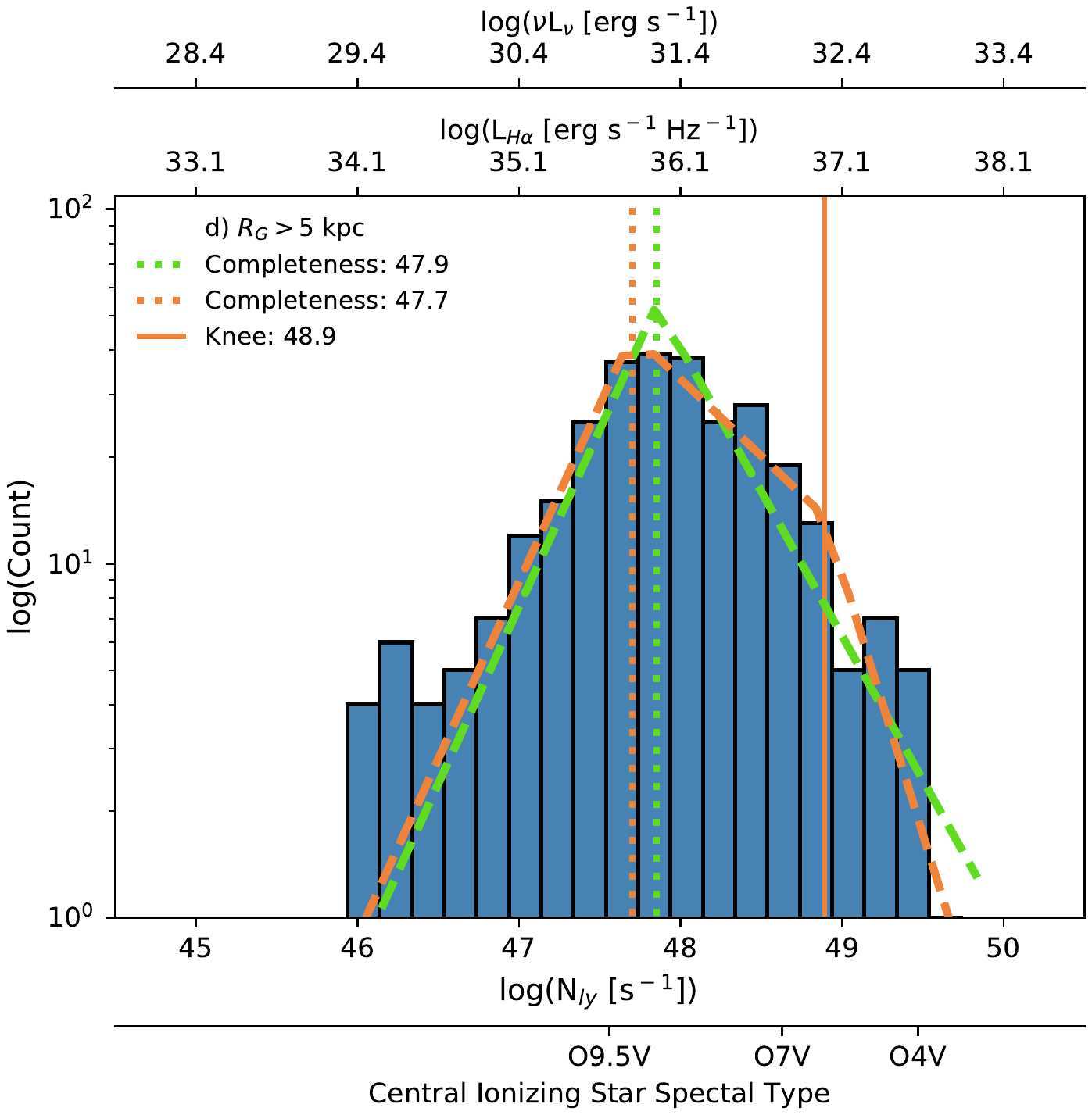}}\qquad
  \subfloat{\label{fig:vgps_small}%
    \includegraphics[scale=0.385,trim={3.75cm 6.25cm 3.5cm 6.5cm}, clip]{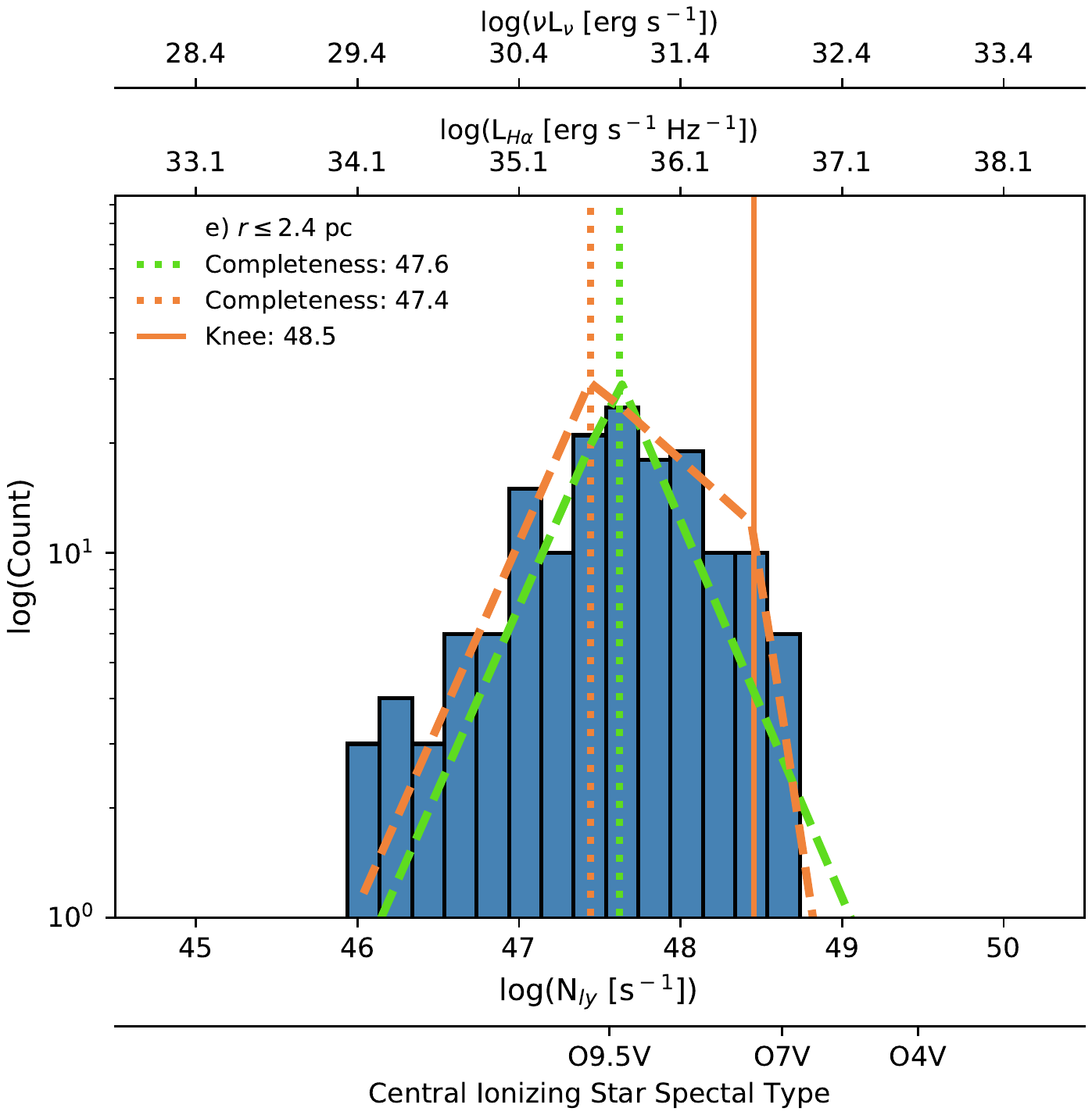}}\qquad
  \subfloat{\label{fig:vgps_large}%
    \includegraphics[scale=0.385,trim={3.75cm 6.25cm 3.5cm 6.5cm}, clip]{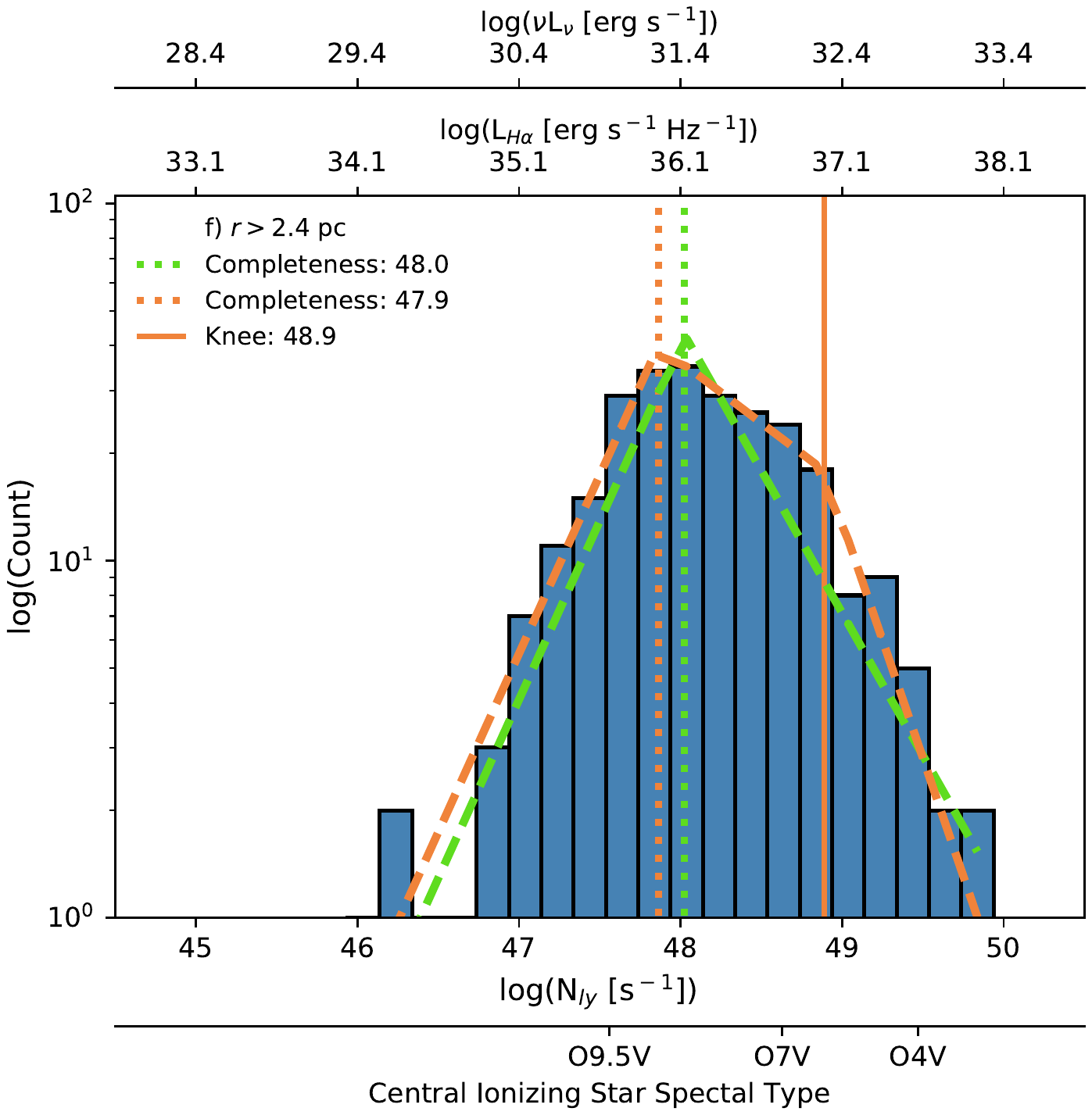}}\\
  \subfloat{\label{fig:vgps_arm}%
    \includegraphics[scale=0.385,trim={3.75cm 6.25cm 3.5cm 6.5cm}, clip]{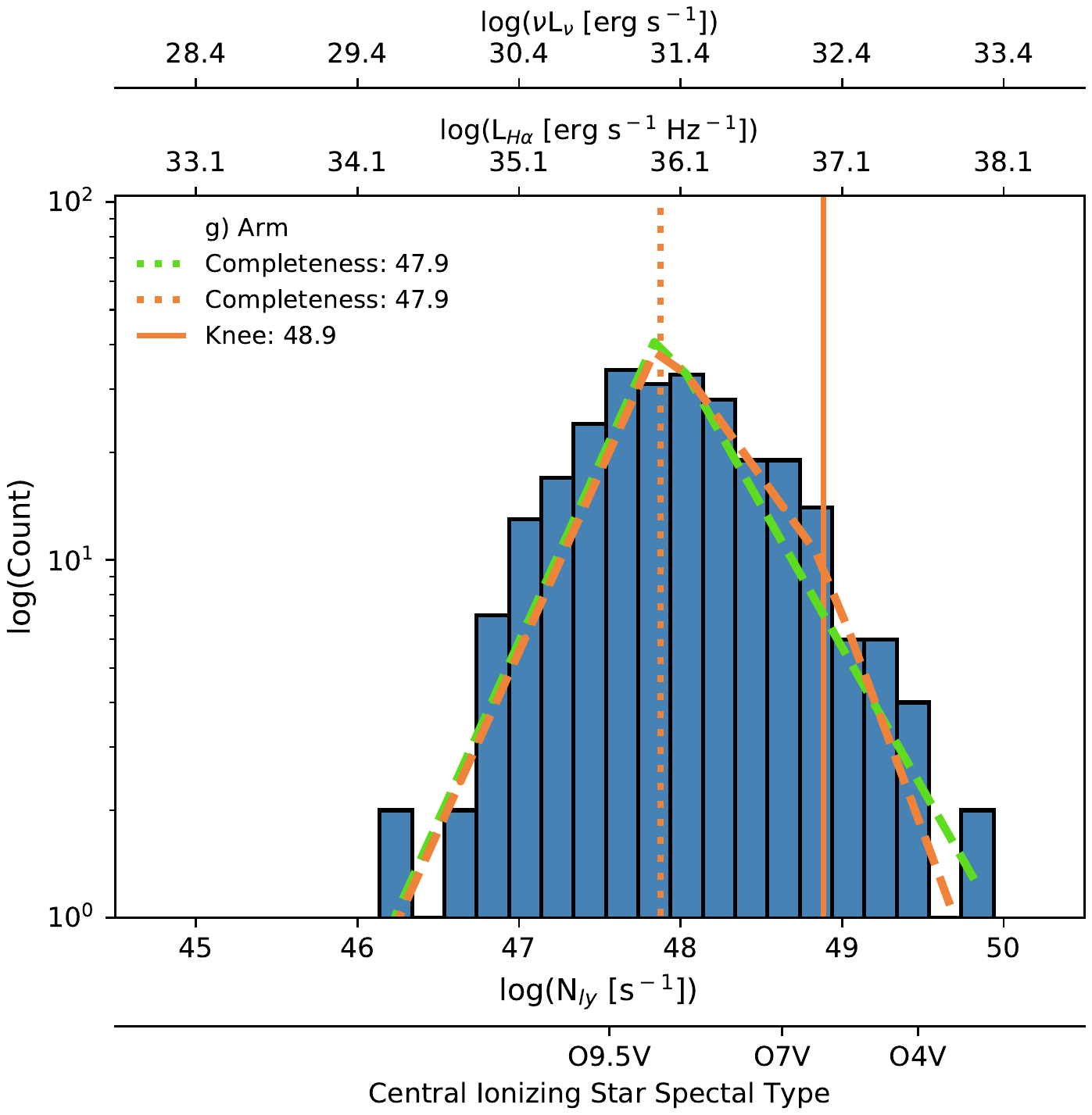}}\qquad
  \subfloat{\label{fig:vgps_interarm}%
    \includegraphics[scale=0.385,trim={3.75cm 6.25cm 3.5cm 6.5cm}, clip]{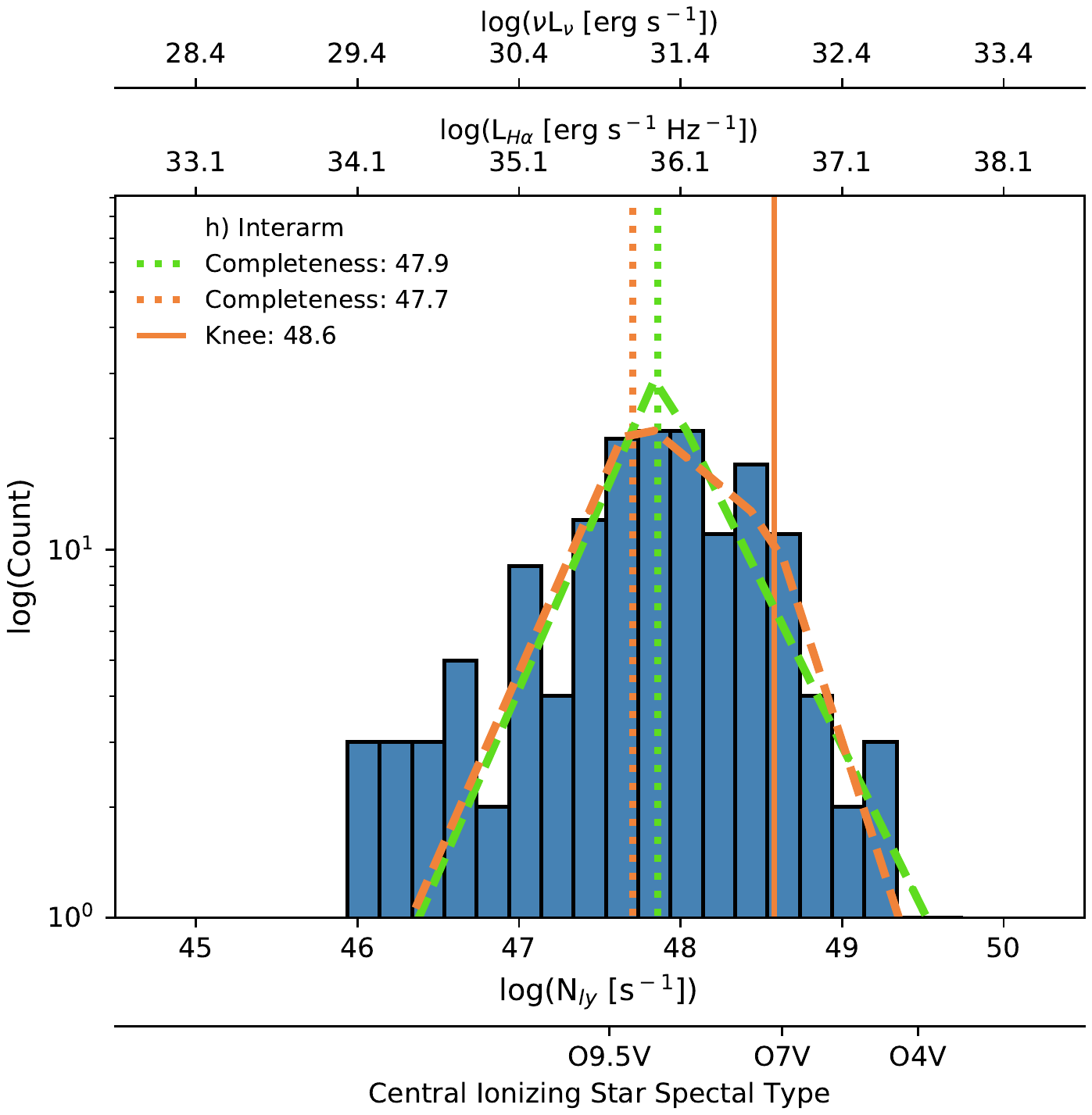}}
\caption{Single and double power law fits to the catalog-only $21\,\cm$ VGPS data subsets: $d_\sun \leq 7.75$ \kpc (panel \subref*{fig:vgps_neardist}), $d_\sun > 7.75$ \kpc (panel \subref*{fig:vgps_fardist}), $\rgal \leq 5$ \kpc (panel \subref*{fig:vgps_nearrgal}), $\rgal > 5$ \kpc (panel \subref*{fig:vgps_farrgal}), $r \leq 2.4 \pc$ (panel \subref*{fig:vgps_small}), $r > 2.4 \pc$ (panel \subref*{fig:vgps_large}), arm (panel \subref*{fig:vgps_arm}), and interarm (panel \subref*{fig:vgps_interarm}).}
\label{fig:vgps2}
\end{sidewaysfigure*}

\newpage
\clearpage

\subsection{Power Law Indices, Knees, and Completeness Limits of the Monte Carlo-Generated Luminosity Distributions}
\label{appsubsec:sim_summary}

Here we show graphical summaries of the median power law indices, knee luminosities, and completeness limit luminosities for each data subset at each wavelength derived from the Monte Carlo-generated luminosity distributions. We also include the MADs for each value, though in many cases they are smaller than the size of the plotted markers. $\alpha$ is the single power law index as defined in Equation \ref{eq:singlaw}. $\alpha_{1}$ and $\alpha_{2}$ are the double power law indices as defined in Equation \ref{eq:double_law}.

\begin{sidewaysfigure*}[h]
\centering
  \subfloat{\label{fig:complete_alpha}%
    \includegraphics[scale=0.45,trim={3cm 7.75cm 3cm 8cm}, clip]{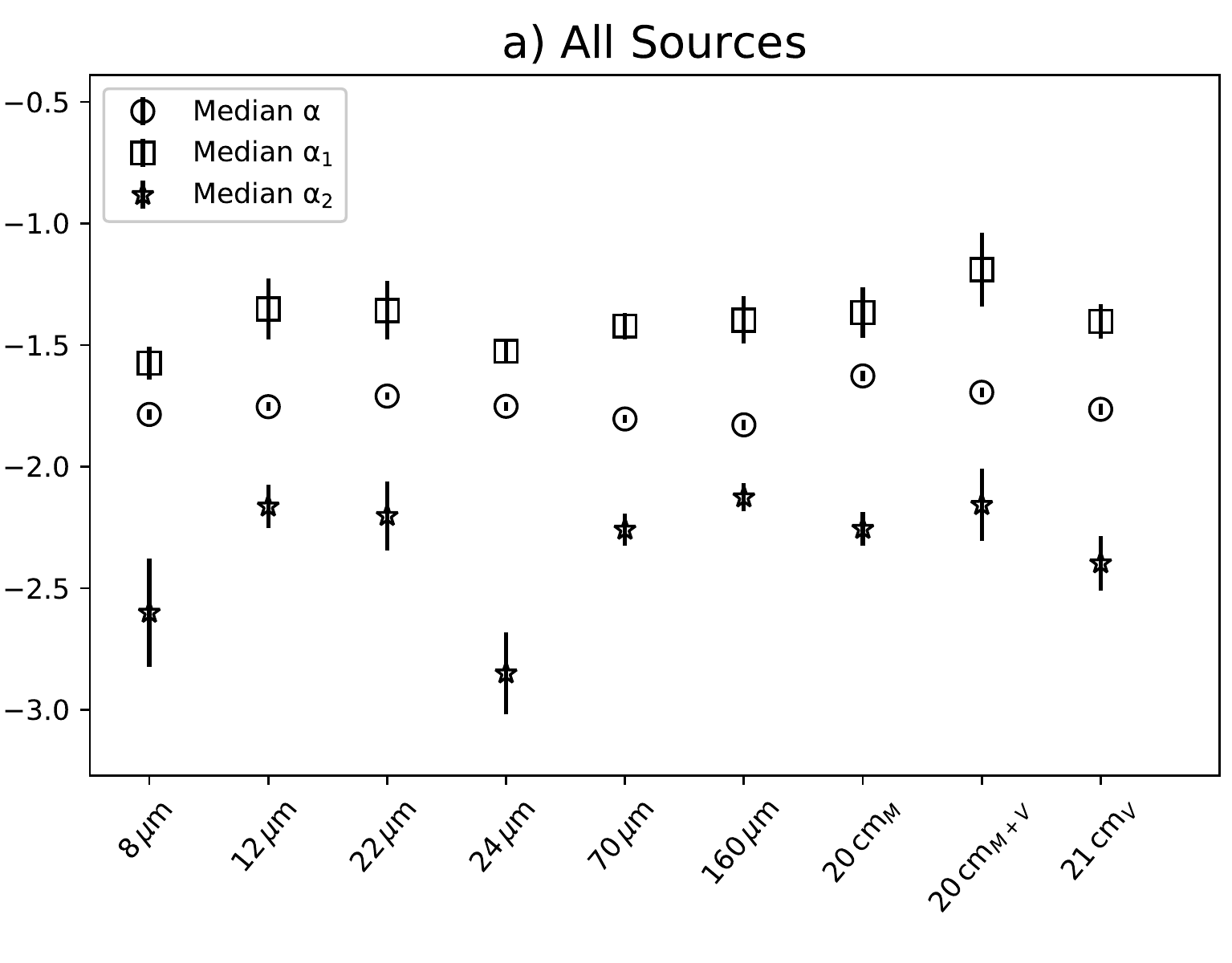}}\qquad
  \subfloat{\label{fig:neard_alpha}%
    \includegraphics[scale=0.45,trim={3cm 7.75cm 3cm 8cm}, clip]{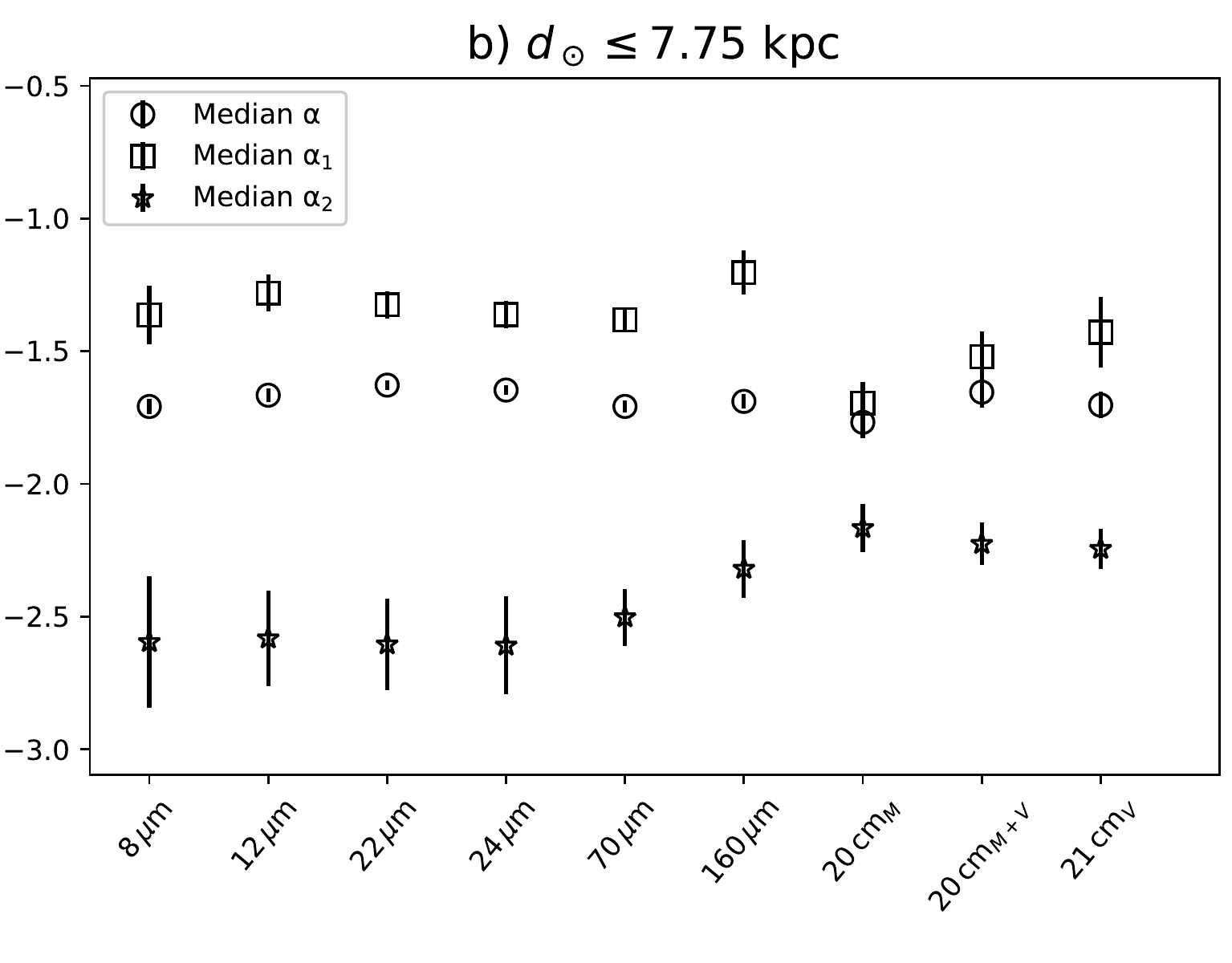}}\qquad
  \subfloat{\label{fig:fard_alpha}%
    \includegraphics[scale=0.45,trim={3cm 7.75cm 3cm 8cm}, clip,trim={3cm 7.75cm 3cm 8cm}, clip]{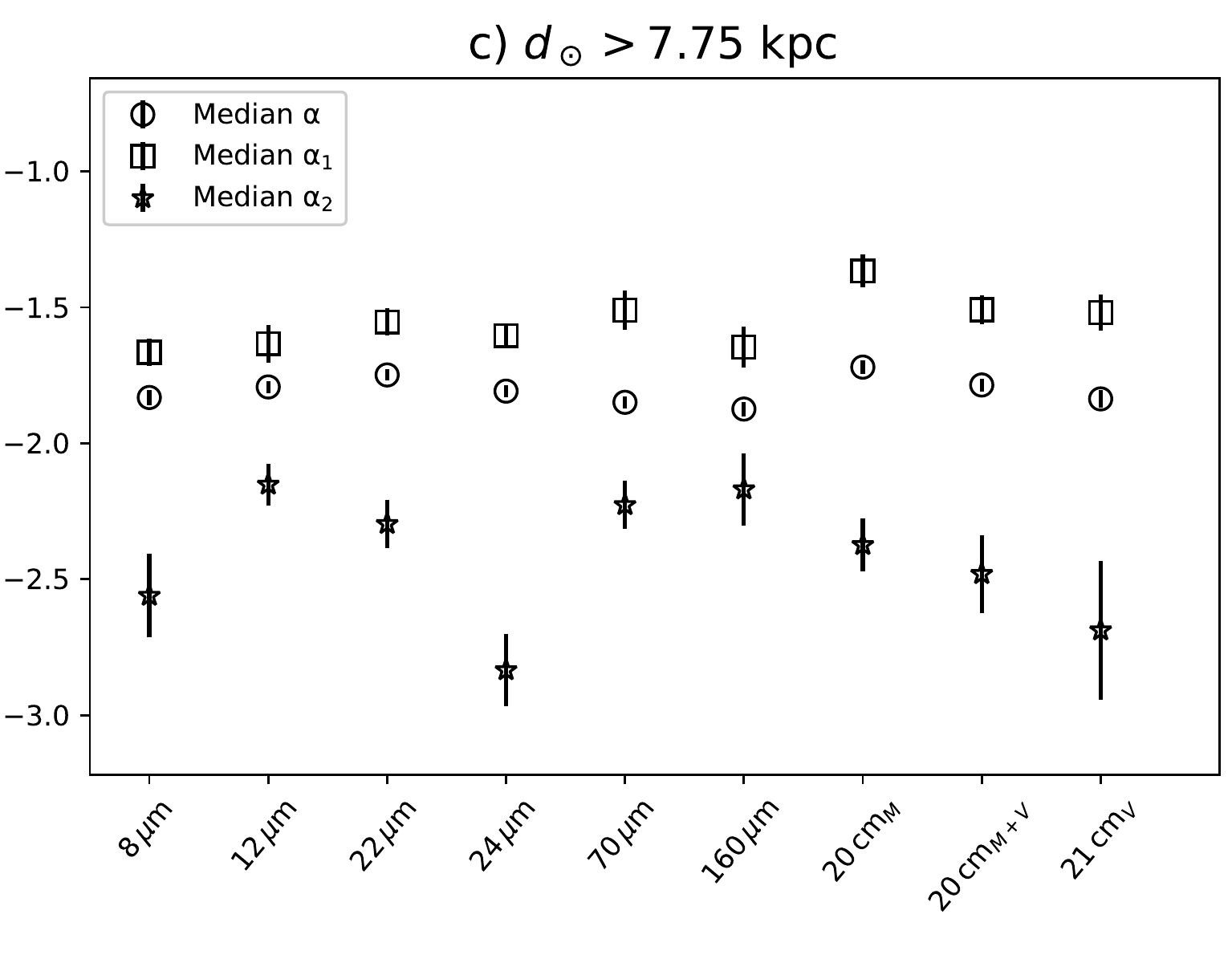}}\\
  \subfloat{\label{fig:nearrgal_alpha}%
    \includegraphics[scale=0.45,trim={3cm 7.75cm 3cm 8cm}, clip]{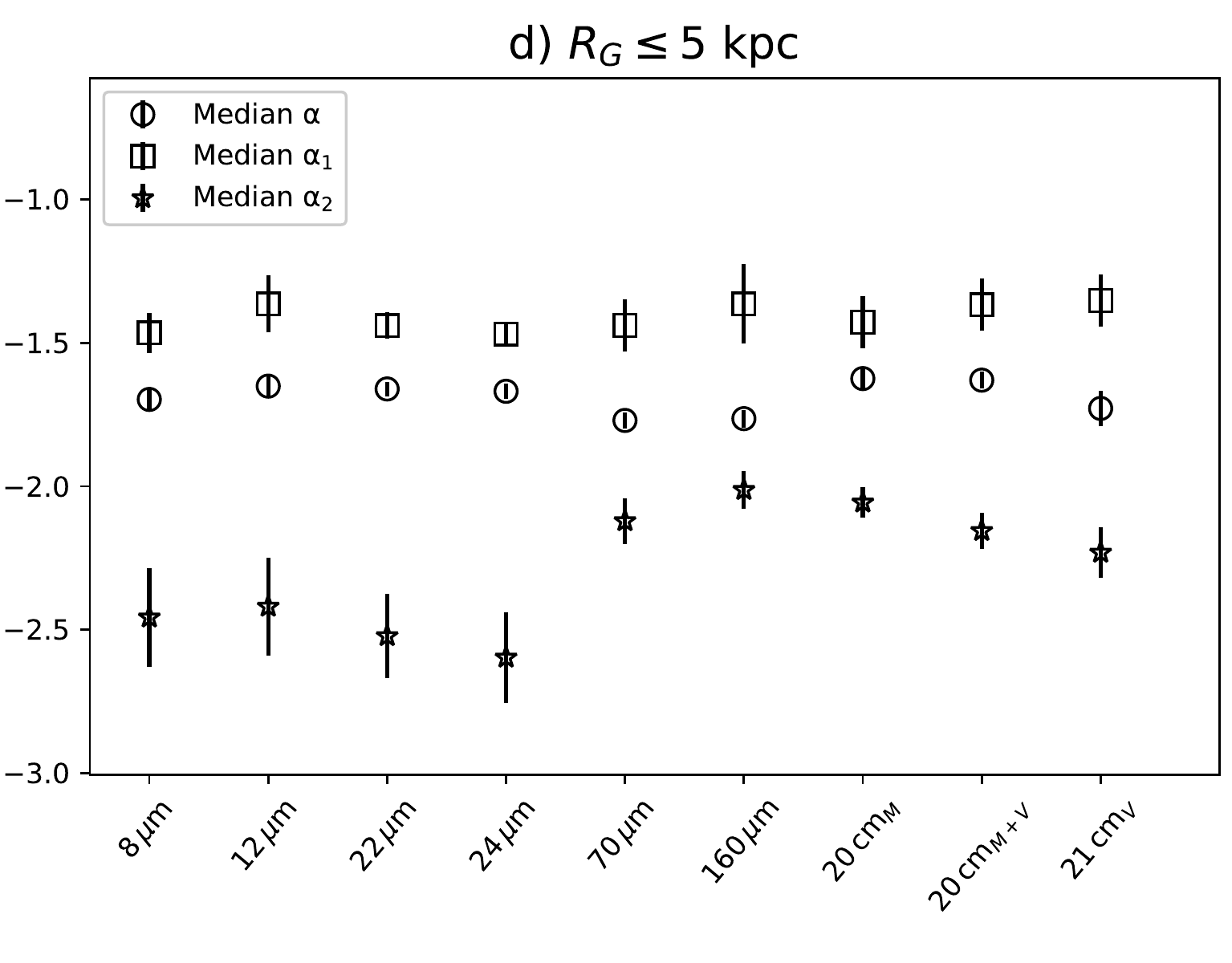}}\qquad
  \subfloat{\label{fig:farrgal_alpha}%
    \includegraphics[scale=0.45,trim={3cm 7.75cm 3cm 8cm}, clip]{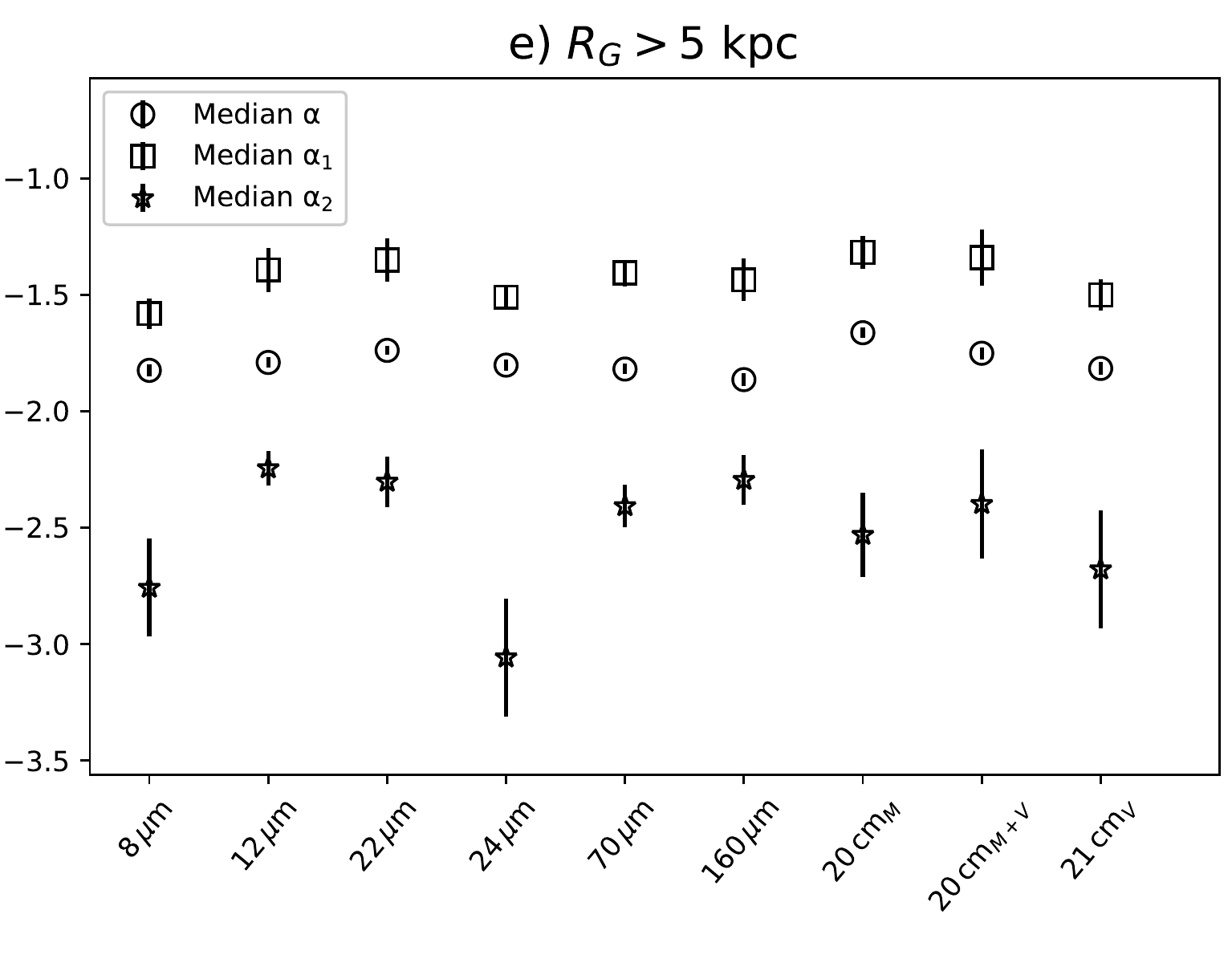}}\qquad
  \subfloat{\label{fig:small_alpha}%
    \includegraphics[scale=0.45,trim={3cm 7.75cm 3cm 8cm}, clip]{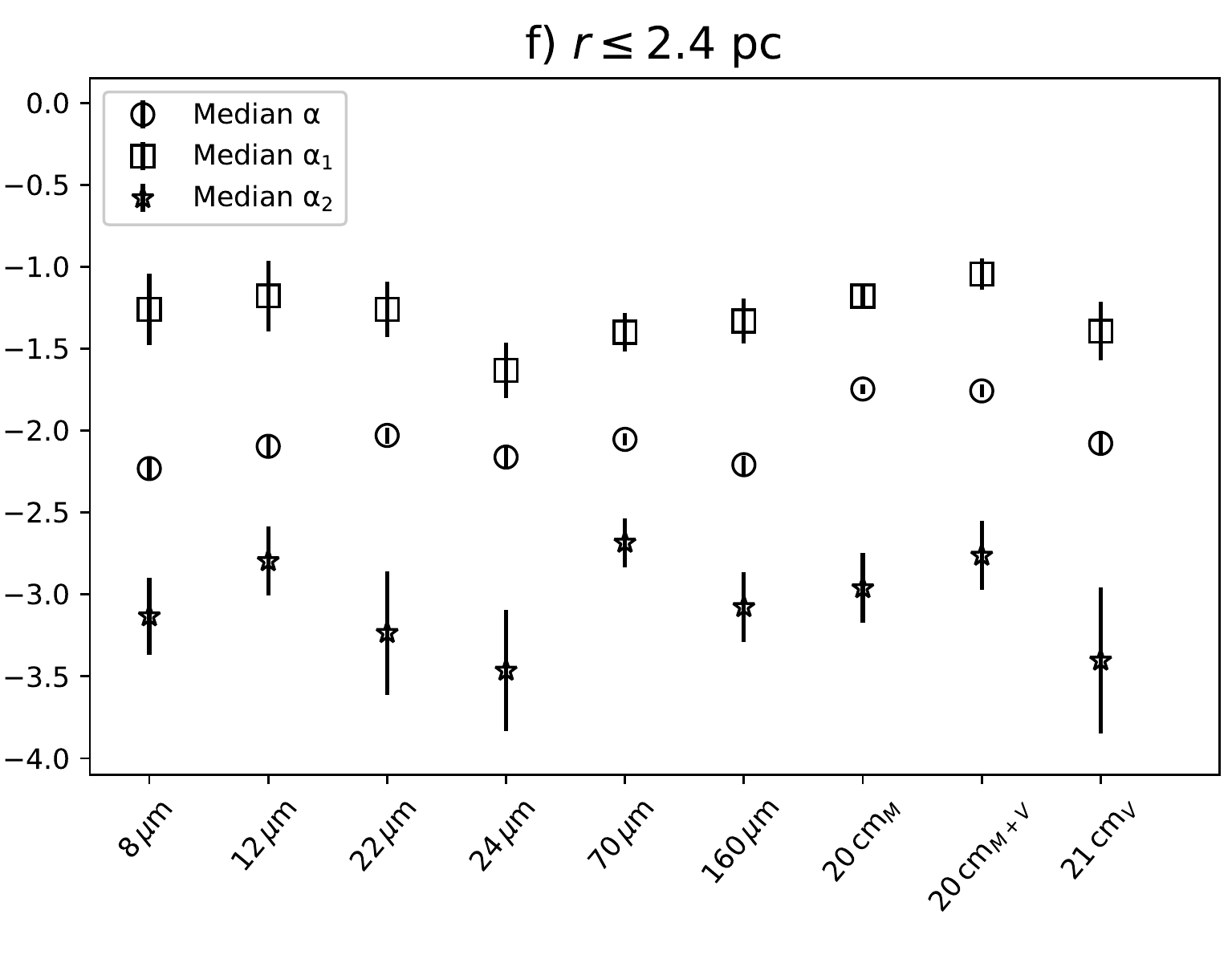}}\\
  \subfloat{\label{fig:large_alpha}%
    \includegraphics[scale=0.45,trim={3cm 7.75cm 3cm 8cm}, clip,trim={3cm 7.75cm 3cm 8cm}, clip]{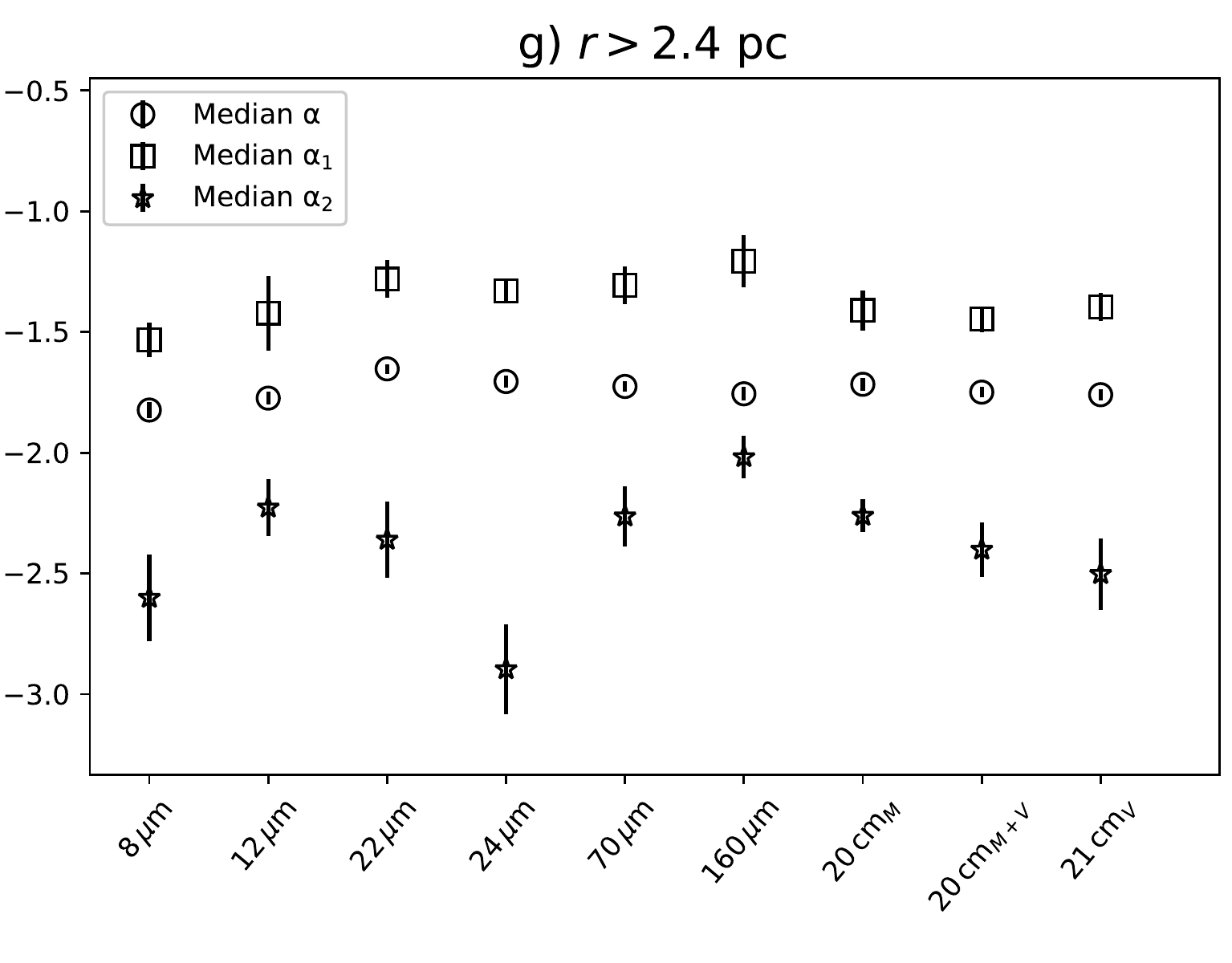}}\qquad
  \subfloat{\label{fig:arm_alpha}%
    \includegraphics[scale=0.45,trim={3cm 7.75cm 3cm 8cm}, clip]{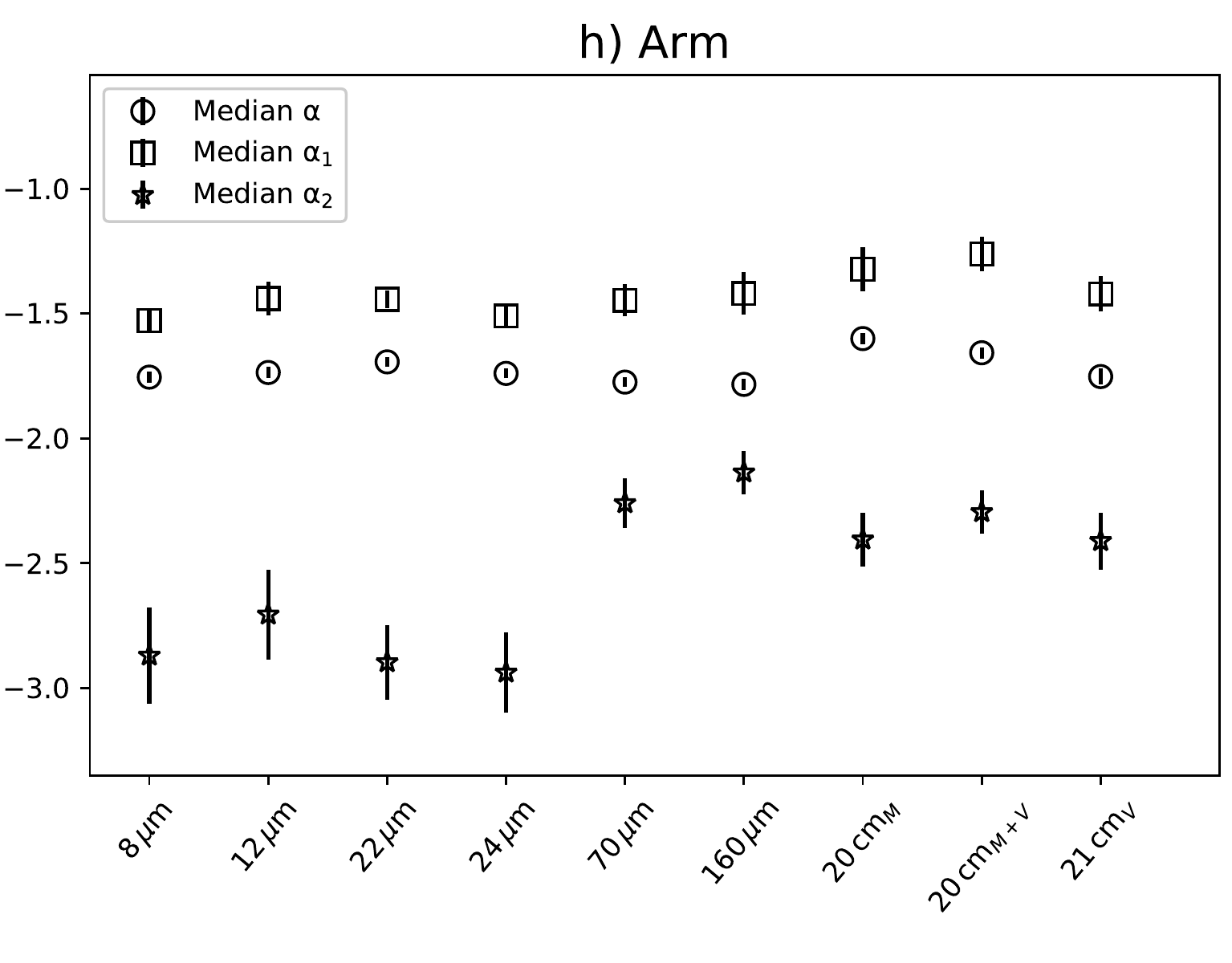}}\qquad
  \subfloat{\label{fig:interarm_alpha}%
    \includegraphics[scale=0.45,trim={3cm 7.75cm 3cm 8cm}, clip]{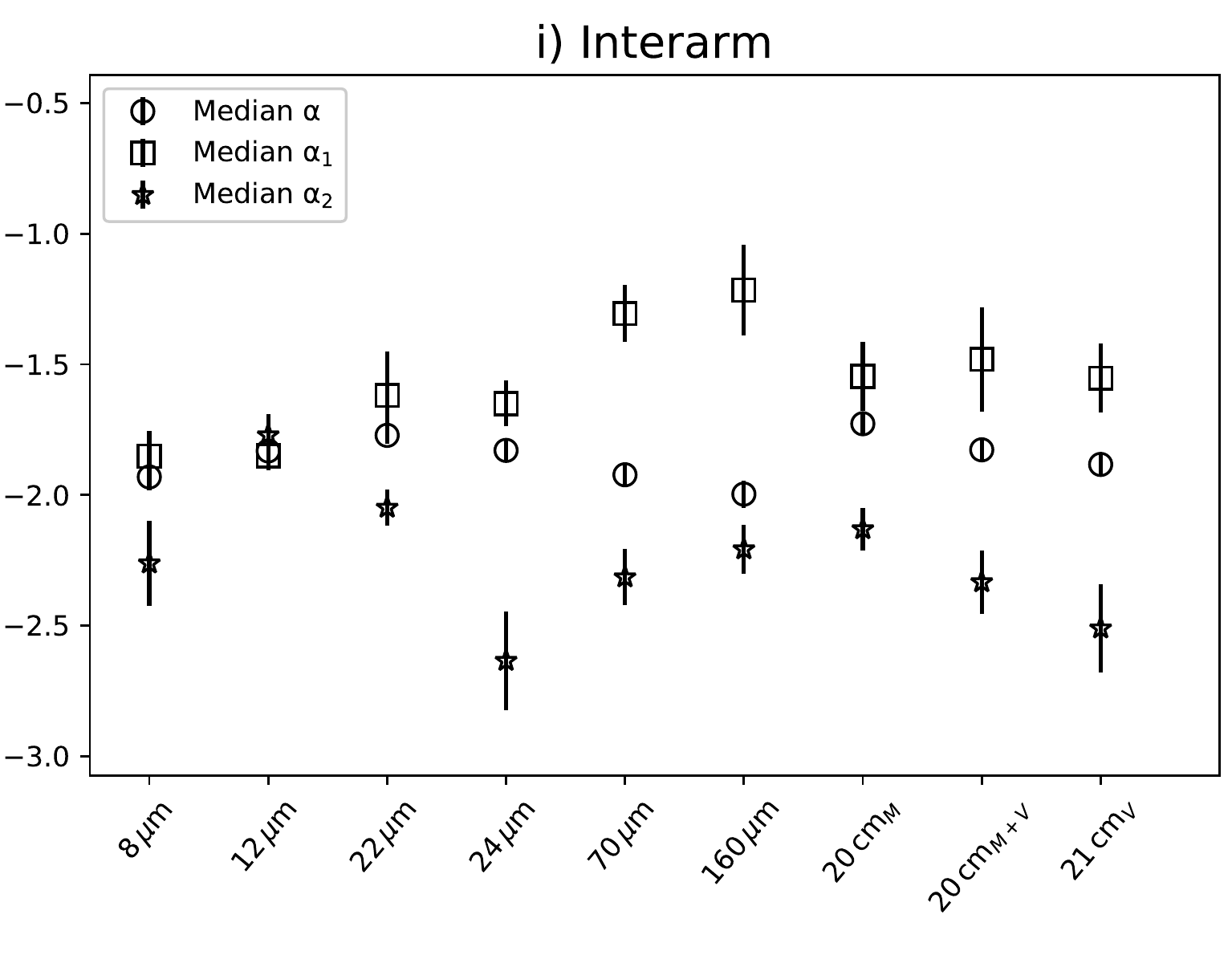}}
\caption{Median single and double power law indices and MADs from the Monte Carlo-generated luminosity distributions: all sources (panel \subref*{fig:complete_alpha}), $d_\sun \leq 7.75$ \kpc (panel \subref*{fig:neard_alpha}), $d_\sun > 7.75$ \kpc (panel \subref*{fig:fard_alpha}), $\rgal \leq 5$ \kpc (panel \subref*{fig:nearrgal_alpha}), $\rgal > 5$ \kpc (panel \subref*{fig:farrgal_alpha}), $r \leq 2.4 \pc$ (panel \subref*{fig:small_alpha}), $r > 2.4 \pc$ (panel \subref*{fig:large_alpha}), arm (panel \subref*{fig:arm_alpha}), and interarm (panel \subref*{fig:interarm_alpha}).}
\label{fig:alphacomp2}
\end{sidewaysfigure*}

\begin{sidewaysfigure*}[h]
\centering
  \subfloat{\label{fig:complete_knee}%
    \includegraphics[scale=0.45,trim={3cm 7.75cm 3cm 8cm}, clip]{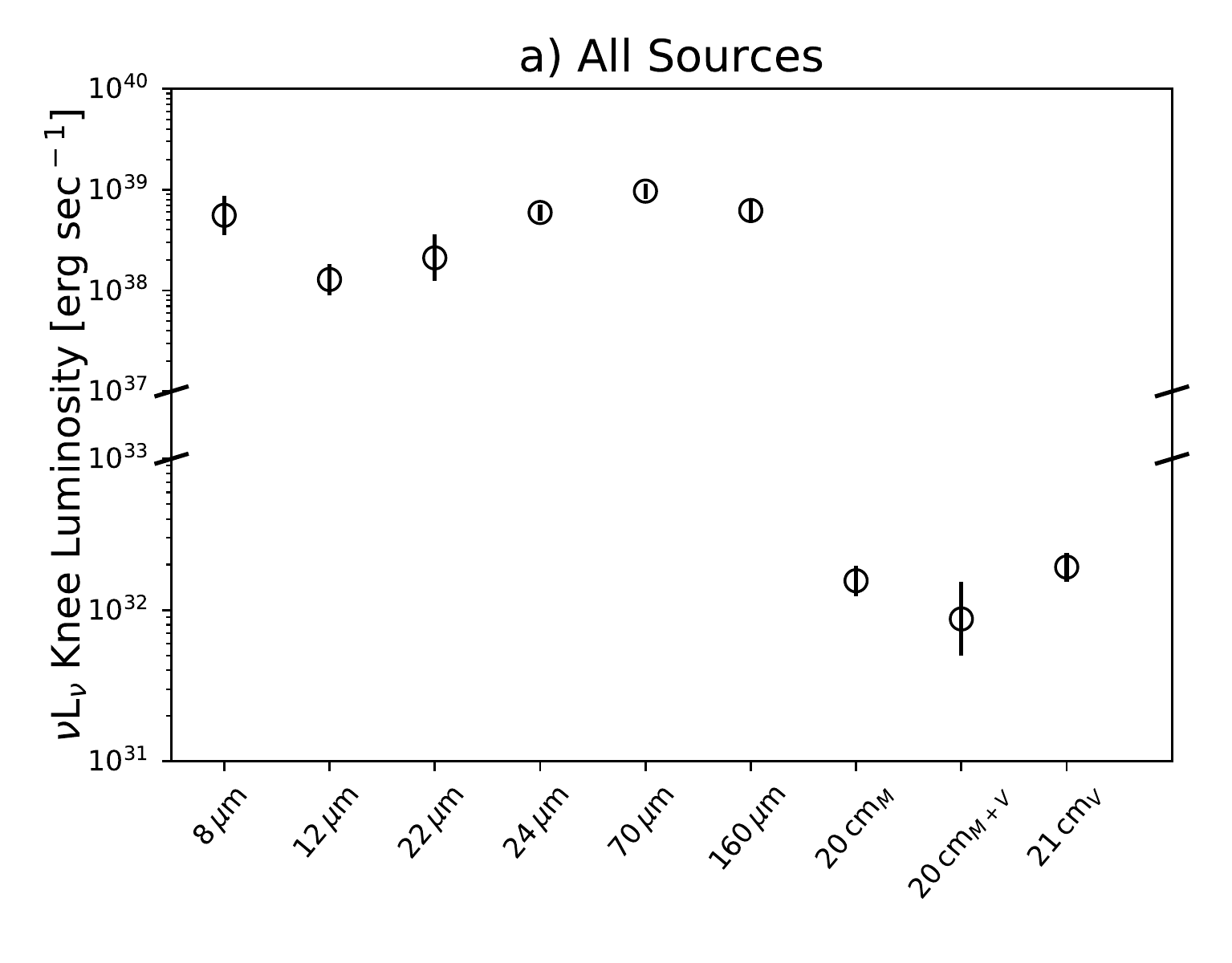}}\qquad
  \subfloat{\label{fig:neard_knee}%
    \includegraphics[scale=0.45,trim={3cm 7.75cm 3cm 8cm}, clip]{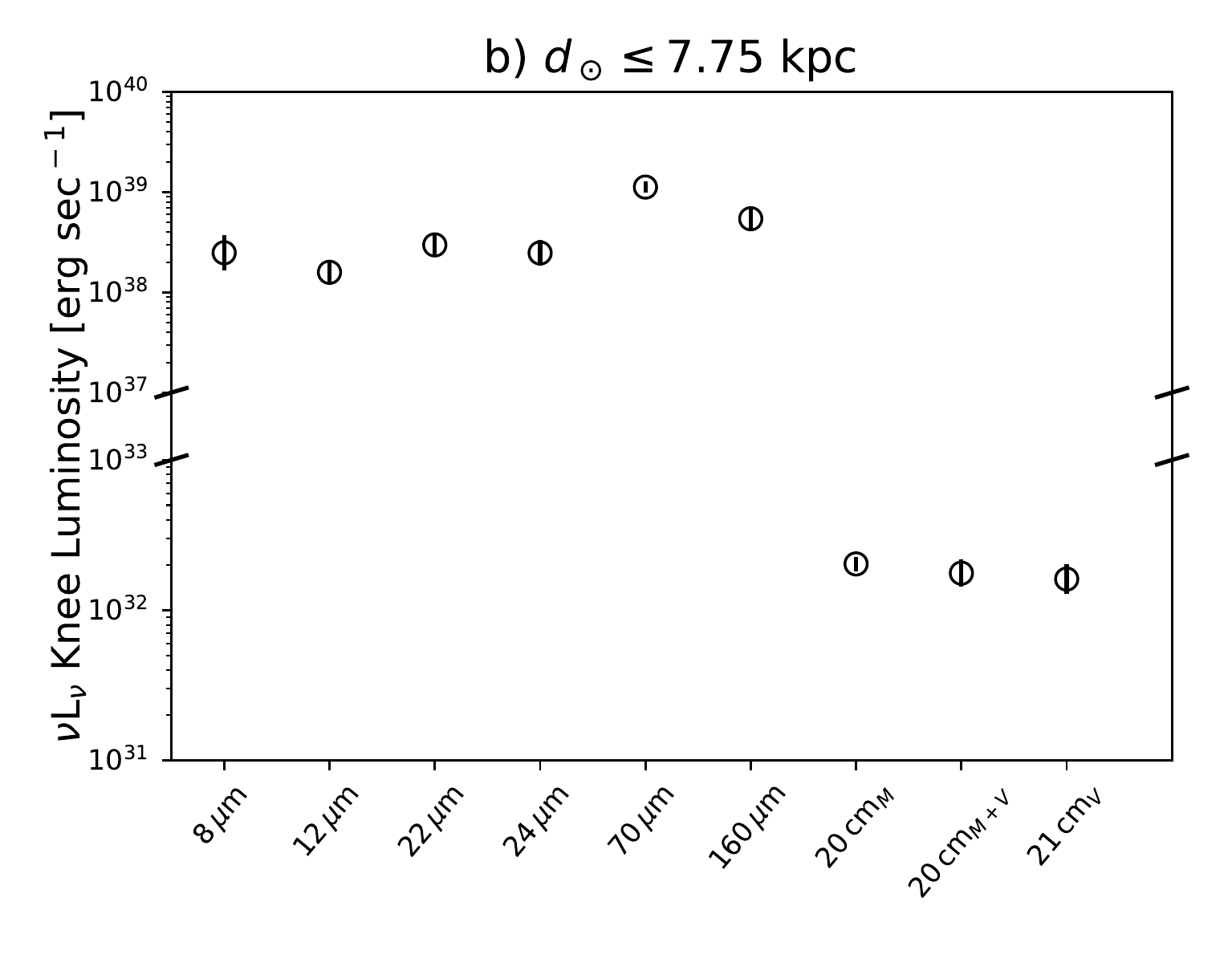}}\qquad
  \subfloat{\label{fig:fard_knee}%
    \includegraphics[scale=0.45,trim={3cm 7.75cm 3cm 8cm}, clip]{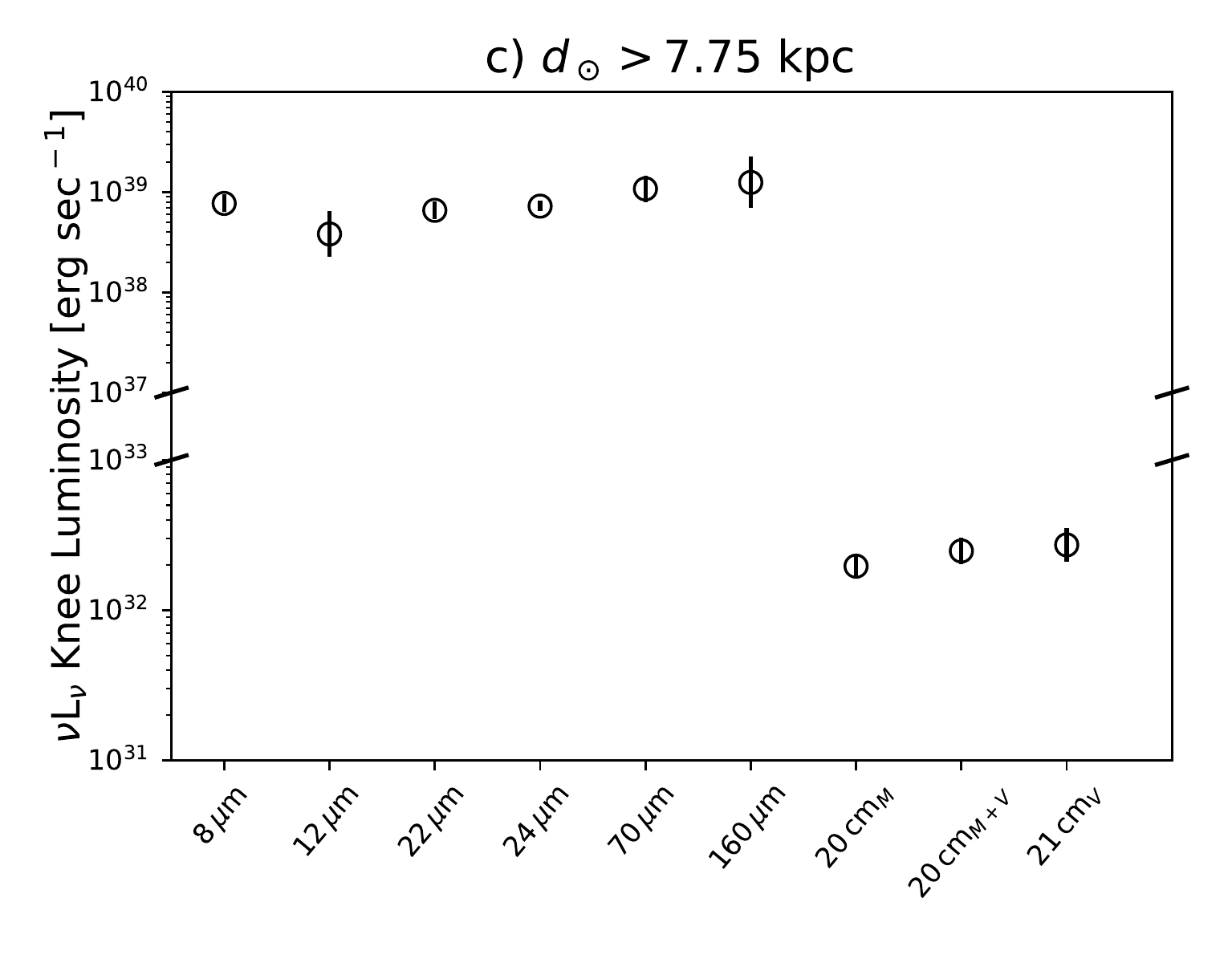}}\\
  \subfloat{\label{fig:nearrgal_knee}%
    \includegraphics[scale=0.45,trim={3cm 7.75cm 3cm 8cm}, clip]{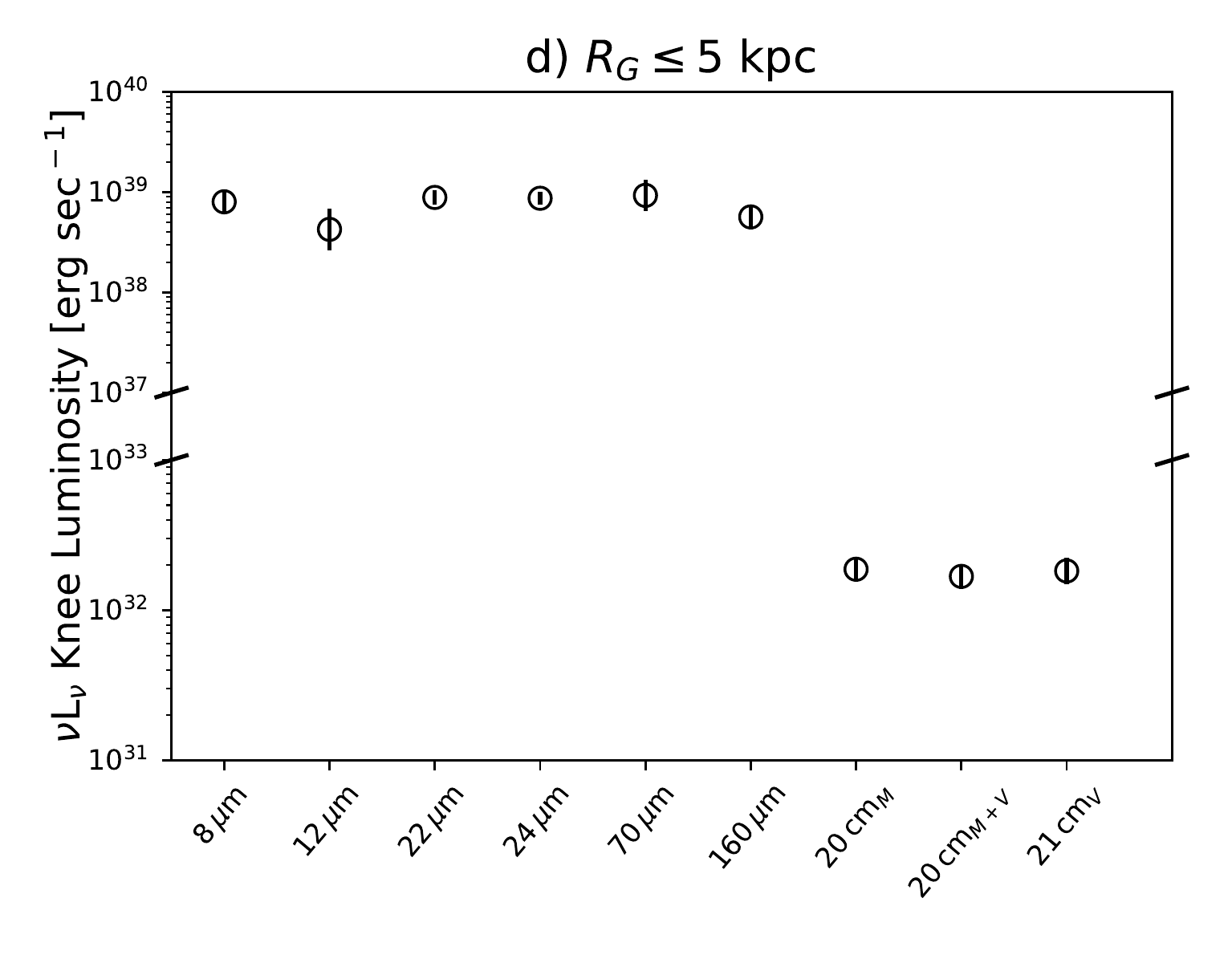}}\qquad
  \subfloat{\label{fig:farrgal_knee}%
    \includegraphics[scale=0.45,trim={3cm 7.75cm 3cm 8cm}, clip]{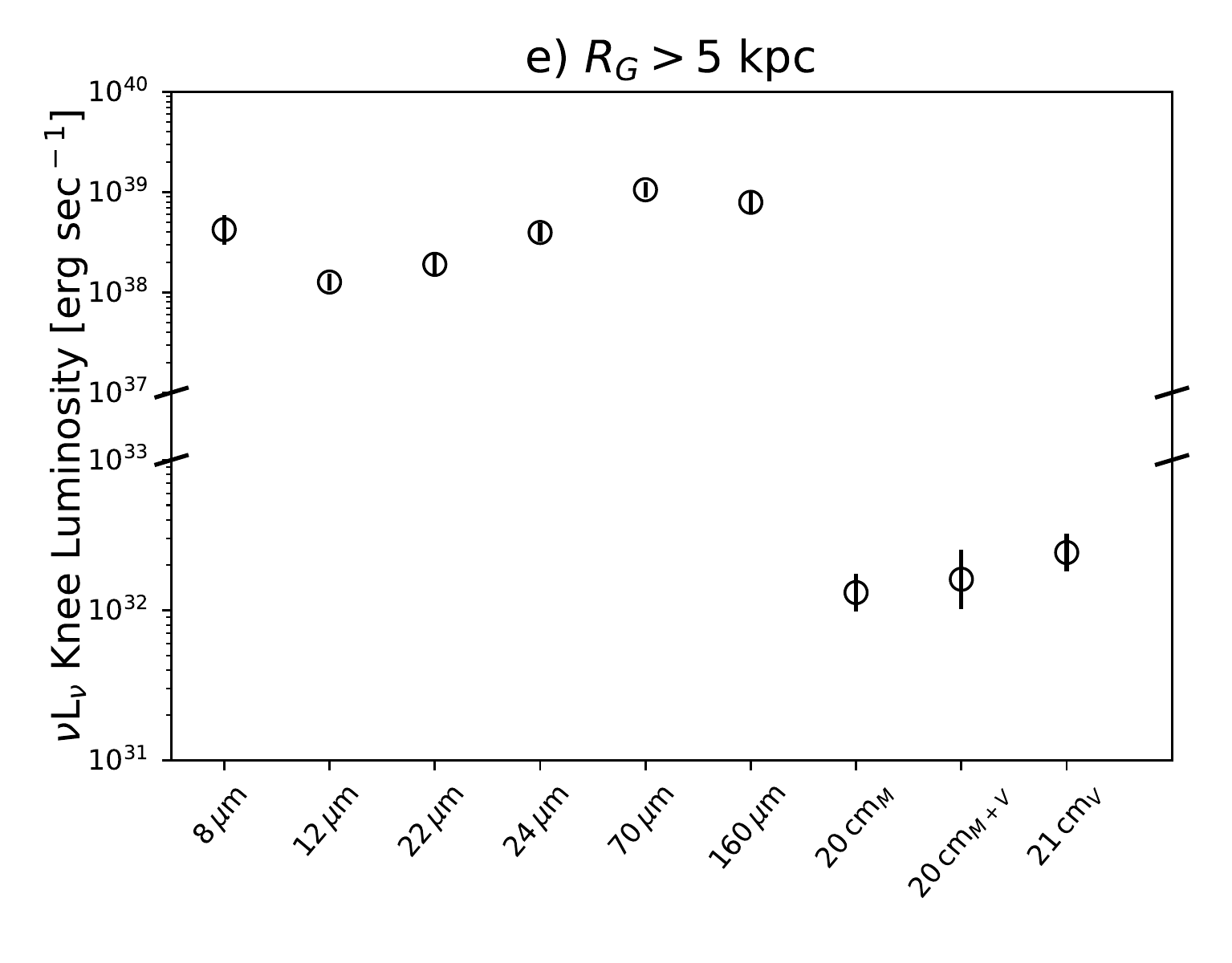}}\qquad
  \subfloat{\label{fig:small_knee}%
    \includegraphics[scale=0.45,trim={3cm 7.75cm 3cm 8cm}, clip]{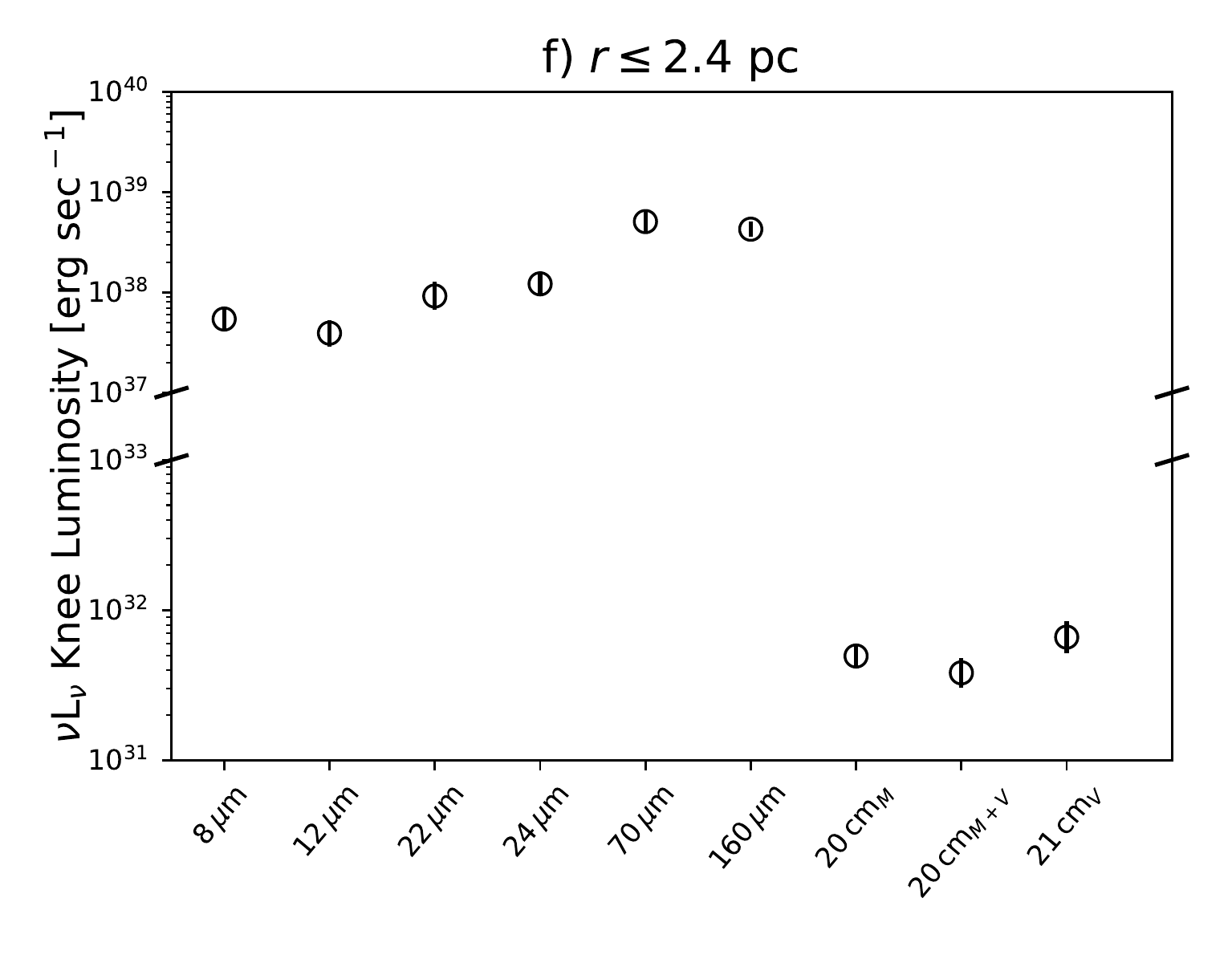}}\\
  \subfloat{\label{fig:large_knee}%
    \includegraphics[scale=0.45,trim={3cm 7.75cm 3cm 8cm}, clip]{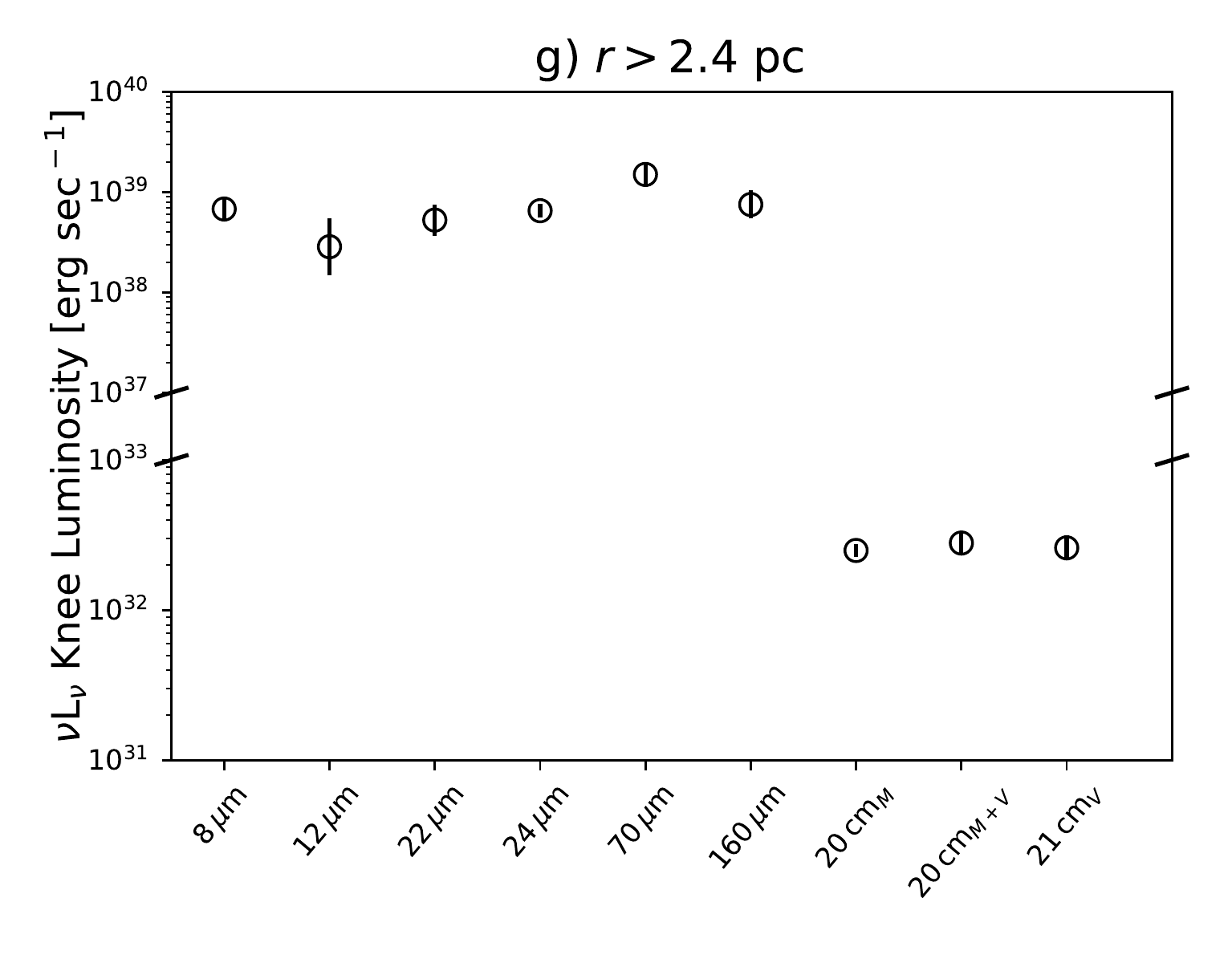}}\qquad
  \subfloat{\label{fig:arm_knee}%
    \includegraphics[scale=0.45,trim={3cm 7.75cm 3cm 8cm}, clip]{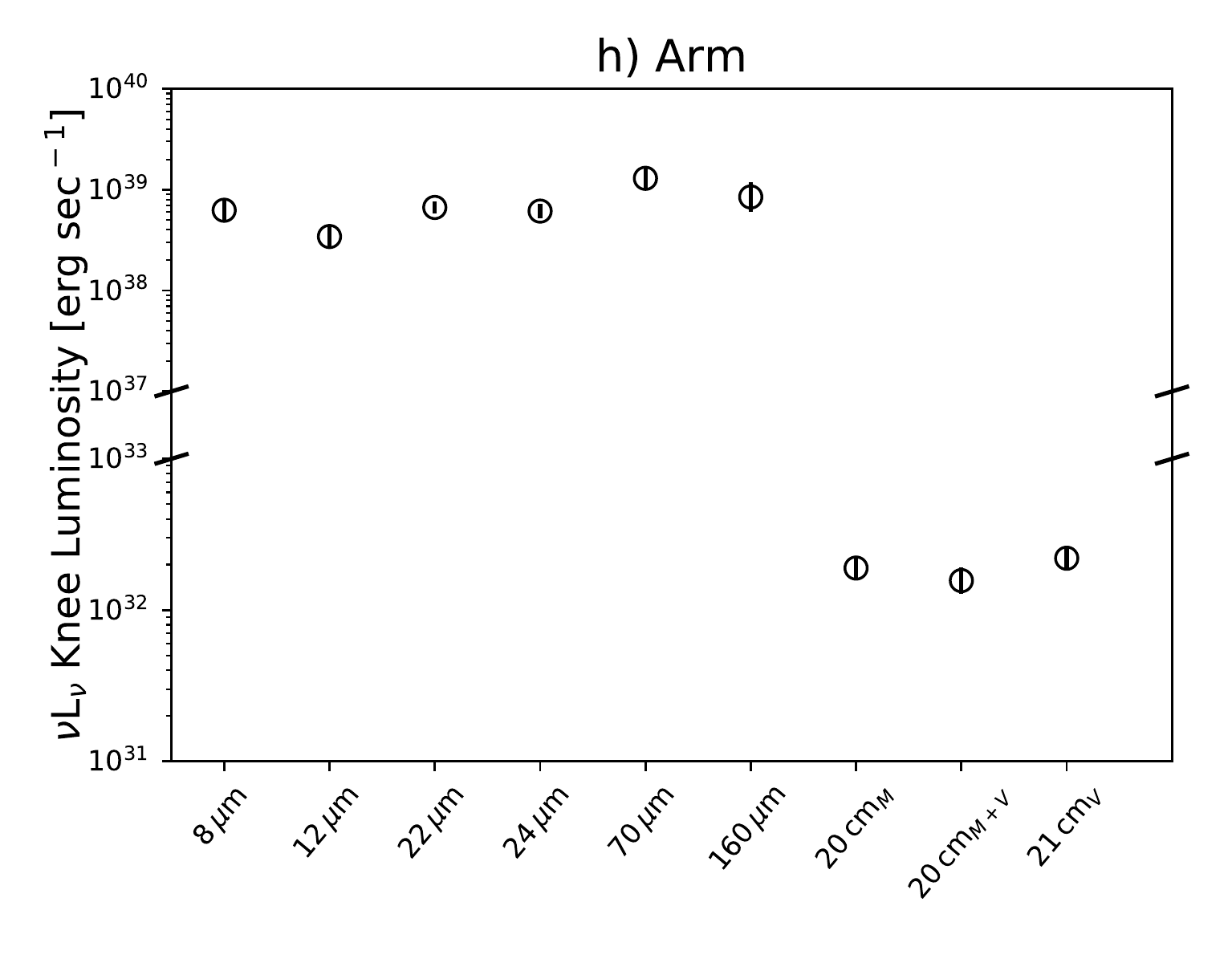}}\qquad
  \subfloat{\label{fig:interarm_knee}%
    \includegraphics[scale=0.45,trim={3cm 7.75cm 3cm 8cm}, clip]{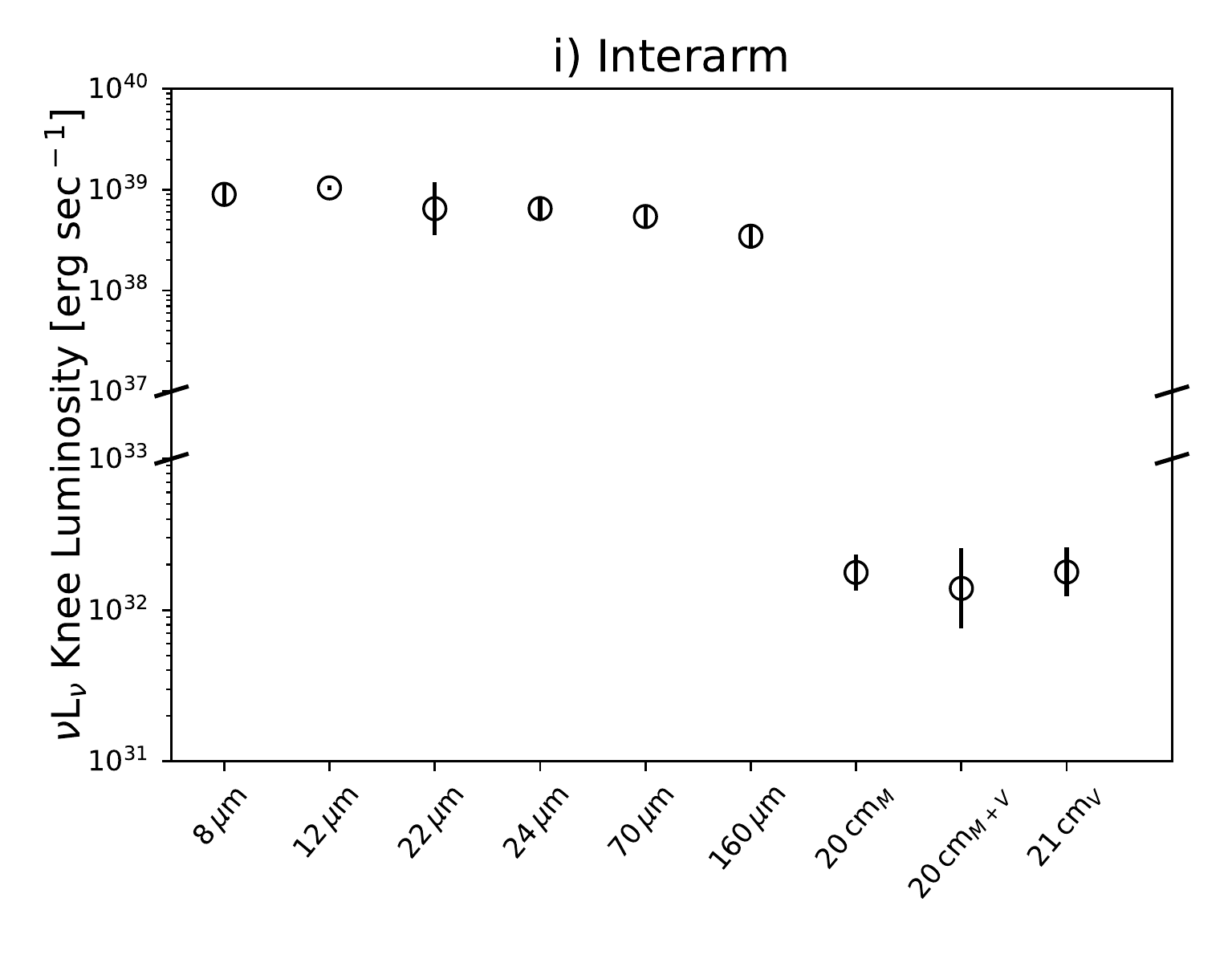}}
\caption{Median knees and MADs from the Monte Carlo-generated luminosity distributions: all sources (panel \subref*{fig:complete_knee}), $d_\sun \leq 7.75$ \kpc (panel \subref*{fig:neard_knee}), $d_\sun > 7.75$ \kpc (panel \subref*{fig:fard_knee}), $\rgal \leq 5$ \kpc (panel \subref*{fig:nearrgal_knee}), $\rgal > 5$ \kpc (panel \subref*{fig:farrgal_knee}), $r \leq 2.4 \pc$ (panel \subref*{fig:small_knee}), $r > 2.4 \pc$ (panel \subref*{fig:large_knee}), arm (panel \subref*{fig:arm_knee}), and interarm (panel \subref*{fig:interarm_knee}).}
\label{fig:kneecomp2}
\end{sidewaysfigure*}

\begin{sidewaysfigure*}[h]
\centering
  \subfloat{\label{fig:complete_limit}%
    \includegraphics[scale=0.45,trim={3cm 7.75cm 3cm 8cm}, clip]{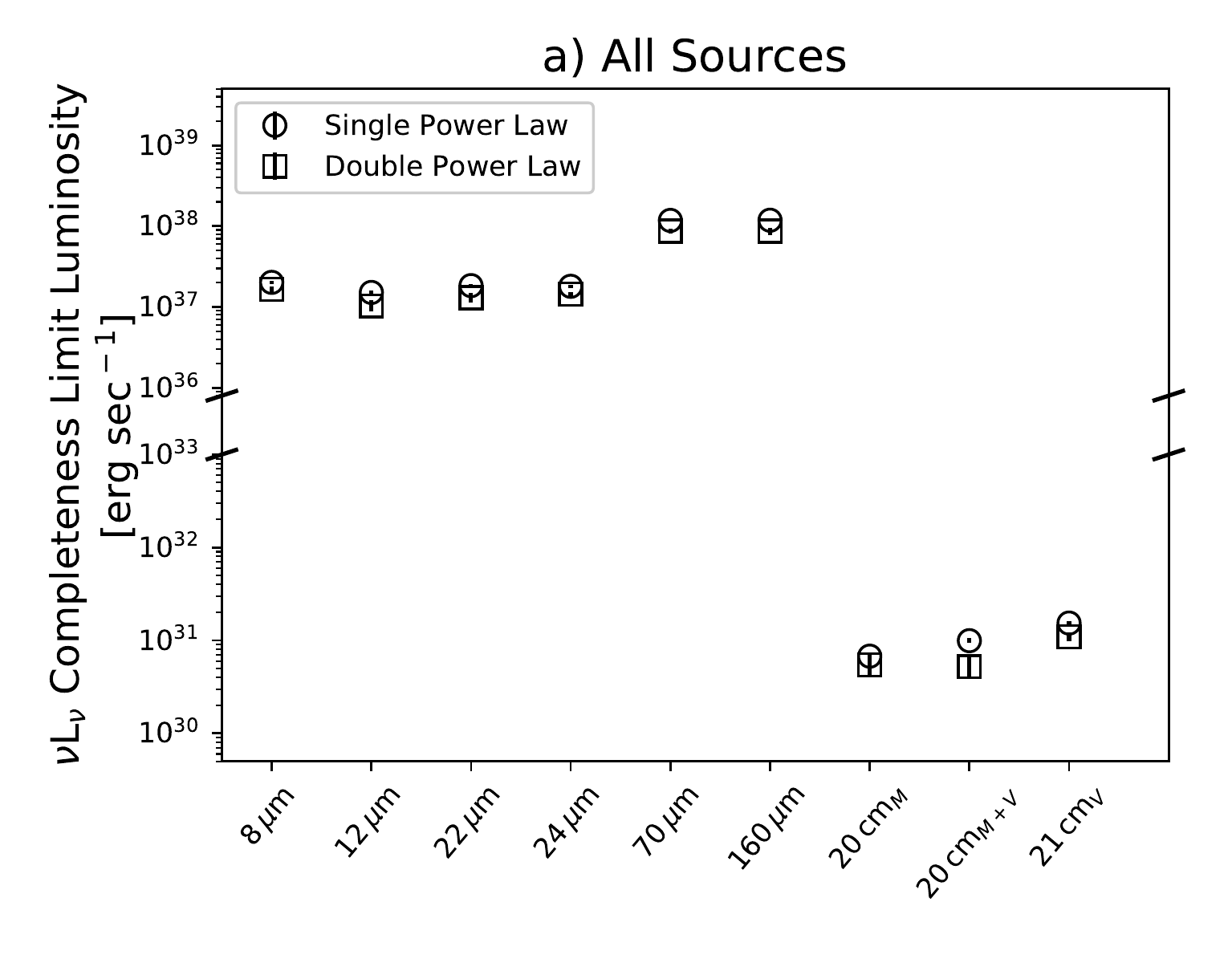}}\qquad
  \subfloat{\label{fig:neard_limit}%
    \includegraphics[scale=0.45,trim={3cm 7.75cm 3cm 8cm}, clip]{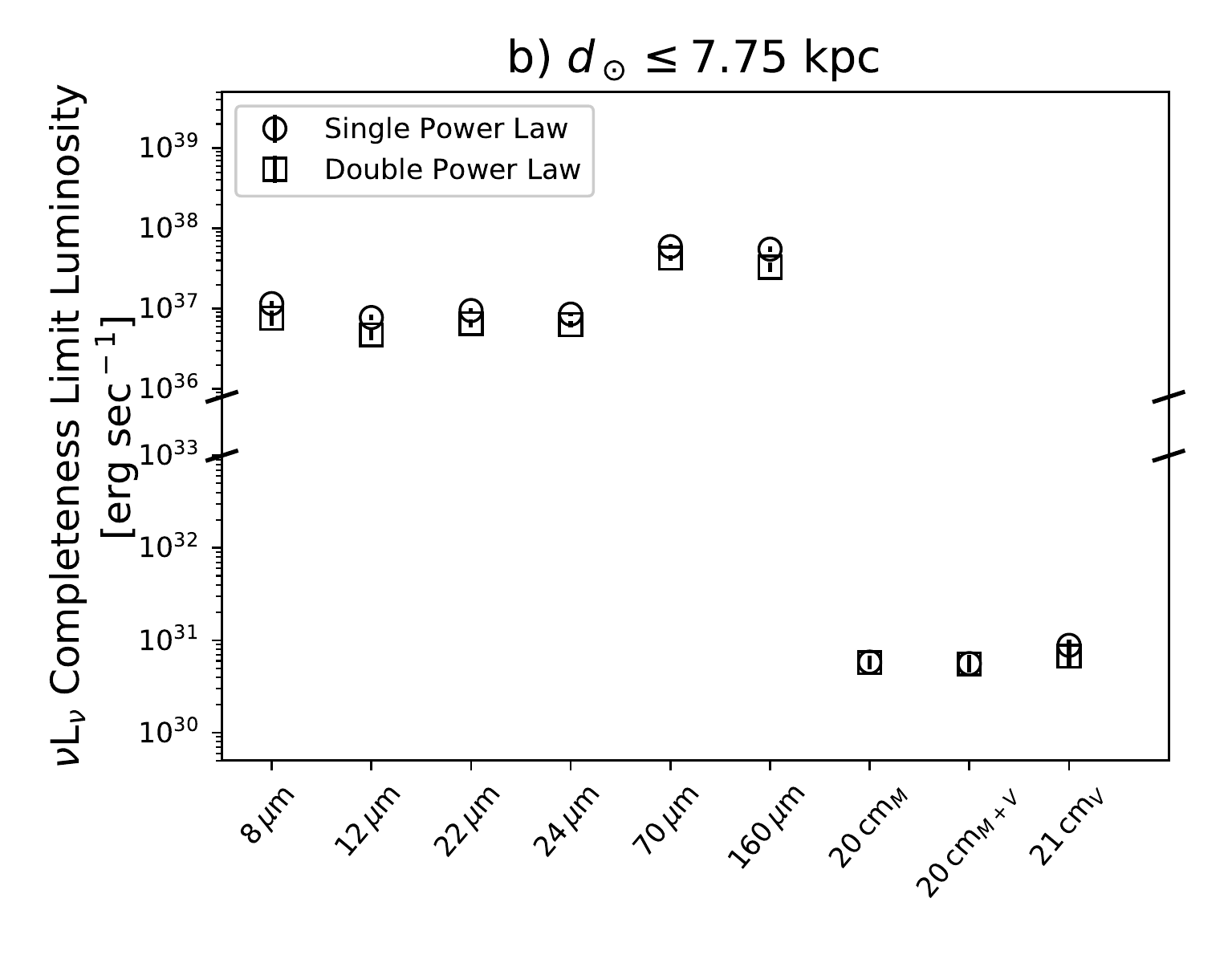}}\qquad
  \subfloat{\label{fig:fard_limit}%
    \includegraphics[scale=0.45,trim={3cm 7.75cm 3cm 8cm}, clip]{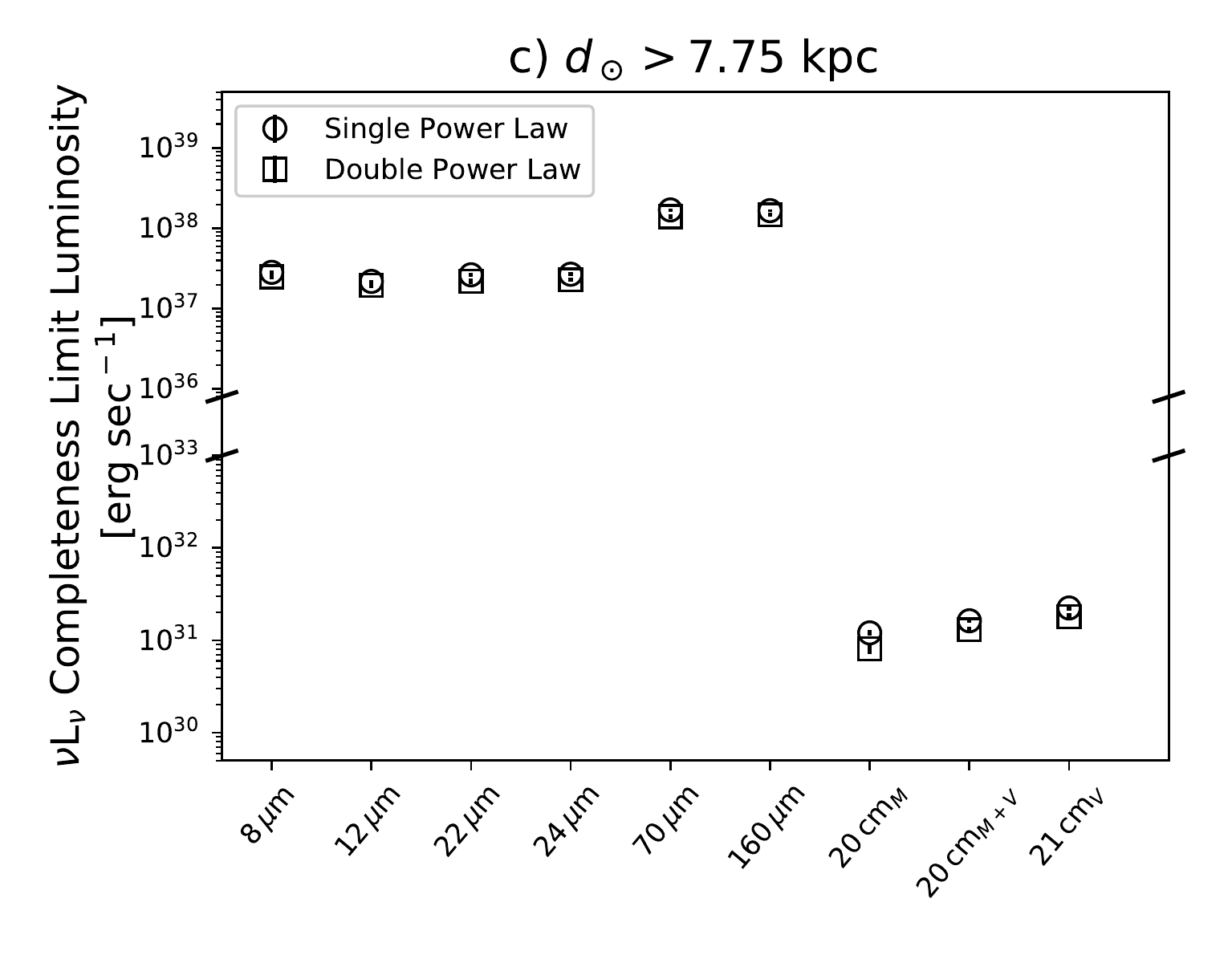}}\\
  \subfloat{\label{fig:nearrgal_limit}%
    \includegraphics[scale=0.45,trim={3cm 7.75cm 3cm 8cm}, clip]{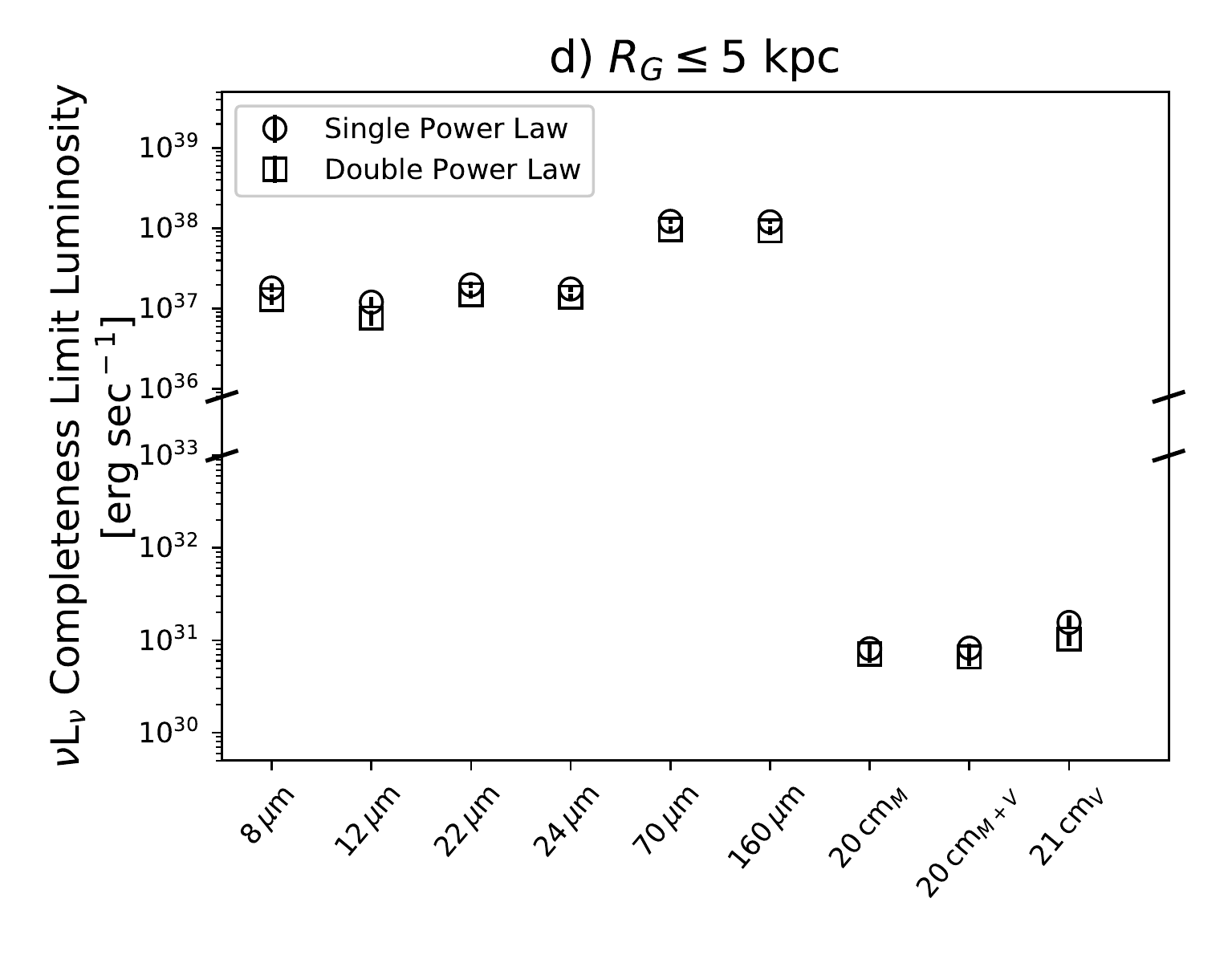}}\qquad
  \subfloat{\label{fig:farrgal_limit}%
    \includegraphics[scale=0.45,trim={3cm 7.75cm 3cm 8cm}, clip]{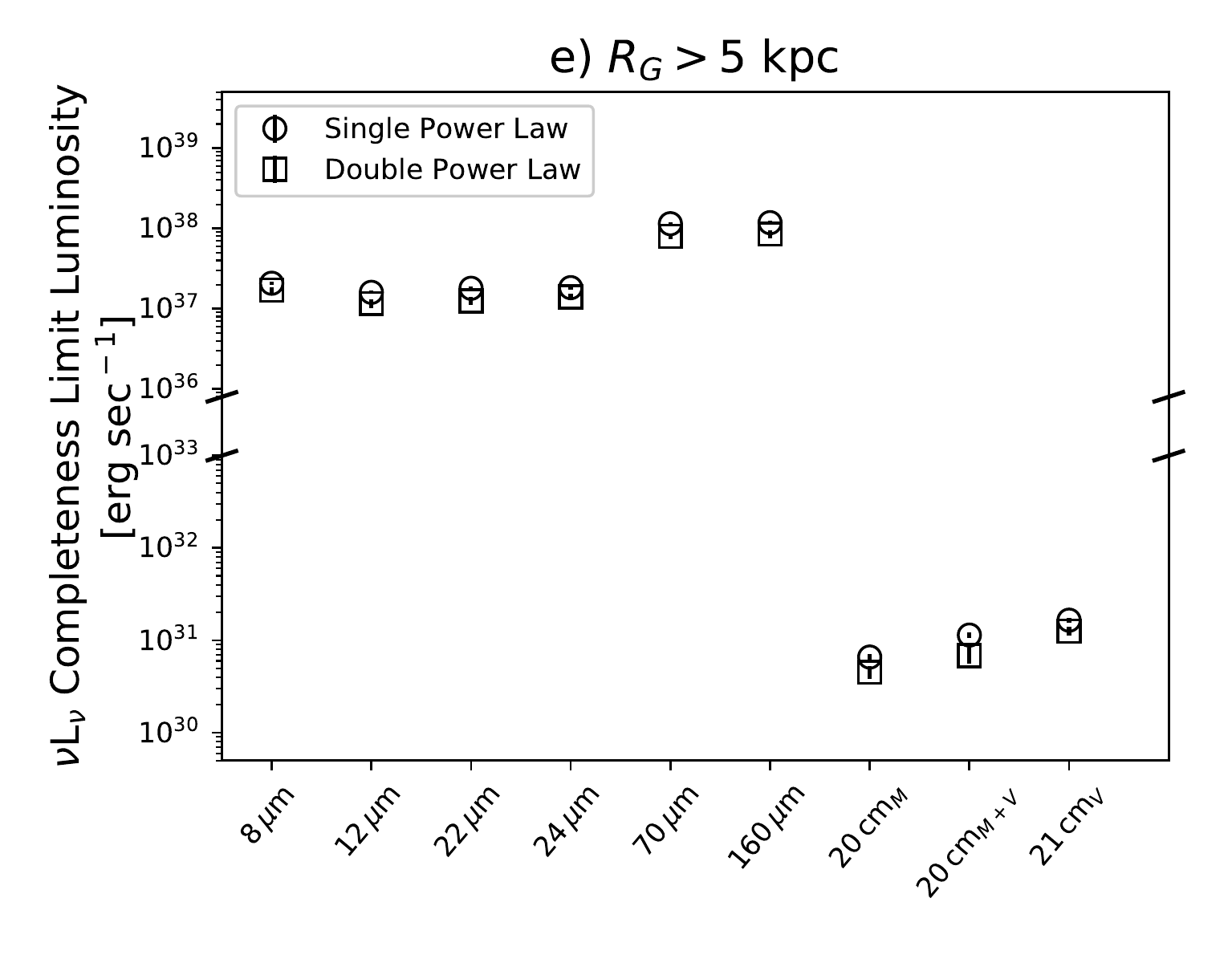}}\qquad
  \subfloat{\label{fig:small_limit}%
    \includegraphics[scale=0.45,trim={3cm 7.75cm 3cm 8cm}, clip]{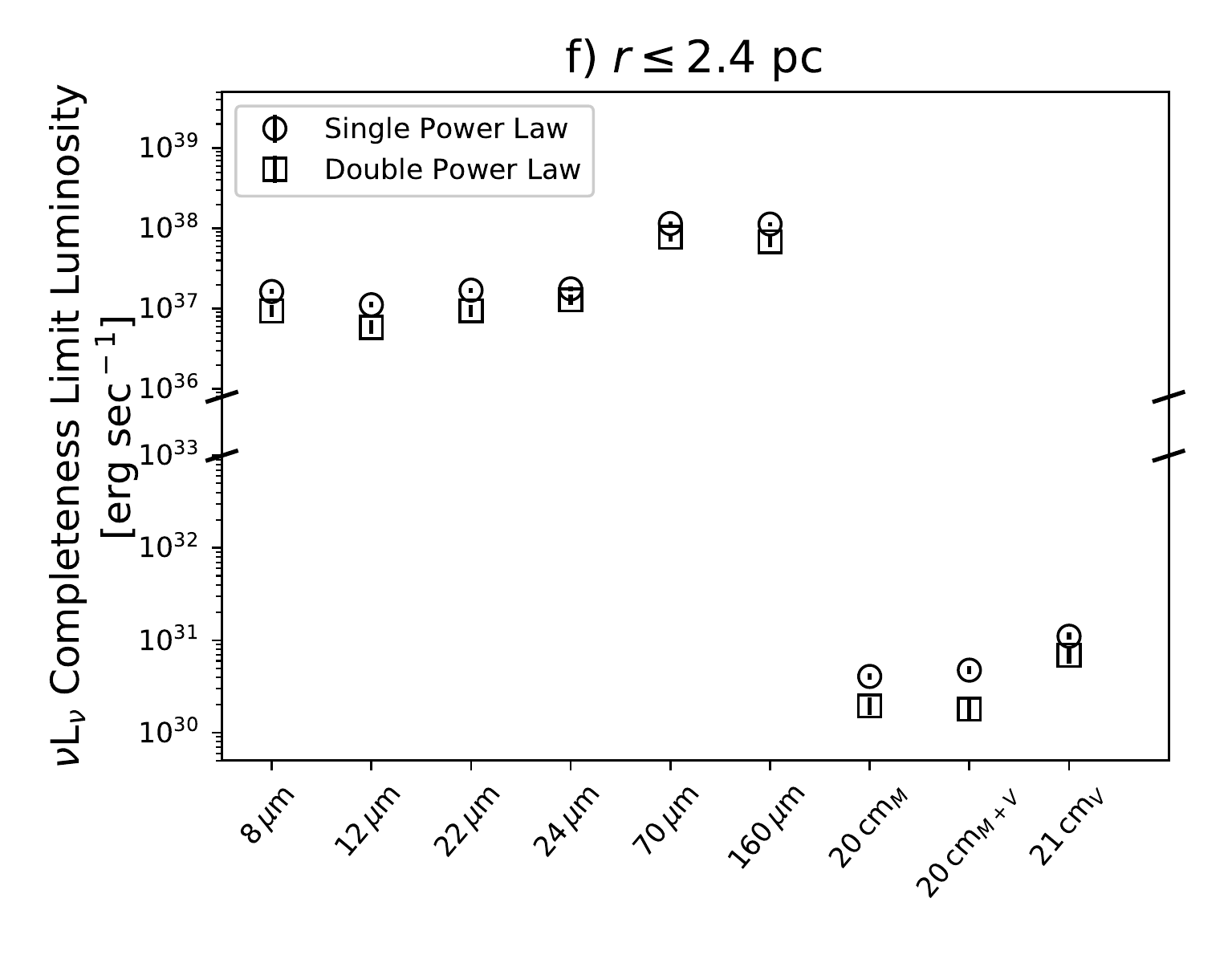}}\\
  \subfloat{\label{fig:large_limit}%
    \includegraphics[scale=0.45,trim={3cm 7.75cm 3cm 8cm}, clip]{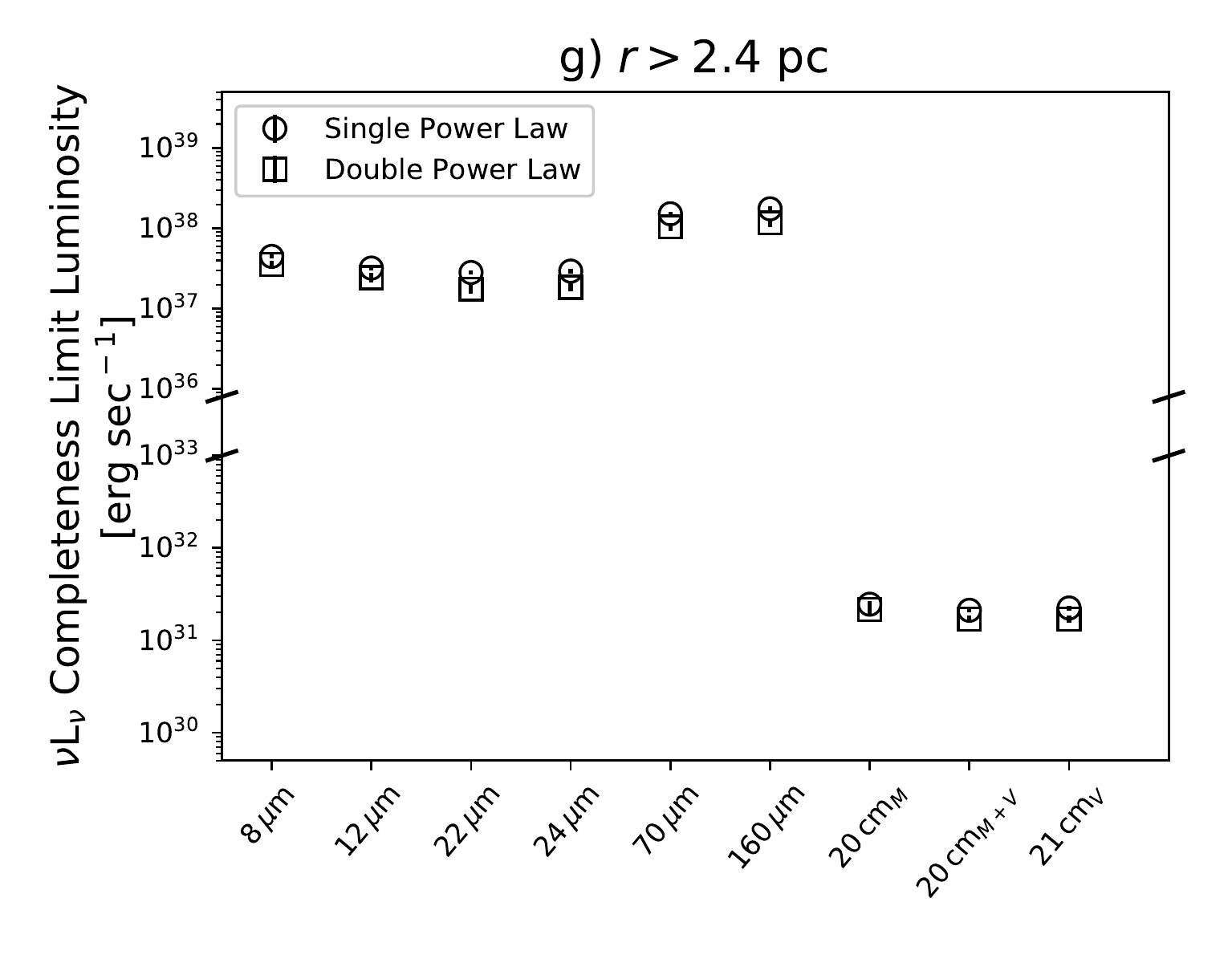}}\qquad
  \subfloat{\label{fig:arm_limit}%
    \includegraphics[scale=0.45,trim={3cm 7.75cm 3cm 8cm}, clip]{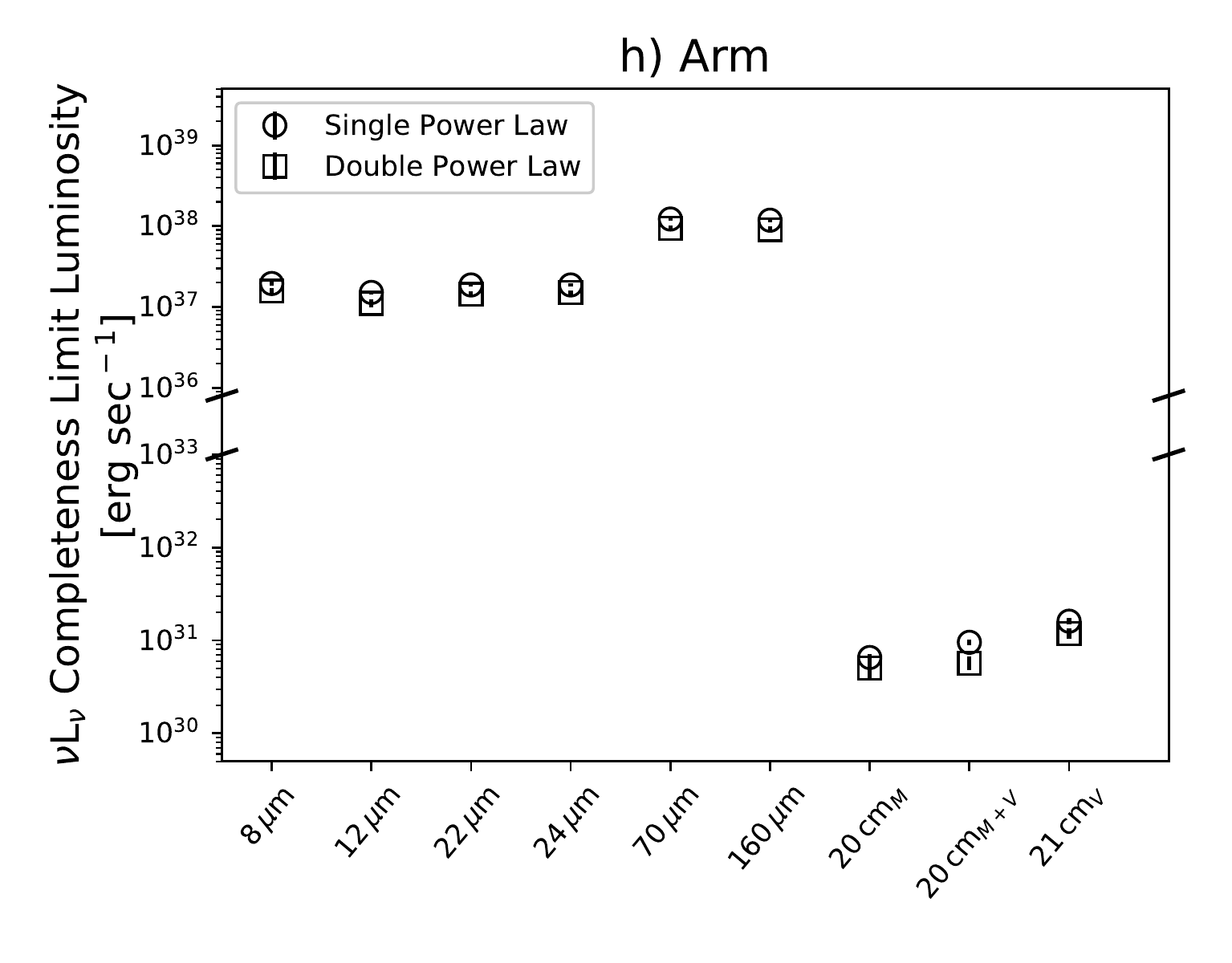}}\qquad
  \subfloat{\label{fig:interarm_limit}%
    \includegraphics[scale=0.45,trim={3cm 7.75cm 3cm 8cm}, clip]{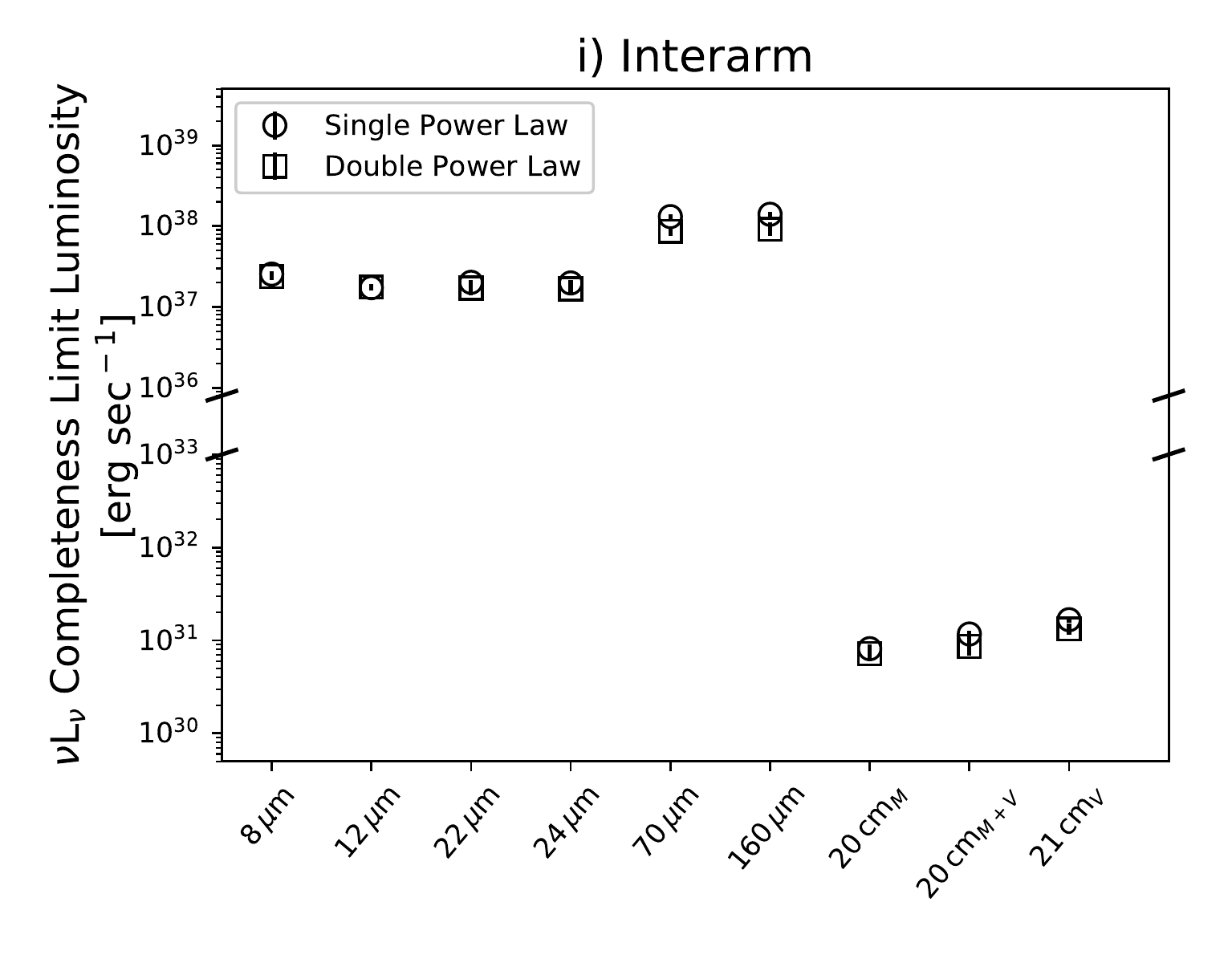}}
\caption{Completeness limits and MADs from the Monte Carlo-generated luminosity distributions: all sources (panel \subref*{fig:complete_limit}), $d_\sun \leq 7.75$ \kpc (panel \subref*{fig:neard_limit}), $d_\sun > 7.75$ \kpc (panel \subref*{fig:fard_limit}), $\rgal \leq 5$ \kpc (panel \subref*{fig:nearrgal_limit}), $\rgal > 5$ \kpc (panel \subref*{fig:farrgal_limit}), $r \leq 2.4 \pc$ (panel \subref*{fig:small_limit}), $r > 2.4 \pc$ (panel \subref*{fig:large_limit}), arm (panel \subref*{fig:arm_limit}), and interarm (panel \subref*{fig:interarm_limit}).}
\label{fig:limitcomp2}
\end{sidewaysfigure*}

\clearpage

\subsection{Blending Analysis}
\label{appsubsec:blend}

Here we compare the median power law indices, knee luminosities, and completeness limit luminosities of the unblended data for each subset at each wavelength to those of the blended data. $\alpha$ is the single power law index as defined in Equation \ref{eq:singlaw}. $\alpha_{1}$ and $\alpha_{2}$ are the double power law indices as defined in Equation \ref{eq:double_law}.

\begin{sidewaysfigure*}[h]
\centering
  \subfloat{\label{fig:glimpse_alpha}%
    \includegraphics[scale=0.45,trim={3cm 7.75cm 3cm 8cm}, clip]{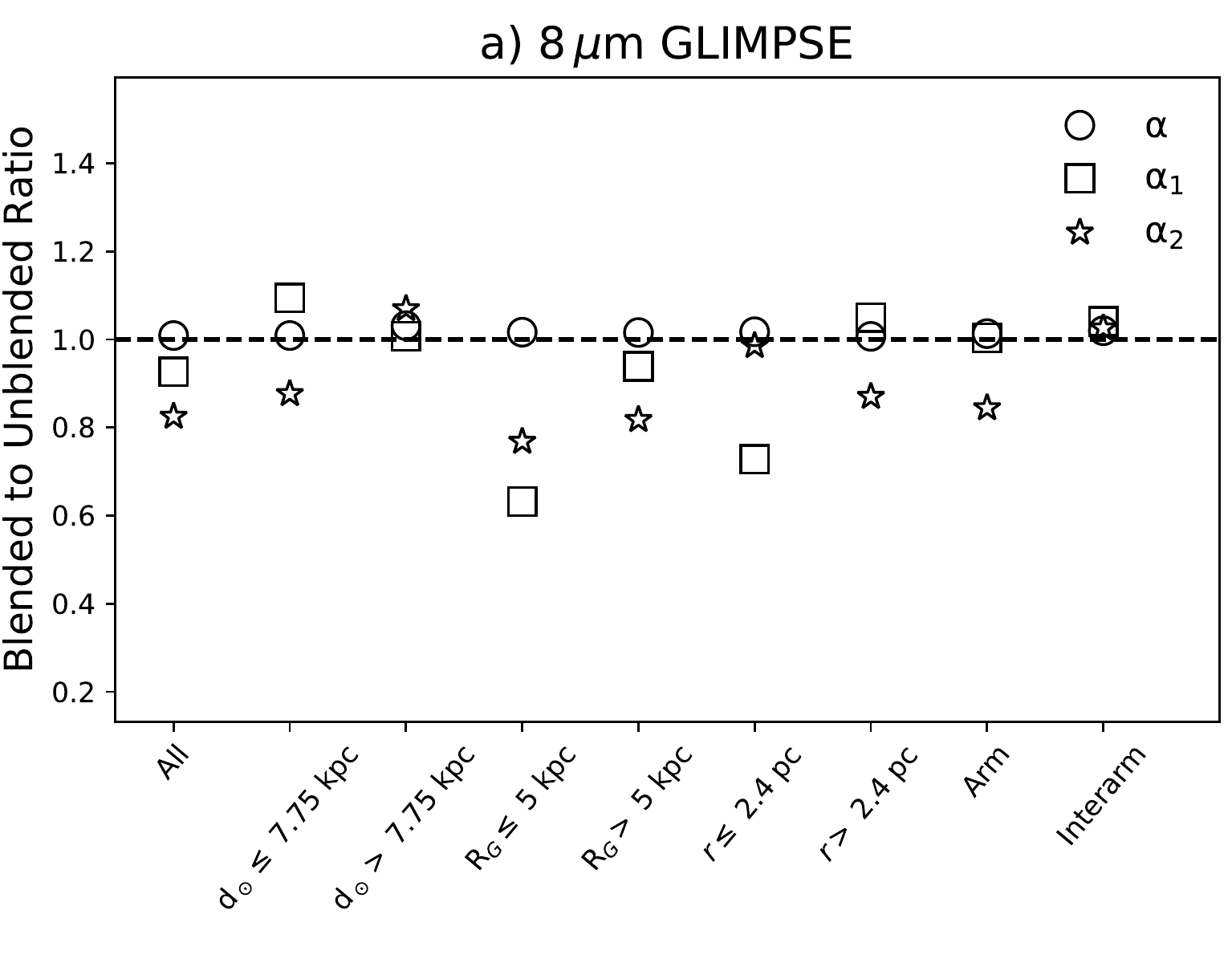}}\qquad
  \subfloat{\label{fig:wise3_alpha}%
    \includegraphics[scale=0.45,trim={3cm 7.75cm 3cm 8cm}, clip]{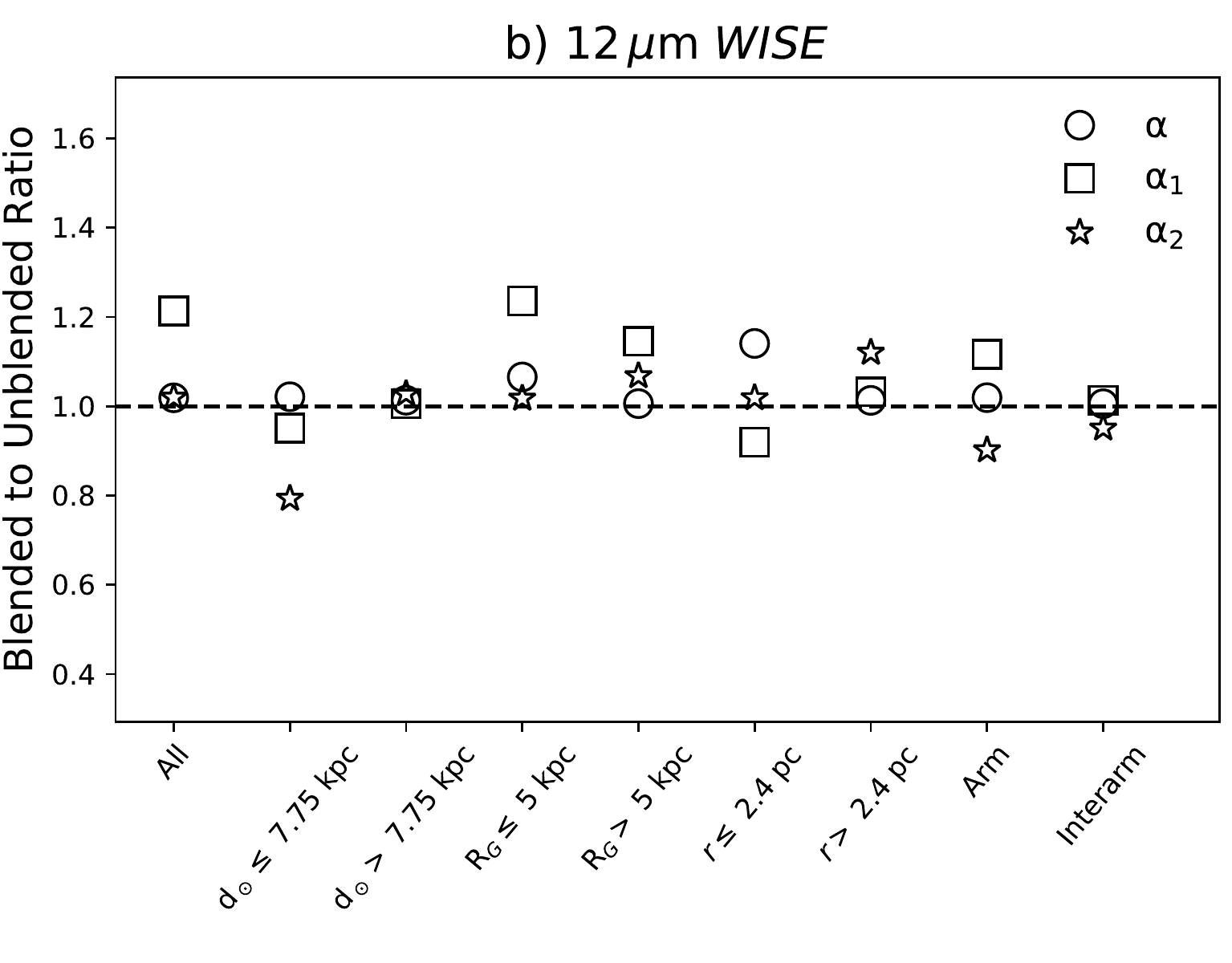}}\qquad
  \subfloat{\label{fig:wise4_alpha}%
    \includegraphics[scale=0.45,trim={3cm 7.75cm 3cm 8cm}, clip]{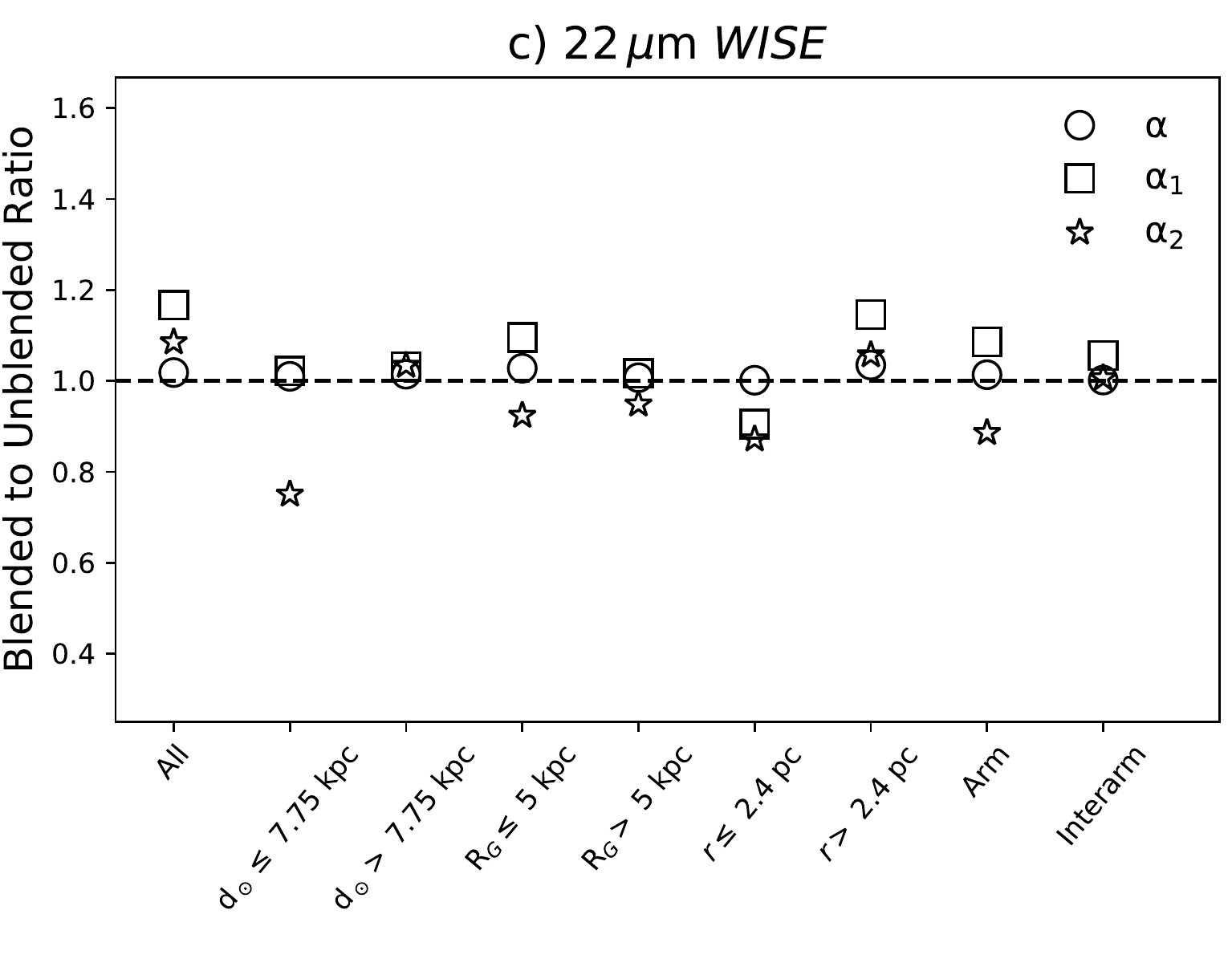}}\\
  \subfloat{\label{fig:mipsgal_alpha}%
    \includegraphics[scale=0.45,trim={3cm 7.75cm 3cm 8cm}, clip]{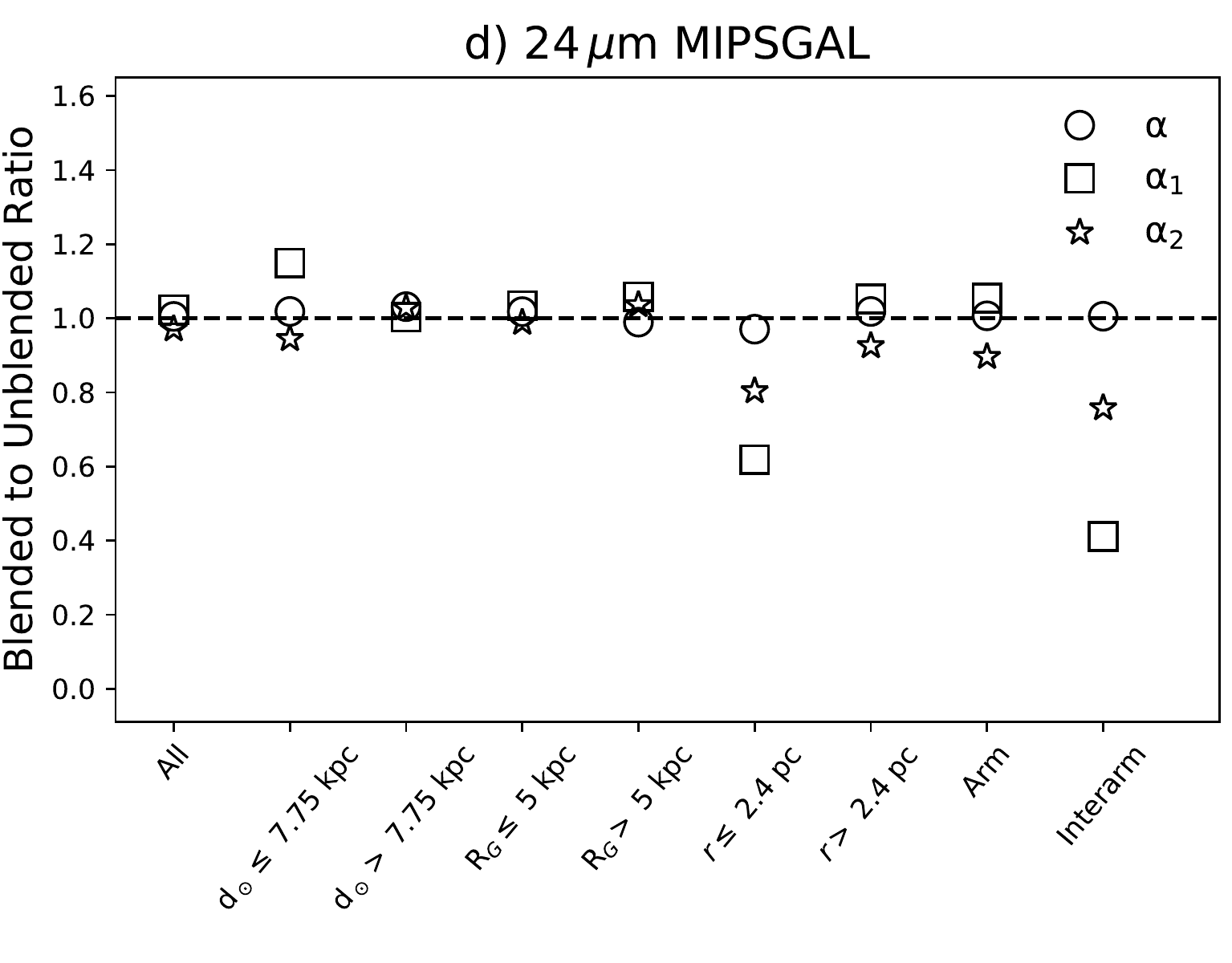}}\qquad
  \subfloat{\label{fig:higal70_alpha}%
    \includegraphics[scale=0.45,trim={3cm 7.75cm 3cm 8cm}, clip]{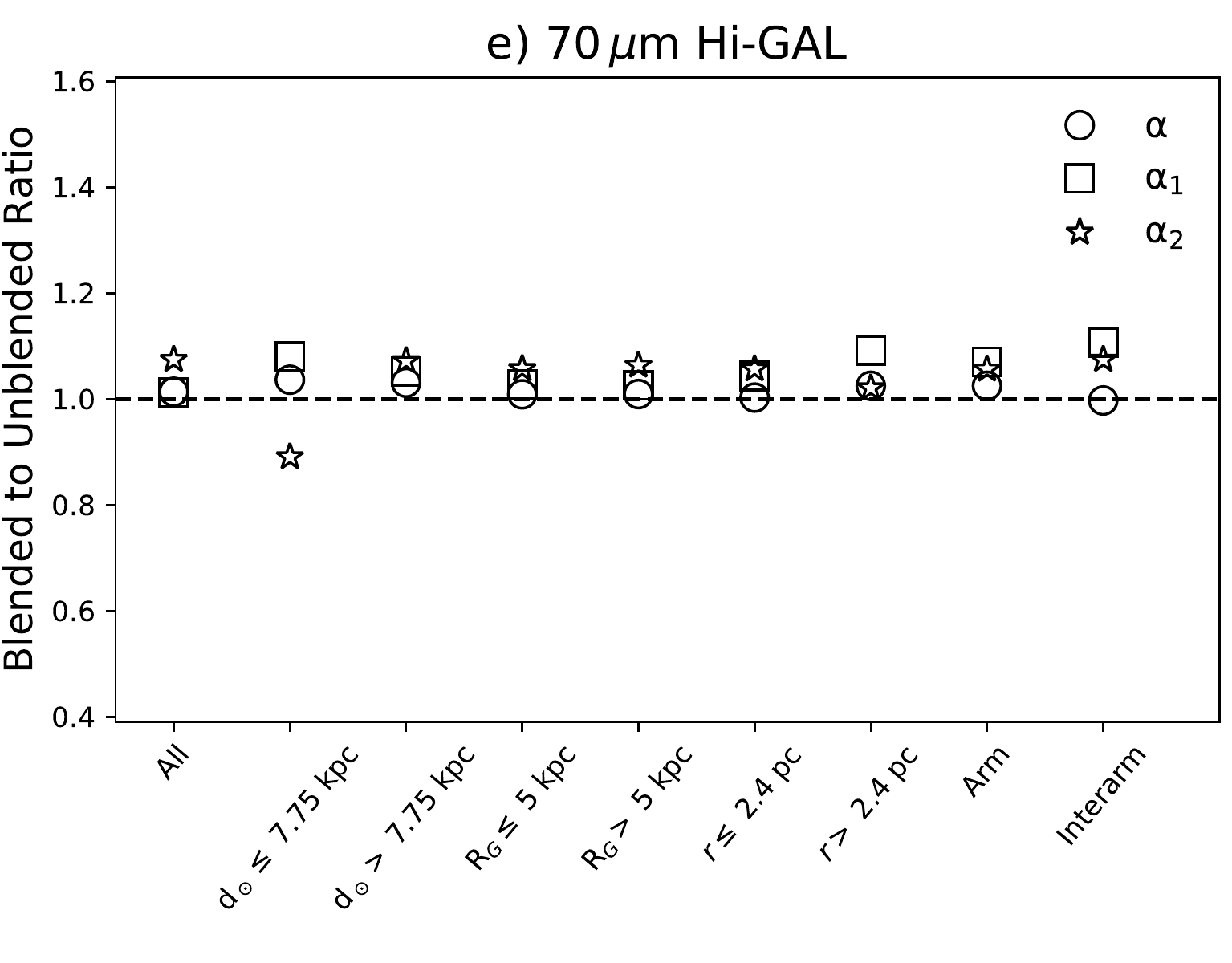}}\qquad
  \subfloat{\label{fig:higal160_alpha}%
    \includegraphics[scale=0.45,trim={3cm 7.75cm 3cm 8cm}, clip]{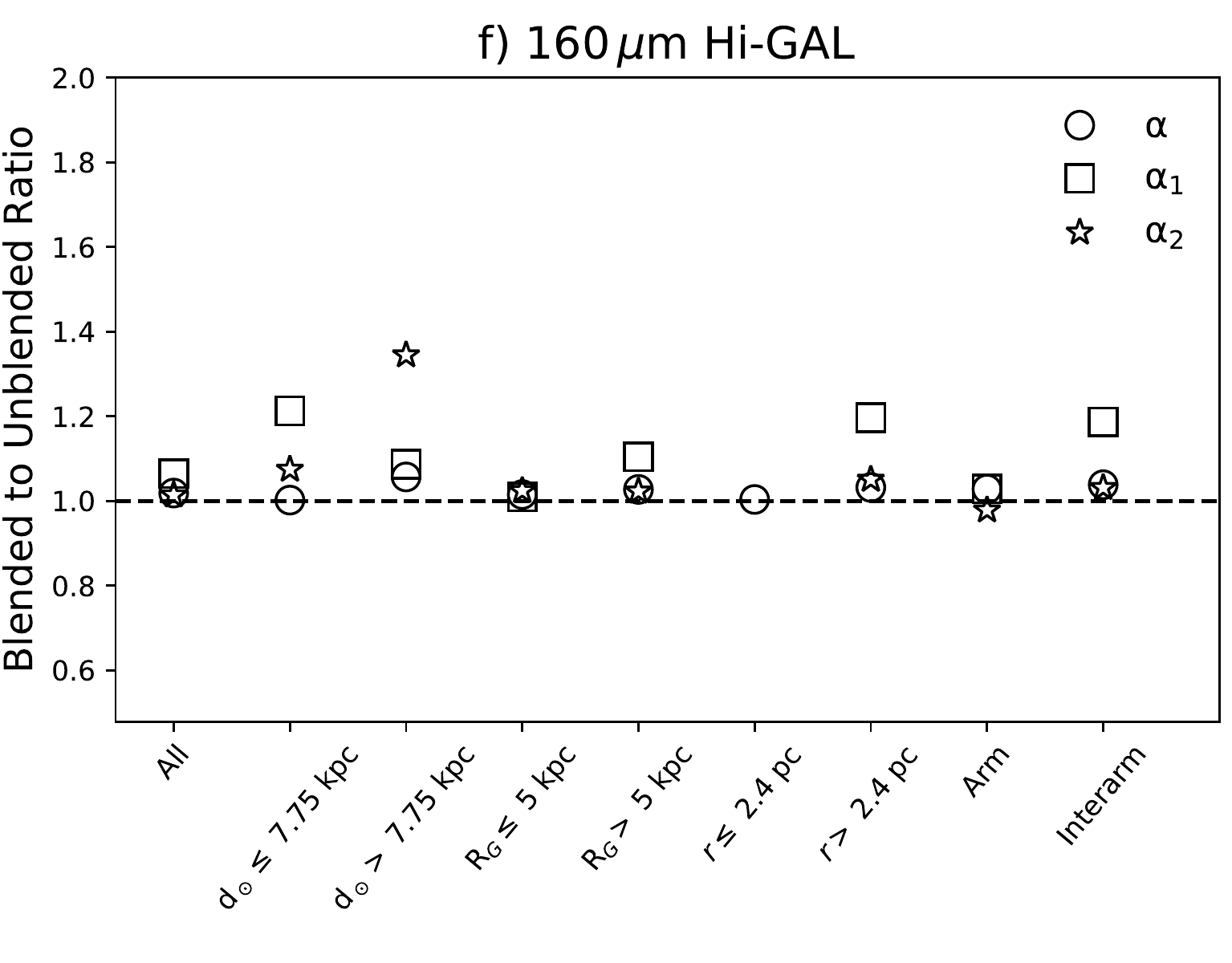}}\\
  \subfloat{\label{fig:magpis_alpha}%
    \includegraphics[scale=0.45,trim={3cm 7.75cm 3cm 8cm}, clip]{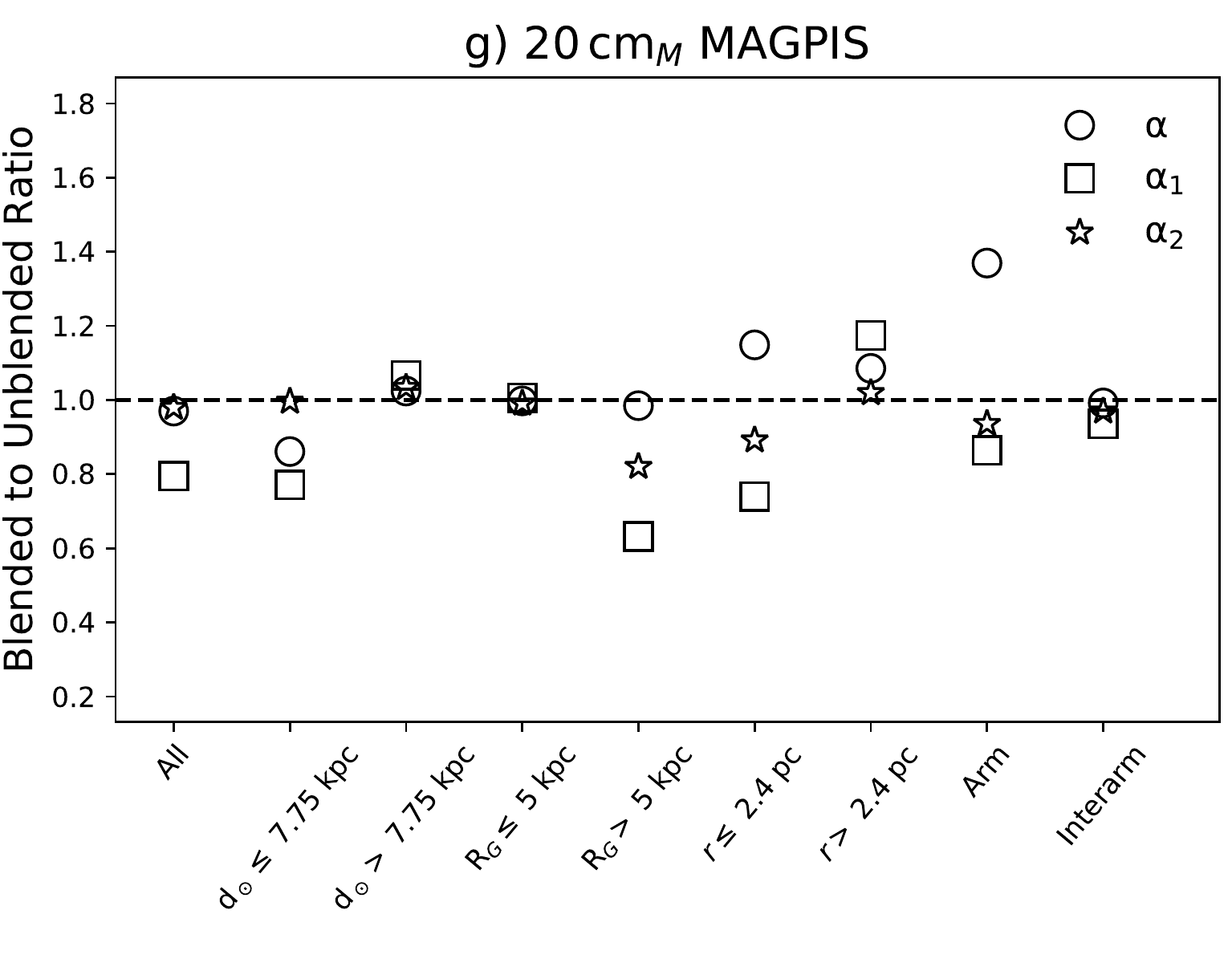}}\qquad
  \subfloat{\label{fig:magpis_vgps_alpha}%
    \includegraphics[scale=0.45,trim={3cm 7.75cm 3cm 8cm}, clip]{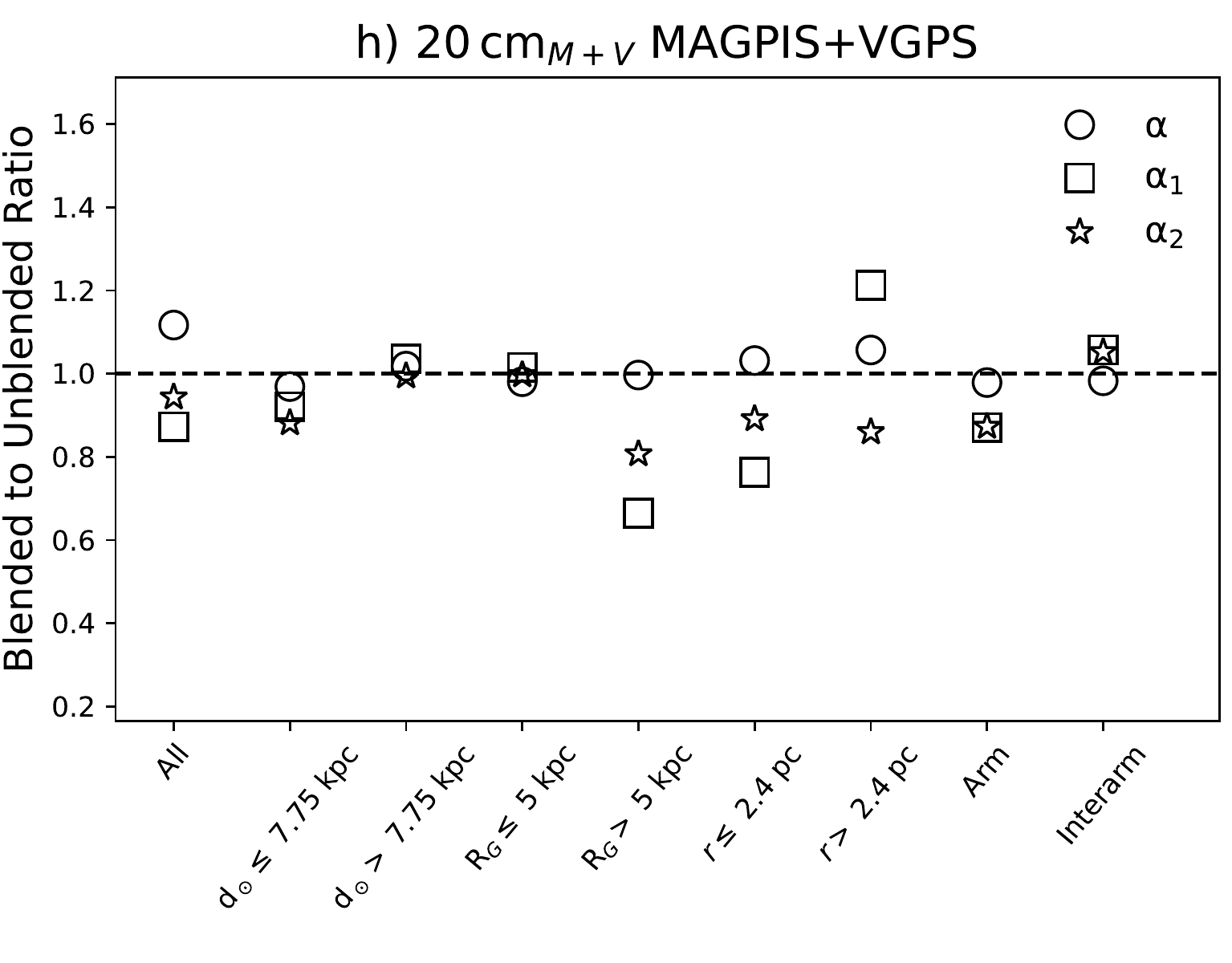}}\qquad
  \subfloat{\label{fig:vgps_alpha}%
    \includegraphics[scale=0.45,trim={3cm 7.75cm 3cm 8cm}, clip]{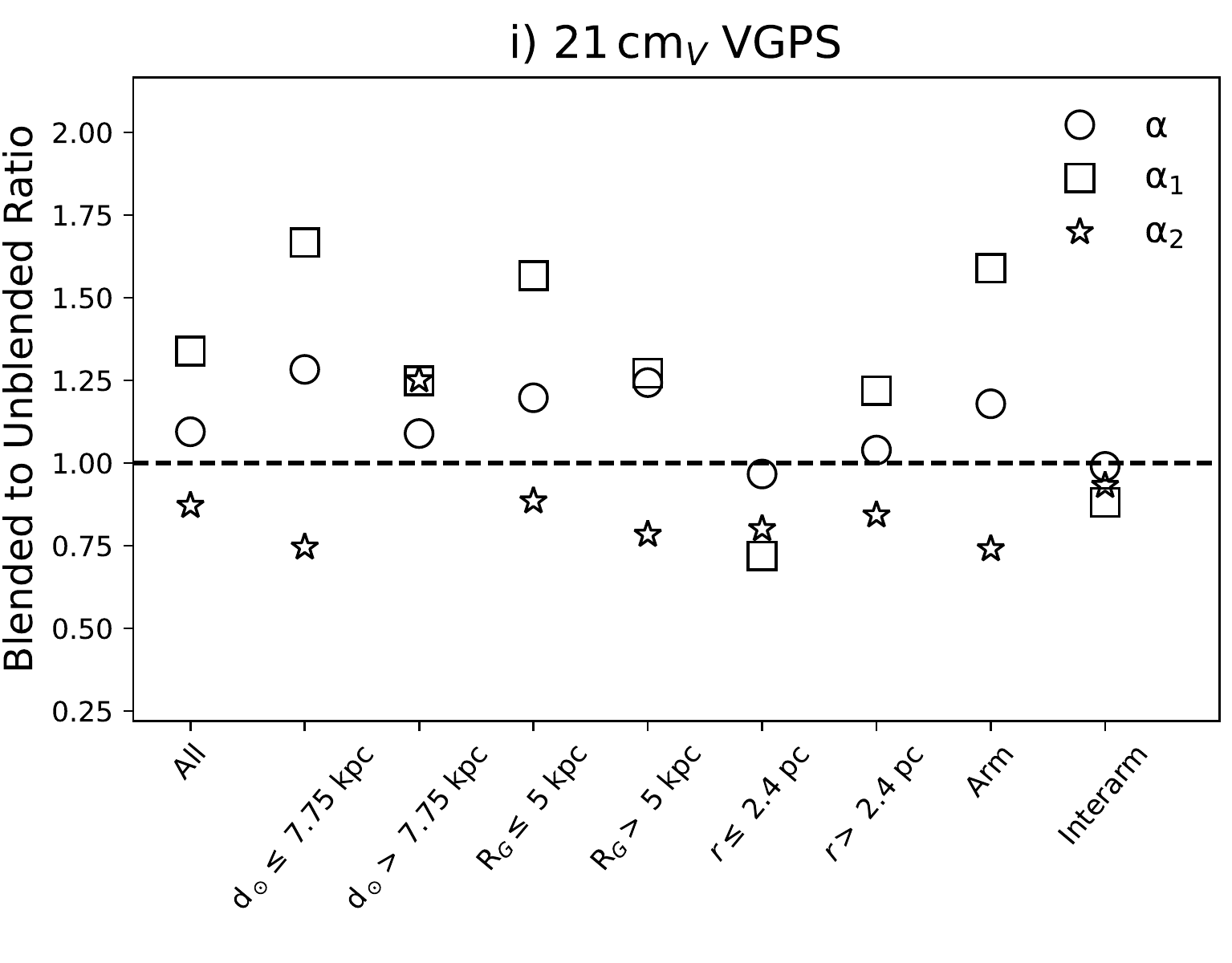}}
\caption{Comparison of unblended and blended single and double power law indices for $8\,\microns$ GLIMPSE (panel \subref*{fig:glimpse_alpha}), $12\,\microns$ \textit{WISE} (panel \subref*{fig:wise3_alpha}), $22\,\microns$ \textit{WISE} (panel \subref*{fig:wise4_alpha}), $24\,\microns$ MIPSGAL (panel \subref*{fig:mipsgal_alpha}), $70\,\microns$ Hi-GAL (panel \subref*{fig:higal70_alpha}), $160\,\microns$ Hi-GAL (panel \subref*{fig:higal160_alpha}), $20\,\cm$ MAGPIS (panel \subref*{fig:magpis_alpha}), $21\,\cm$ MAGPIS+VGPS (panel \subref*{fig:magpis_vgps_alpha}), and $21\,\cm$ VGPS (panel \subref*{fig:vgps_alpha}). The vertical axis has been constrained for clarity in panel \subref*{fig:higal160_alpha}; the double power law indices of the small physical size subset lie outside the displayed vertical range.}
\label{fig:alphacomp_blend2}
\end{sidewaysfigure*}

\begin{sidewaysfigure*}[h]
\centering
  \subfloat{\label{fig:glimpse_knee}%
    \includegraphics[scale=0.45,trim={3cm 7.75cm 3cm 8cm}, clip]{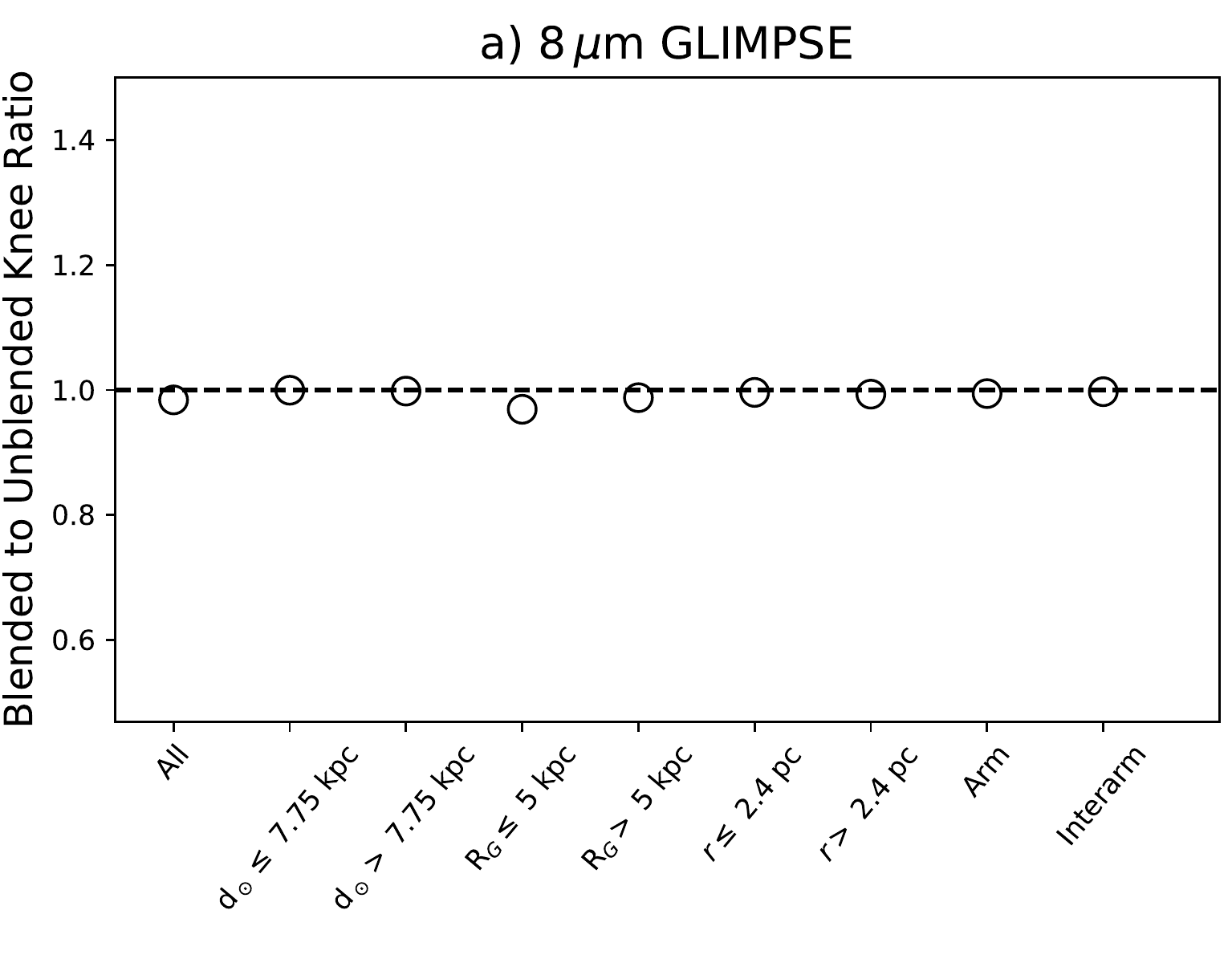}}\qquad
  \subfloat{\label{fig:wise3_knee}%
    \includegraphics[scale=0.45,trim={3cm 7.75cm 3cm 8cm}, clip]{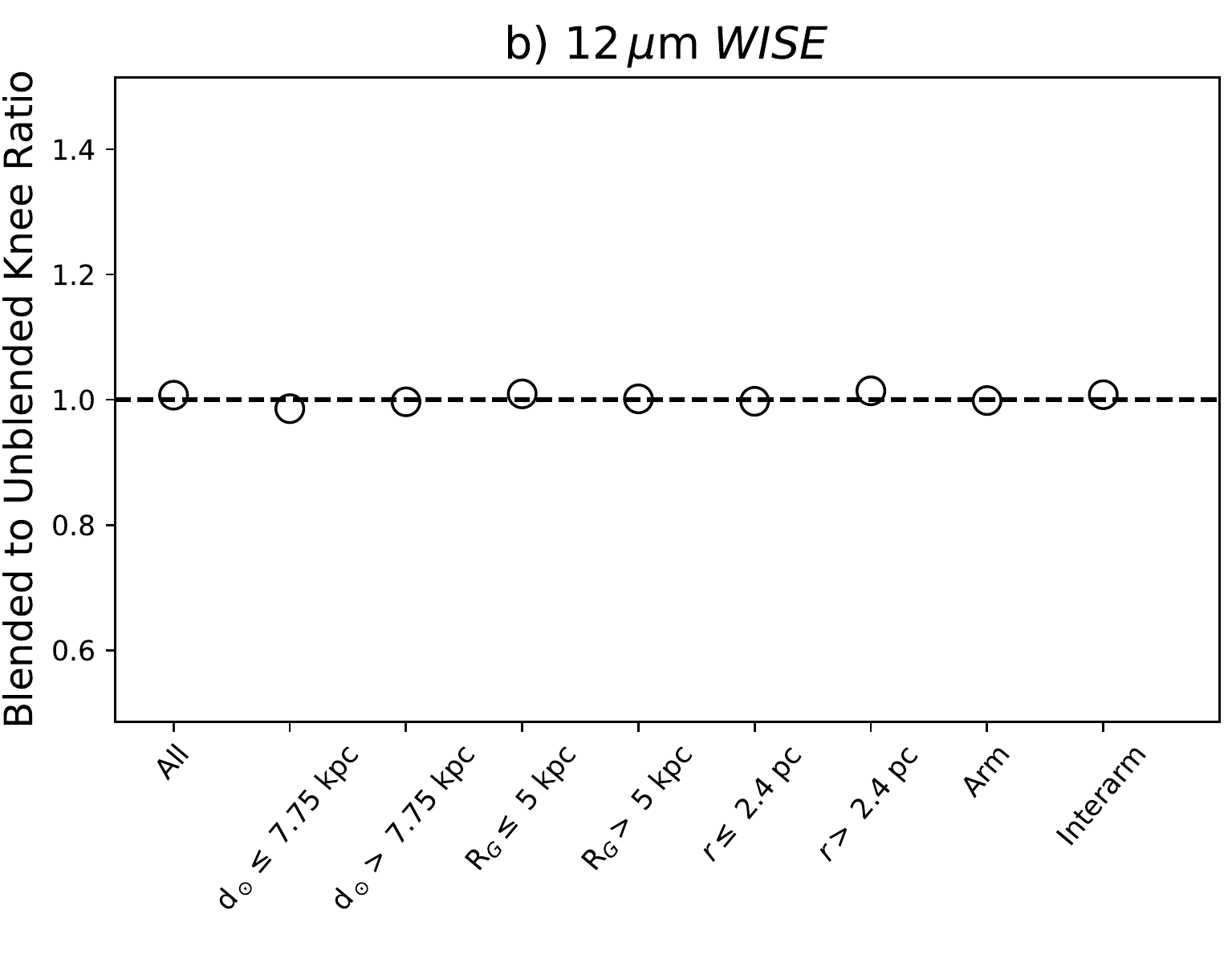}}\qquad
  \subfloat{\label{fig:wise4_knee}%
    \includegraphics[scale=0.45,trim={3cm 7.75cm 3cm 8cm}, clip]{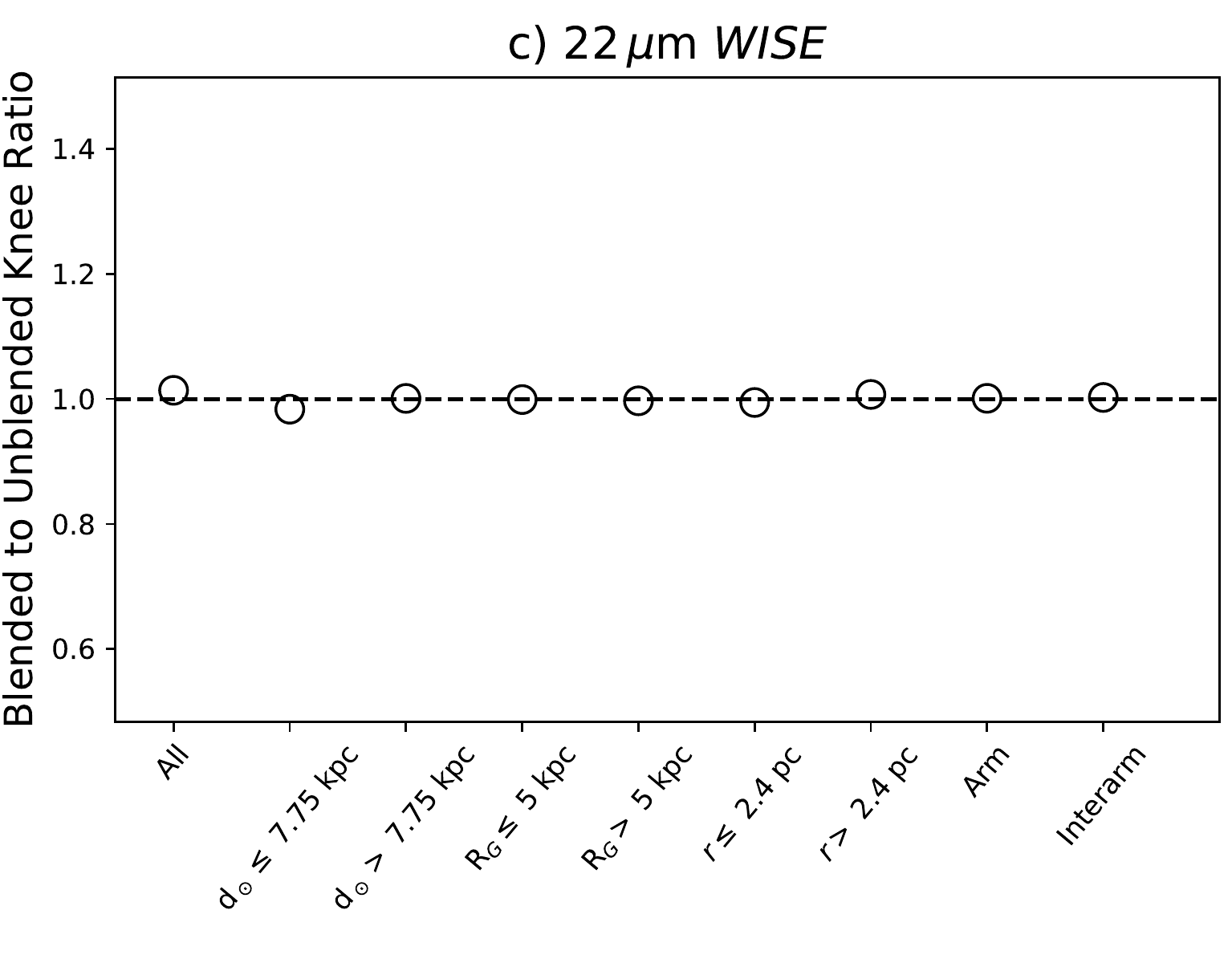}}\\
  \subfloat{\label{fig:mipsgal_knee}%
    \includegraphics[scale=0.45,trim={3cm 7.75cm 3cm 8cm}, clip]{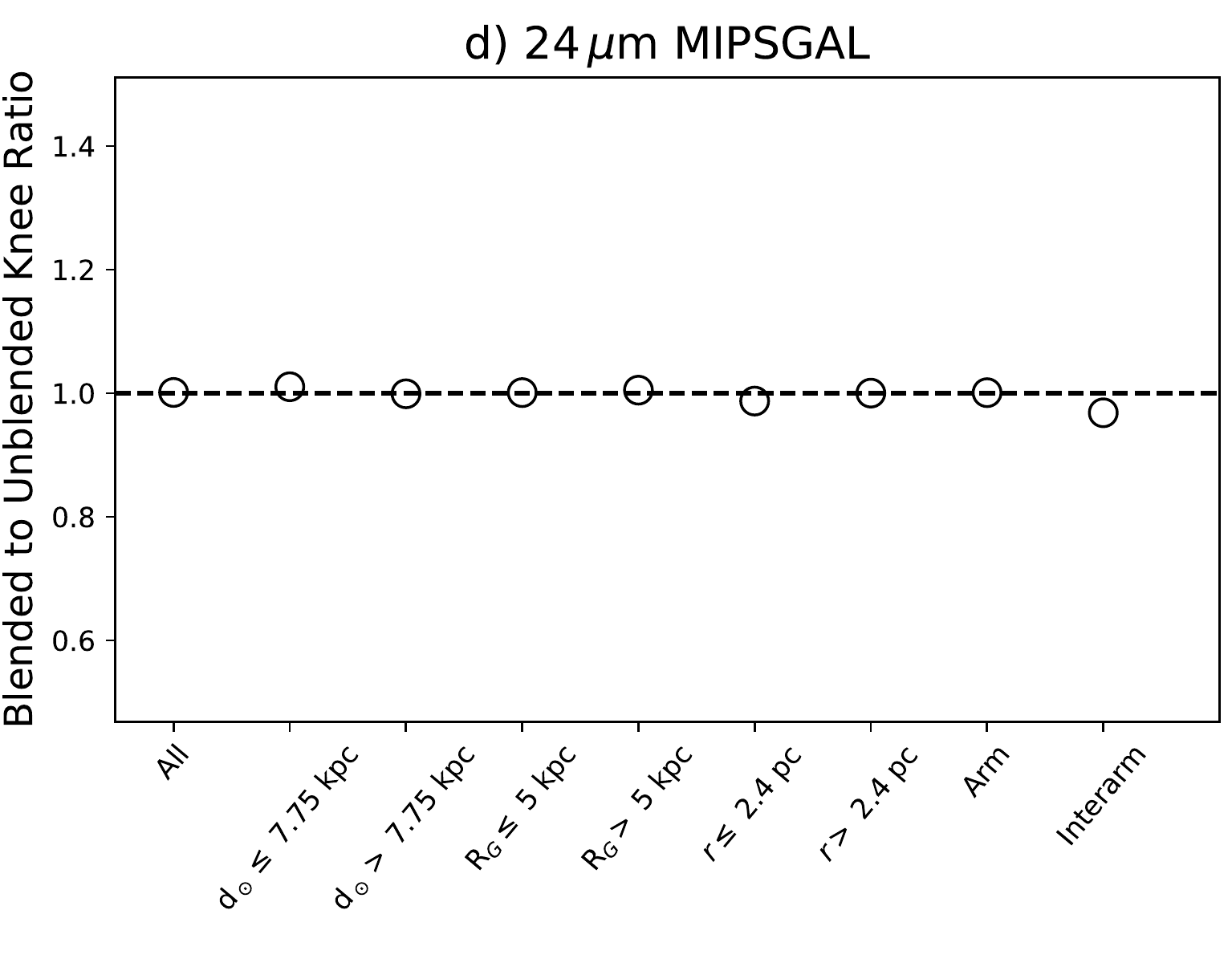}}\qquad
  \subfloat{\label{fig:higal70_knee}%
    \includegraphics[scale=0.45,trim={3cm 7.75cm 3cm 8cm}, clip]{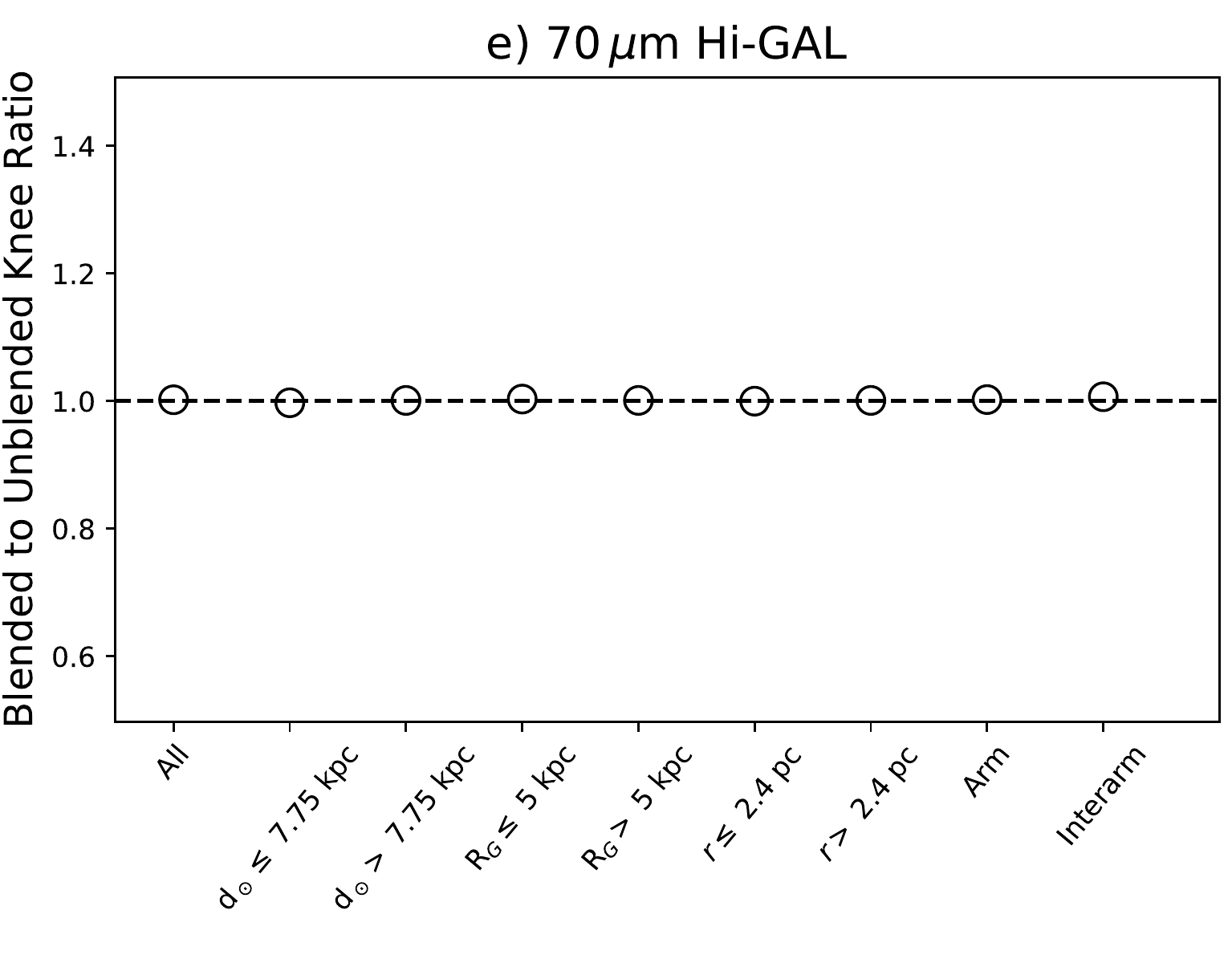}}\qquad
  \subfloat{\label{fig:higal160_knee}%
    \includegraphics[scale=0.45,trim={3cm 7.75cm 3cm 8cm}, clip]{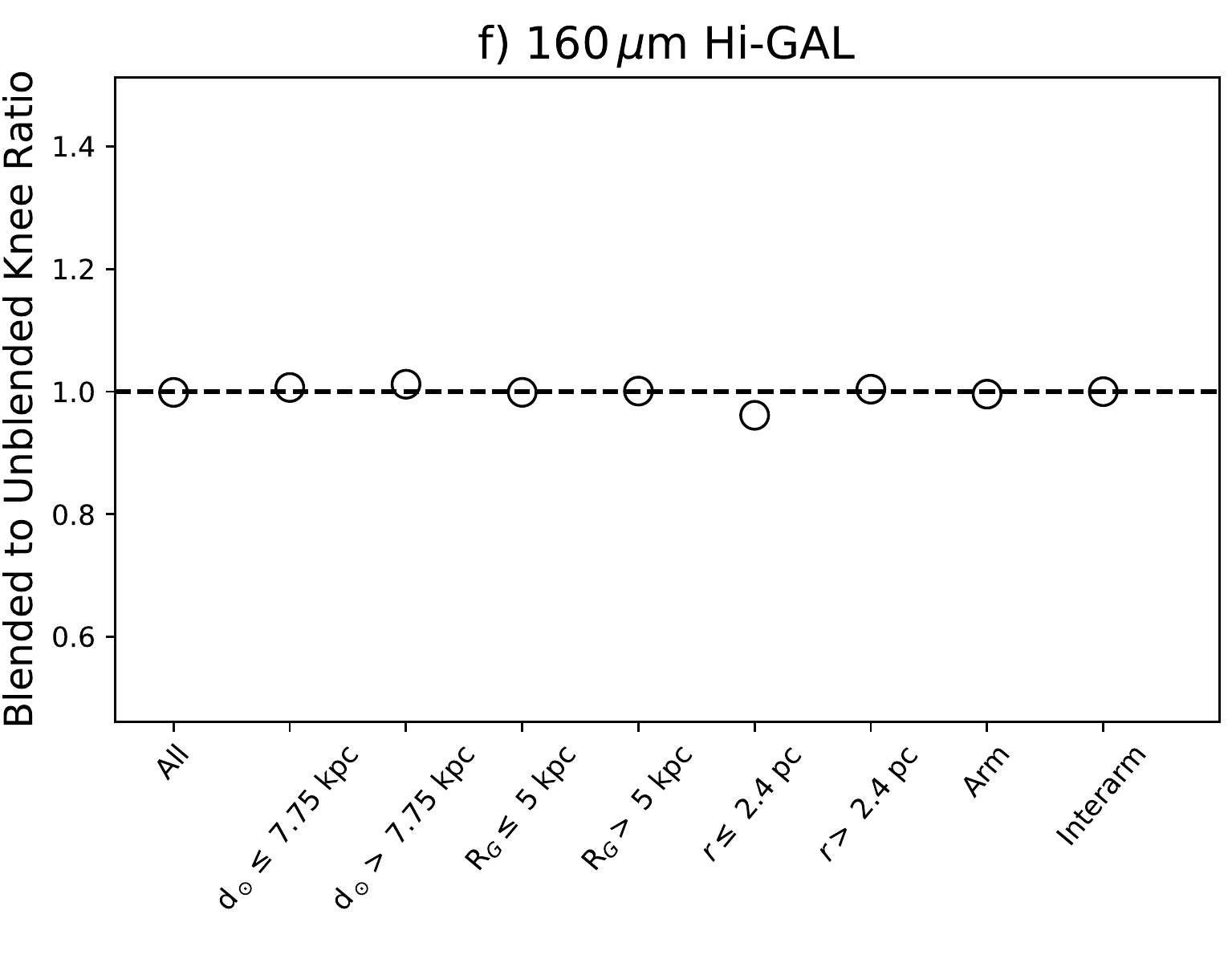}}\\
  \subfloat{\label{fig:magpis_knee}%
    \includegraphics[scale=0.45,trim={3cm 7.75cm 3cm 8cm}, clip]{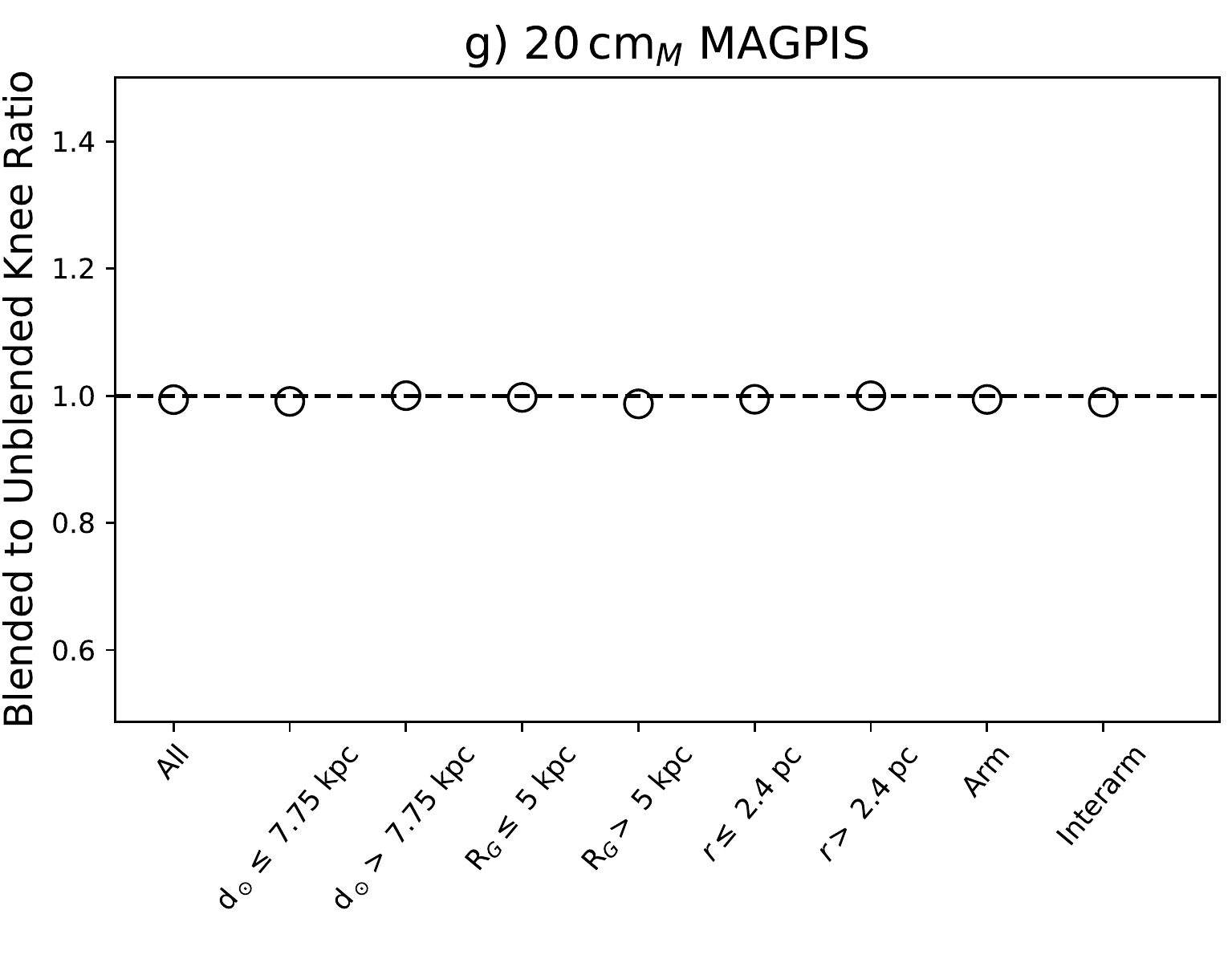}}\qquad
  \subfloat{\label{fig:magpis_vgps_knee}%
    \includegraphics[scale=0.45,trim={3cm 7.75cm 3cm 8cm}, clip]{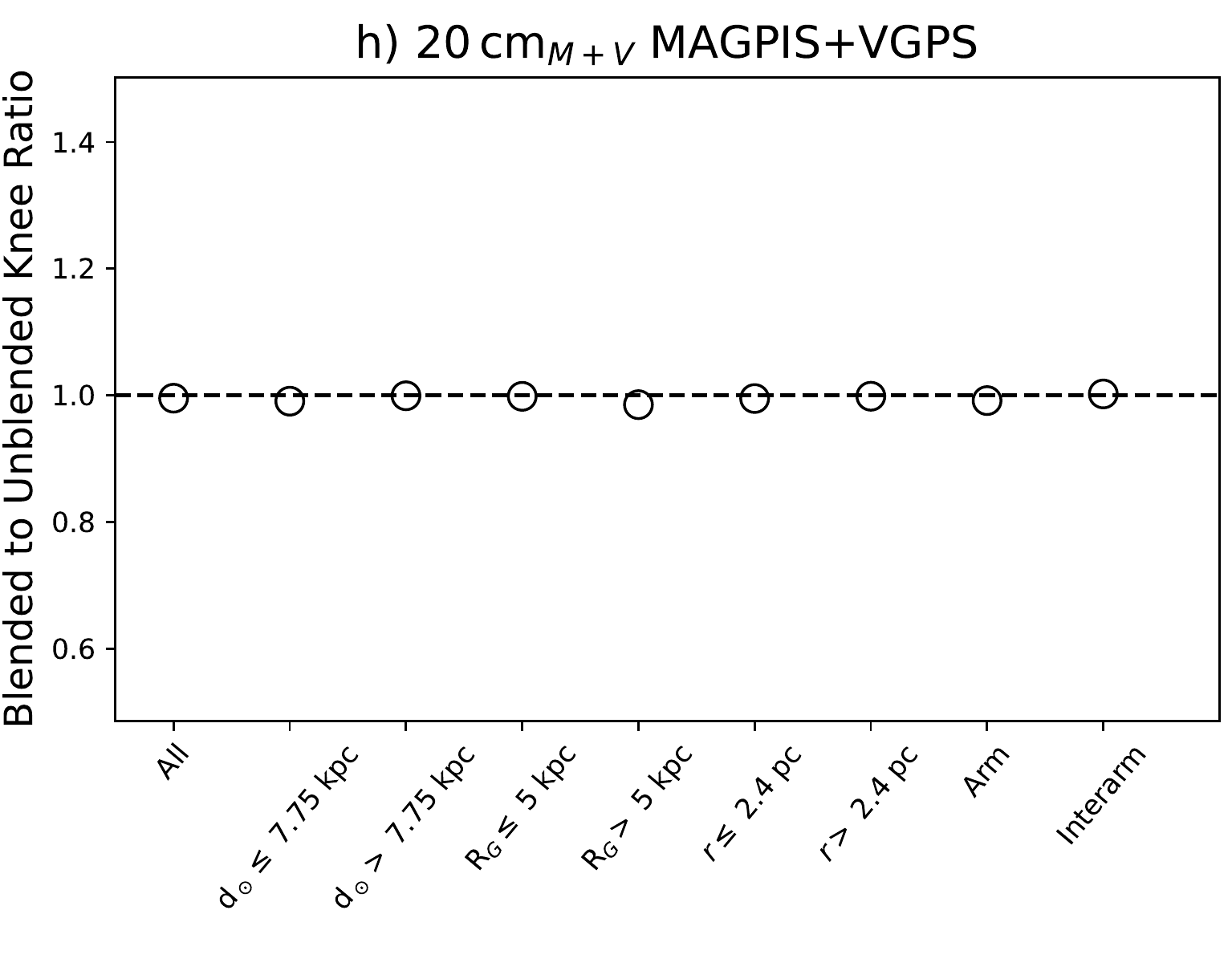}}\qquad
  \subfloat{\label{fig:vgps_knee}%
    \includegraphics[scale=0.45,trim={3cm 7.75cm 3cm 8cm}, clip]{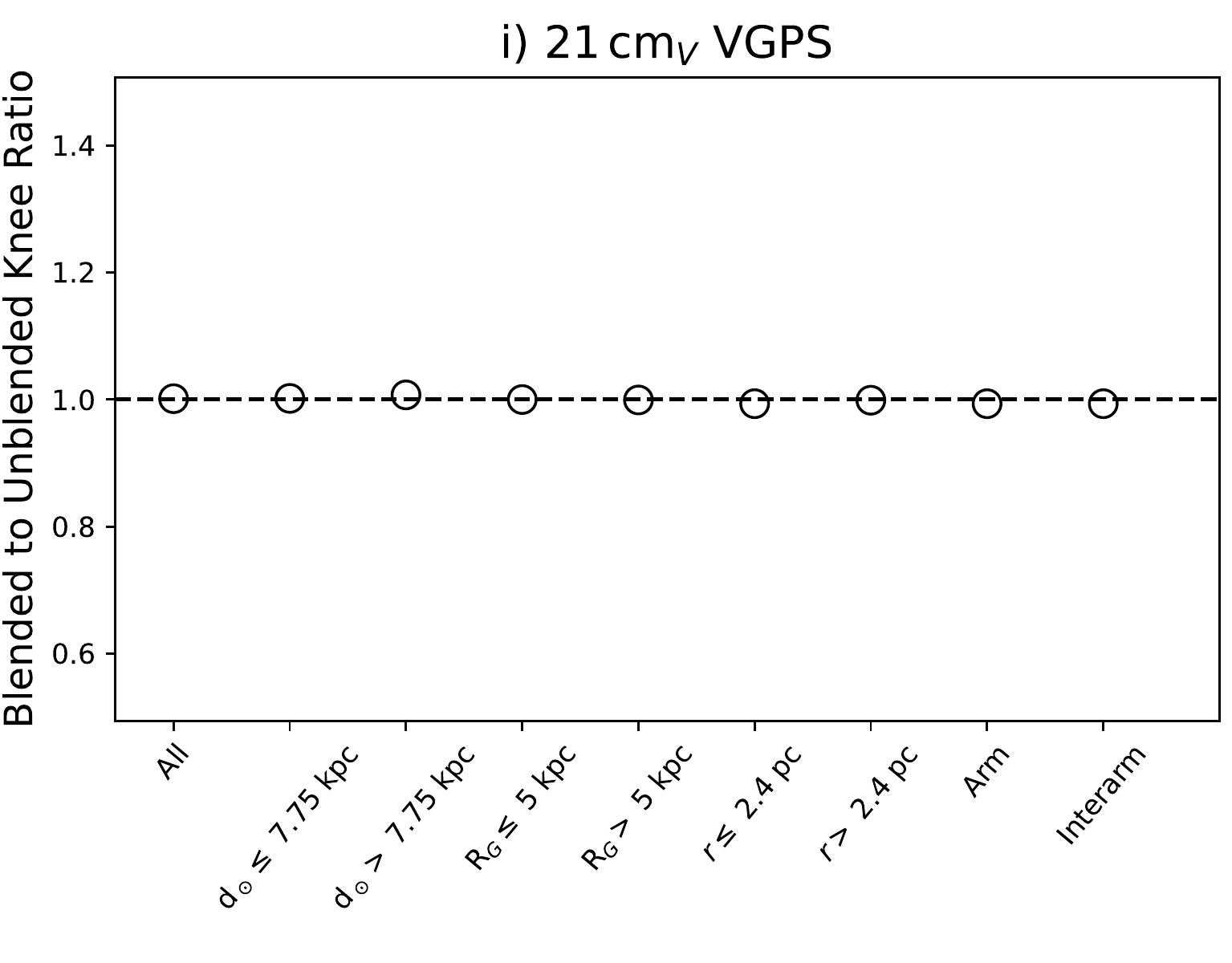}}
\caption{Comparison of unblended and blended knee values for $8\,\microns$ GLIMPSE (panel \subref*{fig:glimpse_knee}), $12\,\microns$ \textit{WISE} (panel \subref*{fig:wise3_knee}), $22\,\microns$ \textit{WISE} (panel \subref*{fig:wise4_knee}), $24\,\microns$ MIPSGAL (panel \subref*{fig:mipsgal_knee}), $70\,\microns$ Hi-GAL (panel \subref*{fig:higal70_knee}), $160\,\microns$ Hi-GAL (panel \subref*{fig:higal160_knee}), $20\,\cm$ MAGPIS (panel \subref*{fig:magpis_knee}), $21\,\cm$ MAGPIS+VGPS (panel \subref*{fig:magpis_vgps_knee}), and $21\,\cm$ VGPS (panel \subref*{fig:vgps_knee}).}
\label{fig:kneecomp_blend2}
\end{sidewaysfigure*}

\begin{sidewaysfigure*}[h]
\centering
  \subfloat{\label{fig:glimpse_limit}%
    \includegraphics[scale=0.45,trim={3cm 7.75cm 3cm 7.8cm}, clip]{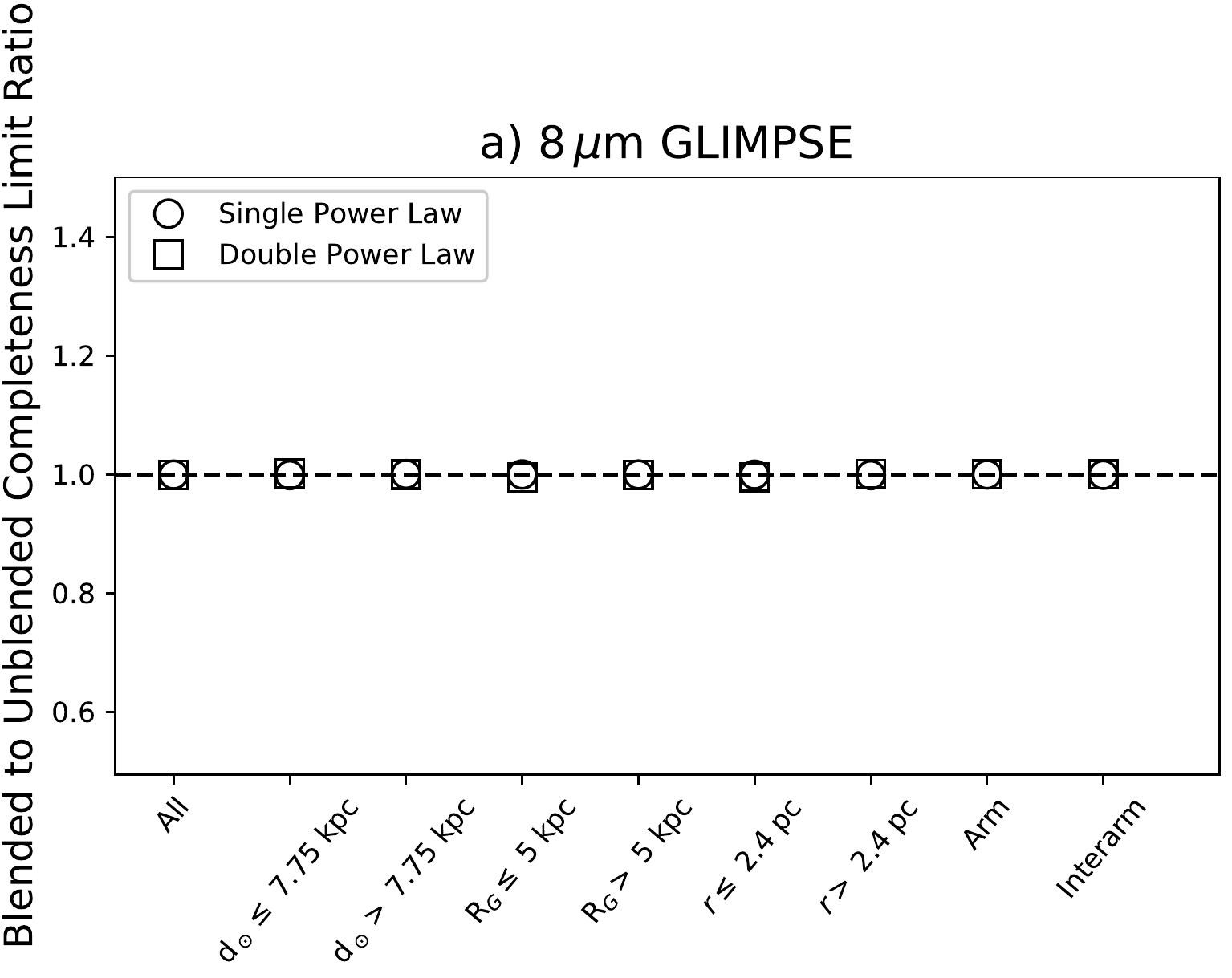}}\qquad
  \subfloat{\label{fig:wise3_limit}%
    \includegraphics[scale=0.45,trim={3cm 7.75cm 3cm 7.8cm}, clip]{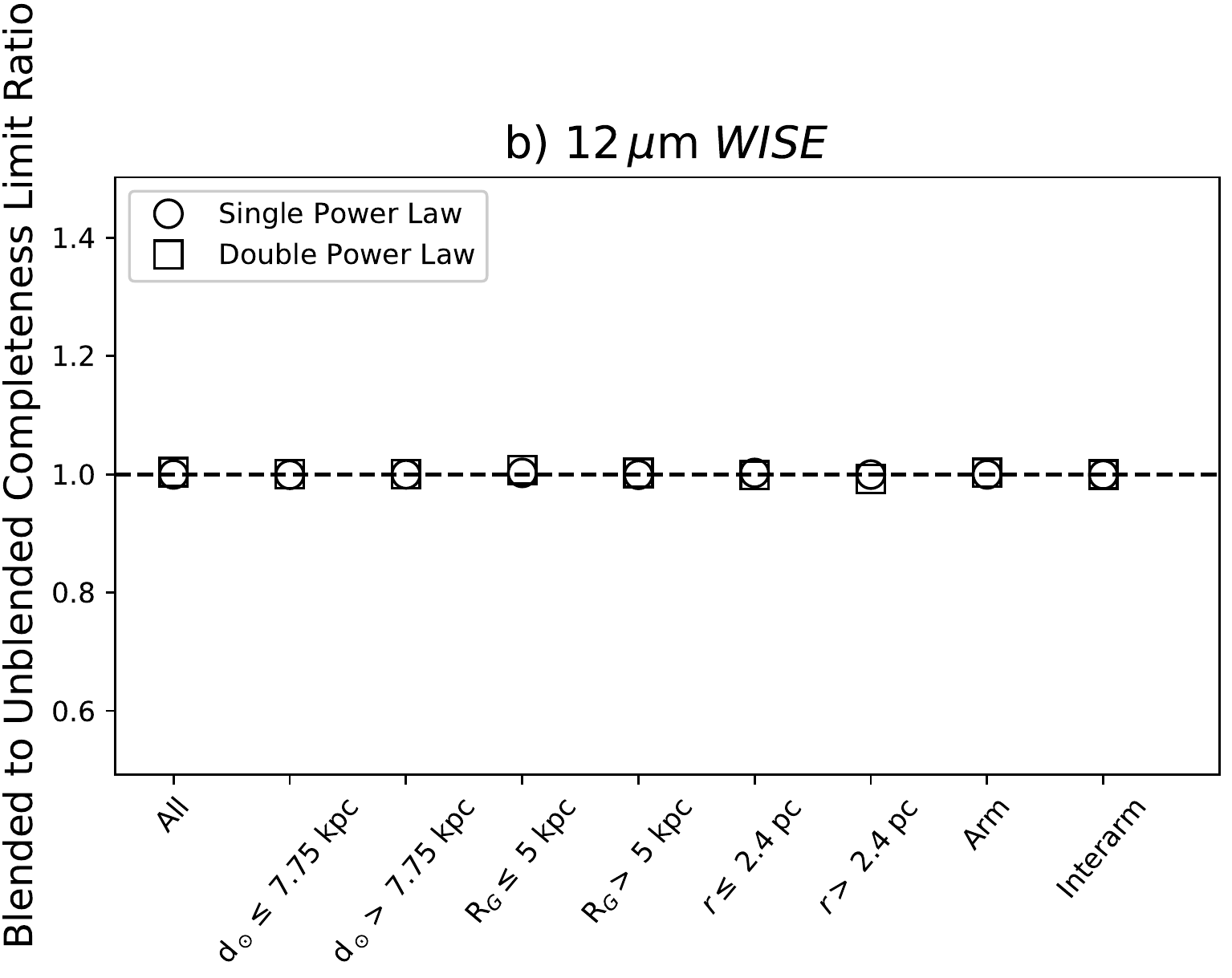}}\qquad
  \subfloat{\label{fig:wise4_limit}%
    \includegraphics[scale=0.45,trim={3cm 7.75cm 3cm 7.8cm}, clip]{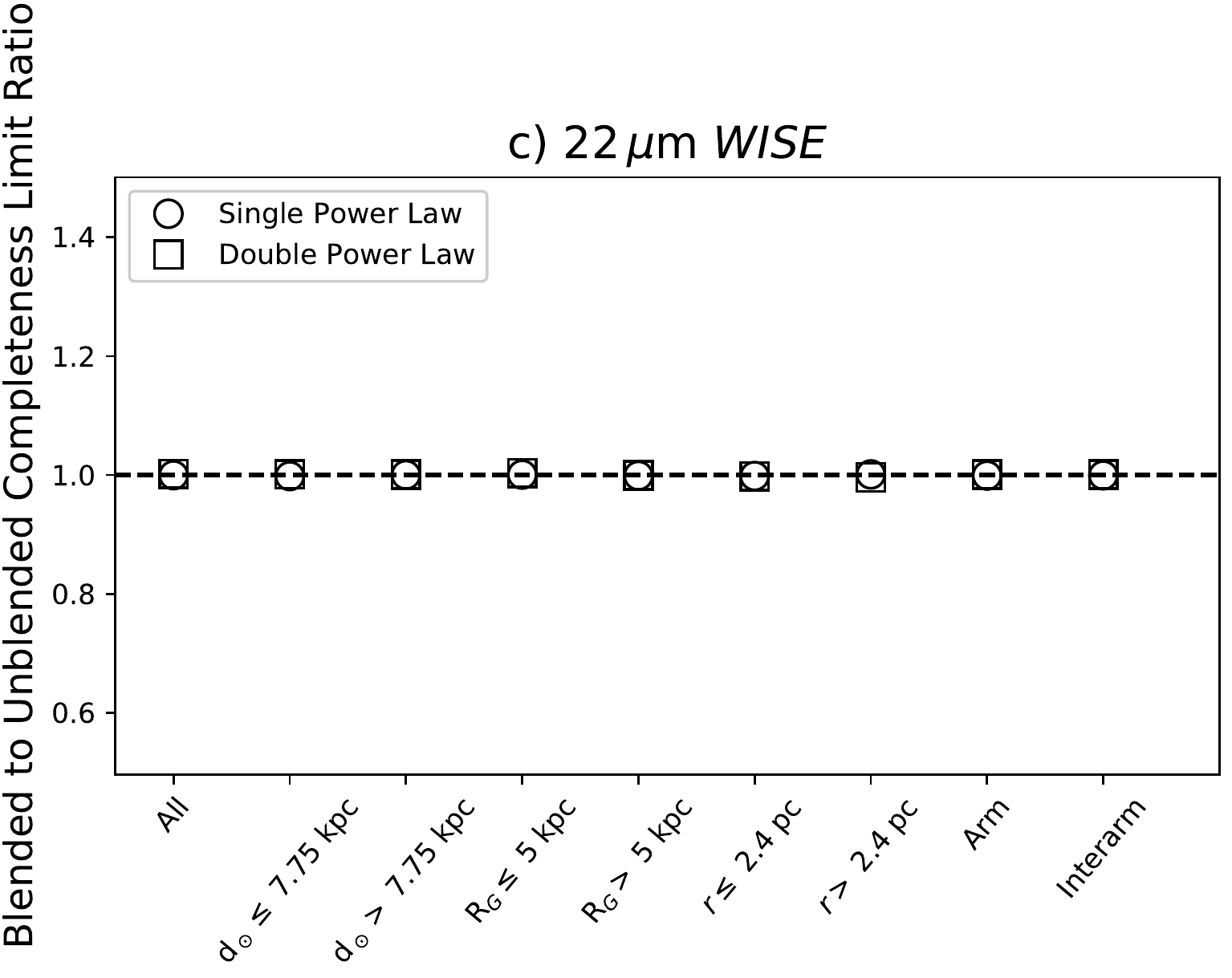}}\\
  \subfloat{\label{fig:mipsgal_limit}%
    \includegraphics[scale=0.45,trim={3cm 7.75cm 3cm 7.8cm}, clip]{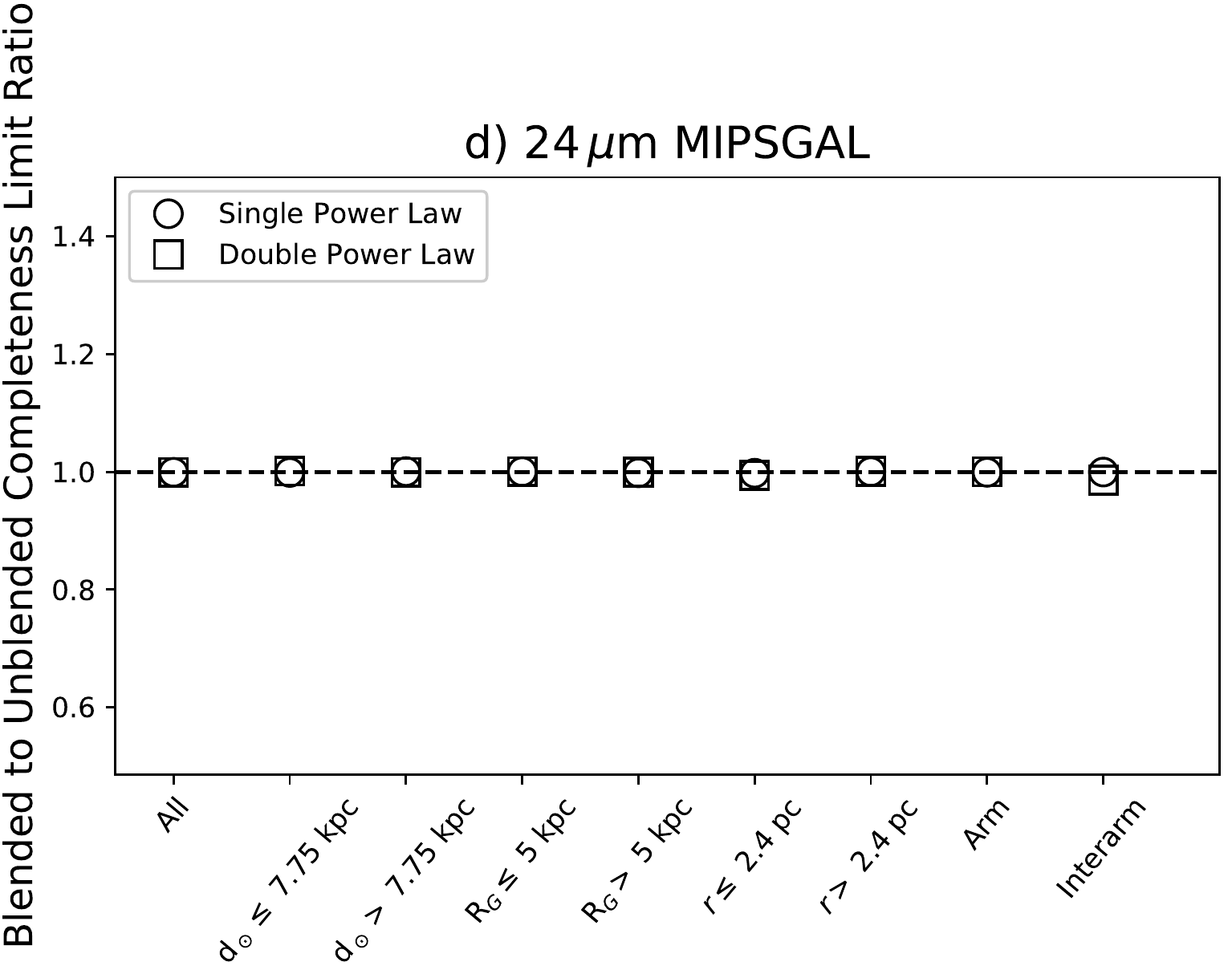}}\qquad
  \subfloat{\label{fig:higal70_limit}%
    \includegraphics[scale=0.45,trim={3cm 7.75cm 3cm 7.8cm}, clip]{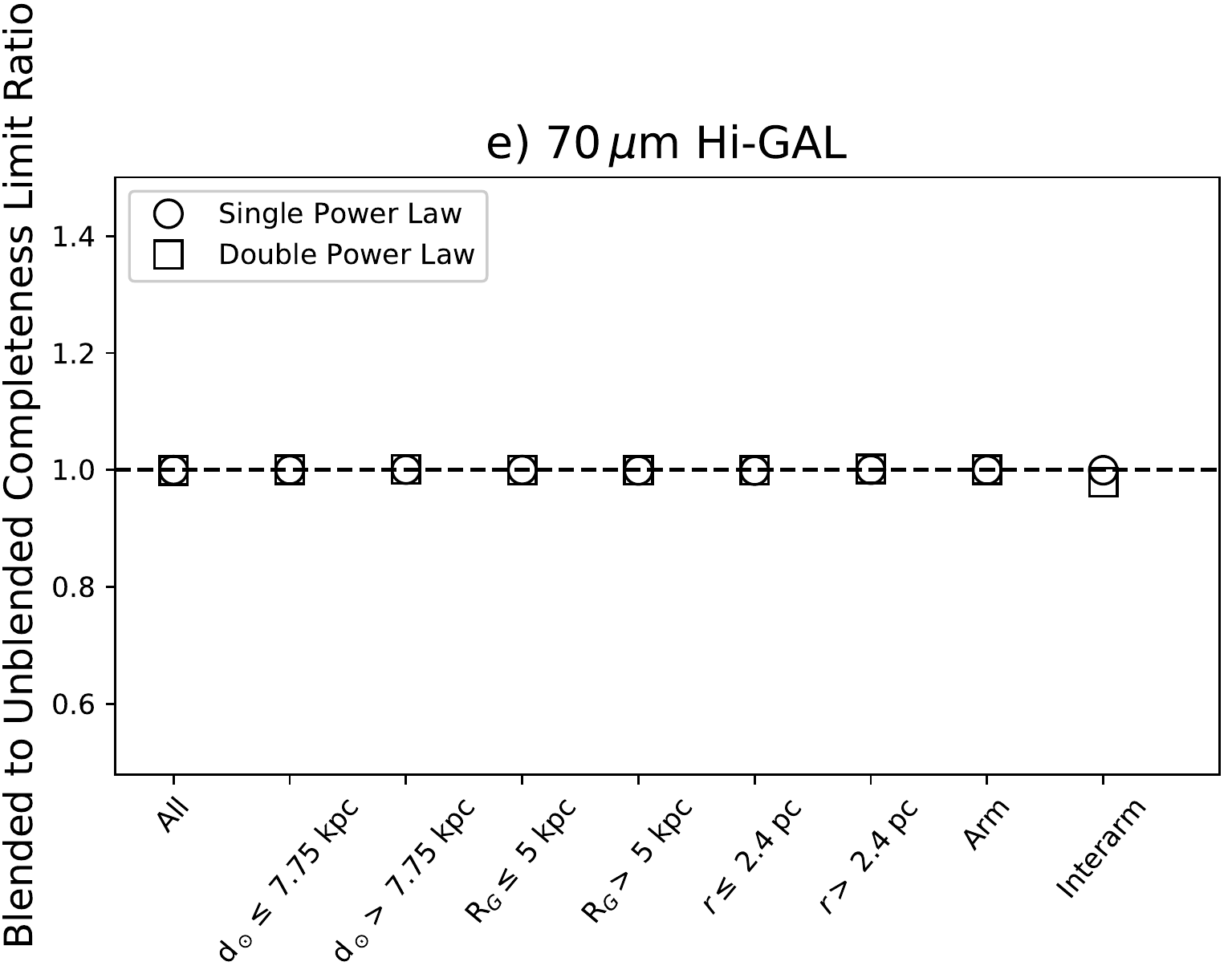}}\qquad
  \subfloat{\label{fig:higal160_limit}%
    \includegraphics[scale=0.45,trim={3cm 7.75cm 3cm 7.8cm}, clip]{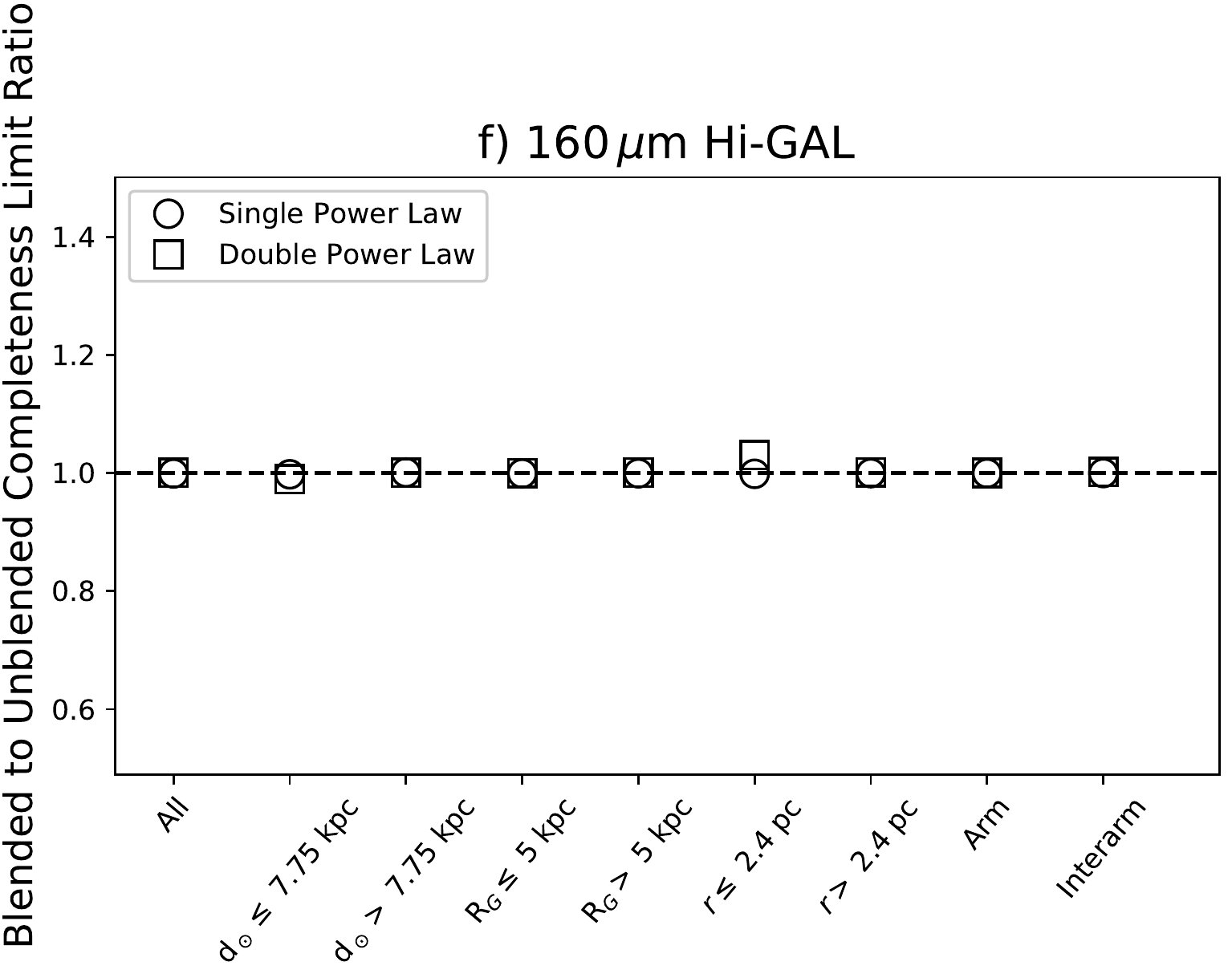}}\\
  \subfloat{\label{fig:magpis_limit}%
    \includegraphics[scale=0.45,trim={3cm 7.75cm 3cm 7.8cm}, clip]{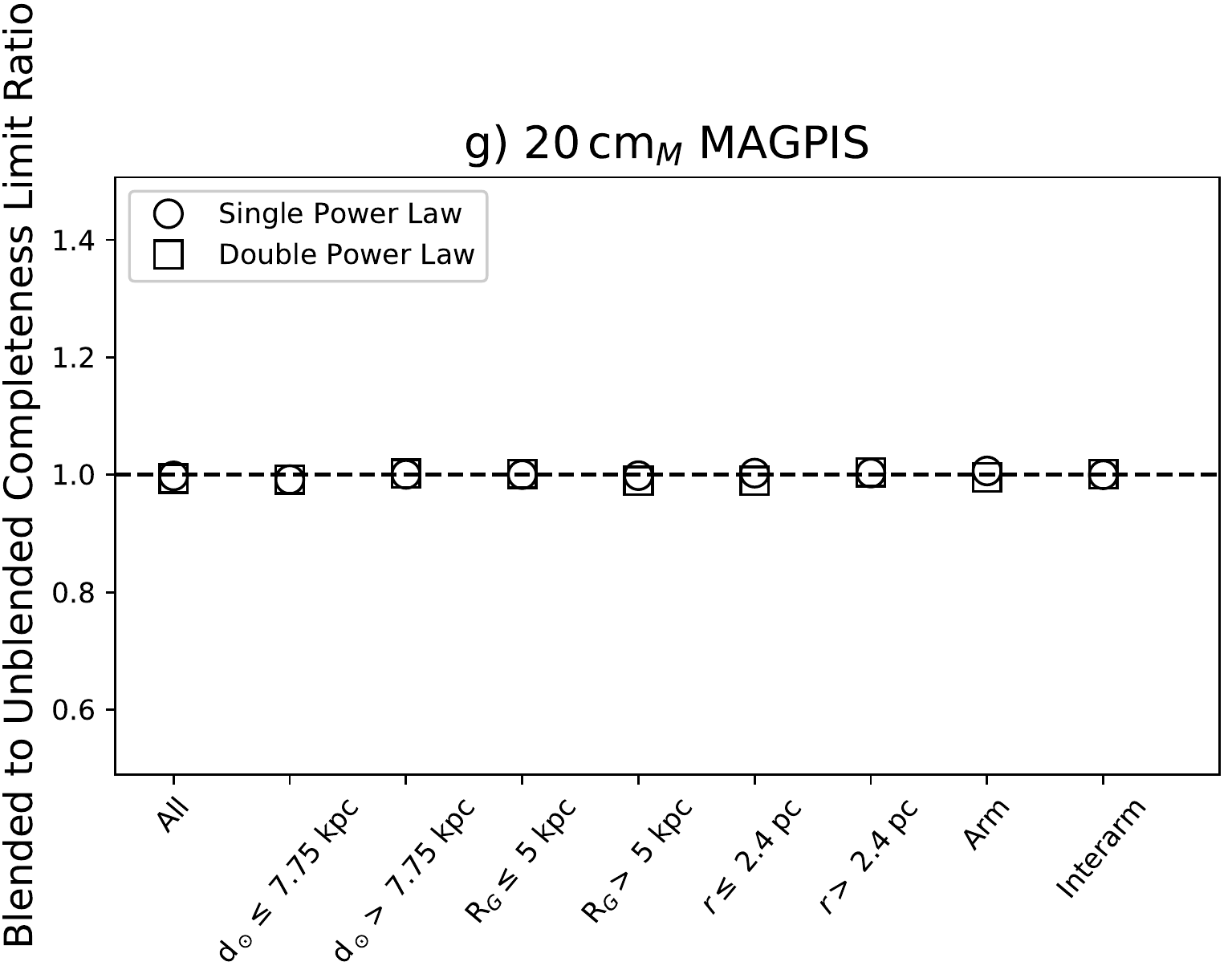}}\qquad
  \subfloat{\label{fig:magpis_vgps_limit}%
    \includegraphics[scale=0.45,trim={3cm 7.75cm 3cm 7.8cm}, clip]{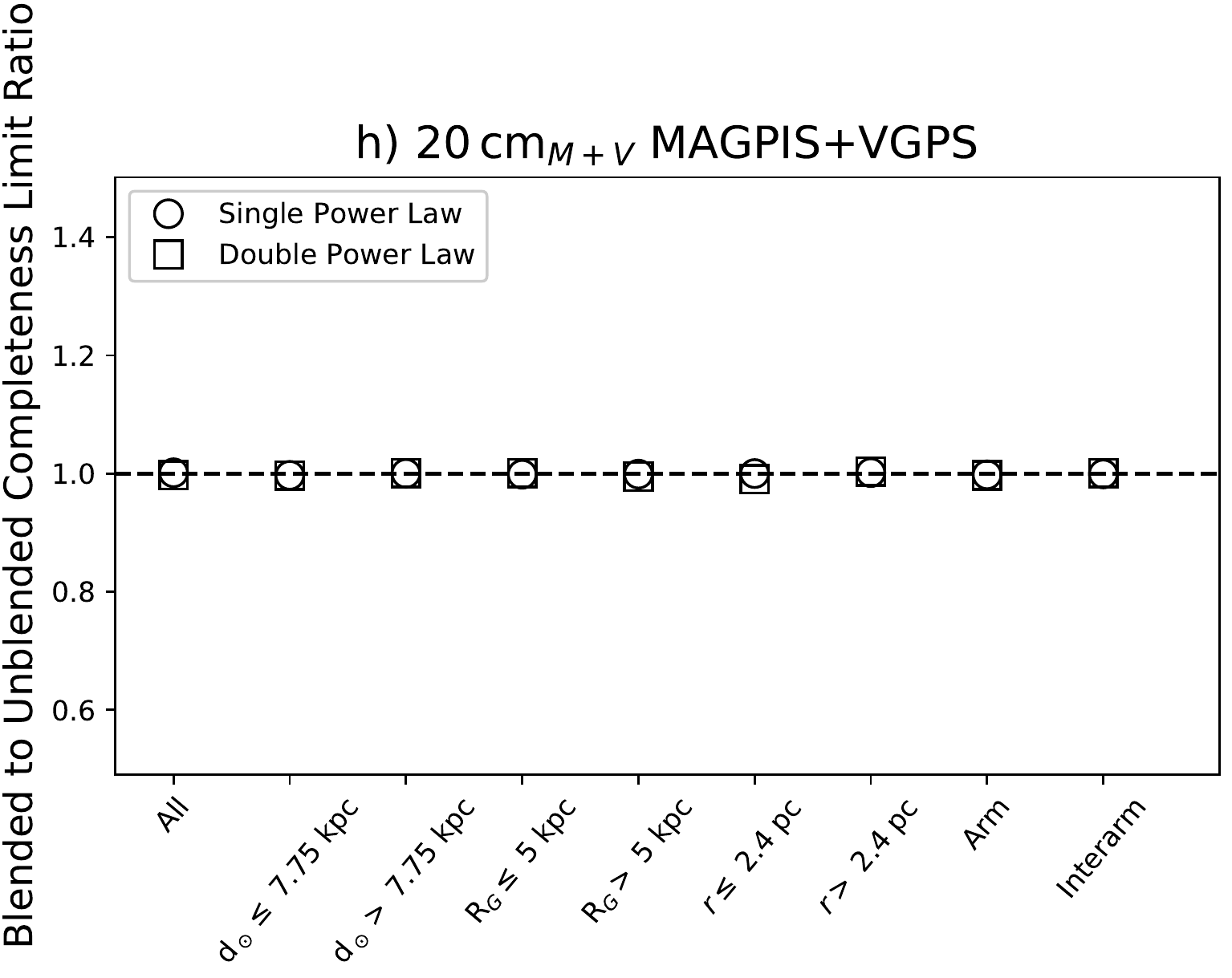}}\qquad
  \subfloat{\label{fig:vgps_limit}%
    \includegraphics[scale=0.45,trim={3cm 7.75cm 3cm 7.8cm}, clip]{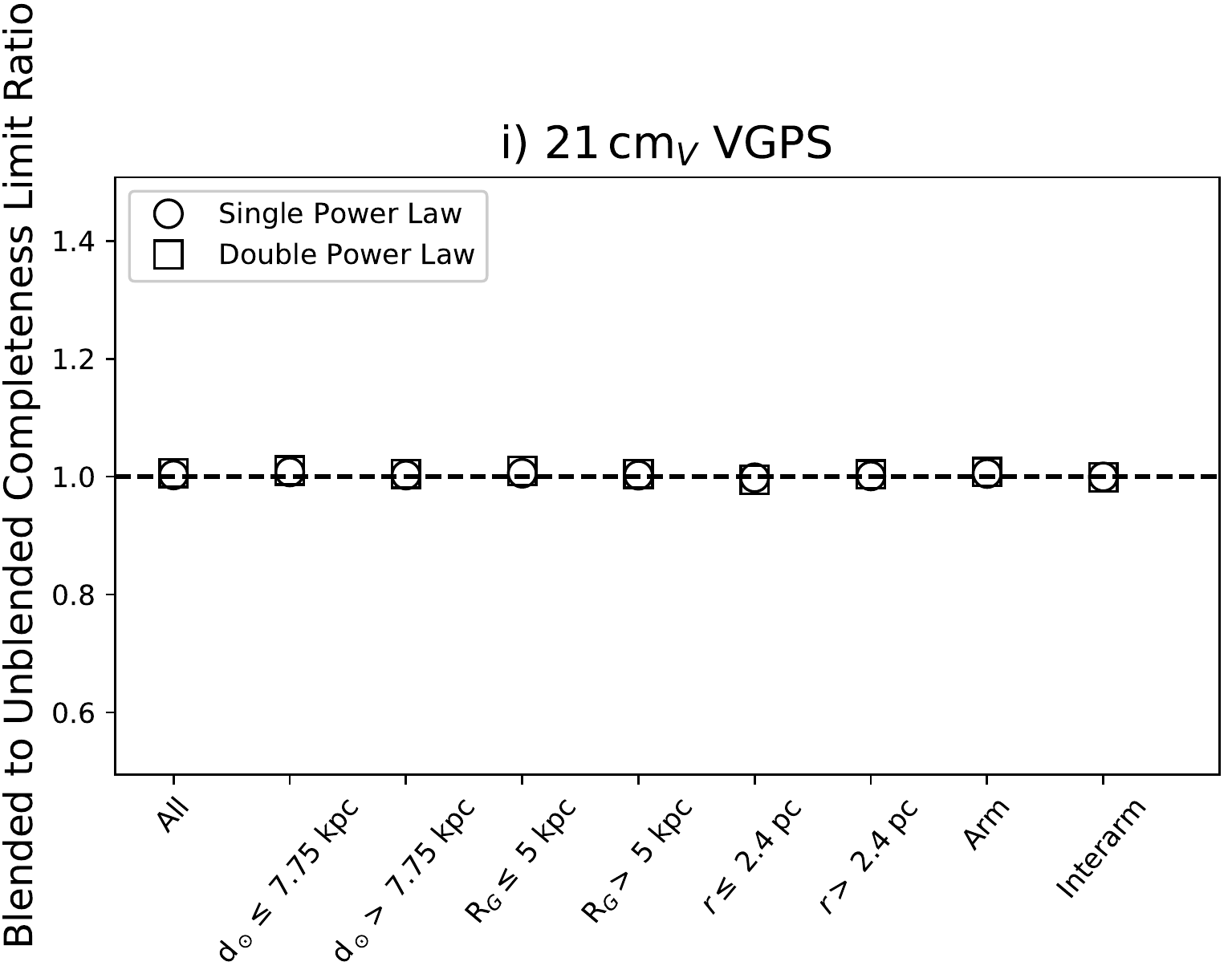}}
\caption{Comparison of unblended and blended completeness limits for $8\,\microns$ GLIMPSE (panel \subref*{fig:glimpse_limit}), $12\,\microns$ \textit{WISE} (panel \subref*{fig:wise3_limit}), $22\,\microns$ \textit{WISE} (panel \subref*{fig:wise4_limit}), $24\,\microns$ MIPSGAL (panel \subref*{fig:mipsgal_limit}), $70\,\microns$ Hi-GAL (panel \subref*{fig:higal70_limit}), $160\,\microns$ Hi-GAL (panel \subref*{fig:higal160_limit}), $20\,\cm$ MAGPIS (panel \subref*{fig:magpis_limit}), $21\,\cm$ MAGPIS+VGPS (panel \subref*{fig:magpis_vgps_limit}), and $21\,\cm$ VGPS (panel \subref*{fig:vgps_limit}).}
\label{fig:limitcomp_blend2}
\end{sidewaysfigure*}

\end{document}